\documentclass[oneside,final, letterpaper]{ucr}

\usepackage[dvipsnames]{xcolor}
\usepackage{amsmath,amssymb, graphicx}
\usepackage[utf8]{inputenc}

\usepackage{subcaption}

\usepackage{paralist} % also compact lists
 \usepackage{booktabs}
 \usepackage{comment}
 \usepackage{verbatim} % for comment
\usepackage{balance}
\usepackage{mathtools}
\DeclarePairedDelimiter\ceil{\lceil}{\rceil}

\usepackage{amsmath}
\usepackage{tikz}
\usepackage{textcomp}  %for degree sign
\usepackage{wrapfig}
\usepackage[font=small,skip=0pt]{caption}
\usepackage[utf8]{inputenc}
\usepackage[english]{babel}
\usepackage{fancyhdr}
\usepackage{lastpage}
\newcommand*\circled[1]{\tikz[baseline=(char.base)]{
            \node[shape=circle,draw,inner sep=0.4mm] (char) {#1};}}

\newcommand{\new}[1]{{{#1}}}

\newcommand{\cmnt}[1]{}
\newcommand{\ali}[1]{{#1}}
\newcommand{\chk}[1]{{#1}}
\renewcommand\footnoterule{\kern-3pt \hrule width 2in \kern 2.6pt}

\usepackage{amsmath,amssymb, graphicx}
\usepackage{comment}

\usepackage{mathtools}
\usepackage{tikz}
\usepackage[utf8]{inputenc}
\usepackage[english]{babel}
\newtheorem{thm}{Theorem}
\newtheorem{corollar}{Corollary}[thm]
\newtheorem{defn}{Definition}
\def\contra{\tikz[baseline, x=0.22em, y=0.22em, line width=0.032em]\draw (0,2.83)--(2.83,0) (0.71,3.54)--(3.54,0.71) (0,0.71)--(2.83,3.54) (0.71,0)--(3.54,2.83);}
\begin{document}

% Declarations for Front Matter

\title{Thermal Safety and Real-Time Predictability on Heterogeneous Embedded SoC Platforms}
\author{Seyed Mehdi Hosseini Motlagh}
\degreemonth{December}
\degreeyear{2020}
\degree{Doctor of Philosophy}
\chair{Prof. Hyoseung Kim}
\othermembers{Prof. Nael Abu-Ghazaleh\\
Prof. Daniel Wong\\
Prof. Shaolei Ren}
\numberofmembers{4}
\field{Computer Science}
\campus{Riverside}

\maketitle
\copyrightpage{}
\approvalpage{}

\degreesemester{Fall}

\begin{frontmatter}

% \begin{acknowledgements}
\begin{acknowledgements}
I would like to thank my research advisor, Dr. Hyoseung Kim, for his guidance and support during all these years. I am greatly indebted to him for his constant
inspiration, encouragement, and patience throughout my doctoral study. His exceptional
commitment to research and strong demand for excellence have guided me this
far and his valuable feedback contributed greatly to the improvement and progress
of my research and dissertation. I feel extremely lucky that I met him
and that he let me be a part of his research team.

I thank the rest of my dissertation committee members: Dr. Neal Abu-Ghazaleh,
Dr. Daniel Wong, and Dr. Shaolei Ren for the discussions and directions for the future work.  Their insightful comments and constructive criticism helped me to improve the dissertation in many ways.

I want to thank the people that more closely helped me with my research. I want
to thank  Dr. Christian R. Shelton for his help and support and collaboration during the PhD years. I would like to thank  Ali Ghahreman-Nezhad  for the successful teamwork throughout the years. I am beholden to Dr. Farshad Khunjush for his trust, help and guidance. I would not have ended up in the US if it wasn’t for him. Thanks to the researchers that I worked with in Google, Danijela Mijailovic, Yao Qin, Pablo Ruiz Junco, and particularly my managers, Sreekumar Kodaraka and David Ruiz.

In addition, I thank my friends and fellow students at University of California, Riverside for numerous discussions related to my research work. I sincerely thank all the current and previous members in Real-Time Embedded and Networked systems (RTEN) Lab, including  Hyunjong Choi, Yecheng Xiang, Mohsen Karimi, Yidi Wang, Abdulrahman Bukhari, and Daniel Enright.

I want to thank my friends all around the world, who contributed to make my
PhD years an awesome journey. In particular, I owe special thanks to Ali Mokari Amiri, Nima Azizi, Mahdi Aminian, Pouya Haghighat, Ehsan Mohandesi, and Omid Jahromi. I am grateful to my Iranian friends at UC Riverside:  Seyed Hossein Mostafavi, Joobin Gharibshah, Abbas Roayaei Ardakany,  Amir Feghahati, and Amir-Hossein Nodehi Sabet. 

Last but not least, I would like to express my gratitude to my parents and my family members for their continuous support and encouragement.  Without their tremendous sacrifices and great love, I would not have chance for successfully accomplishing my PhD.

\end{acknowledgements}
% \end{acknowledgements}

\begin{dedication}
\null\vfil
{\large
\begin{center}
To my parents for all the support.
\end{center}}
\vfil\null
\end{dedication}

\begin{abstract}

Recent embedded systems are designed with high-performance System-on-Chips (SoCs) to satisfy the computational needs of complex applications widely used in real life, such as airplane controllers, autonomous driving automobiles, medical devices, drones, and hand-held devices. Modern SoCs integrate multi-core CPUs and various types of accelerators including graphics processing units (GPUs), digital signal processing (DSP), video encoding, and decoding units. The performance gain of such SoCs comes at the cost of high power consumption, which in turn leads to high heat dissipation. Uncontrolled heat dissipation is one of the main sources of interference that can adversely affect the reliability and real-time performance of safety-critical applications. The mechanisms currently available to protect SoCs from overheating, such as frequency throttling or core shutdown, may exacerbate the problem as they cause unpredictable delay and deadline misses. Dynamic changes in ambient temperature further increase the difficulty of solving this problem.

This dissertation addresses the challenges caused by thermal interference in real-time mixed-criticality systems (MCSs) built with heterogeneous embedded SoC platforms. We propose a novel thermal-aware system framework with analytical timing and thermal models to guarantee safe execution of real-time tasks under the thermal constraints of a multi-core CPU/GPU integrated SoC. For mixed-criticality tasks, the proposed framework bounds the heat generation of the system at each criticality level and provides different levels of assurance against ambient temperature changes. In addition, we propose a data-driven thermal parameter estimation scheme that is directly applicable to MCSs built with commercial-off-the-shelf multi-core processors to obtain a precise thermal model without using special measurement instruments or access to proprietary information. The practicality and effectiveness of our solutions have been evaluated using real SoC platforms and our contributions will help develop systems with thermal safety and real-time predictability.
\end{abstract}

\tableofcontents
\listoffigures
\listoftables
\end{frontmatter}

\def\myb[#1]{\boldsymbol{\bf{#1}}}
\chapter{Introduction}

% Ensuring continuous operation with high assurance in the physical environment 
% remains a significant challenge to cyber-physical systems (CPS). This is particularly important for safety-critical applications with real-time mixed-criticality components, e.g., automotive, aerospace, manufacturing, and defense systems, where even occasional timing failures of high-criticality components can lead to catastrophic consequences. Various types of thermal inference contribute to the difficulty of this problem. Operating SoC in high operating temperature causes several effects. An increase in temperature increases the power consumption of the embedded systems which leads to rapid battery drain. Besides,  operating in high temperature reduces the system reliability substantially~\cite{srinivasan2004impact}.  For instance,  10\degree C to 15\degree C increase in temperature doubles the probability of failure of the underlying electronic devices~\cite{viswanath2000thermal}. Thermal violation in implantable medical devices can lead to physical harm~\cite{chandarli2015response}. Therefore, bounding the maximum temperature is an important issue, especially when real-time requirements have to be satisfied. In this dissertation, we focus on the challenges caused by thermal interference in mixed-criticality systems (MCSs) on modern multi-core platforms integrated with GPUs. 
Rising demands for data and computationally-intensive applications such as deep learning
and big-data driven applications entail the request for higher performance processors, e.g., multi-core CPUs, Graphics Processor Units (GPUs), and Tensor Processing Units (TPUs) as well as more aggressive computing technologies such as 3D chips. With this rapidly increasing power trend, the problem of thermal management in systems is becoming acute. My research focuses on enabling thermal-aware real-time computing in next-generation cyber-physical systems (CPS), such as self-driving cars driving across the country, drones flying over volcanic areas, and medical devices for humans, where the systems have to timely accomplish their missions even under harsh thermal conditions. This is an important problem because the two requirements, real-time and thermal, are fundamental to the reliable operation of such systems, but hard to be holistically analyzed and optimized by the current state of the art. Solving this problem will bring significant benefits like improving system safety and reliability, adding more intelligent features, and reducing costs for high-volume products. This dissertation presents novel analytical methodologies and system software techniques to analyze and safely guarantee both requirements. 

One of major factors that plays a role in temperature increase of modern systems-on-chips (SoCs) is heat generation due to complex computations. There exist two reasons that directly or indirectly result in temperature increase: 1) the increased number of accelerators and CPU cores, and 2) the high power consumption of each subsystems. Nowadays, different types of accelerators, such as TPUs, FPGAs, DSPs, GPUs, video processing units, and also multiple heterogeneous CPU cores are integrated into SoCs to satisfy various needs of complex applications. The power consumption of active units to perform computation generates heat. There exists heat conduction between these units, which causes temperature increase on other units even when they are idle. Aside from it, the need for high computation which can inevitably necessitate a high clock frequency and a large number of transistors further raises the power consumption of each computing unit.  The power consumption is converted into heat, resulting in a significant increase in temperature due to performing computation on active computing units. Moreover, an increase in temperature has a considerable impact on the growth of leakage current which leads to a rise in static power consumption. An increase in power consumption levels up the device temperature, thereby resulting  in again an increase in power consumption. This detrimental loop causes``thermal runaway''~\cite{ahmed2016necessary}.

Ambient temperature is one of the key factors that affect the cooling process of mixed-criticality CPS.
% Most embedded devices are designed to work in various environmental conditions in practice like harsh ambient temperature which affect the cooling process.
Designing of cooling packages such as heat sinks and cooling fans can be costly~\cite{tiwari1998reducing}. There is also an extra challenge in cooling solutions through packaging only to enable thermal safety under various thermal conditions. Active/passive cooling packages attached to SoCs dissipate the heat generated due to computation. These solutions try to transfer heat from SoCs to the environment. Fans and heat sinks are an example of active and passive cooling packages respectively that blow air to chip  or maximize the hot surface area in contact with the cooling medium surrounding it, e.g., transferring heat to ambient air by heat convection. These mechanisms are only effective when SoCs are exposed to low-temperature ambient air. In harsh ambient temperature, heat flux is reduced substantially, meaning heat convection to the air becomes little,  because the temperature difference between the operating chip and the air is negligible. Therefore, the harsh ambient temperature reduces the the effectiveness of the cooling process while SoCs still generate heat to perform computation. This condition can occur frequently in practice which may lead to burnout of SoC. In addition, for many high-performance embedded systems like medical implants or smartphones, that require thermal safety, packaging solutions are expensive, bulky and inapplicable.

To protect the processor chip from thermal damage, a set of policies are defined in the thermal governor of the OS for different scenarios. When the chip temperature crosses a trip point, a predefined cooling scenario is performed such as frequency throttling or shutting down of CPU cores~\cite{choi2004fine, herbert2007analysis, wang2009temperature}. Such thermal countermeasures, however, lead to timing unpredictability in real-time mixed-criticality systems (MCS) since the deadlines of tasks could be unexpectedly violated by reduced processing speed or temporarily unavailable CPU cores. A single delayed response in the pedestrian detection in a self-
driving car can lead to irreparable harm. A delay in the control system of the Patriot
missiles, used to protect Saudi Arabia during the Gulf War led to injuries and enormous
economic damage\footnote{L’Espresso, Vol. XXXVIII, No. 14, 5 April 1992, p. 167.}.

This unwanted performance degradation may also affect the timing predictability of integrated accelerators as well. In this dissertation, we focus on integrated GPUs because the  unique  characteristics  of  GPU operations, e.g., kernel execution on the GPU and interactions between  CPU  and  GPU  cores  for  data  transfer, introduce new  challenges  to  thermal-safety  design  and  timing assurance. For instance, miscellaneous operations for GPU segments require CPU intervention. The intermittent performance degradation of CPU cores adds up additional delay for completion of real-time tasks because GPUs are required to spend extra time for miscellaneous operations while other tasks are waiting for access to shared GPUs. This delay due to CPU slowdown leads to performance unpredictability of real-time GPU-using tasks on multi-core SoC platforms. 

Despite its importance, the thermal parameter estimation of commercial off-the-shelf (COTS) processors still remains a challenging problem. The practical use of thermally-safe mechanisms remains largely limited due to the fact that it is extremely difficult to obtain a precise thermal model of commercial-processors without using special measurement instruments or access to proprietary information, such as the power traces of micro-architectural units and detailed floorplan maps. 

In this dissertation, we focus on challenges arising from the thermal interference on heterogeneous multi-core SoC platforms. We develop novel frameworks supported by analytical time and thermal models to bound the maximum operating temperature of CPU cores with the assurance of performance predictability in mixed-criticality domains. Additionally, we propose a data-driven thermal parameter estimation scheme that is directly applicable to MCSs built with commercial-off-the-shelf multi-core processors to obtain a precise thermal model without using special measurement instruments or access to proprietary information. The main thesis supported by this dissertation is as follows:\\

\begin{minipage}{5.4in}
\textbf{Thesis Statement:} The timing predictability of real-time tasks in dynamic thermal conditions is achievable on multi-core GPU-integrated SoC platforms by designing a novel thermal-aware system framework with the support of analytical timing and thermal models.
\end{minipage}
\\

\section{Thesis Outline and Contributions}
The aim of my PhD work is to test the hypothesis that thermal-aware real-time computing
on heterogeneous multi-core CPU and GPU systems can be achieved by novel analytical
and software support. To do so, my work develops an ensemble framework for real-time
tasks to support various policies under different circumstances while providing analytical
foundations to check both thermal and temporal safety. This dissertation includes the following
several components:
\subsection{Chapter 2: Thermal-Aware Servers for Real-Time Tasks on Multi-Core GPU-Integrated Embedded Systems}
In Chapter 2, we propose a thermal-aware CPU-GPU framework to handle both real-time CPU-only and GPU-using tasks. Our framework satisfies thermal-safety under the given thermal constraint on CPU and GPU cores and offers bounded response time to real-time tasks. The framework also introduces two novel mechanisms for GPU requests to reduce task response time.  

\smallskip\noindent\textbf{Contributions.} The contributions of this work are as follows:
\begin{itemize}
\item We propose a thermal-aware CPU-GPU server framework for multi-core GPU-integrated real-time systems. We characterize different timing penalties and present a protocol for CPU and GPU thermal servers.
\item For real-time predictability on a GPU, we propose a GPU server design with a variant of the sporadic server policy, where the GPU segments of tasks execute with no thermal violation.
\item We propose an enhancement to the waiting queue of the GPU server to mitigate the pessimism of a priority-based queue.
\item We introduce a \textit{miscellaneous operation time reservation} mechanism for deferrable and sporadic CPU servers to reduce CPU-GPU handover delay and remote blocking time.
\item We extensively analyze the thermal safety and task schedulability of CPU and GPU servers with various budget replenishment policies.
\end{itemize}

\subsection{Chapter 3: On Dynamic Thermal Conditions in Mixed-Criticality Systems}
 The key contribution of our work in Chapter 3 is showing that the problem of thermal-aware real-time scheduling can be decomposed into thermal schedulability (how much CPU budget is usable under thermal constraints) and timing schedulability (if tasks are schedulable using given budget).
 
 Under harsh ambient temperature, a system may not be able to utilize 100\% of CPU time even if the CPU runs at the minimum possible frequency with active/passive cooling packages.  In such a condition, the only option left to ensure timing and thermal guarantees of critical tasks is to secure cooling time by suspending less critical tasks. Our work addresses this problem. 
 
 \smallskip\noindent\textbf{Contributions.} The contributions of this work are as follows:
\begin{itemize}
%\item We introduce an ambient temperature-aware framework for mixed-criticality multi-core systems, where thermal-aware servers are used to bound heat generation and a criticality mode change is triggered by ambient temperature.
\item We show that the problem of thermal-aware real-time scheduling can be decomposed into thermal schedulability (how much CPU budget is usable under thermal constraints) and timing schedulability (if tasks are schedulable using given budget). Our thermal schedulability achieves the simplicity in timing analysis by ensuring that the budget is guaranteed to be made available for any execution patterns without violating thermal constraints.

\item We extensively analyze the thermal safety of a multi-core system and bound the maximum operating temperature that the system can reach. At a specific ambient temperature level, we characterize the worst-case thermal behavior of a system \cmnt{by  using the notion of idle servers} and also determine the minimum time for the system to transition from one criticality level to a lower level.
\item \new{We introduce the notion of \textit{idle thermal servers} that allow bounding the maximum operating temperature caused by multiple preemptive active servers scheduled dynamically on a multi-core processor for a given  mixed-criticality taskset.} 

\end{itemize}

\subsection{Chapter 4: Data-driven Thermal Parameter Estimation for COTS-based Mixed-criticality Systems}
In Chapter 4, we propose a fast and accurate scheme to estimate the thermal parameters of COTS multi-core processors for real-time MCS. Our scheme requires only a small number of temperature traces from on-chip thermal sensors which are widely available in today's processors. Our scheme also improves the accuracy of thermal parameters through the ensemble of measurements from different frequency levels and execution patterns. 

\smallskip\noindent\textbf{Contributions.} The contributions of this work are as follows:
\begin{itemize}
    \item We present a thermal estimation scheme that has low computational cost by design. Given that steady-state profiles are much compact than transient-state profiles, our scheme first estimates the thermal parameters of a given system using only steady-state profiles, and then uses transient-state data for calibration purpose. 
    
    \item We characterize various sources of errors in thermal parameter estimation, and reduce their negative effects through the multiple refinement stages of our scheme. Our scheme also enables locating errors in the temperature profiles.
    \item Our scheme can identify the relative distance between CPU cores and produce an estimated chip floorplan from temperature profiles. It can also estimate the relative power consumption for a given  workload on each CPU core.
    \item We present techniques to further improve the accuracy of thermal parameters by exploiting the ensemble of measurement data obtained at various frequency and workload settings.
\end{itemize}
 %usually intro
\chapter{Thermal-Aware Servers for Real-Time Tasks on Multi-Core GPU-Integrated Embedded Systems}
%\boldmath
The recent trend in real-time applications raises the demand for powerful embedded systems with GPU-CPU integrated systems-on-chips (SoCs). This increased performance, however, comes at the cost of power consumption and resulting heat dissipation. Heat conduction interferes the execution time of tasks running on adjacent CPU and GPU cores. The violation of thermal constraints causes timing unpredictability to real-time tasks due transient performance degradation or permanent system failure. In this chapter, we propose a thermal-aware server framework to safely upper bound the maximum temperature of GPU-CPU integrated systems running real-time sporadic tasks. Our framework supports variants of real-time server policies for CPU and GPU cores to satisfy both thermal and timing requirements. In addition, the framework incorporates two mechanisms, miscellaneous-operation-time reservation and pre-ordered scheduling of GPU requests, which significantly reduce task response time. We present analysis to design thermal-server budget and to check the schedulability of CPU-only and GPU-using sporadic tasks. The thermal properties of our framework have been evaluated on a commercial embedded platform. Experimental results with randomly-generated tasksets demonstrate the performance characteristics of our framework with different configurations.

% \begin{IEEEkeywords}
% IEEEtrtn, journal, \LaTeX, chapter, template.
% \end{IEEEkeywords}

% \IEEEpeerreviewmaketitle
\section{Introduction}

% \hyoseung{the flow of the intro will be: 1) temperature is important, 2) there are prior studies, especially thermal servers, 3) but they have several limitations when used for today's embedded SoCs, 4) therefore, we propose this work.}

% \hyoseung{Paragraph 2: We should appreciate prior work such that the notion of thermal servers has been proposed earlier and they have shown to be effective in uniprocessor or homogeneous CPU cores. But due to several limitations, they cannot be used in recent CPU-GPU SoCs. Thus, we propose a new scheme.}

% \hyoseung{Paragraph 3: limitations of existing work, specifically on thermal servers: no consideration of CPU-GPU SoCs. The use of GPU will lead to both thermal and temporal violations. Give a brief description on these problems.}
%\textbf{parag1:temperature increase causes severl problem }
High temperature in embedded systems with modern systems-on-chips (SoCs) causes several major issues. 
%Temperature rise due to an increase in transistor density in SoCs~\cite{kuroda2001cmos} raises major concerns  especially in real-time embedding systems. 
An increase in temperature has a considerable impact in the growth of leakage current which leads to rise in static power consumption. An increase in power consumption levels up the device temperature, thereby resulting  in again an increase in power consumption. 
This detrimental loop causes not only rapid battery drain but also  ``thermal runaway''~\cite{ahmed2016necessary}.  Furthermore, studies show that operating in high temperature reduces the system reliability substantially~\cite{srinivasan2004impact}.  For instance,  10\textdegree C to 15\textdegree C increase in temperature doubles the probability of failure of the underlying electronic devices~\cite{viswanath2000thermal}. Thermal violation not only reduces the reliability in real-time implantable medical devices, but also causes physical harm~\cite{chandarli2015response}. Therefore, bounding the maximum temperature is an important issue, especially when real-time requirements have to be satisfied. 

% \textbf{parag2: the power managment in todday integrated cpu gpu platfor is challenging}\\
Thermal management on today's multi-core CPU-GPU integrated platforms with real-time requirements is a challenging problem. Dynamic Thermal Management (DTM) is triggered when thermal violation occurs so that it forces frequency throttling or shutdown of the SoC for cooling purpose. This unwanted performance degradation leads to timing unpredictability in task execution, and real-time tasks may miss their deadlines. Thermal violation avoidance in uni-processor systems has been studied extensively in the literature of real-time systems~\cite{ chandarli2015response, 4032360}  but it cannot be directly used for multi-core GPU-integrated devices due to heat conduction between processor units. Dynamic Voltage Frequency Scaling (DVFS) techniques to mitigate the heat and power dissipation of processors also has been widely studied in the literature~\cite{ 4032360, chen2009proactive}. However, aside from a considerable reduction in system reliability over time due to continuous frequency changes~\cite{iranfar2017thespot, lasance2003thermally,  xiang2010system}, not all embedded devices support DVFS, especially for integrated GPUs.  

% \textbf{parag3:why multicore thermal schems cannot apply for cpu gpu systems}
Despite the popularity of integrated GPUs in modern multi-core SoCs, state-of-the-art approaches are incapable of simultaneously addressing thermal management and real-time schedulability issues. On the one hand, the GPU access segment of a real-time task has been modeled as a critical section to ensure schedulability~\cite{GPUSync, elliott2012globally, patel2017}. These approaches, however, are oblivious of thermal constraints so that the system may suffer from heat dissipation and intermittent performance drops by DTM. On the other hand, there are previous studies~\cite{Ahmed2017, ahmed2011minimizing}  introducing the concept of thermal servers for real-time uni-processor and multi-core platforms. However, their schemes are not ready to use for a multi-core SoC with an integrated GPU. The unique characteristics of GPU operations, e.g., kernel execution on the GPU and interactions between CPU and GPU cores for data transfer, introduce new challenges to thermal server design and schedulability analysis. To the best of our knowledge, there is no prior work that offers both thermal safety and real-time schedulability in  multi-core GPU-integrated embedded systems.

In this chapter, we propose a thermal-aware CPU-GPU framework to handle both real-time CPU-only and GPU-using tasks. Our framework enhances the notion of thermal servers to satisfy the given thermal constraint on CPU and GPU cores and to offer bounded response time to real-time tasks.
The framework also introduces two mechanisms, miscellaneous-operation-time reservation and pre-ordered waiting queue for GPU requests, to reduce task response time. We will show with experimental results that our framework is effective in satisfying both thermal and temporal requirements. 

\smallskip\noindent\textbf{Contributions.} The contributions of this chapter are as follows:
\begin{itemize}
\item We propose a thermal-aware CPU-GPU server framework for multi-core GPU-integrated real-time systems.

 We characterize different timing penalties and present a protocol for CPU and GPU thermal servers.
\item For real-time predictability on a GPU, we propose a GPU server design with a variant of the sporadic server policy, where the GPU segments of tasks execute with no thermal violation.
\item We propose an enhancement to the waiting queue of the GPU server to mitigate the pessimism of a priority-based queue.
%Accordingly, a \textit{hybrid} scheduler is proposed to choose one of the implementations based on the schedulability analysis result during run-time;
\item We introduce a \textit{miscellaneous operation time reservation} mechanism for deferrable and sporadic CPU servers to reduce CPU-GPU handover delay and remote blocking time.
\item We extensively analyze the thermal safety and task schedulability of CPU and GPU servers with various budget replenishment policies.
\end{itemize}

% This chapter is organized as follows. Section~\ref{sec:rel} explains related work. Thermal preliminaries are explained in Section~\ref{sec:background}. We introduce the system model, task model and the limitations of existing thermal servers for GPU-needed taskset in Section~\ref{sec:model}. Accordingly, we address the limitations in design of thermal servers for both CPU and GPU under different replenishment policies with an overrun option in Section~\label{sec:chap2_frame}.  Server design and schedulability analysis are presented in Section~\ref{sec:analysis}. We evaluate our proposed framework throughout experimental results in Section~\ref{sec:eval}. Finally, we conclude the chapter and outline our future work. 
\section{Related Work}
\label{sec:chap2_rel}

Real-time GPU management has been studied to handle GPU requests with the goal of improving timing predictability~\cite{kato2011timegraph, kato2011rgem, kato2012gdev, zhou2015gpes}. To guarantee the schedulability of GPU-using real-time tasks, synchronization-based approaches~\cite{GPUSync,elliott2012globally,patel2017} that model GPU access segments as critical sections have been developed. The work in \cite{8046309,kim2018server} introduces a dedicated GPU-serving task as an alternative to the synchronization-based approaches. While these prior studies have established many fundamental aspects on real-time task schedulability with GPUs, thermal violation issues have not been considered. 
%Kim et al.~\cite{7010477} proposed the overrun option for small global critical sections in virtual machine environment to continue task execution in global shared resources hoping to reduce the remote blocking time. 
%Thermal concern is also out of the scope for these studies. 

There exist extensive studies on bounding the maximum temperature in real-time uni-processor systems~\cite{chandarli2015response, Youngmoon2018} and non-real-time heterogeneous multi-core systems~\cite{7092527, Prakash, 7746768, 7372657, 7904613}. In \cite{chandarli2015response}, the authors proposed a novel scheme that bounds the maximum temperature by analyzing task execution and cooling phases in uni-processor real-time systems. Real-time thermal-aware resource management for uni-processor systems has been developed with the consideration of varying ambient temperature and diverse task-level power dissipation~\cite{Youngmoon2018}. For non-real-time heterogeneous systems, Singla et. al~\cite{7092527} proposed a dynamic thermal and power management algorithm to adjust the frequency of GPU and CPU as well as number of active cores by computing the power budget. Prakash et al.~\cite{Prakash} proposed a control-theory based dynamic thermal management technique for mobile games.
% to regulate the frequency of both CPU and GPU in order to avoid thermal violation.  
Gong et al.~\cite{7904613} presented a thermal model on a real-life heterogeneous mobile platform. In~\cite{7372657} and~\cite{7746768}, the authors proposed a proactive frequency scheduling for heterogeneous mobile devices to maintain their performance. However, these studies cannot be directly applied to real-time multi-core GPU-integrated systems where the GPU is shared among tasks.

% By proving an upper bound on the worst-case response time of a task, they introduced a scheduling algorithm and provided its optimality. 
The notion of periodic thermal-aware servers was proposed in \cite{ahmed2011minimizing} for uni-processors. In this work, the optimal server utilization has been proved and the budget replenishment period is determined by a heuristic algorithm. Similar to the thermal server, the notion of cool shapers was proposed in~\cite{kumar2011cool} to satisfy the maximum temperature constraint by throttling task execution. The notion of hot tasks was introduced in~\cite{Huang2014} to partition lengthy tasks into several chunks to avoid continuous task execution and thermal violation while maximizing throughput. Although all of these aforementioned studies have brought valuable contributions, they have been proposed for uni-processor platforms.  In contrast, our framework addresses the temperature bounding problem for real-time tasks with GPU segments running on modern CPU-GPU integrated SoCs. 

Recently, the authors of~\cite{DSouza2017ThermalIO} introduced  a novel technique for periodic tasks executing on multi-core platform. This technique introduces an Energy Saving (ES) task that runs with the highest priority and captures the sleeping time of CPU cores. The technique can be seen as an alternative to a thermal server because the ES task effectively models the budget-depleted duration of a thermal server. The authors of~\cite{Ahmed2017} proposed thermal-isolation servers that avoid the thermal interference among tasks in temporal and spatial domains with  thermal composability. These techniques, however, cannot address the challenges of scheduling GPU-using real-time tasks.

\section{Background on Thermal Behavior}
\label{sec:chap2_background}

In this section, we briefly introduce the thermal model used in this chapter. It depends on power consumption, heat dissipation, and the conductive heat transfer between adjacent power-consuming resource components, which includes CPU cores, CPU peripherals, GPU, caches, and other IPs.

% \smallskip\noindent\textbf{Power model.} The total power consumption of CMOS circuits is modeled as the summation of dynamic and static powers ($P(t) = P_D(t) + P_S(t)$)\cite{4484694}. Static power $P_S$ depends on the semiconductor technology and the operating temperature caused by current leakage. Therefore, static power presents when a power resource is switched on. Static power can be modeled as a linear function of temperature. Hence, static power is $P_S(t) = k_1 \theta(t) + k_2 $, where $k_1$ and $k_2$ are technology-dependent system constants, and $\theta$ is the operating temperature \cite{4212027}. Dynamic power $P_D(t)$ is the amount of power consumption due to the processor operating frequency $f$, modeled as $P_D = k_0 f^s$, where $f$ and $k_0$ are system constants that depend on the semiconductor technology. It is worth noting that in this chapter, our framework runs processor cores at fixed operating frequency to obtain predictable worst-case task execution time. Consequently, the total power is the function of $\theta(t)$ because the other factors such as frequency remain invariant during execution. 

\smallskip\noindent\textbf{Uni-processor thermal model.} With respect to the power function~\cite{4212027, 4484694}, the thermal model of a uni-processor is modeled as an RC circuit in the literature~\cite{4484694, schor2012worst, Skadron}. According to the Fourier's Law~\cite{wang2010schedulability} and considering $t_0 = 0$, the temperature of a core $\theta(t)$ after $t$ time units operating at a fixed clock frequency is given by:\footnote{Details are available in~\cite{DSouza2017ThermalIO} and~\cite{Ahmed2017}.}
%can be derived by solving the following differential equation\footnote{Details are available in~\cite{DSouza2017ThermalIO} and~\cite{Ahmed2017}.}:
\begin{equation}
     \theta(t) = \alpha + (\theta(t_0)-\alpha) e^{\beta t}
     \label{eq:chap2_tempActive}
\end{equation}
where $\alpha > 0$ and $\beta < 0 $ are constants. 

Most operating systems in embedded platforms transition the processor to the sleep state when there is no task execution. Therefore, the power consumption can be assumed to be negligible in such a case. The thermal function in the sleep state (\textit{aka} cooling phase) where the frequency is switched off can be modeled as:
\begin{equation}
    \theta(t) = \theta(t_0) e^{\beta t}.
    \label{eq:chap2_tempCool}
\end{equation}
 because in the cooling phase, $\alpha = 0$.
 
\smallskip\noindent\textbf{Heterogeneous multi-core thermal model.} With the presence of multiple cores and other power-consuming resources, there is heat dissipation not only to ambient but also between nodes. In this chapter, we only consider CPU and GPU cores as power-consuming nodes because other IPs consume much less power than them, thereby causing negligible thermal effects. 

In such a CPU-GPU integrated system with the lateral thermal conductivity between processing cores, the temperature of a core depends not only on its current temperature and power consumption, but also on those of adjacent cores.  Prior work~\cite{6629322} showed that the temperature of each core at time $t+\Delta$ can be modeled with an acceptable accuracy as follows:
\[ \Theta(t+\Delta) = A \times P(t+\Delta) + \Gamma \times \Theta(t)  \]
where $A$ is  $m \times m$ matrix, and $\Gamma$, $P$ and $\Theta$ are $m \times 1$ matrices, respectively. One can interpret the thermal model as a function of the current temperature and heat produced by execution and the thermal conductivity between cores. 
Hence, according to the thermal composability characteristics of heat transfer~\cite{Ahmed2017} with the initial temperature of $\theta(t_0)$, the temperature after $t$ time units is given by:
\begin{equation}
\theta_i(t_0+t) = \alpha + (\theta(t_0)-\alpha) e^{\beta t} +  \sum_{j=1}^{m} \gamma_{ij} \theta_j(t_0+t).
    \label{eq:chap2_tempCompo}
\end{equation} 
\section{System Model}
\label{sec:chap2_model}
In this section, we describe the thermal-aware server as well as the task models used in this chapter and explain the procedure of a kernel launch on a GPU. Then, we characterize the scheduling penalties that arise from the use of a GPU with thermal-aware servers.
%in a multi-core system and the limitation of the current thermal-server schemes when they are used in a GPU-enabled system.

\subsection{Computing Platform}
We consider a temperature-constrained embedded platform equipped with an integrated CPU-GPU SoC. The SoC has multiple CPU cores and one GPU core, each running at a fixed clock frequency. Note that such SoC design with a single GPU is popular in today's embedded processors, such as Samsung Exynos 5422 and NVIDIA TX2. The assumption on the fixed operating frequency is particularly suitable for the GPU as DVFS capabilities are not widely supported on embedded on-chip GPUs. The thermal behavior of CPU and GPU cores follows the model described in Section III. For simplicity, we assume that the amount of temperature generated by other SoC components, such as peripherals and caches, is either negligible or acceptably small.

\subsection{Thermal-Aware Servers}
%The notion of thermal-aware periodic resource interfaces, called {\em thermal servers}, has been investigated for real-time systems with thermal constraints \cite{Ahmed2017,AHMED2011226}. The intuition behind the thermal server is to guarantee that a taskset assigned to a core accomplish without thermal interference of other nodes. Under this approach, each real-time task is statically assigned to one server and runs within its assigned server. 

%For the ease of presentation, 
We consider one thermal-aware server for each CPU and GPU core.\footnote{Running multiple thermal servers on each CPU/GPU core is left for future work.}
%\footnote{as long as the sum of their budget does not exceed the value provided by our analysis given in Sec.~\ref{sec:server}.}
Each server is statically associated with one core and does not migrate to another core at runtime. However, unlike prior work \cite{Ahmed2017,ahmed2011minimizing}, we do not limit the server to follow only the polling server policy. We will show in the later section that this flexibility brings significant benefit in the schedulability of tasks accessing the GPU. 

To bound the temperature of each core, its corresponding server $v_i$ is modeled as $v_i = (C_i^v,T_i^v) $
where $C_i^v$ is the maximum execution budget and $T_i^v$ is the budget replenishment period of $v_i$. For brevity, we will use ${v_g = (C^g, T^g)}$ to denote the GPU server and ${v_c = (C^c,T^c)}$ for the CPU server. For budget replenishment policies, we consider \textit{polling}~\cite{sha1986solutions}, \textit{deferrable}~\cite{strosnider1995deferrable}, and \textit{sporadic servers}~\cite{sprunt1989aperiodic}. Under the polling server policy, the corresponding server activates periodically and executes ready tasks until its budget is depleted. The budget is fully replenished at the start of the next period. If there is no task ready, the remaining budget is immediately depleted. In contrast, under the deferrable server, any unused budget is preserved until the end of the period. Hence, a task can execute at any time during the server period while the budget is available. The sporadic server also preserves remaining budget, but replenishes the budget sporadically; only the amount of budget consumed is replenished after $T^v$ time units from the time when that budget is used. Let $J^v$ denote the task release jitter relative to the server release. The value of  $J^v$ is $T^v$ under the polling server policy and $T^v - C^v$ under the deferrable and sporadic server policies~\cite{bernat1999new}.

\subsection{Task Model}
This work considers sporadic tasks with implicit deadlines under \textit{partitioned fixed-priority preemptive scheduling}, which is widely used in many real-time systems. Each task $\tau_i$ has been statically allocated to one CPU core (thus to the server of that core) with a unique priority. Tasks are labeled in an increasing order of priority, i.e., $i<j$ implies $\tau_i$ has lower priority than $\tau_j$. Without loss of generality, each task can contain at most one GPU segment, but it can be easily extended to multiple GPU segments. A task $\tau_i$ is modeled as ${\tau_i = ((C_{i,1},E_{i},C_{i,2}), T_i,s_i)}$, where $C_{i,1}$ and $C_{i,2}$ are the worst-case execution time (WCET) of the normal execution segments of task $\tau_i$, and $E_i$ is the worst-case time of the GPU access segment. The normal execution segments run entirely on the CPU core and the GPU segment involves GPU operations. Let $T_i$ denote the the minimum inter-arrival time of $\tau_i$ and $s_i$ indicate whether $\tau_i$ has a GPU segment, i.e., $s_i=1$ means a GPU-using task. In case $\tau_i$ executes only on the CPU, $s_i$, $E_i$ and $C_{i,2}$ are all zero. Thus, the accumulated sum of the WCETs of $\tau_i$ is denoted as
\[  C_i = s_i \times (C_{i,1}+E_i+C_{i,2}) + (1-s_i)\times C_{i,1}. \]
Furthermore, $V(\tau_i)$ represents the CPU server where $\tau_i$ is assigned. Tasks are considered as fully compute-intensive and independent from each other during normal segment execution. 
%This implies that there is no other interference between task executions. 
The only  resource shared among tasks is the GPU and it is modeled as a critical section protected by a suspension-based mutually-exclusive lock (mutex). Note that this approach follows the well-established locking-based real-time GPU access schemes~\cite{patel2017,GPUSync,elliott2012globally}. We will later present how pending GPU requests are queued in our proposed framework. 

\subsection{GPU Execution Model}
\begin{figure}[t]
\centering
\includegraphics[scale=0.40]{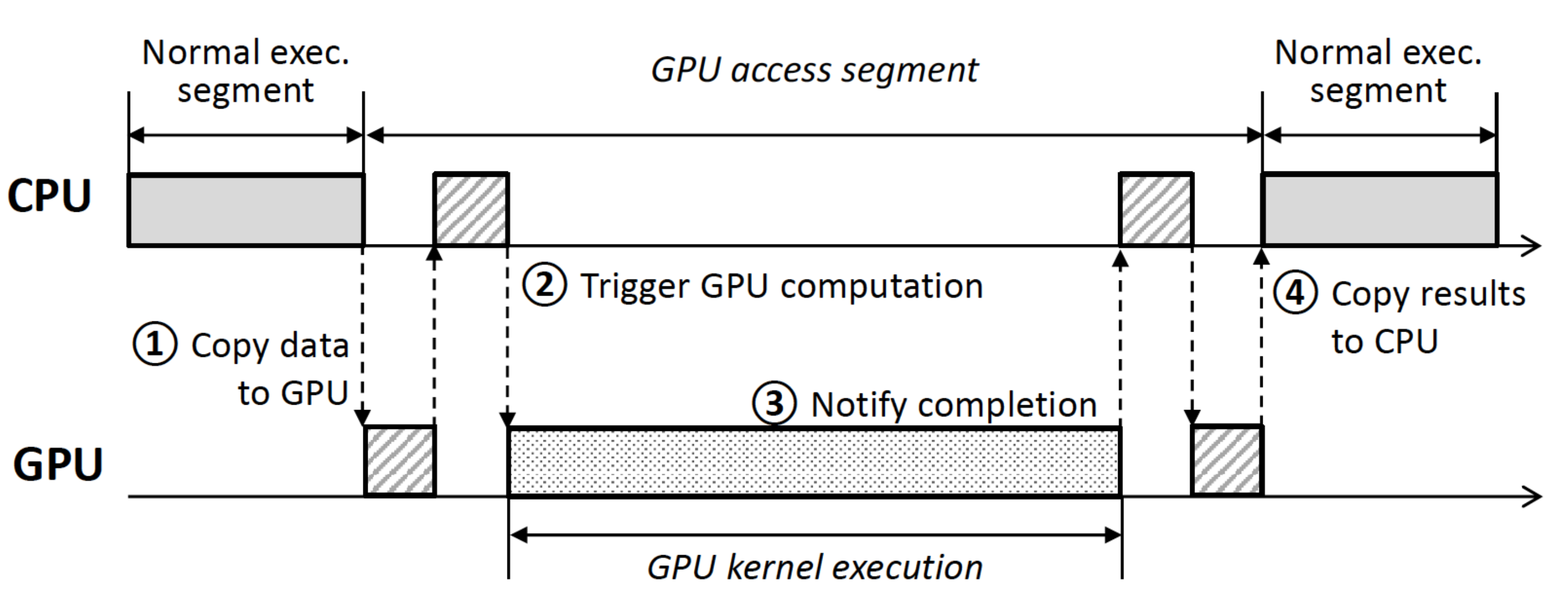}
\caption{Task execution with a GPU segment.}
\label{fig:chap2_gpu_model}
%\vspace{-10.00mm}
\end{figure}
The GPU has its own memory region, which is assumed to be sufficient enough for the tasks under consideration. 
We do not consider the concurrent execution of GPU requests from different tasks because of the resulting unpredictability in kernel execution time~\cite{patel2017,otterness2017evaluation}.
%and ``co-scheduled GPU programs from different programs are not truly concurrent, but are multi-programmed instead"~\cite{otterness2017evaluation}. 
Once a task acquires the GPU lock, its GPU segment is handled through the following steps (see Fig.~\ref{fig:chap2_gpu_model}): 
\begin{enumerate}
    \item \textit{Data Transfer to the GPU:} The task first copies data needed for the GPU computation, from CPU memory to GPU memory. This can be done by Direct Memory Access (DMA), which requires minimal CPU intervention. If the GPU uses a unified memory model, this step can be omitted.
%An acknowledgement signal is then send to the host; Accordingly, the node issues a computation signal to the GPU to launch the kernel;
    \item   \textit{Kernel Launch: } Kernel launches on the GPU. Meanwhile, the task on the CPU side self-suspends and waits for the GPU computation to complete. 
    \item \textit {Kernel Notification Signal:} The GPU signals the CPU to notify the completion of kernel execution.
    \item \textit{Data Transfer to the CPU:} The task wakes up and transfers the results from  GPU memory to  CPU memory.
\end{enumerate}
\cmnt{Finally, the task continues its normal execution segment. }
It is worth noting that a GPU kernel {\em cannot self-suspend} on the GPU in the middle of execution.\footnote{Although GPU kernel preemption is available on some recent GPU architectures, e.g., Nvidia Pascal~\cite{pascal}, to the best of our knowledge, the self-suspension of a kernel is not supported in any of today's GPU architectures.}
%\footnote{To the best of our knowledge, there is no API to suspend and resume GPU kernel execution in CUDA and OpenCL at the time of writing.} 
On the other hand, since there is no CPU intervention during kernel execution, the task on the CPU side {\em self-suspends} to save CPU cycles. As a result, other tasks have a chance to execute or the CPU core sleeps during the kernel execution. 
For a task $\tau_i$, the total time for the above four steps consists of two major parts:
\begin{itemize}
    \item   Miscellaneous operations that require CPU intervention. Let $M_{i,1}$ denote the time for data transfer before the kernel execution and $M_{i.2}$ denote that after the kernel execution.
    \item  Pure GPU kernel operations that do not require any CPU intervention denoted as $K_i$.
\end{itemize}
In such aspect, the GPU segment time of a task $\tau_i$ is modeled as $E_i= M_{i,1} + K_i + M_{i,2}$. As the GPU is non-suspendable during kernel execution, the GPU server should have enough budget larger than or equal to $E_i$. The CPU server only needs to have budget larger than $M_{i,1}$ or $M_{i,2}$. %We will thoroughly address the prospective limitations in our framework and apply them during task schedulability analysis. 

\subsection{Challenges of Thermal-Aware Servers with an Integrated GPU}
\label{sec:chap2_challenges}
Aside from the blocking delays coming from the locking-based GPU access approach, e.g., local and remote blocking to acquire the GPU lock~\cite{patel2017,GPUSync,elliott2012globally}, there are other new challenges faced by thermal-aware servers in a CPU-GPU integrated system. 
\begin{itemize}
    \item \textit{Server budget depletion}: Task execution in a server is scheduled with respect to the  available budget. When the server budget is depleted, a task has to wait until the budget is replenished.
    \item \textit{Mutual budget availability}: 
    %In multi-core synchronization protocols, there is less likely to have the server notion for the shared resources other than CPU. However, here 
    If a task $\tau_i$ issues a GPU request, both CPU and GPU servers must have enough budget to handle this request. It is worth noting that server budget needs for the CPU and GPU servers are different ($M_{i,1}$ or $M_{i,2}$ vs. $E_i$).
    \item \textit{CPU-GPU handover delay}: Even if both servers are designed to have enough budget, each server may have to wait for the other's budget to be replenished when their interactions are needed, e.g., data transfer between the CPU and the GPU.
    %\new{Suppose that there is a polling server on the CPU core and a deferrable server on the GPU. Even if the result is ready on the GPU, the CPU does not respond to the GPU signal until the beginning of its next replenishment period.} %Therefore, data handover synchronization has to be well addressed.
    \item \textit{Back-to-back heat generation}: In case of the deferrable server, some tasks can use up all the budget at the end of the budget replenishment period, and at the very beginning of the next period, some other tasks can start to consume the replenished budget. This causes the server to run longer than its budget and generate heat in a back-to-back manner. 
    %This phenomenon causes the CPU and/or GPU runs for longer time; Hence thermal violation occurs.
\end{itemize}

%\subsubsection{Non-preemptive shared GPU}
%As discussed earlier, each GPU request contains various operations which may or may not require CPU intervention.
% For the sake of execution time predictability, GPU request has to be implemented as ``busy-waiting" on the CPU under most real-time synchronization protocols and their analyses, such as MPCP, FMLP and OMLP. 
%However, considering the shared GPU as a critical section brings up some other challenges as:
%\begin{itemize}
    % \item \textit{busy waiting}: There is no chance for other tasks on the host node to execute when CPU intervention is not needed. Therefore, the task whose kernel is launched can be self-suspended and other tasks have chance to execute their normal execution segments. Consequently, it leads to a better server budget utilization on the CPU cores.  
    %\item \textit{non-preemption}: Since GPU is a non-preemption resource, the GPU server has to be designed in a way to handle the whole GPU request without any interruption.
    % \item \textit{server based mutual resource}: Aside from thermal violation consideration, the server on the GPU increases the remote blocking time overhead.
%\end{itemize}

% \textbf{GOAL:} In this work, we target bounding the remote blocking time of a task as a function of GPU segments length. Furthermore, we introduce our proposed server based framework to bound the maximum temperature on multi-core CPU integrated with GPU platforms.

\section{Framework Design}
\label{sec:chap2_frame}
In this section, we present our proposed framework. We first give a protocol for CPU and GPU thermal servers, and then explain thermal server design to address the challenges discussed in the previous section. We lastly describe a \textit{miscellaneous operation time reservation} mechanism to trade-off between server budget and GPU waiting time. 

% \hyoseung{Give a forward reference to the next section for analysis of the framework.}

\subsection{Thermal-Aware CPU-GPU Server Protocol}
Our framework simultaneously bounds the worst-case blocking time for GPU access and the maximum temperature of CPU and GPU cores. The thermal server designed with our framework isolates the thermal conductive effects of each compute node to other nodes. To achieve these properties, we establish the following rules in our framework.
% \hyoseung{Let us not say that we use MPCP. Since there is only one shared resource, we don't need the whole MPCP but only its subset features for GPU access control, e.g., mutual exclusion, priority boosting.}
% \begin{figure}[t]
% \centering
% \subfloat [][]{\includegraphics[width=0.28\textwidth]{Chapter2/figs/frmwrk/nobusy_wait.pdf}} %\vspace{-4.00mm}\\
% \subfloat [][]{\includegraphics[width=0.20\textwidth]{Chapter2/figs/frmwrk/busy_wait}}
% \caption{Example task scheduling with an unlimited GPU server budget and a CPU server following the polling server policy. a) No busy waiting: $\tau_1$ finishes at 22.  b) Busy waiting: $\tau_1$ finishes at 14.} 
% \label{fig:chap2_busy_wait}
% %\vspace{-10.00mm}
% \end{figure}

\smallskip\noindent\textbf{Shared GPU Server}
\begin{enumerate}
    \item Pending GPU requests are inserted to a priority queue that orders the requests based on the priorities of the corresponding tasks. This rule assures that the GPU request of a high-priority task is blocked by at most one lower-priority request in the queue.
    \item To handle the GPU request of a task $\tau_i$, there must exist at least $E_i$ budget available on the GPU server. This rule is due to the preemptive and non-self-suspending characteristics of the GPU. 
    \item If there is an insufficient amount of budget to launch a GPU segment, the GPU is locked until having enough budget for that GPU segment. This rule assures that starvation does not happen and the critical section of a higher-priority task waits for just a single critical section of a lower-priority task. With these rules, remote blocking time can be bounded. Later, we will propose an enhancement to these rules.

\end{enumerate}

\begin{figure}[ht]
\centering
\begin{subfigure}{.55\textwidth}
  \centering
  % include first image
  \includegraphics[width=\linewidth]{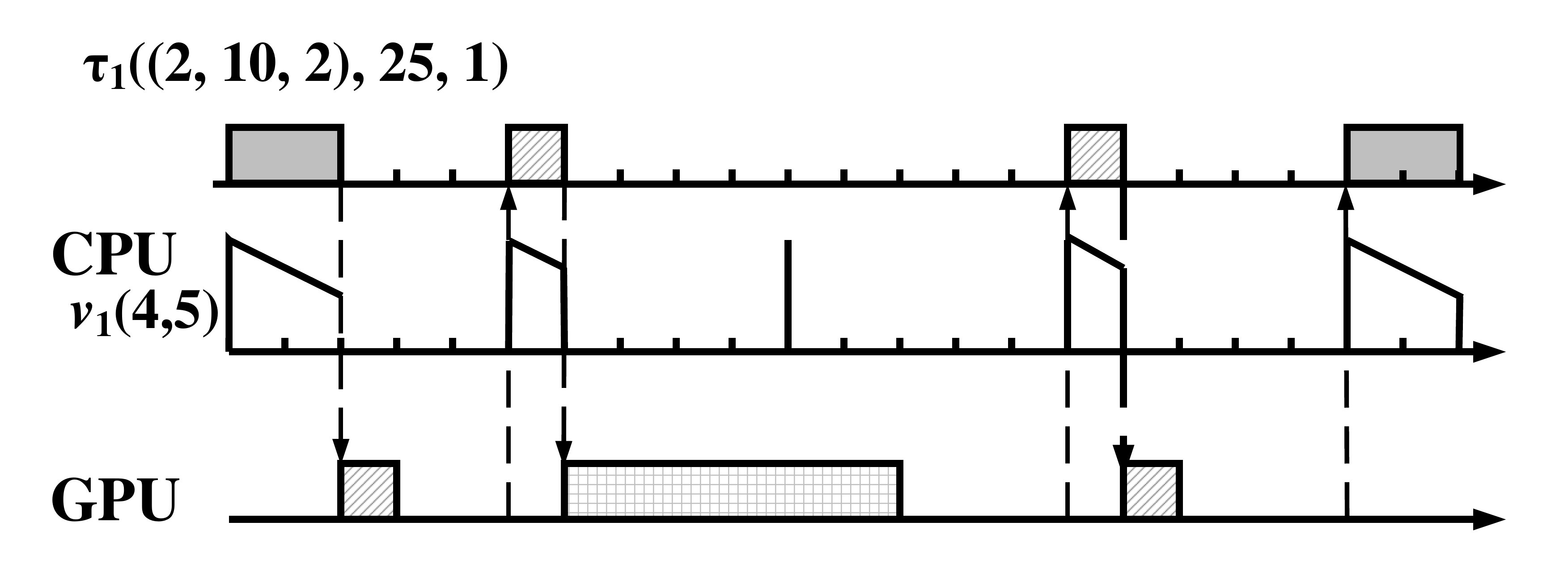}
  \caption{}
  \label{fig:chap2_busy_waitA}
\end{subfigure}
\begin{subfigure}{.43\textwidth}
  \centering
  % include second image
  \includegraphics[width=\linewidth]{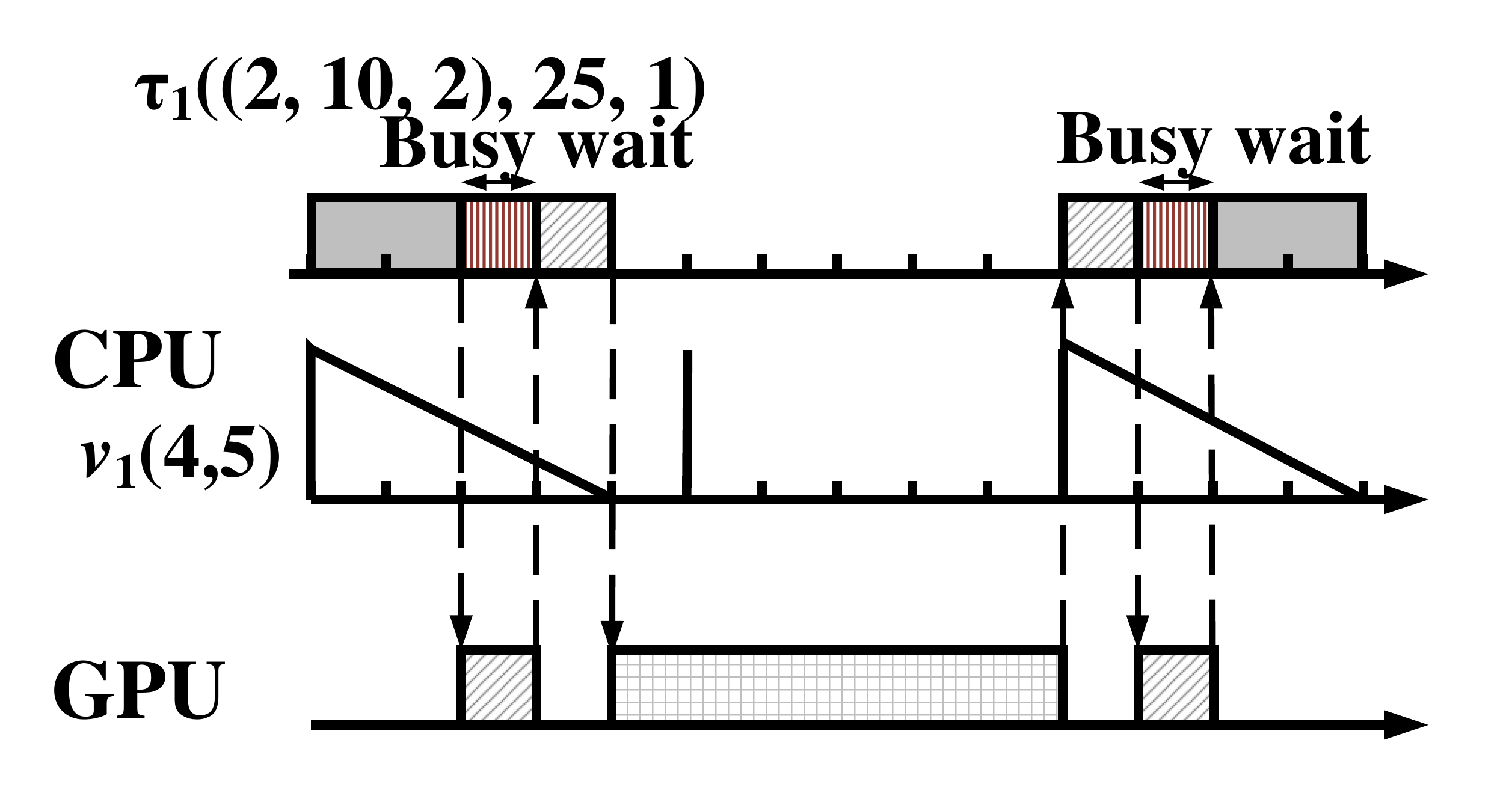}
   \caption{}
  \label{fig:chap2_busy_waitB}
\end{subfigure}
\caption{Example task scheduling with an unlimited GPU server budget and a CPU server following the polling server policy a) No busy waiting: $\tau_1$ finishes at 22. b) Busy waiting: $\tau_1$ finishes at 14.}
\label{fig:chap2_busy_wait}
%\vspace{-10.00mm}
\end{figure}

\smallskip\noindent\textbf{CPU Core Server}
\begin{enumerate}
    % \item If there is no job launching on the shared GPU and there is enough budget on the GPU request and at least $M_i^s$ available budget on the core server, the GPU is granted to the task $\tau_i$ from the top of waiting list.
    \item Servicing a GPU request boosts the priority of the corresponding task to the highest-priority level. As a result, the normal segments of higher-priority tasks are blocked until the GPU segment completes. If there is not enough budget on the corresponding CPU server (e.g., $M_{i,1}$) or on the GPU server (e.g., $E_i$), no other task can execute on the CPU or the GPU.
    \item During data transmission to/from the GPU, the CPU server ``busy-waits". The reasoning behind this rule is to reduce the total response time of a task. Fig. \ref{fig:chap2_busy_wait} illustrates task scheduling with and with or without this rule on a CPU server using the polling server policy. As one can see in this example, without this rule, it takes one additional replenishment period for transferring data to/from the CPU, because during processing the transferring request on the GPU, the CPU server has no other workload to execute; hence it deactivates until the beginning of the next replenishment period. %In fact, this is the reason of using the notation of $M_{i,1}$ and $M_{i,2}$.
    % \item During pure kernel launching where there is no CPU intervention needed, the corresponding task self-suspends. 
\end{enumerate}
\subsection{GPU Server Design}
% Hereby, we discuss the type of the GPU server and its reasoning. Suppose that \new{both the GPU and the CPU servers work under the polling policy}. Because of their \new{distinctive} specifications, they have \new{the} different budget and replenishment period. As discussed, the CPU only needs to have \new{an} enough budget to execute either $M_{i,1}$ or $M_{i,2}$ \new{segment}. Aside from \new{the} overhead time of CPU synchronization due to \new{the} periodicity of \new{the} replenishment, the worst case of {the} critical section executes three times of the host node replenishment period  because each part of $M_{i,1}$, $K_i$ and $M_{i,2}$ executes at beginning of each GPU server. The GPU server has to be long enough to support the largest parts of all tasks. Therefore, it incurs large waiting time for launching small kernel. Synchronization for data handover exacerbates the whole GPU section time (especially for launching small kernel). 
We now discuss budget replenishment policies for the GPU server and their implications. 
%Because of their distinctive hardware specifications of the GPU and the CPU, and also the non-preemptivity of a GPU kernel execution, the different budget and replenishment period are designated to the GPU and CPU servers. 
One may consider both the CPU and GPU servers following the polling server policy. 
In this case, the CPU-GPU handover delay can be at least three complete replenishment periods of a CPU server.
For instance, consider a task holding a GPU lock and its kernel being executed on the GPU. Once the kernel execution completes, the GPU has to wait for the task on the CPU to transfer the results and release the lock. It is possible that at this time, the CPU server budget has been already depleted. Thus, the GPU has to wait for the next period of the CPU server, and this kind of extra delay happens for each sub-segment of the CPU segment, i.e., $M_{i,1}$, $K_i$ and $M_{i,2}$. 
%Thus, the lock remains held and tasks on other servers have to wait for the server budget replenishment of the lock-holding task.  As a result, it incurs extra remote blocking time to tasks on other cores.    
%Aside from the CPU-GPU server coordination delay due to the budget depletion and the replenishment period, similar to the CPU core server design, it would take at least three CPU replenishment periods to launch each GPU kernel because of the coordination between the CPU and the GPU servers. 
%Note that CPU and GPU servers may have different budget and period. 
Moreover, since operations on the GPU are non-preemptive and non-suspendable, the GPU server budget has to be large enough that at least one entire GPU sub-segment of $M_{i,1}$, $K_i$ or $M_{i,2}$ can complete its execution within the same period. However, having a large replenishment period would exacerbate the response time of GPU segments especially for small kernels.

One may consider busy waiting between GPU sub-segments so as to fill their execution gaps because such gaps may deactivate the GPU's polling server. This approach, however, not only requires over-provisioning of the GPU budget, but also produces a considerable amount of heat. In the worst case, the extra heat generation due to the busy-waiting on the GPU server may continue over two periods of the CPU server because the CPU server with the polling server policy can be deactivated between GPU sub-segments.
%\new{Moreover, since the GPU is shared among multiple CPU cores, busy-waiting does not appear to be practically feasible.}

% 2 large period chunk. 
% \begin{figure}[t]
% \centering
% \subfloat [][]{\includegraphics[width=0.5\textwidth]{Chapter2/figs/frmwrk/def.pdf}}\vspace{-4.00mm}\\
% \subfloat [][]{\includegraphics[width=0.5\textwidth]{Chapter2/figs/frmwrk/sprdc.pdf}}
% \caption{Example task scheduling with the GPU server under a) the deferrable server policy and b) the sporadic server policy. For both tasks $\tau_1$ and $\tau_2$, $E_i=8$ and $M_{i,1}=M_{i,2}=2$.}
% \label{fig:chap2_thermal_bck2bck}
% %\vspace{-7.00mm}
% \end{figure}

\begin{figure}[t]
\centering
\begin{subfigure}{.85\textwidth}
  \centering
  % include first image
 \includegraphics[width=\textwidth]{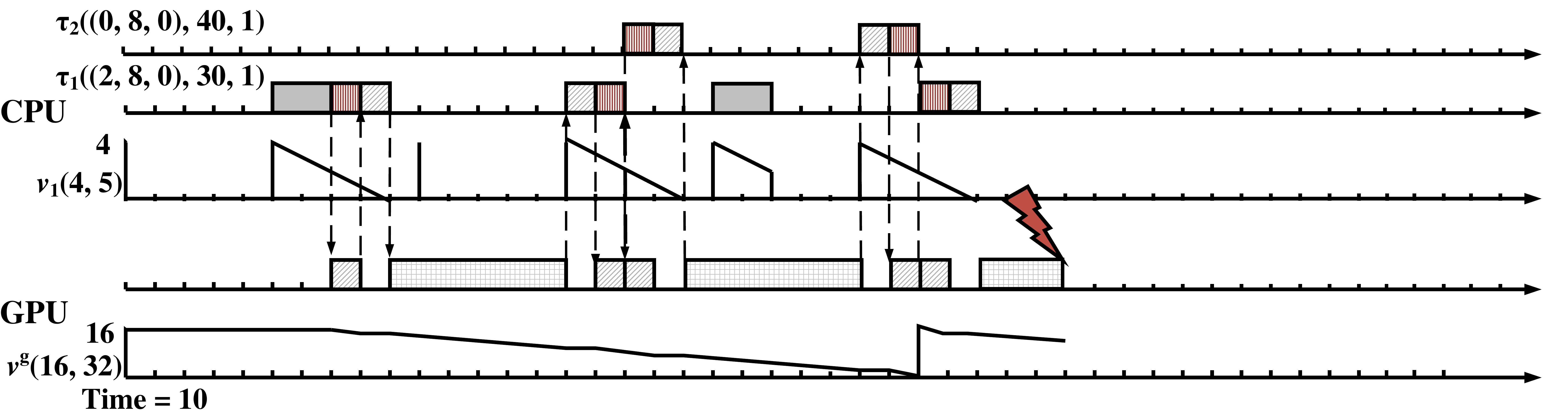}
  \caption{}
  \label{fig:chap2_thermal_bck2bckA}
\end{subfigure}
\newline
\begin{subfigure}{.85\textwidth}
  \centering
  % include second image
  \includegraphics[width=\textwidth]{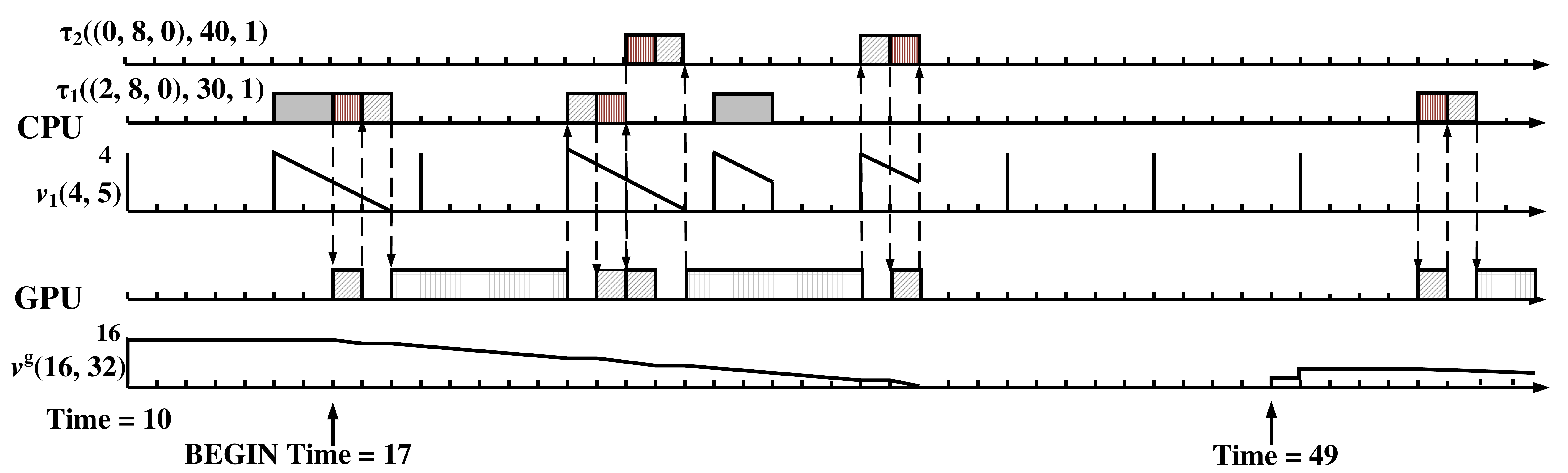}
  \caption{}
  \label{fig:chap2_thermal_bck2bckB}
\end{subfigure}
\caption{Example task scheduling with the GPU server under a) the deferrable server policy and b) the sporadic server policy. For both tasks $\tau_1$ and $\tau_2$, $E_i=8$ and $M_{i,1}=M_{i,2}=2$.}
\label{fig:chap2_thermal_bck2bck}
%\vspace{-10.00mm}
\end{figure}

One may suggest the deferrable server policy for the GPU to mitigate the heat generation issue and to minimize the response time of a GPU segment. This approach, however leads to the thermal back-to-back execution phenomenon. Fig.~\ref{fig:chap2_thermal_bck2bck}a illustrates an example of possible drawbacks of the deferrable server chosen for the GPU. The first jobs of $\tau_1$ and $\tau_2$ arrive at the latest moments in the first GPU server period, and the second job of $\tau_1$ arrives at the beginning of the second GPU server period. Although the tasks are schedulable by the given server budgets, it causes burst heat generation by back-to-back execution, which can potentially lead to thermal violation. In order to avoid this, the budget of the deferrable server for the GPU has to be halved to avoid thermal violation but  it can drastically lower task schedulability.

% \begin{figure}[t]
% \centering
% \includegraphics[width=0.5\textwidth]{figs/frmwrk/sprdc.pdf}
% \caption{b) The previous example of synchronization using the proposed sporadic GPU server}
% \label{fig:chap2_gpu_sporadic}
% %\vspace{-10.00mm}
% \end{figure}
In contrast, the sporadic server policy can take the merits of both the polling server and deferrable server policies. If the sporadic server policy is used for the GPU, a GPU segment can execute at any time as long as there is enough budget, and back-to-back heat generation does not occur. Fig.~\ref{fig:chap2_thermal_bck2bck}b illustrates the previous example with the sporadic server on the GPU. As can be seen, unlike the deferrable server case, the budget replenishment of the GPU server is one period apart from its consumption time, thereby preventing potential thermal violation. 
The sporadic server on the GPU is also practically effective because the GPU server needs to have a relatively large budget with a long replenishment period due to its non-preemptive nature whereas the CPU server typically has a short replenishment period to reduce task response time. 
% \textbf{MEHDI:} define what you mean about sporadic server. define the type of sporadic server that u use. how do you know it is really good. what are drawback of using sporadic server for gpu. why we do not need to use for CPU

In summary, due to the aforementioned reasons, our framework specifically uses the sporadic server policy for the GPU, while all the three policies (polling, deferrable, and sporadic) are allowed for CPU cores. We also set the budget of the GPU sporadic server to be at least as large as one complete GPU segment of any task, i.e., $\max(E_i)$, because this can reduce the response time of a GPU segment and remote blocking time. The detailed analysis on these delays will be presented in Section~\ref{sec:chap2_analysis}.

The thermal server for the GPU is a resource abstraction managed on the CPU side, similar to other real-time GPU management schemes~\cite{patel2017,8046309,kim2018server,GPUSync,elliott2012globally}. One can implement the GPU thermal server as part of GPU drivers or application-level APIs.

\subsection{Miscellaneous Operation Time (MOT) Reservation}
\label{sec:chap2_Overrun}
In order to reduce the CPU-GPU handover delay, we propose an MOT reservation mechanism. 
With this mechanism, a small portion of the CPU {server budget is} reserved only for miscellaneous operations in a GPU segment, e.g., transferring data to/from the GPU. The MOT reservation is feasible with the deferrable and sporadic server policies but not with the polling server policy because the polling server is unable to keep unused budget by design. Although this MOT reservation reduces the amount of CPU budget for regular task execution, it guarantees that the GPU does not need to wait for the budget replenishment of the CPU server during the data transmission phase of a GPU segment.
It can also reduce the remote blocking time of other tasks. 
This reserved budget for MOT has to be  the largest amount of the CPU intervention time in all GPU requests. The trade off between the MOT reservation and the reduced budget for regular task execution will be extensively investigated in the evaluation. 

\begin{figure}[t]
\centering
\begin{subfigure}{.85\textwidth}
  \centering
  % include first image
\includegraphics[width=\textwidth]{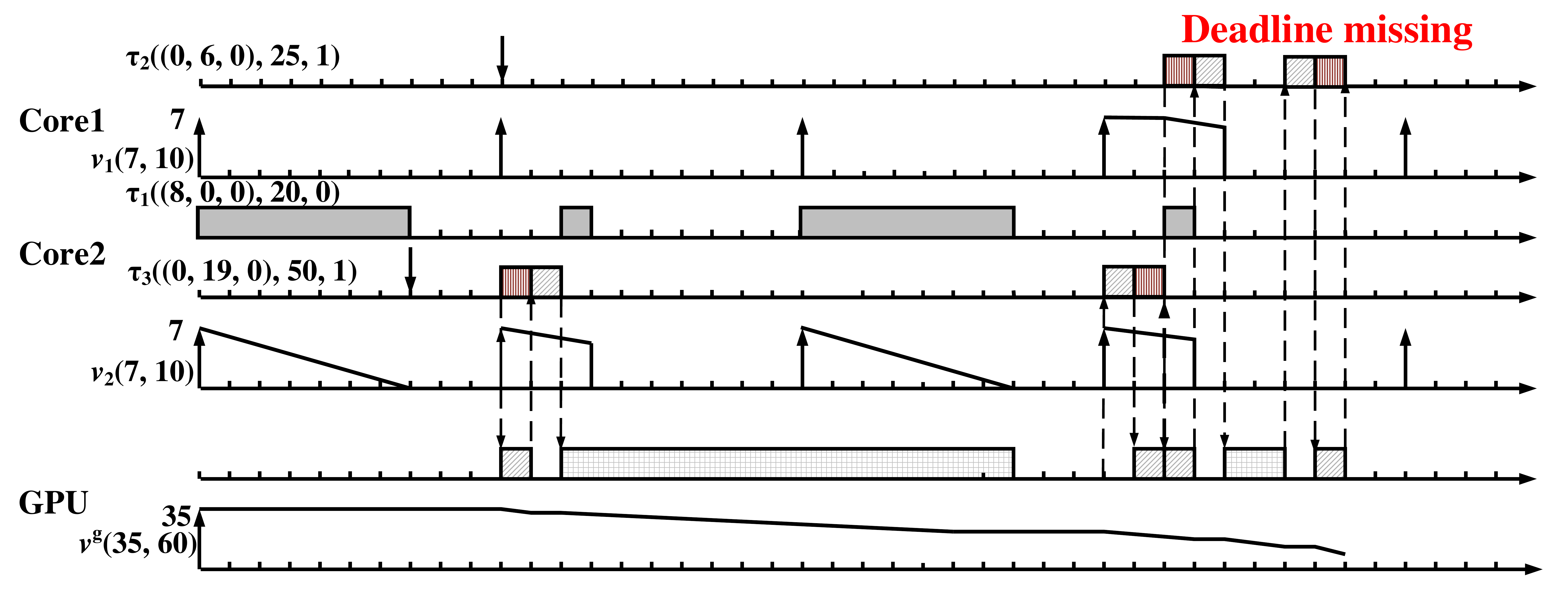}
  \caption{}
  \label{fig:chap2_overrunA}
\end{subfigure}
\newline
\begin{subfigure}{.85\textwidth}
  \centering
  % include second image
 \includegraphics[width=\textwidth]{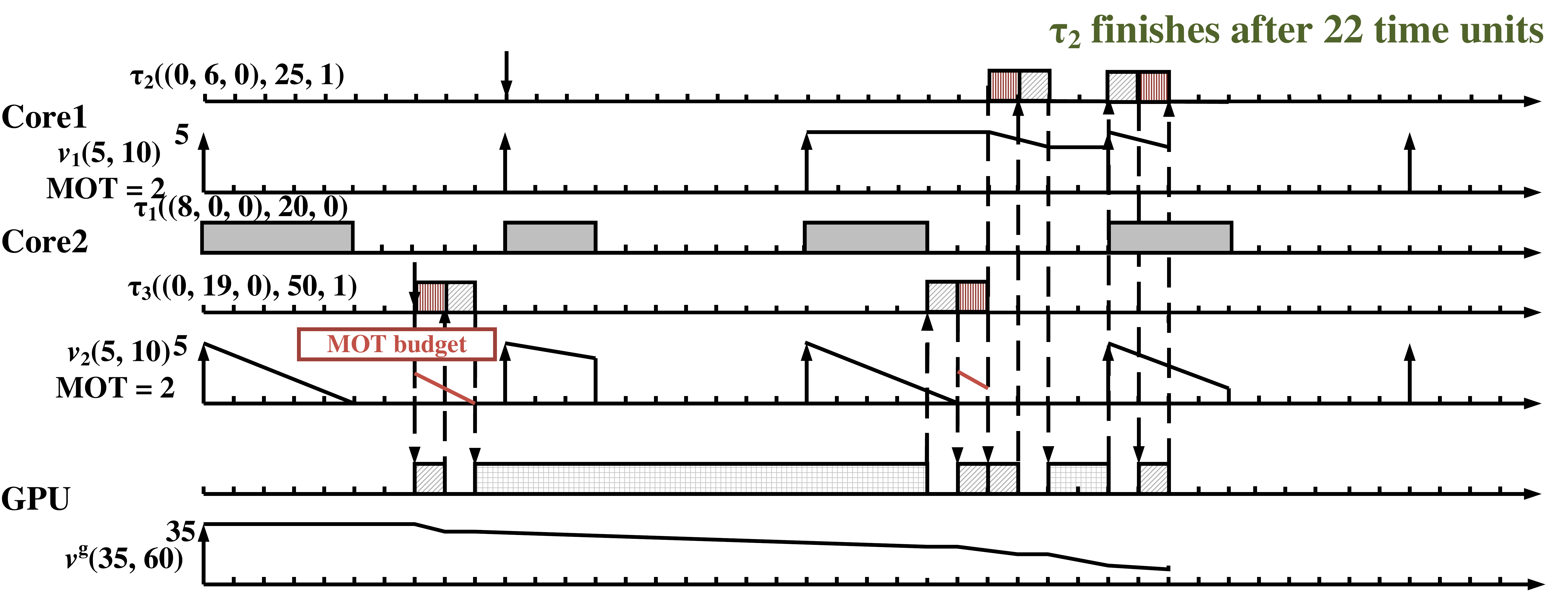}
  \caption{}
  \label{fig:chap2_overrunB}
\end{subfigure}
\caption{Data transfer in a GPU request a) without and b) with the MOT reservation mechanism.}
\label{fig:chap2_overrun}
%\vspace{-10.00mm}
\end{figure}

Fig.~\ref{fig:chap2_overrun} illustrates an example to highlight the benefit of the MOT reservation mechanism. As one can see in Fig.~\ref{fig:chap2_overrun}a, $\tau_3$ has to wait until 10 to transfer its data to the GPU because $\tau_1$ already has consumed all budget of $v_2$. Similarly, although the result of the GPU kernel of $\tau_3$ is ready at 27, its result begins to be transferred at 30. These delays cause the remote blocking of $\tau_2$ lasts for 22 time units although there exists an available amount of the server budget on $v_1$. Designating 2 time units as the MOT budget (see Fig.~\ref{fig:chap2_overrun}b) leads $\tau_3$ to transfer its data to the GPU at 7 and finishes its kernel launch at 27; hence $\tau_2$ initiates its kernel launch accordingly. Consequently, designating some amount of the server budget as MOT reservation leads to reduction in the remote blocking of a GPU-using task from other CPU core.

% \begin{figure}[t]
% \centering
% \subfloat [][]{\includegraphics[width=0.5\textwidth]{Chapter2/figs/frmwrk/non-over.pdf}}\\
% \subfloat [][]{\includegraphics[width=0.5\textwidth]{Chapter2/figs/frmwrk/ovr.pdf}}
% \caption{Data transfer in a GPU request a) without and b) with the MOT reservation mechanism.}
% \label{fig:chap2_overrun}
% %\vspace{-10.00mm}
% \end{figure}

\section{Thermal and Schedulability Analysis}
\label{sec:chap2_analysis}
In this section, we present the schedulability analysis of our framework. We first design our thermal-aware servers for multi-core GPU-integrated  platforms to avoid thermal violation. Then, we analyze  task schedulability with and without the MOT reservation mechanism. 

\subsection{Design of Server Budget}
\label{sec:chap2_server}
In our framework, task execution is performed within thermal-aware servers. As discussed, the notion of servers is to isolate each compute node from others in terms of the thermal aspect. Accordingly, under any circumstance of task execution, the thermal violation avoidance has to be guaranteed in the server design. The specifications of servers are independent of their running tasks (except for the design of the MOT reservation),  whereas task schedulability does depend on them. In our proposed framework, introducing the thermal-aware servers for both the CPU and the GPU makes the physical characteristics of underlying platforms be transparent of the task schedulability test.

\subsubsection{Single-Core Platforms}
First, we calculate the ``maximum" budget that a server can have while limiting the temperature not  to exceed the given thermal constraint in a single-core platform under a given replenishment period. In the worst case of a polling server, the server exhausts all of its budget at the beginning of its period and then sleeps until the beginning of the next replenishment period. Let $t_{wk}$ and $t_{slp}$ denote the active time (i.e., the budget-consuming phase) and the sleeping time (i.e., the cooling phase) of a CPU core server, respectively. Hence, the server period $T$ is \begin{equation} 
\label{eq:chap2_1}
T = t_{wk} + t_{slp}.
\end{equation}
In the steady state of the system, we are interested in bounding the server's maximum temperature. According to  Eq.~\ref{eq:chap2_tempActive}, \[ \alpha + (\theta_s - \alpha ) e^{\beta t_{wk}} \leq \theta_M\]
where $ \theta_s$ and $\theta_M$ are the steady state temperature and the thermal constraint, respectively. Therefore,
% \[ (\theta_s - \alpha ) e^{\beta t_{wk}} \leq   \theta_M -\alpha \]
\begin{equation}  
\label{eq:chap2_2}
e^{\beta t_{wk}} \geq  \frac{ \theta_M -\alpha}{\theta_s - \alpha } \Longrightarrow t_{wk} \leq \frac{1}{\beta} \ln{\frac{ \theta_M -\alpha}{\theta_s - \alpha }}.
 \end{equation}
On the other hand, in the cooling phase, according to Eq.~\ref{eq:chap2_tempCool} to respect the steady state, $ \theta_M e^{\beta t_{slp}} = \theta_s$.
Hence, 
\begin{equation} \label{eq:chap2_3}
t_{slp} = \frac{1}{\beta} \ln{\frac{\theta_s}{\theta_M}}.
 \end{equation}
 By substituting Eqs.~\ref{eq:chap2_2} and \ref{eq:chap2_3} by Eq.~\ref{eq:chap2_1}, we have
\[ \frac{1}{\beta}\ln{ \frac{ \theta_M -\alpha}{\theta_s - \alpha }} + \frac{1}{\beta} \ln{\frac{\theta_s}{\theta_M}} \leq T
\]
\[  \frac{ \theta_M -\alpha}{\theta_s - \alpha } \times \frac{\theta_s}{\theta_M} \leq e^{\beta T }.
\]
Therefore, the worst-case steady state temperature at the beginning of each period is
\begin{equation} \label{eq:chap2_4}
\theta_s = \frac{\alpha \theta_{M} e^{\beta T}}{\theta_{M} (e^{\beta T}-1)+\alpha}.
\end{equation}
Accordingly, the maximum budget for the period $T$ is
\begin{equation} \label{eq:chap2_5}
t_{wk} = T - \frac{1}{\beta} \ln{\frac{\theta_s}{\theta_M}}.
\end{equation}
Consequently, for a given replenishment period, a server ${v_i = (T - \frac{1}{\beta} \ln{\frac{\theta_s}{\theta_M}}, T)}$ can bound the maximum temperature to $\theta_M$.

As one can figure out from the analysis, the maximum budget converges because of $\alpha$. This means that after some point, an increase in the replenishment period has no effect on the maximum feasible budget. As discussed earlier through our framework design, the budget has to be considered as $\frac{t_{wk}}{2}$ for a deferrable server due to the \textit{thermal back-to-back} phenomenon. This phenomenon does not occur in a sporadic server, and it can use the computed budget as is.

\subsubsection{Homogeneous Multi-core Platforms}
The worst case for the budget of polling servers on a multi-core CPU happens when all of them exhaust their budget completely. Therefore, according to Eq. \ref{eq:chap2_tempCompo} for the composability characteristics of the heat transfer, we have
 \[    \alpha + (\theta_s - \alpha) e^{\beta t_{wk}} +  \sum_{\begin{subarray}{l}j=1\\ j \neq i \end{subarray}}^{m} \gamma_{i,j} \theta^j_M \leq \theta^i_M\]
where $\theta^i_M$ is the maximum temperature for the $i$th node. In the worst case, every core may reach its maximum temperature at the same time. Hence,
\[  \underbrace{(1 +  \sum_{\begin{subarray}{l}j=1\\ j \neq i \end{subarray}}^{m} \gamma_{i,j})}_{\lambda_i} [ \alpha + (\theta_s - \alpha) e^{\beta t_{wk}}] \leq \theta^{i}_{M}. \]
However, the geographic location of cores on the chip results in different values of the conduction coefficients although it remains symmetric (i.e., $\gamma_{i,j} = \gamma_{j,i}$). Denoting ${\lambda  = \max_{1 \leq i \leq m}{\lambda_i}}$, $\theta_M$ is given by $\theta_M = \lambda [ \alpha + (\theta_s - \alpha) e^{\beta t_{wk}}] $. Similar to the single-core analysis given in the previous subsection, the steady state temperature is 
\begin{equation} \label{eq:chap2_10}
\theta_s = \frac{\alpha \frac{\theta_{M}}{ \lambda} e^{\beta T}}{\frac{\theta_{M}}{\lambda} (e^{\beta T}-1)+\alpha}
\end{equation}
 Therefore, the server budget for each compute node $i$ is
 \begin{equation} \label{eq:chap2_11}
C^c = t_{wk} = T - \frac{1}{\beta} \ln{\frac{\theta_s \times \lambda}{\theta_M}} .
\end{equation}

\subsubsection{Heterogeneous Multi-Core  GPU-Integrated Platforms}
Hereby, we will determine the budget of servers in the presence of an integrated GPU. Similar to the homogeneous multi-core platform, the worst case happens when all servers exhaust their budgets completely. There is also a GPU segment execution on the GPU which causes extra heat dissipation. Since the GPU runs at a different frequency and its architecture is different from the CPU cores, its heat generation parameters differ from those of the CPU cores. Hence, 

% Besides, in \new{the} presence of \new{the} GPU, we assume that \new{the} GPU launches a kernel during the \new{the whole} period of CPU \new{servers}. The reasoning of this assumption is that the GPU is a non-preemptive resource, and in its worst case, it can execute for \new{the} entire \new{CPU} core period. Hence, 

\[\left\{
                \begin{array}{ll}
                  \theta^i_M = \lambda_i [ \alpha + (\theta_s - \alpha) e^{\beta t_{wk}}]  +  \gamma_{i,g}[\alpha^g + (\theta^g_s - \alpha^g) e^{\beta^g t^g_{wk}}] \\
                 
                \theta^g_M = \alpha^g + (\theta^g_s - \alpha^g) e^{\beta^g t^g_{wk}} +  \underbrace{\sum_{j=1}^{m} \gamma_{g,j}}_{\gamma^g} [ \alpha + (\theta_s - \alpha) e^{\beta t_{wk}}]  
                \end{array}
              \right. 
\]
\begin{equation}
\label{eq:chap2_double}
\Longrightarrow\left\{
                \begin{array}{ll}
                  \lambda \theta^i_M e^{\beta t_{slp}} + \gamma_{i,g} \theta^g_M e^{\beta^g t^g_{slp} }= \theta_s \\
                \theta^g_M e^{\beta^g t^g_{slp} }+ \gamma^g \theta^i_M e^{\beta t_{slp}} = \theta^g_s
                \end{array}
              \right.             
\end{equation}where the symbols with a superscript $g$ represent the corresponding parameters of the GPU. 
\subsubsection{Miscellaneous Operation Time Reservation}
To reduce the remote locking time, some portion of the CPU server budget can be reserved for data transferring from/to the GPU that needs CPU intervention. The MOT budget has to be large enough to handle the longest data transferring time, therefore 
\begin{equation} \label{eq:chap2_overrun}
C^c  =  t_{wk}  - \max_{\forall \tau_i}{(M_{i,1}, M_{i,2})}
\end{equation}

It is noteworthy that acquiring a lock on the GPU happens when there is enough budget on the GPU server to execute the whole GPU request; hence, no budget reservation is needed on the GPU side.

\subsection{ Task Schedulability Analysis}
The thermal analysis in the previous subsection gives the maximum budget of each CPU/GPU server that satisfies the thermal constraint of the system. Hereby, we present the schedulability analysis of a task $\tau_i$ in our framework.

Before introducing our analysis, we review the existing response time test for independent tasks with no thermal constraints and no shared GPU under hierarchical scheduling~\cite{saewong2002analysis}, which is 
\begin{equation}
\label{eq:chap2_hierchical}
 W_i^{n+1}   =  C_i + \sum_{\begin{subarray}{h} \tau_h \in V(\tau_i) \\
       h > i \end{subarray} }\ceil*{\frac{W_i^n  +J^c}{T_h}}C_h + \ceil*{\frac{W^{n}+C^c}{T^c}} (T^c - C^c)
\end{equation}
where $J^c$ is the jitter of a task running in a server (see Section~\ref{sec:chap2_model}) and $W^0 = C_i$. The recursion terminates successfully when $W^{n+1}= W^n$ and  fails when $W^{n+1}> T_i$. This equation considers the budget depletion of a CPU server. The first term is the amount of CPU time used by the task $\tau_i$, the second term captures the back-to-back execution of each higher-priority task $\tau_h$,  and the third term captures the amount of interference that the server can generate due to the periodic budget replenishment. However, this equation cannot be used directly for the thermal constraint problem with a  shared GPU.
 
Our analysis extends the existing response time test by considering the factors discussed in Section V:  (i) local blocking time, (ii) remote blocking time, (iii) back-to-back execution due to remote blocking, (iv) mutual budget availability, (v) CPU-GPU handover delay, (vi) multiple priority inversions, and (vii) CPU and GPU server budget co-depletion.  We take into account the factors (iv) and (v) together in the analysis of the handover delay, the factor (vi) as part of the local blocking time analysis, and the factor (vii) as part of the remote blocking time analysis. By considering all these factors, the following recurrence equation bounds the worst-case response time of a task $\tau_i$: 
% [ comment: the original response time task for independent tasks with no thermal constraints under hierarchical scheduling [ref paper 31]] and explain each term briefly. For the total amount of CPU time used by the task under analysis and its higher prioty tasks, the last term captures (T^c-C^C) the amount of interference that the server can generate.
% \begin{equation}
    %%%%%%  resizing equation in one line
%   \resizebox{.9\hsize}{!}{$ W_i^{n+1} = C_i + B_i^l + B_i^r +\sum_{\begin{subarray}{l} \tau_h \in V(\tau_i) \\
%      h > i \end{subarray}}\ceil*{\frac{W_i^n + J_h +B^r_h}{T_h}}C_h +  \ceil*{\frac{W_i^{n}+C_k^v}{T_k^v}} (T_k^v - C_k^v)$}
  \begin{align}
& W_i^{n+1}   =  C_i + B_i^l + B_i^r +  H_i^{gc}  + \ceil*{\frac{W_i^{n}+C^c- s_i (H_i^{gc} +  K_i)}{T^c}} (T^c - C^c)  + \nonumber \\
 &\sum_{\begin{subarray}{l} \tau_h \in V(\tau_i) \\
       h > i \end{subarray}}\ceil*{\frac{W_i^n +J^c +(W_h - C_h) - s_i  (H_i^{gc} +  E_i)}{T_h}}C_h  
  \end{align}
% \end{equation}
where $C_i$ is the worst-case execution time of $\tau_i$, $B_i^l$ is the local blocking time, $B_i^r$ is the remote blocking time, $H^{gc}_i$ is the CPU-GPU handover delay. We will later discuss each of them in details. The recurrence equation terminates when $W^{n+1}_i = W^n_i$, and the task $\tau_i$ is schedulable if its response time does not exceed its implicit deadline (i.e., $W^n_i <= T_i$).

% The latter term captures extra  delay  a task can experience after acquiring the GPU lock. When a task transfers data from/to the share GPU but there is not enough budget available in the sporadic GPU server for its kernel execution, the task has to wait at most the jitter of GPU server ($T^g-C^g$). After the GPU budget is replenished, if the host core server is inactive, the task has to wait for additional $T^c$ for replenishment of the CPU core server. Similarly, at most $T^c$ amount of time is needed for the CPU core server be activated to transfer data from GPU to the host core. 

The second line of the equation captures the delay due to the server budget depletion on the CPU side (an extension of the last term of Eq.~\ref{eq:chap2_hierchical}). It is worth noting that during the pure GPU kernel execution of $\tau_i$ ($K_i$), no CPU budget is consumed by $\tau_i$. Any other tasks can execute on the CPU core if there is a remaining budget. The task $\tau_i$ only consumes the CPU server budget when it executes  normal segments or the miscellaneous operations (e.g. data copy) of GPU segments. Therefore, $H^{gc}_i $ and $K_i$, which already exist in $W_i^n$ for a GPU-using task $\tau_i$, are excluded such that only the CPU-consuming parts of this task is affected by the CPU server budget depletion. Any additional delay due to CPU budget replenishments during GPU segment execution will be captured by $H^{gc}_i$.

The last line captures the preemption time by the normal execution segments of higher-priority tasks (an extension of the second term of Eq.~\ref{eq:chap2_hierchical}). The fact that a higher-priority task $\tau_h$ can only preempt the normal execution segments of $\tau_i$ leads to the deduction of $H^{gc}_i$ and $E_i$ from $W_i^n$. There can be at most one additional job of $\tau_h$ that has arrived during $\tau_i$'s GPU segment execution and interfere $\tau_i$'s normal segments. This holds true if $\tau_h$ is schedulable, i.e., $W_h\le T_h$, because other arrivals of $\tau_h$ during $\tau_i$'s GPU segment will finish their executions while $\tau_i$ is self-suspended for its kernel execution on the GPU. The interference from this additional carry-in job of $\tau_h$ is taken into account by modeling it as a dynamic self-suspending model and adding $W_h-C_h$ to $W_i^n$~\cite{bletsas2018errata}. 

Now, we present the detailed analysis of the delay factors used in our response time test.
\subsubsection{\textbf{CPU-GPU Handover Delay}}
\begin{figure}[t]
\centering
\includegraphics[width=0.8\textwidth]{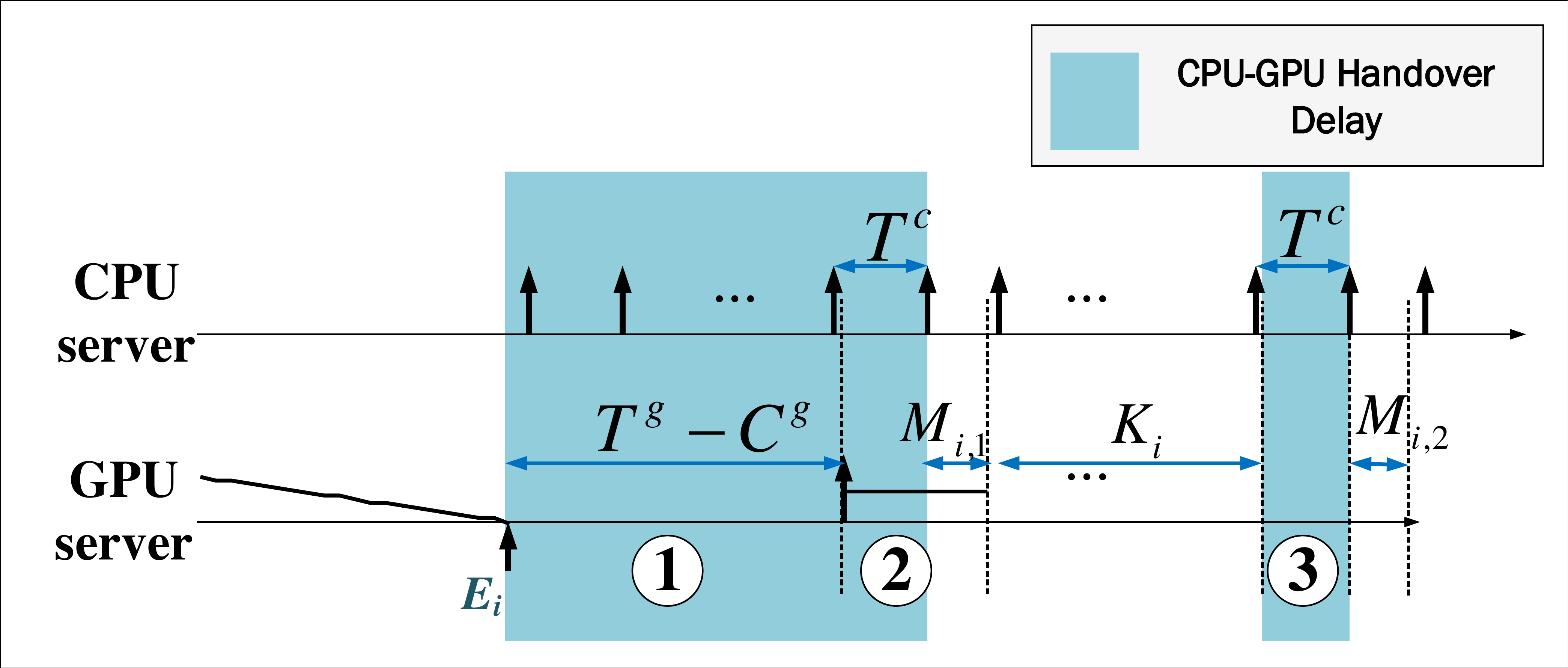}
\caption{The worst-case scenario for CPU-GPU handover delay with a CPU polling server. When a GPU request with $E_i$ is chosen from the GPU waiting queue for execution, it experiences the delay highlighted in blue boxes.}
\label{fig:chap2_handover}
\end{figure}

 It captures an extra delay a task  can experience after acquiring the GPU lock because of the factors (iv) and (v).  Fig. \ref{fig:chap2_handover} illustrates this type of delay decomposed into three parts when the polling server policy is used for a CPU server. \circled{1} When there is not enough budget available in the GPU server for the execution of a GPU segment, the task has to wait at most the jitter of the GPU server ($T^g-C^g$). \circled{2} After the GPU budget is replenished, if the CPU server is inactive, the task has to wait for additional $T^c$ time units for the next period of the CPU server. \circled{3} After the completion of kernel execution on the GPU, at most $T^c$ time units are needed for the CPU server to be activated in order to transfer the results back from the GPU to the CPU. Hence, 
\begin{equation}
\label{eq:chap2_handover}
    H_i^{gc} = s_i  (T^g - C^g + 2  T^c).
\end{equation}
For a CPU server with the deferrable and sporadic server policies, \circled{2} and \circled{3} change to the jitter of the CPU server (i.e., $T^c - C^c$) because a GPU segment needs to wait for the replenishment of its corresponding CPU server's budget. The handover delay is then given by
\begin{equation}
\label{eq:chap2_handover_deferrable}
{H_i^{gc} = s_i  [T^g - C^g + 2  (T^c-C^c)]}.
\end{equation}

\subsubsection{\textbf{Local Blocking}}
  It occurs when a task $\tau_i$ is blocked by the lower-priority task on the same core. As in the case of MPCP~\cite{rajkumar1990real, rajkumar1988real}, a task can be blocked by each of its lower-priority task $\tau_l$'s GPU segment at most once due to the priority boosting of our framework. To obtain a tight bound, we analyze local blocking time from two different perspectives.
  
  On the one hand, the worst-case local blocking time of a task $\tau_i$ happens when each normal execution segment of $\tau_i$ is blocked by the GPU segment of each lower-priority task $\tau_l$ with the amount of ${E_l- K_l = M_{i,1}+M_{i,2}}$, which is the maximum CPU time used by the GPU segment of $\tau_l$. Hence, the total local blocking time of a task $\tau_i$ is
  \begin{equation}
  {(1+s_i) \sum_{ l<i \And \tau_l \in V(\tau_i) \And s_i>0} E_l-K_l}
  \end{equation}
  where $(1+s_i)$ indicates the number of normal execution segments of $\tau_i$. It is worth noting that under the RM policy, the total blocking time can be bounded by just
  \begin{equation}
  {\sum_{ l<i \And \tau_l \in V(\tau_i) \And s_i>0} E_l-K_l}
  \end{equation}
   because each task has only one GPU segment and the period of any lower-priority task $\tau_l$ is larger than that of $\tau_i$, which leads to only one blocking time from each $\tau_l$ during the job execution of $\tau_i$.

On the other hand, the worst-case local blocking time of $\tau_i$ can also be bounded by the amount of GPU budget available during one period of $\tau_i$. The reasoning behind this approach is that lower-priority tasks cannot execute GPU segments more than the available budget on the GPU. Thus, the maximum total blocking time of $\tau_i$ is bounded by ${(1+s_i)(\ceil{\frac{T_i}{T^g}}+1) C^g}$ where ``+1" is due to the carry-in effect. 

Using these two approaches, the total local blocking time of $\tau_i$ is bounded by 
\begin{equation}
% B_i^{l} = \sum_{\begin{subarray}{l} l<i \\ \tau_l \in V(\tau(i)) \\ S_i>0\end{subarray}}s_l \left[(T^g - C^g)+E_l + (T^v_k - C^v_k)\right].
 \begin{split}
B_i^l  =  %2 \times (T^v_k - C^v_k) + 
(s_i+1) \cdot \min \left(\left(\ceil*{\frac{T_i}{T^g}}+1\right) C^g , \sum_{\begin{subarray}{l} l<i \\ \tau_l \in V(\tau_i) \\ s_i>0\end{subarray}} E_l-K_l\right).
 \end{split}
\end{equation}
%  It is noteworthy that a lower priority task can only accomplish its kernel execution when there exist enough budgets on both CPU and GPU servers. The GPU services a kernel request from the head of its waiting queue only if there is enough remaining budget that the whole kernel can be executed within a single period. The term $T^g - C^g$ which is the waiting time of the sporadic GPU server for replenishment is not necessary because if there is not enough GPU budget, the higher priority executes immediately.  There is also no need to consider the CPU server depletion since this term is already captured in calculation of worst-case response time.  

\subsubsection{\textbf{Remote Blocking}}
% Remote blocking of a task happens when either a lower priority task or any higher priority task of other cores launches their kernel on GPU. Consequently, the corresponding task is blocked until they release their lock on the GPU. Therefore, the remote blocking time $B_i^r$ of a task $\tau_i$ is given by 
 It occurs when the GPU segment of a task is blocked in the GPU waiting queue due to other GPU requests. 
Recall that the GPU segments of tasks are ordered in a priority queue according to their tasks' original priorities. The response time of a GPU segment of $\tau_i$ is given by ${W'_i = H_i^{gc} + E_i }$. The reasoning is that after a task acquires the GPU lock, it has to wait $H^{gc}_i$  for the handover delay of data transferring and mutual server synchronization (Eq.~\ref{eq:chap2_handover}). There is no other delay than $H_i^{gc}$ added to the GPU segment length $E_i$ because our framework sets the GPU budget to be large enough to perform any GPU segment in one GPU period and boosts the priority of the task executing a GPU segment.  Hence, the remote blocking time of $\tau_i$ is bounded by the following recurrence equation:
\begin{equation}
B_{i}^{r,n+1} =  \max_{\begin{subarray}{l} l<i  \\ s_i>0\end{subarray}}{W'_l} + \sum_{\begin{subarray}{l} h>i \\ s_i>0\end{subarray}}\bigg(\ceil*{\frac{B_{i}^{r,n}}{T_h}}+1\bigg). W'_{h}.
\end{equation}
where the base is ${B_{i}^{r,0} =  \max_{\begin{subarray}{l} l<i  \\ s_i>0\end{subarray}}{W'_l}}$. The first term of the equation captures the waiting time for acquiring the GPU lock due to the currently-running of one GPU segment of a lower- priority task and the second term represents the waiting time for the GPU segments of higher-priority tasks. This analysis is pessimistic because it assumes that the GPU budget is exhausted after the completion of each GPU segment and the GPU segment of the next task has to wait for the amount of $H^{gc}$. It is noteworthy that because of the  non-preemptive GPU resource, the GPU server budget has to be large enough that the GPU segment of any task is able to execute without interruption which leads to enormous remote blocking time due to a considerable data handover  delay. Let $\Gamma^g$ denote the set of GPU-using tasks in a taskset. For the lowest-priority GPU-using task, ${|\Gamma^g-1| \times (T^g-C^g)}$ is the amount of delay only from the GPU server budget replenishment of higher-priority GPU-using tasks. To mitigate this issue, we will present another design of the waiting queue for GPU requests in the later part of this section.

\subsection{Improvement}
\subsubsection{ \textbf{Miscellaneous Operation Time Reservation Policy}}
Recall our MOT reservation mechanism presented in Section~\ref{sec:chap2_Overrun}. When enabled, it ensures that there is always an enough amount of budget for miscellaneous operations (e.g. data copy from/to GPU); thus, the GPU does not have to wait until the start of the next CPU budget replenishment. 
%As discussed in Section ~\ref{sec:chap2_Overrun}, some budget portion is reserved on \new{the} CPU core server according to Eq.~\ref{eq:chap2_overrun} to stipulate enough budget for transferring data. 
Hence, the CPU-GPU handover delay with the MOT mechanism is
\begin{equation}
    H_i^{gc} = s_i  (T^g - C^g).
\end{equation}

The improvement in the handover delay has also a profound impact on the remote blocking delay by reducing the worst-case response time of a GPU segment. However, it is worth noting that for deferrable CPU servers, their budget needs to be halved as discussed earlier in Section~\ref{sec:chap2_server} to avoid the thermal back-to-back phenomenon. \begin{comment}In the next section, we will evaluate the influence of this option from different aspects on schedulability percentage. \end{comment}

\subsubsection{\textbf{Remote Blocking Enhancement}}
To address the problem of enormous remote blocking time due to CPU-GPU handover delay, we propose an alternative approach to the GPU waiting queue. This approach implements the queue based on a variant of the first-come first-served (FCFS) policy with a pre-defined bin-packing order. To be more precise, a bin-packing heuristic is employed to determine the number of bins, where the size of each bin is the GPU budget and the length of a GPU  segment is the size of an item to be packed. 
The total number of items in the bins is $|\Gamma^g|$ and the number of bins is related to the waiting time for a GPU segment. 
The reason for employing the FCFS policy is to avoid starvation of jobs with large period because under this policy, jobs with shorter period get serviced only once at any time.  
If a small-period job arrives meanwhile, it waits until the rest of waiting jobs get serviced and after finishing all other jobs, it gets serviced according to its position in the bins. Since in this approach all tasks have the same amount of waiting time, it leads to a significant reduction in the waiting time of low-priority GPU-using tasks but a moderate increase in that of high-priority ones. It is worth reminding that the replenishment period of the GPU server is typically much larger than that of the CPU server. The remote blocking time for a GPU-using task $\tau_i$ under the polling server policy for CPU cores is
\begin{equation}
B_{i}^{r} = s_i \left[ (|bins|+1) T^g + 2  (|\Gamma^g|-1)  T^c \right].
\end{equation}
  In this approach, missing activation points of the CPU polling server can still happen in the transferring time of data from/to the CPU server and due to the FCFS characteristics of the queue, the total amount is ${2  (|\Gamma^g|-1)  T^c}$. The remote blocking time of $\tau_i$ under the deferrable and sporadic CPU server policies without the MOT mechanism is
\begin{equation}  
  {B^r_{i} = s_i \left(|bins|+1\right) T^g + 2  (|\Gamma^g|-1) (T^c -C^c) }.
\end{equation}
  This is because in the worst case, the GPU server waits for the amount of CPU server jitter. The remote blocking time with the MOT mechanism is 
\begin{equation}  
  {B^r_{i} = s_i \left(|bins|+1\right) T^g}.
\end{equation}

 To this end, our framework takes a \textit{hybrid} scheduling scheme that chooses one of the proposed queue implementations which successfully passes the schedulability analysis. 
\section{Evaluation}
\label{sec:chap2_eval}
 This section gives the experimental evaluation of our framework. First, we explain our implementation on a real platform. Then, we explore the impact of proposed approaches on task schedulability with randomly-generated tasksets based on  practical parameters.
\subsection{Implementation}
We did our experiments on an ODroid-XU4 development board \cite{ODROIDXU4} equipped with a Samsung Exynos5422 SoC. There exist two different CPU clusters of little Cortex-A7 and  big Cortex-A15 cores, where each cluster consists of four homogeneous cores. There exists an integrated Mali-T628 GPU on the chip which supports OpenCL 1.1. Built-in sensors with sampling rate of 10 Hz with the precision of 1\textdegree~C are on each big CPU core and also the GPU to measure the temperature\footnote{There are no temperature sensors on little cores since the power consumption and heat generation of the little cluster is considerably low.}. The DTM throttles the frequency of the big CPU cluster to 900 MHz when one of its cores reaches the pre-defined maximum temperature. During experiments, the CPU fan is always either turned off or on at its maximum speed and the CPU is set to run at its maximum frequency. %The temperature threshold when there is no workload on the GPU is 95\textdegree~C and when the CPU fan is forced to off, we observed that the DTM throttles the frequency at 84\textdegree~C. 

We stressed the CPU cores of the big cluster and the GPU with different settings by executing \texttt{sgemm} program of the Mali SDK benchmark \cite{Mali} to measure the system parameters used in Section~\ref{sec:chap2_analysis}. We observed that without launching any kernel on the GPU, because of the heat conduction, the temperature on the GPU rises from 40\textdegree~C (the ambient temperature) to 70\textdegree~C. However, the GPU has less thermal effect on CPU cores due to the low heat dissipation. The kernel execution on the GPU in the presence of CPU workloads raises the CPU temperature by 5-10\textdegree~C. 

\begin{figure}[ht]
\centering
\begin{subfigure}{.45\textwidth}
  \centering
  % include first image
\includegraphics[width=\textwidth]{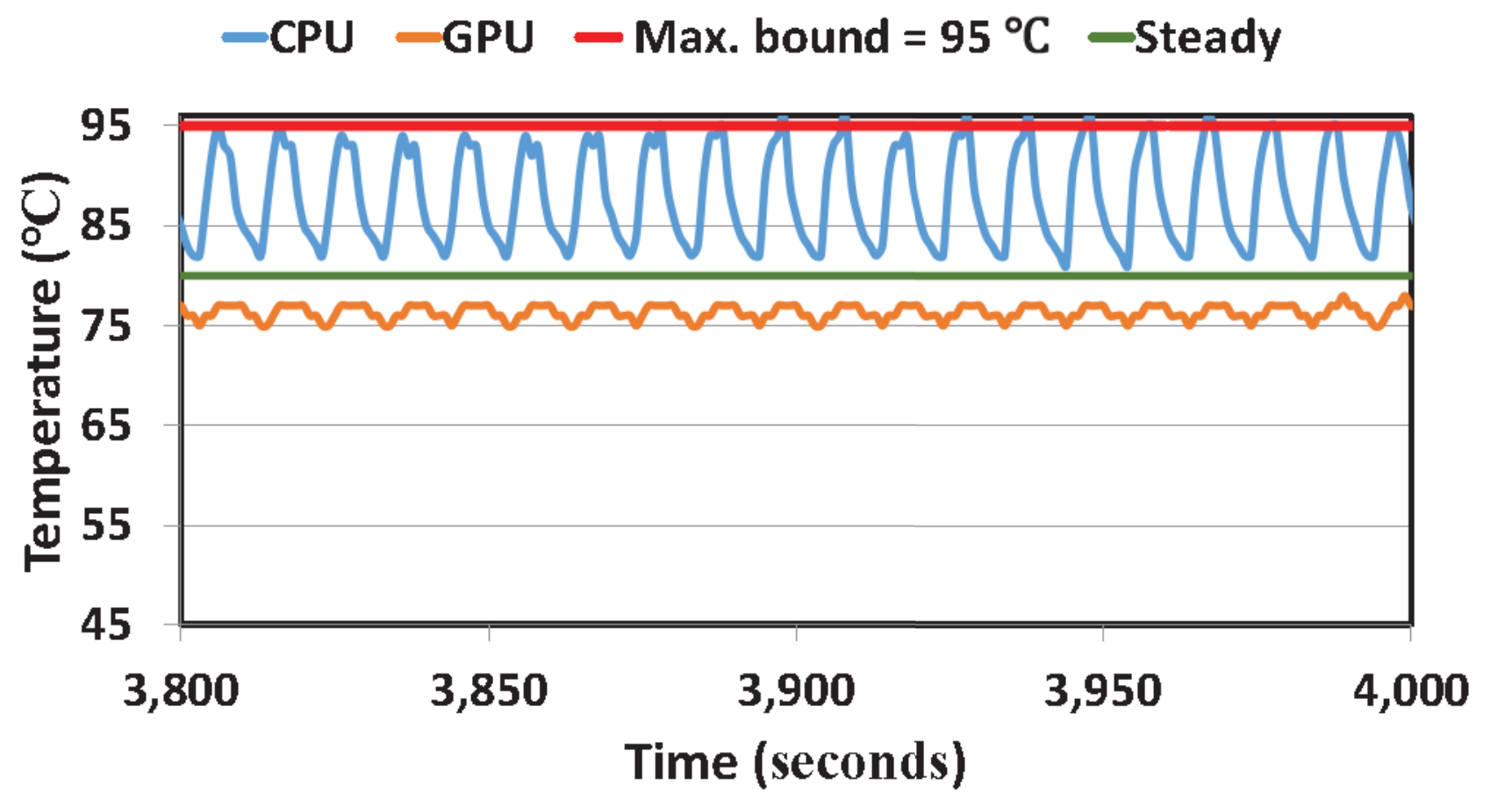}
  \caption{}
  \label{fig:chap2_boardA}
\end{subfigure}
\begin{subfigure}{.45\textwidth}
  \centering
  % include second image
\includegraphics[width=\textwidth]{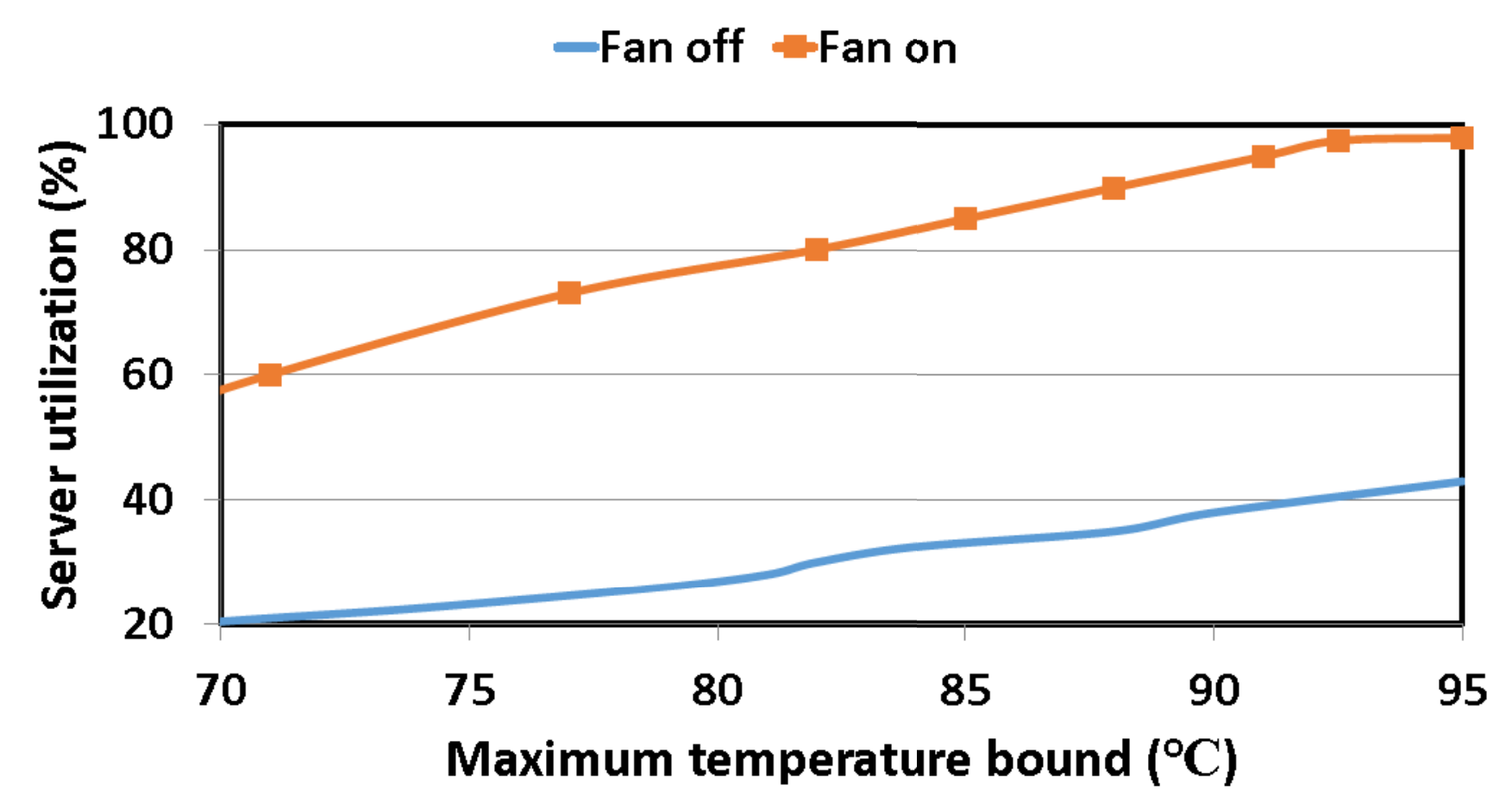}
  \caption{}
  \label{fig:chap2_boardB}
\end{subfigure}
\begin{subfigure}{.45\textwidth}
  \centering
  % include third image
\includegraphics[width=\textwidth]{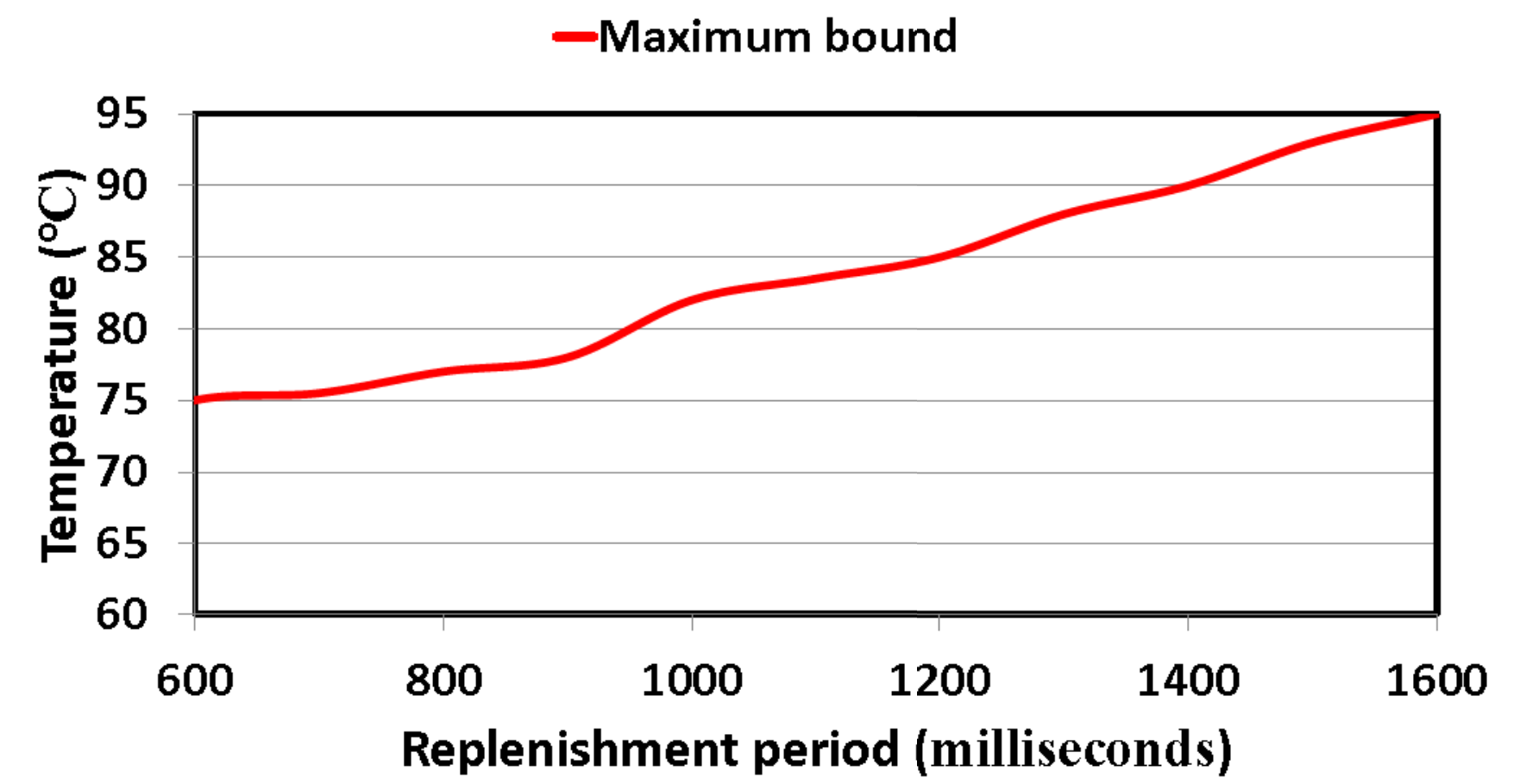}
  \caption{}
  \label{fig:chap2_boardC}
\end{subfigure}
\caption{a) Server design with the given maximum temperature of 95\textdegree C when the CPU fan is off.  b) Server utilization w.r.t the given maximum temperature. c) The maximum observed temperature w.r.t the server replenishment period when the CPU fan is off and CPU utilization is 30\%.}
\label{fig:chap2_board}
%\vspace{-10.00mm}
\end{figure}

Fig.~\ref{fig:chap2_board}a illustrates the result of the implementation of the CPU polling server with the replenishment period of 1 second\footnote{We conducted the experiment with large values of replenishment period because of the coarse granularity of the sampling rate of the on-board temperature sensors.} and the maximum temperature bound of 95\textdegree~C when the CPU fan is off. As one can see, the CPU temperature oscillates between 78\textdegree~C to 95\textdegree~C after a long time of the system operating in the steady state.

Figures~\ref{fig:chap2_board}b depicts the server utilization with respect to the maximum temperature bound. The maximum achievable utilization of the CPU server is only 43\% when the fan is off and almost 95\% when the fan is on. The gap in server utilization between these two cases (fan on/off) decreases as the value of the maximum temperature bound reduces. The steady state temperatures under different maximum temperature bounds remain almost the same regardless of whether the fan is on/off. 

With our proposed thermal-aware server design, it is possible to bound the operating temperature to any thermal constraint. It is worth noting that the same server utilization value can give different temperature bounds depending on the value of replenishment period used. Fig.~\ref{fig:chap2_board}d shows the temperature bounds at the invariant server utilization of 30\% and the replenishment period in the range of $[600, 1600]$ milliseconds. The length of 1600 milliseconds is the maximum feasible server replenishment period at the server utilization of 30\% that the board can reach when the fan is off. This is because with a larger period value, the CPU exceeds its maximum temperature bound during the active phase.

% \begin{figure}[t]
% \centering
% \subfloat [][]{\includegraphics[width=0.33\textwidth]{Chapter2/figs/results/board_serverutil.pdf}}
% \subfloat [][]{\includegraphics[width=0.33\textwidth]{Chapter2/figs/results/board_serverutilization.pdf}}
% \subfloat [][]{\includegraphics[width=0.33\textwidth]{Chapter2/figs/results/board_30period.pdf}}
% % \caption{a) Server design with the given maximum temperature of 95\textdegree~C when the CPU fan is off. b) Server utilization w.r.t the given maximum temperature. c) The maximum observed temperature w.r.t the server replenishment period when the CPU fan is off and CPU utilization is 30\%. }
% \label{fig:chap2_board}
% \end{figure}
 With these results, we have shown that it is possible to satisfy the maximum temperature constraint based on our proposed server design. Next, we will use the measured parameters in our analysis and discuss its effect in taskset schedulability.

\subsection{Schedulability Experiments}
\textbf{Task Generation.} We randomly generate 10,000 tasksets for each experimental setting. The base parameters given in Table \ref{tab:chap2_spec} and the measured parameters from the board are used for the taskset generation and the server design, respectively. It is worth noting that the GPU parameters are in compliance with the case study of prior work \cite{kato2011timegraph, kato2012gdev, 8046309, 7010477}. Server budgets are determined according to the maximum temperature need by applying the equations of Section~\ref{sec:chap2_server} and the measured system parameters. The number of tasks in each taskset is determined based of the uniform distribution in the range of [8, 20]. Then, the utilization of the taskset is partitioned randomly for these tasks in a way that no task has the utilization more than the CPU server utilization. The total WCET of each task (i.e., $C_i$) is calculated based on the task's utilization and its randomly-chosen period. If the task $\tau_i$ is a CPU-only task, the whole $C_i$ is assigned to $C_{i,1}$ otherwise $C_i$ is divided into $E_i$, $C_{i,1}$ and $C_{i,2}$, according to the random ratio of the GPU segment length to the normal WCET. In this phase, if $E_i$ is more than the GPU server budget, another random ratio is generated. Then, $E_i$ is partitioned randomly into the miscellaneous time ($M_{i,1}+M_{i,2}$) and the pure kernel execution time ($k_i$) according to the ratio of miscellaneous operations given in Table~\ref{tab:chap2_spec}. The accumulated miscellaneous-operation time is randomly divided into $M_{i,1}$ and $M_{i,2}$. Finally, tasks are assigned to CPU cores by using the worst-fit decreasing (WFD) heuristic for load balancing across cores. When the MOT reservation is used, the MOT budget is determined by the  maximum of $M_{i,1}$ and $M_{i,2}$ for all tasks. 
\begin{table}[h!]
\centering
 \caption{ Base parameters for taskset generation}
  \label{tab:chap2_spec}
 \begin{tabular}{c | c} 
 \hline
 Parameters & Values \\ [0.5ex] 
 \hline\hline
Number of CPU cores & 4 \\ 
Number of tasks & [8, 20]\\
Taskset utilization & [0.4, 1.6] \\
Task period and deadline & [30, 500] ms \\
Percentage of GPU-using tasks & [10, 30] \% \\
Ratio of GPU segment len. to normal WCET & [2, 3]:1  \\
Ratio of misc. operations in GPU segment $\frac{M_{i,1} + M_{i,2}}{E_i}$ & [10, 20]\% \\
Server Period & 10 ms \\
GPU Period &  20 ms \\
     [1ex] 
 \hline
 \end{tabular}
\end{table}

\begin{figure}[ht]
\centering
\begin{subfigure}{.45\textwidth}
  \centering
  % include first image
\includegraphics[width=\textwidth]{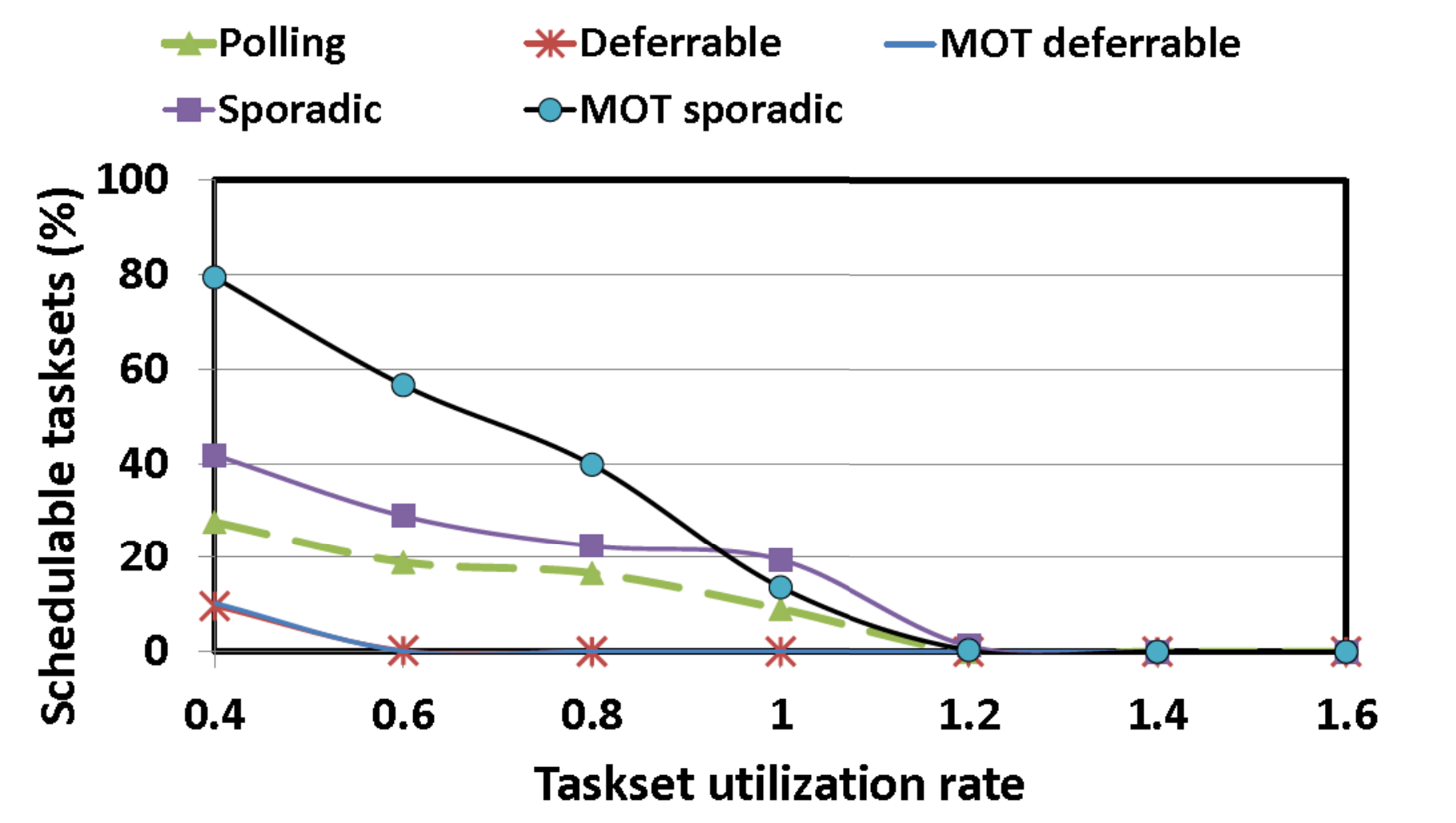}
  \caption{}
  \label{fig:chap2_analtasksetA}
\end{subfigure}
\begin{subfigure}{.45\textwidth}
  \centering
  % include second image
\includegraphics[width=\textwidth]{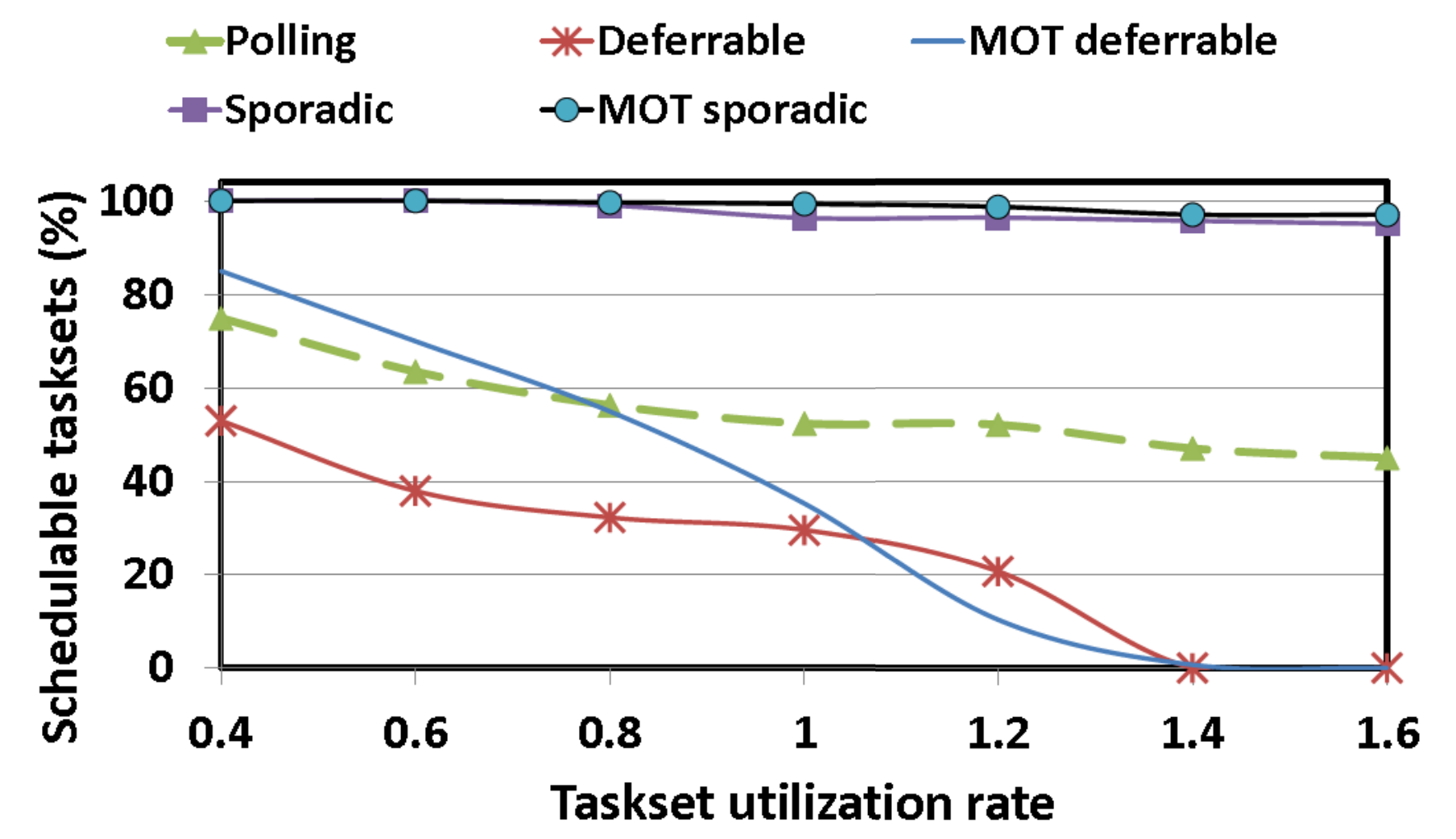}
  \caption{}
  \label{fig:chap2_analtasksetB}
\end{subfigure}
\newline
\begin{subfigure}{.45\textwidth}
  \centering
  % include third image
\includegraphics[width=\textwidth]{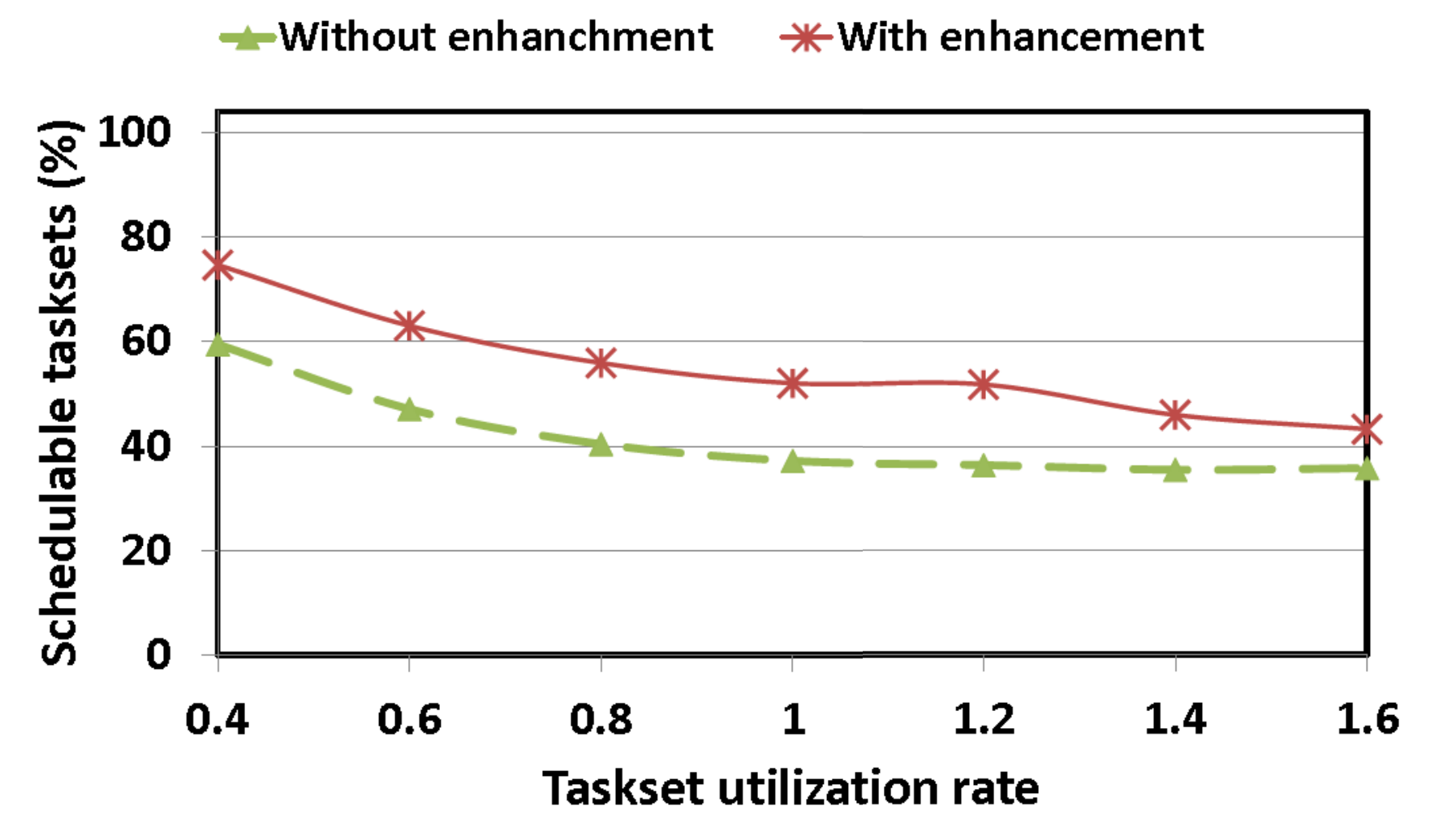}
  \caption{}
  \label{fig:chap2_analtasksetC}
\end{subfigure}
\begin{subfigure}{.45\textwidth}
  \centering
  % include fourth image
\includegraphics[width=\textwidth]{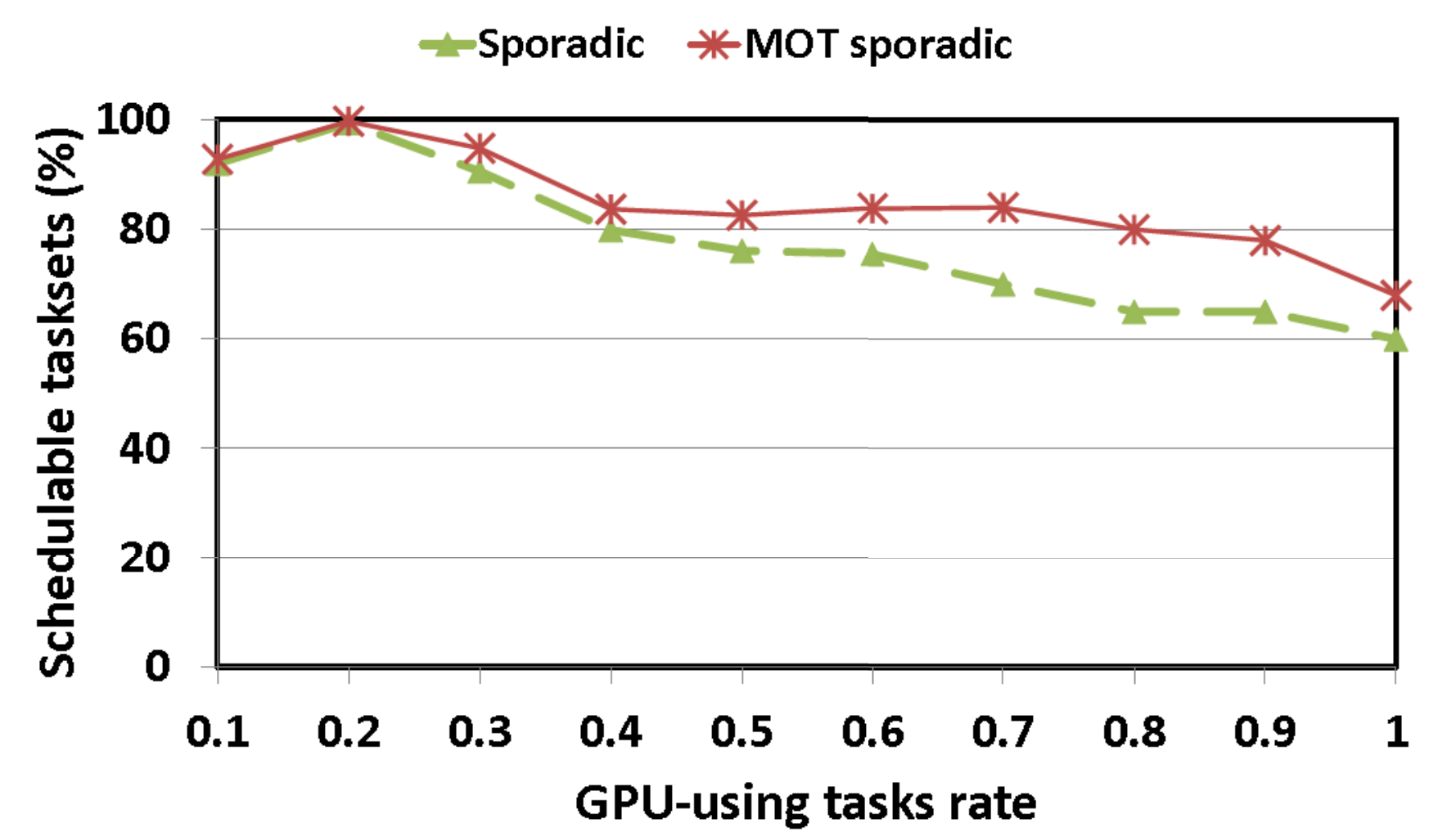}
  \caption{}
  \label{fig:chap2_analtasksetD}
\end{subfigure}
\newline
\begin{subfigure}{.5\textwidth}
  \centering
  % include third image
\includegraphics[width=\textwidth]{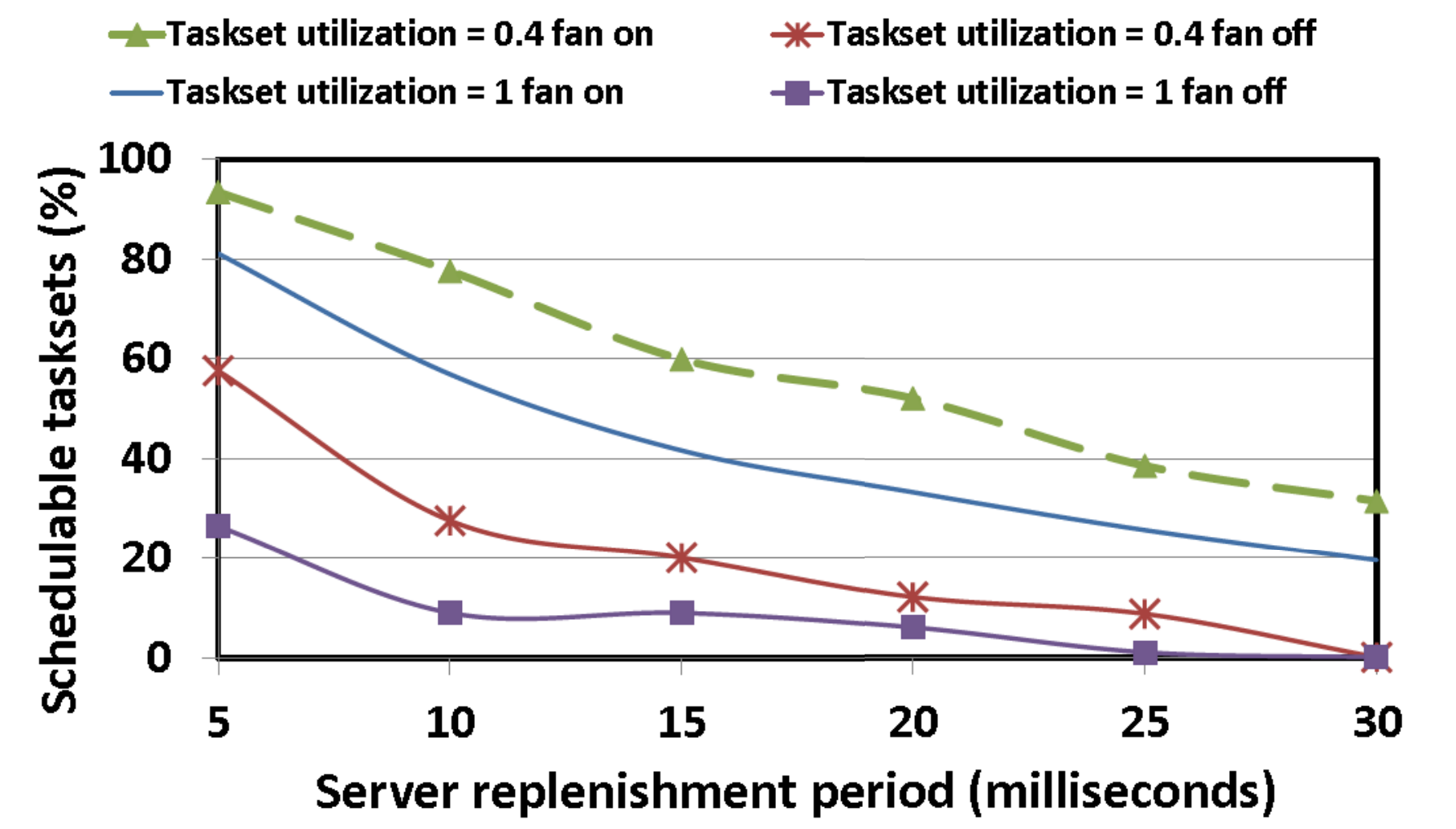}
  \caption{}
  \label{fig:chap2_analtasksetE}
\end{subfigure}
\caption{a) and b) The percentage of schedulable tasksets under the given maximum temperature constraint of 95\textdegree~C when the CPU fan is off and on, respectively. c) Taskset schedulability results with and without the remote blocking enhancement under the polling server policy while the CPU fan is off. d) The percentage of schedulable tasksets w.r.t the ratio of GPU-using tasks.}
\label{fig:chap2_analtaskset}
%\vspace{-10.00mm}
\end{figure}

\textbf{Results.}
Figures~\ref{fig:chap2_analtaskset}a-b depict the percentage of the schedulable tasksets when the CPU fan is on or off with different taskset utilization. The CPU sporadic servers outperform the other CPU server policies as expected because their replenishment budget are as large as the polling server and the remote blocking and CPU-GPU handover delays are as low as those in the deferrable server. Compared to the polling server, the deferrable server with MOT yields a higher percentage of schedulable tasksets especially when taskset utilization is low. Designating some portion of CPU server budget under the deferrable replenishment policy results in improvement in taskset schedulability. However, the percentage of schedulable taskset under the deferrable server drops sharply as taskset utilization increases because of the insufficient amount of server budget (which is just the half of the polling/sporadic server budget). It is worth noting that when the CPU fan is off, there is a large gap between the two sporadic servers with and without MOT at low taskset utilization. This is due to the advantage of the MOT mechanism that also reduces enhanced remote blocking delay.

\cmnt{the rate of schedulable tasksets under deferrable policy with/without overrun option is very low because the budget for polling server is drastically low; hence the budget for deferrable server which is half of its corresponding polling server is negligible. When temperature demands rises, the deferrable has more chance to schedule some tasksets but even in this condition their budgets are still low. }

% \begin{figure}[t]
% \centering
% \subfloat [][]{\includegraphics[width=0.3\textwidth]{Chapter2/figs/results/ANAL_off.pdf}}
% \subfloat [][]{\includegraphics[width=0.3\textwidth]{Chapter2/figs/results/ANAL_on.pdf}}
% \subfloat [][]{\includegraphics[width=0.3\textwidth]{Chapter2/figs/results/ANAL_enhance_polling.pdf}}\\
% \subfloat [][]{\includegraphics[width=0.3\textwidth]{Chapter2/figs/results/ANAL_Gtasks_sprodic.pdf}}
% \subfloat [][]{\includegraphics[width=0.3\textwidth]{Chapter2/figs/results/ANAL_period_poll.pdf}}
% %  \caption{a) and b) The percentage of schedulable tasksets under the given maximum temperature constraint of 95\textdegree~C when the CPU fan is off and on, respectively. c) Taskset schedulability results with and without the remote blocking enhancement under the polling server policy while the CPU fan is off. d) The percentage of schedulable tasksets w.r.t the ratio of GPU-using tasks. }
% \label{fig:chap2_analtaskset}
% \end{figure}

Fig.~\ref{fig:chap2_analtaskset}c shows the effect of the proposed remote blocking enhancement under the polling policy. As one can see, the remote blocking reduces substantially due to our proposed remote blocking enhancement by up to 20\% especially when there exists less workload in the system.  Fig.~\ref{fig:chap2_analtaskset}d depicts the impact of the rate of GPU-using tasks on schedulable taskset rate under the CPU sporadic server policy with and without MOT. As the number of GPU-using tasks increases, taskset schedulability goes decreases as more tasks contend for the shared GPU.

Next, we investigate the effect of server periods on the schedulability rate. In this experiment, the temperature constraint is fixed to the maximum level and the server period is varied from 5 to 30 milliseconds (see Fig.~\ref{fig:chap2_analtaskset}e) under the polling server policy. As expected, because of the large CPU-GPU handover delay, the percentage of schedulable tasksets drops significantly as the server replenishment period increases. 
\section{Case Study}
We have implemented a prototype of our framework and conducted a case study on ODroid-XU4. We show that without our framework, a real-time task experiences unexpectedly large delay due to the thermal violation, but with our framework, the operating temperature is safely bounded within a desired range.

In the case study, we used three types of applications: a real-time GPU-using task, and non-real-time CPU-only and GPU-using tasks. The non-real-time CPU-only application is run on each of the four big CPU cores with the lowest priority. The non-real-time GPU-using matrix multiplication task is run on one big CPU core with the medium priority. This task randomly generates two matrices and performs the matrix multiplication repetitively using the GPU-accelerated Mali OpenCL library. The size of matrix is set to $512\times512$ in order not to cause unnecessarily long waiting time to the real-time task. On the other big CPU core, the highway workzone recognition application for autonomous driving~\cite{lee2013kernel} is run as the real-time GPU-using task with the highest priority. This task is configured to process 8 frames per second and has one GPU segment based on OpenCL. A video consisting of around 800 frames is given as input to this task. To avoid unexpected delay in data fetching, video frames are preloaded into memory as a vector during the initialization of the task and the loaded frames are repeatedly processed at runtime. After the initialization, all tasks are signaled to start their execution together and the CPU fan is turned off during the experiment. Other tasks in the system, including system maintenance and monitoring processes, are assigned to the little cluster cores. The thermal-aware servers are implemented based on the polling server policy with the budget replenishment period of 10 milliseconds. 

\begin{figure}[ht]
\centering
\begin{subfigure}{.48\textwidth}
  \centering
  % include first image
\includegraphics[width=\textwidth]{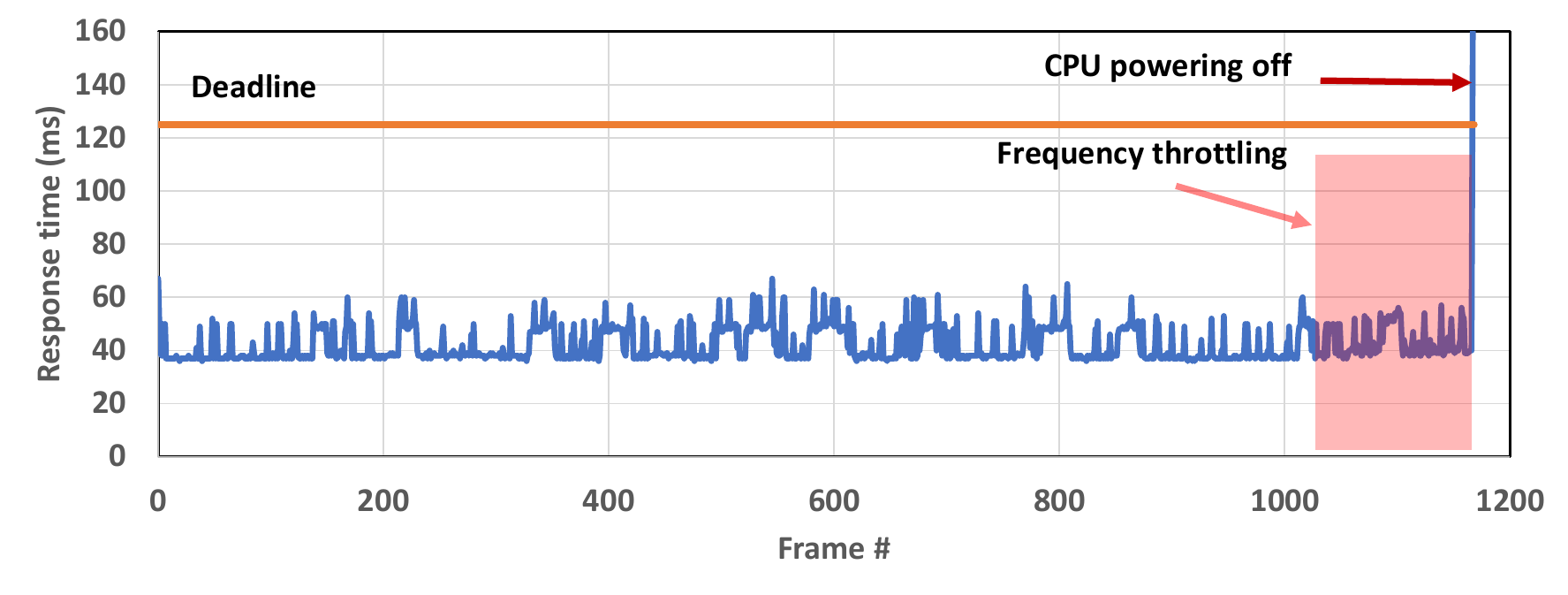}
  \caption{}
  \label{fig:chap2_casestudyA}
\end{subfigure}
\begin{subfigure}{.48\textwidth}
  \centering
  % include second image
\includegraphics[width=\textwidth]{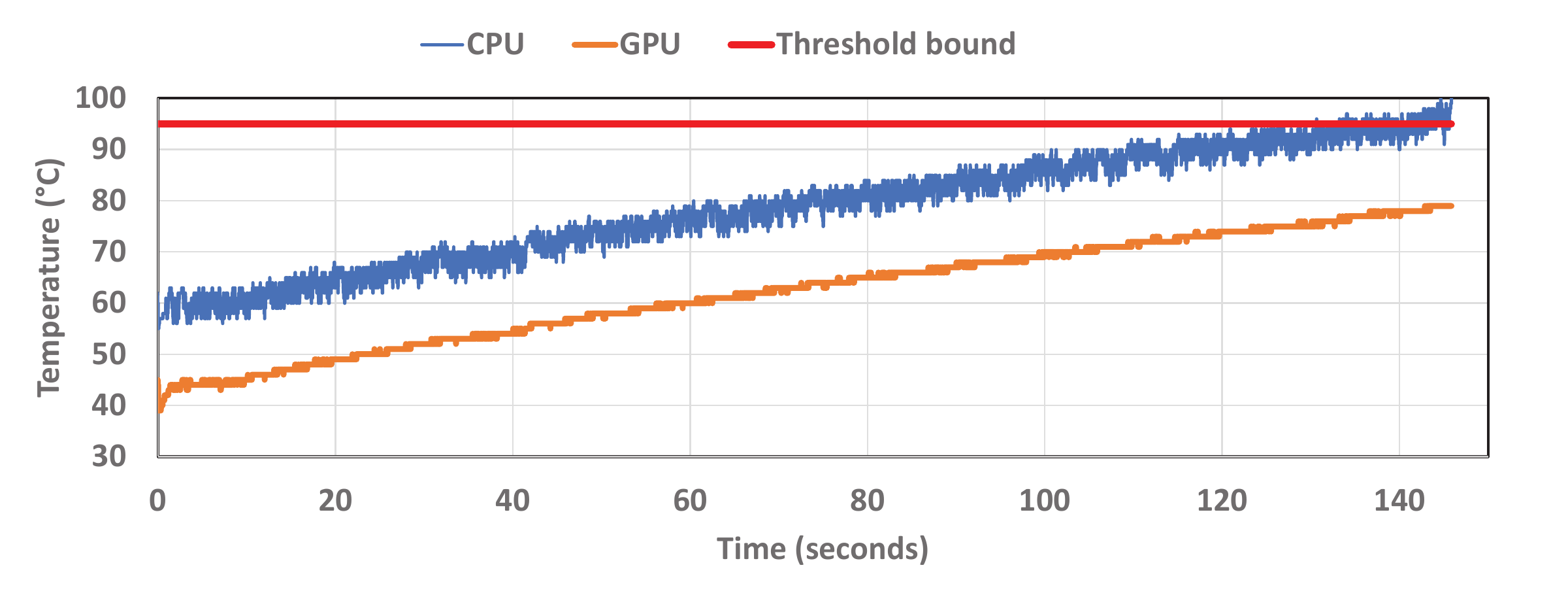}
  \caption{}
  \label{fig:chap2_casestudyB}
\end{subfigure}
\newline
\begin{subfigure}{.48\textwidth}
  \centering
  % include third image
\includegraphics[width=\textwidth]{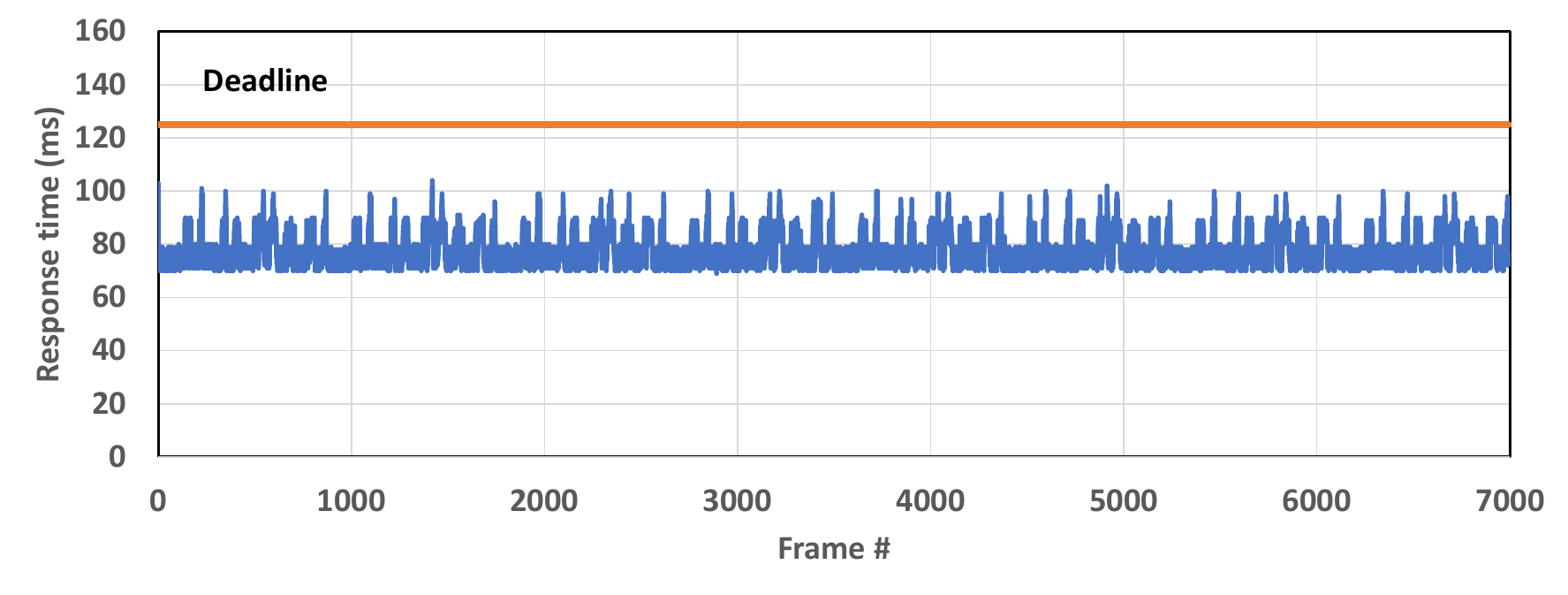}
  \caption{}
  \label{fig:chap2_casestudyC}
\end{subfigure}
\begin{subfigure}{.48\textwidth}
  \centering
  % include fourth image
\includegraphics[width=\textwidth]{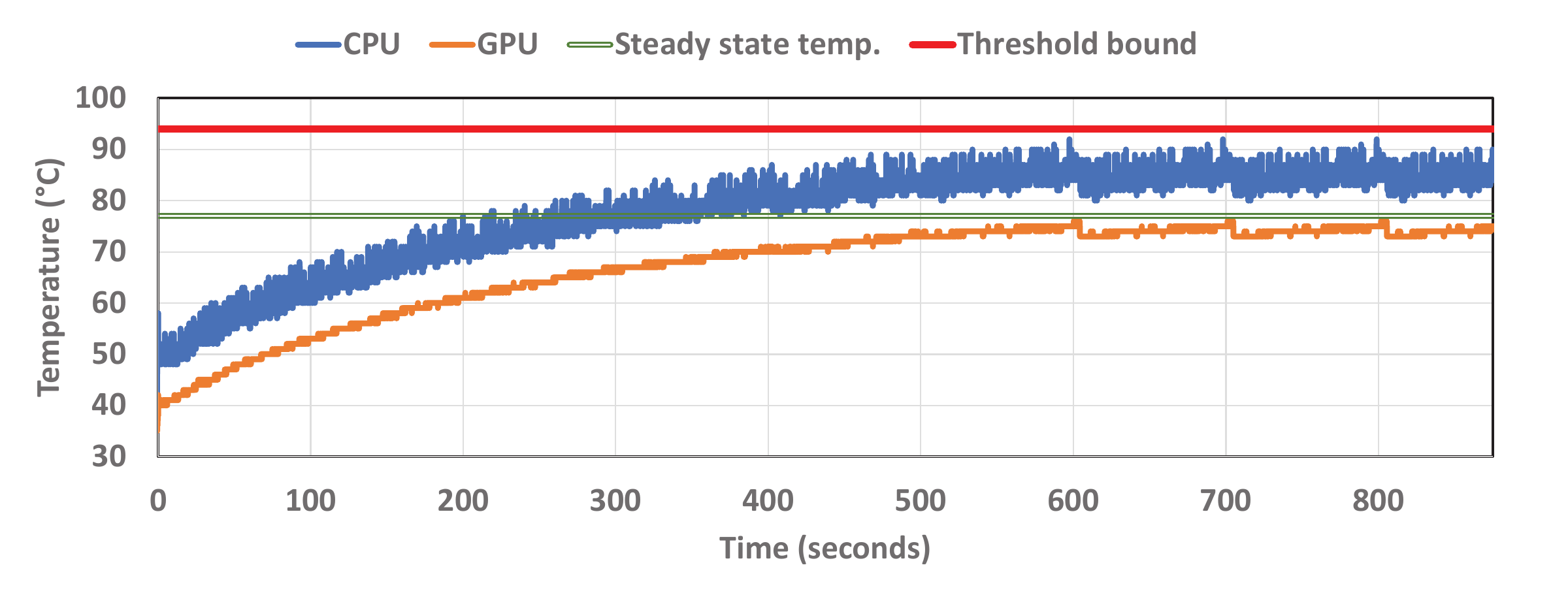}
  \caption{}
  \label{fig:chap2_casestudyD}
\end{subfigure}
\caption{a) Response time of the real-time workzone recognition task without our framework. b) Temperature of CPU and GPU without our framework. c) Response time of the real-time task with our framework d) Temperature of CPU and GPU with our framework.}
\label{fig:chap2_casestudy}
%\vspace{-10.00mm}
\end{figure}

Two scenarios are used to show the effectiveness of our proposed framework. The first one is the baseline system with no thermal-aware server. Fig.~\ref{fig:chap2_casestudy}a shows the response time of each frame (job) of the real-time workzone recognition task, and Fig.~\ref{fig:chap2_casestudy}b depicts the temperature measurements during the experiment. As one can see, after processing around 1040 frames, the DTM was triggered and it started throttling the CPU frequency from 2.0 GHz to 900 MHz since the CPU temperature had reached the threshold of 95\textdegree~C. However, the frequency throttling did prevent the operating temperature from rising on both the CPU and the GPU, which caused the OS to power off the big CPU cluster temporarily. This experiment took less than three minutes. In the second scenario, all tasks execute within the thermal-aware servers. As illustrated in Fig.~\ref{fig:chap2_casestudy}c, the observed response time of each frame was larger compared to the first scenario due to the budget of servers, but all frames were processed before the deadline. Fig.~\ref{fig:chap2_casestudy}d depicts that it took around 500 seconds to reach the steady state temperature. Moreover, the operating temperature was tightly bounded by the thermal threshold in any circumstance. This result shows the thermal safety and accuracy of our framework in practical settings.

% \begin{figure}[t]
% \centering
% \subfloat [][]{\includegraphics[width=0.43\textwidth]{Chapter2/figs/casestudy/cs_time100.pdf}}
% \subfloat [][]{\includegraphics[width=0.43\textwidth]{Chapter2/figs/casestudy/cs_temp100.pdf}}\\
% \subfloat [][]{\includegraphics[width=0.43\textwidth]{Chapter2/figs/casestudy/cs_time033.pdf}}
% \subfloat [][]{\includegraphics[width=0.43\textwidth]{Chapter2/figs/casestudy/cs_temp033.pdf}}
% % \caption{a) Response time of the real-time workzone recognition task without our framework. b) Temperature of CPU and GPU without our framework. c) Response time of the real-time task with our framework d) Temperature of CPU and GPU with our framework. }
% \label{fig:chap2_casestudy}
% \end{figure}

\section{Summary}

In this chapter, we proposed a novel thermal-aware framework to bound the maximum temperature of CPU cores as well as an integrated GPU while guaranteeing real-time schedulability. Our framework supports various server policies and provides analytical foundations to check both thermal and temporal safety. Experimental results show that each CPU server policy provided by our framework is effective in bounding the maximum temperature. We proposed the miscellaneous operation time reservation mechanism for the CPU servers in order to improve task schedulability by reducing the CPU-GPU handover delay. We also introduced a remote blocking enhancement technique that employs the bin-packing strategy to reduce the remote blocking caused by other tasks.

\chapter{On Dynamic Thermal Conditions in Mixed-Criticality Systems}
%\boldmath
The rising demand for powerful embedded systems to support modern complex real-time applications signifies the on-chip temperature challenges. 
 Heat conduction between CPU cores interferes in the execution time of tasks running on other cores. The violation of thermal constraints causes timing unpredictability to real-time tasks due to transient performance degradation or permanent system failure. Moreover, dynamic ambient temperature affects the operating temperature on multi-core systems significantly. 
 
  In this chapter, we propose a thermal-aware server framework to safely upper-bound the maximum operating temperature of multi-core mixed-criticality systems. With the proposed analysis on the impact of ambient temperature, our framework manages mixed-criticality tasks to satisfy both thermal and timing requirements. We present techniques to find the maximum ambient temperature for each criticality level to guarantee the safe operating temperature bound. We also analyze the minimum time required for a criticality mode change from one level to another. The thermal properties of our framework have been evaluated on a commercial embedded platform. A case study with real-world mixed-critical applications demonstrates the effectiveness of our framework in bounding operating temperature under dynamic ambient temperature changes.

% \begin{IEEEkeywords}
% IEEEtrtn, journal, \LaTeX, chapter, template.
% \end{IEEEkeywords}

% \IEEEpeerreviewmaketitle
\section{Introduction}

Ensuring continuous operation with high assurance in the physical environment 
remains a significant challenge to cyber-physical systems (CPS). This is particularly important for safety-critical applications with real-time mixed-criticality components, e.g., automotive, aerospace, manufacturing, and defense systems, where even occasional timing failures of high-critical\new{ity} components can lead to catastrophic consequences. Various types of unexpected changes in the physical environment may affect the system behavior and contribute to the difficulty of this problem. 

Ambient temperature is one of the key factors that affect many mixed-criticality CPS applications. For instance, in automobiles, a report from the National Renewable Energy Laboratory of the US Department of Energy~\cite{farrington2000impact} indicates that cabin air temperature can reach up to and 82$^{\circ}$C in Phoenix, Arizona. \new{The heat generated by the engine worsens the ambient temperature level of nearby electronic control units~\cite{park2010dynamic}.}
\cmnt{The higher ambient temperature due to the affinity to running car engines exacerbates the operating temperature of embedded electronic control units (ECUs)~\cite{park2010dynamic}.}
\cmnt{For the safe operation of automotive computing systems, air conditioning (AC) is required but it cannot always guarantee constant ambient temperature in diverse driving scenarios. Excessive use of AC can also increase carbon monoxide (CO) and nitrogen oxide (NOx) emissions by 71\% and 81\%, respectively, and reduce fuel economy by 22\%~\cite{bevilacqua1999effect}.}
Another example is a fire-containment drone~\cite{toribio2016fire}. Even with a heat protection shield, the drone's computing system starts warning when the ambient temperature reaches 35$^{\circ}$C and becomes nonoperational at 40$^{\circ}$C. This also limits the minimum distance from the drone to a fire hazard.

While dynamic ambient temperature is an important problem, most thermal management schemes for real-time embedded systems~\cite{kumar2011cool,kumar2011system, ahmed2011minimizing,Huang2014,DSouza2017ThermalIO,Ahmed2017} assume fixed, room-level ambient temperature and focus only on the temperature increase caused by the computing system itself. CPS are expected to run in various physical environments; hence, the assumption of room temperature made by prior work limits their practical applicability. There is recent work~\cite{Youngmoon2018} considering dynamic ambient temperature but it assumes a uni-processor single-criticality system.

Under harsh ambient temperature, a system may not be able to utilize 100\% of CPU time even if the CPU runs at the minimum possible frequency with active/passive cooling packages.
The operating temperature of the CPU may still reach the maximum thermal constraint, resulting in temporary system shutdown or permanent hardware damage.
In such a condition, the only option left to ensure timing and thermal guarantees of more critical tasks is to secure cooling time by suspending less critical tasks.
In other words, although DVFS~\cite{fu2010feedback,ma2015improving, 7372657, 7746768, zhang2010thermal} and cooling packages~\cite{chantem2010temperature, elghool2017review, ghahremannezhad2018thermal, ghahremannezhad2019effect} can help tolerate high ambient temperature, it is inevitable to consider only partial operations of the system.

In this work, we aim to design a system that offers different levels of assurance against ambient temperature changes. This is different from the well-known Vestal model~\cite{vestal2007preemptive}, which focuses on varying assurance of execution time, but it shares the spirit of addressing uncertainties in real-time system design. To avoid confusion, we clarify our mixed-criticality model as follows:

% \begin{defn}
% {\bf Thermally mixed-criticality systems} are a type of systems in which lower critical tasks will be suspended in high critical mode to guarantee that the operating temperature of the CPU due to heat dissipation of task execution and also thermal condition of the environment does not exceed the maximum temperature constraint.
% \end{defn}

\begin{defn}
{\bf Thermally mixed-criticality systems} are the systems that assure ambient temperature changes and heat dissipation from lower-criticality task execution do not adversely affect the real-time schedulability of higher-criticality tasks.
\end{defn}

In the thermally mixed-criticality model, ambient temperature plays a key role in determining the maximum amount of workload that can be executed on the CPU. The following questions still remain unanswered by the  state-of-the-art:
\begin{itemize}
	\item  Up to what ambient temperature is the system fully or partially operational? Specifically, for a given criticality level of a mixed-criticality system, can we find the corresponding {\it critical ambient temperature}, under which real-time tasks with that or higher criticality are guaranteed to meet their deadline at the expense of lower-criticality tasks?
	\item  Can we take into account the effect of dynamic ambient temperature along with heat conduction on a modern multi-core processor? 
	\item  If the system moves from a hot to a cold region, how long will it take for the system to cool down and safely resume the operation of low-critical\new{ity} tasks, without violating the processor thermal constraint?
\end{itemize}

This chapter presents a multi-core mixed-criticality scheduling framework with ambient temperature awareness. In our proposed framework, thermal-aware servers are used to bound heat generation at each criticality level and the criticality mode change is triggered by ambient temperature changes.
This is the first work to address the aforementioned limitations and provides analytical guarantees on the timely execution of critical components in dynamic thermal conditions. 

\smallskip\noindent\textbf{Contributions.} The contributions of this chapter are as follows:
\begin{itemize}
%\item We introduce an ambient temperature-aware framework for mixed-criticality multi-core systems, where thermal-aware servers are used to bound heat generation and a criticality mode change is triggered by ambient temperature.
\item \new{We show that the problem of thermal-aware real-time scheduling can be decomposed into thermal schedulability (how much CPU budget is usable under thermal constraints) and timing schedulability (if tasks are schedulable using given budget). Our thermal schedulability achieves the simplicity in timing analysis by ensuring that the budget is guaranteed to be made available for any execution patterns without violating thermal constraints.}

\item We extensively analyze the thermal safety of a multi-core system and bound the maximum operating temperature that the system can reach. At a specific ambient temperature level, we characterize the worst-case thermal behavior of a system \cmnt{by  using the notion of idle servers} and also determine the minimum time for the system to transition from one criticality level to a lower level.
\item \new{We introduce the notion of \textit{idle thermal servers} that allow bounding the maximum operating temperature caused by multiple preemptive active servers scheduled dynamically on a multi-core processor for a given  mixed-criticality taskset.} 
% \item \new{Introducing the notion of \textit{idle servers}}, \chk{we provide a mechanism  to find the maximum ambient temperature for each criticality level}  that the system can tolerate with multiple preemptive periodic servers scheduled dynamically on a multi-core processor for a given  mixed-criticality taskset. 
\item We perform a case study on mixed-criticality applications running on an ODROID-XU4 embedded platform, and evaluate our framework and analysis in different ambient temperature levels.
\end{itemize}

\section{Related Work}
\label{sec:chap3_rel}

% dynamic ambient temperature in non-real time papers
There exist extensive studies on bounding the maximum operating temperature in non-real-time multi-core systems \cite{fu2012feedback,fu2010feedback,ma2015improving, 7372657, 7746768, zhang2010thermal}, most of which propose adjusting CPU clock speed. The authors of~\cite{fu2012feedback,fu2010feedback} proposed a feedback loop that dynamically controls processor temperature by either adapting CPU utilization or scaling frequency to satisfy thermal constraints in varying ambient temperature environment. In~\cite{7372657} and~\cite{7746768}, the authors proposed proactive frequency scheduling to improve overall system performance under different ambient temperatures. Although average-case performance degradation has been discussed in these papers, they cannot be directly applied to our problem. 

%consideration of cold tasks to cool down.
The notion of \textit{hot} and \textit{cold} tasks has been introduced  ~\cite{chantem2010temperature, Huang2014, Youngmoon2018, kumar2011system, jayaseelan2008temperature,   zhang2010thermal} in both real-time and non-real-time systems. They propose scheduling algorithms to interleave hot and cold tasks, adjust the CPU frequency, and force idling time for the CPU to cool down after running hot tasks. In most of them, the scheduling problem is either to find task execution order or to reduce the size of lengthy hot tasks~\cite{Huang2014} in a fixed schedule. However, DVFS may cause a considerable reduction in system reliability over time~\cite{iranfar2017thespot, lasance2003thermally,  xiang2010system} and may be unsupported in some embedded devices.

There exists extensive research on real-time uni-processor systems with stringent thermal constraints~\cite{chen2009proactive, wang2006delay, chen2007minimization, kumar2011cool,ahmed2014temperature, ahmed2013thermal, ahmed2011minimizing, Huang2014, Youngmoon2018, jayaseelan2008temperature}. In uni-processor systems, thermal dependency between workloads is only  \textit{temporal}. However, in multi-core systems, because of heat dissipation between cores, there also exists  \textit{spatial} thermal dependency between the execution of workload on one core and those on other cores. Due to this reason, the work on uni-processor systems cannot be used safely in multi-core real-time systems.

The notion of thermal servers (either by injecting of idle tasks or using thermal-aware servers explicitly) have been proposed in the literature of real-time systems for  uni-processors~\cite{kumar2011cool,kumar2011system, ahmed2011minimizing,Huang2014} and multi-core platforms~\cite{DSouza2017ThermalIO,Ahmed2017}. In particular, the authors of~\cite{DSouza2017ThermalIO} introduced  a novel technique for periodic tasks executing on multi-core platforms. This technique introduces an Energy Saving (ES) task that runs with the highest priority and captures the sleeping time of CPU cores. The technique can be seen as an alternative to a thermal-aware server because the ES task effectively models the budget-depleted duration of a thermal server. The authors of~\cite{Ahmed2017} proposed thermal-isolation servers that avoid the thermal interference among tasks in temporal and spatial domains with thermal composability. Despite their achievements in isolating the thermal-aware design from real-time schedulability analysis, these techniques assume a fixed schedule for idle task or periodic servers, and are inapplicable to dynamically-scheduled (e.g., priority-based) servers in multi-core platforms. Since priority-based preemptive servers are widely used in many real-time systems, such as real-time virtualization~\cite{kim2017predictable,Xi_EMSOFT11}, the restriction imposed by these prior thermal servers is a significant limiting factor.

Matrix exponential is a well-known approach to solve the first-order linear system of differential equation. In~\cite{pagani2015matex}, the authors proposed a technique based on the numerical  Newton–Raphson method to solve the thermal equations in the steady state for each power change. 
Based on this, the work in~\cite{rai2011worst, yang2013real, kumar2011thermally, chantem2009online} finds the worst-case peak temperature by exhaustively searching all possible patterns.
Unlike prior work, 
the unique contribution of our work is that 
we solve the temperature equation for oscillating power signals analytically, by representing them as continuous functions, and analyze the worst-case peak temperature directly for multi-core platforms.
It is worth noting that our work not only proves the worst case for peak temperature but also reduces  computational complexity considerably.

%One of the main contributions of this paper is that in our proposed method, we solve the temperature equation for oscillating power signals analytically, with the advantage of representing them as continuous functions in the steady state. Unlike~\cite{rai2011worst, yang2013real, kumar2011thermally, chantem2009online} which try to find the worst-case peak temperature for a set of arrival patterns by tracing all patterns, we directly analyze the worst-case peak temperature for thermal-aware preemptive servers in multi-core platforms. It is worth noting that due to the use of thermal-aware servers, the computational complexity of our analysis is considerably lower than that of the state-of-art.

% Thermal behavior in multi-core systems when changing thermal behavior can be found in~\cite{pagani2015seboost}. The worst-case peak temperature for a set of arrival patterns modeled by an event arrival curve can be found in~\cite{rai2011worst, yang2013real}. The pattern that has the best/worst thermal behavior for a sequence of on/off (or two speeds) can be found in~\cite{kumar2011thermally, chantem2009online}. The worst-case peak temperature for MPSoC can be calculated in~\cite{schor2012worst, schor2012fast}.

There exists some research focusing on varying ambient temperature in real-time systems~\cite{Youngmoon2018, Huang2014, wang2006delay}. However, there is no discussion in these studies about mixed-criticality systems and the effect of harsh ambient temperature change on the schedulability of critical tasks. 

To the best of our knowledge, there is no prior work on multi-core mixed-criticality systems with the consideration of the effect of ambient temperature change.

% % \input{prelim.tex}
 \section{System Model}
\label{sec:chap3_model}
\subsection{Power Model}\label{sec:chap3_power_model} The total power consumption of CMOS circuits is modeled as the summation of dynamic and static powers~\cite{4484694}, i.e., ${P(t) = P_S(t) + P_D(t)}$. Static power $P_S$ depends on the semiconductor technology and the operating temperature caused by current leakage. 
%Therefore, static power presents when a power resource is switched on. Static power can be modeled as a linear function of temperature. 
Hence, it can be modeled as: $P_S(t) = k_1 \theta(t) + k_2 $, where $k_1$ and $k_2$ are technology-dependent system constants, and $\theta(t)$ is the operating temperature \cite{4212027}. Dynamic power $P_D(t)$ is the amount of power consumption due to the processor operating frequency $f$, modeled as ${P_D = k_0 f^s}$, where $s$ and $k_0$ are system constants that depend on the semiconductor technology.
% It is worth noting that in this chapter, our framework runs processor cores at a fixed operating frequency to obtain deterministic worst-case task execution time. Consequently, the total power is the function of temperature at any time instant $t$ because the other factors such as frequency remain invariant during execution. 

\subsection{Thermal-Aware Mixed-Criticality Servers}
We consider multiple preemptive thermal-aware servers for each CPU core. Each server is statically associated with one core and does not migrate to another core at run-time. A server $v_i$ is modeled as $v_i = (C_i,T_i,L_i)$, where $C_i$ is the maximum execution budget of the server $v_i$, $T_i$ is its budget replenishment period, and $L_i$ is its criticality level. Servers are ordered in an increasing order of priorities, i.e. $i<j$ implies that a server $v_i$ has lower priority than a $v_j$. At the criticality level of $l$, only the servers with criticality level higher than or equal to $l$ (i.e., $L_i \geq l$) are eligible to execute.

For budget replenishment policies, we consider \textit{polling}~\cite{sha1986solutions}, \textit{deferrable}~\cite{strosnider1995deferrable}, and \textit{sporadic servers}~\cite{sprunt1989aperiodic}. 
% Under the polling server policy, the corresponding server activates periodically and executes ready tasks until its budget is depleted. The budget is fully replenished at the start of the next period. If there is no task ready to execute, the remaining budget is immediately depleted. In contrast, under the deferrable server, any unused budget is preserved until the end of the period. Hence, a task can execute at any time during the server period while the budget is available. The sporadic server $v_i$ also preserves remaining budget, but replenishes the budget sporadically; only the amount of budget consumed is replenished after $T_i$ time units from the time when that budget is used. 
Let $J_i$ denote the task release jitter relative to the server release. The value of  $J_i$ is $T_i$ under the polling server policy and $T_i - C_i$ under the deferrable and sporadic server policies~\cite{bernat1999new}.

\subsection{Task Model}
This work considers sporadic tasks with implicit deadlines under \textit{partitioned fixed-priority preemptive scheduling}, which is widely used in many practical systems. Each task $\tau_i$ is statically allocated to one thermal server (thus to the corresponding CPU core of that server) with a unique priority. In each server, tasks are labeled in an increasing order of priority, i.e., $i<j$ implies $\tau_i$ has lower priority than $\tau_j$. There also exist non-real-time tasks running with the lowest priority level in the server and they execute only if there is no real-time task ready to execute and the server has remaining budget.
A task $\tau_i$ is modeled as $\tau_i=(E_i, D_i)$ where $E_{i}$ is the worst-case execution time (WCET) of task $\tau_i$, $D_i$ denotes the minimum inter-arrival time of $\tau_i$ which is implicit deadline of $\tau_i$. %We will also use the following notation:
% \begin{itemize}
%     \item $V(\tau_i)$: the server assigned to a task $\tau_i$
%     \item $P(v_i)$: the CPU core assigned to a server $v_i$.
% \end{itemize}
\cmnt{It is noteworthy that the criticality level of each real-time task $\tau_i$ is at least lower than that of its allocated server $v_j$ (i.e., $CR_i \leq L_j$)\cmnt{ whereas non-real-time tasks can always execute under any criticality level if $v_j$ has enough budget and no real-time task runs on the server.} Therefore, at the criticality level of $l$, in a server $v_j$ whose criticality level is at least $l$, real-time tasks whose criticality level is at least $l$ execute.} 
\new{It is worth mentioning that the simplicity in timing schedulability achieved by our work enables easy adoption of more complex task models. For instance, the analysis for tasks with critical sections under hierarchical scheduling~\cite{7010477,hosseinimotlagh2019thermal} is directly applicable to our work since we ensure periodic resource budget.}

\subsection {Criticality Model}
In this work, there exists a set of $m$ criticality levels ${L=\{l_1,l_2,\dots, l_m\}}$. At criticality mode $l$, only the servers (and tasks within these servers) with criticality level higher than or equal to $l$ are eligible to execute. 
%Those with lower-criticality level are suspended and resume only when the system's criticality mode becomes equal to or less than their criticality level.
Thus, for each criticality level $l$, there exists a subset of servers $V^l \subseteq V$ and a subset of taskset $\Gamma^l \subseteq \Gamma$ that execute.

\begin{defn}
{\bf Critical ambient temperature} of a  criticality level $l$ is the maximum ambient temperature that the  system  can  execute  eligible  servers $v \in V^l$ without  violating  the system's  maximum  temperature  constraint.
\end{defn}
%Each criticality mode is associated with \textbf{{\em critical ambient temperature}}, which is the maximum ambient temperature that the system can execute eligible servers without violating the system's maximum temperature constraint. 
\noindent Details on how criticality mode changes is given in Sec.~\ref{sec:chap3_frame}.
%For instance, if ambient temperature becomes greater than the critical ambient temperature of the current mode $l$, then the mode is changed to $l+1$. If ambient temperature goes below the critical ambient temperature of one lower criticality level $l-1$, the mode is changed to $l-1$ only after {\em shifting time} $t_{shift}$ which we analyze in Sec.~\ref{sec:chap3_thermal_analysis}. 

%add the goal of servers preemption thing and search for thermally scheduablity settiing
% \smallskip
% \noindent\textbf{Goals.} In this work, one of our goals is to develop a framework that analytically determines the value of critical ambient temperature levels corresponding to criticality levels and schedules a mixed-criticality real-time taskset in multi-core systems with thermal constraints. Our another goal is to analyze {\em shifting time} which is the minimum time the system has to wait for a thermally-safe transition from the steady state of one criticality mode to that of one lower criticality mode. 

\section{Framework Design}
\label{sec:chap3_frame}
This section presents the overall design-time and run-time aspects of our framework and how the criticality mode changes. 
%We also present the assumptions made for this work.
With our framework, all tasks in a system run within thermal-aware servers. A criticality mode change is triggered by ambient temperature, which can be obtained from an off-chip temperature sensor.
%\footnote{A light-weight system service can capture ambient temperature from a temperature sensor and send the result to our framework.} 
If the criticality mode changes from a lower criticality level to a higher one, i.e., the critical ambient temperature has been reached, the framework terminates the lower-criticality servers and their tasks immediately. 
%It is worth noting that in the transient phase from a lower criticality level to a higher one, only the running lower-criticality task and its corresponding server are terminated. Therefore, if the running task on the server has the higher criticality level, it continues executing in the transient phase whereas the running task with lower-criticality terminates during its execution. 
This design guarantees that the lower-criticality tasks have no effect on the  \textit{thermal} and \textit{timing} schedulability of tasks running in servers with higher criticality levels. The timing schedulability of tasks refers to the ability to complete their execution by the deadline. Thermal schedulability is defined as follows. 
\begin{defn}
{\bf Thermal schedulability} is to guarantee that under any task execution patterns, the CPU does not exceed the maximum temperature constraint.
\end{defn}

When the ambient temperature changes from a higher criticality level to a lower one, lower-criticality servers and their tasks do not resume immediately. The reasoning is that it takes time for the CPU to cool down to reach the safe temperature level that the increase in workload (due to resuming the lower-criticality tasks) will not lead to a temperature violation. Therefore, in the transition to a lower criticality level, only higher-criticality servers (and also their tasks) still run, and after reaching the safe temperature, then lower-criticality servers (and their tasks) resume. Let \textit{shifting time} denote this delay. We will later determine this delay as a function of physical characteristics and system settings.

\cmnt{Since actual power dissipation is affected by operating temperature as discussed in Sec.~\ref{sec:chap3_power_model}, we consider the peak power dissipation at each criticality level.\footnote{Note that each criticality level has a different critical ambient temperature.}}

%Although a rise in operating temperature leads to an increase in the static power term as discussed in Sec.~\ref{sec:chap3_power_model}, we consider the peak power dissipation at each criticality level for sake of simplicity in our analysis. 
%This assumption leads to having a periodic step signal for power model at each criticality level and can be easily converted to the Fourier series instead of applying the Fourier transform.

%Furthermore, we consider multiple preemptable servers running on each CPU core. To perform our analysis, we determine an equivalent server with the same characteristics on each each. Accordingly the thermal schedulability test of those servers on each criticality level is verified. 

\begin{figure}[th]
\centering
\includegraphics[width=0.72\textwidth]{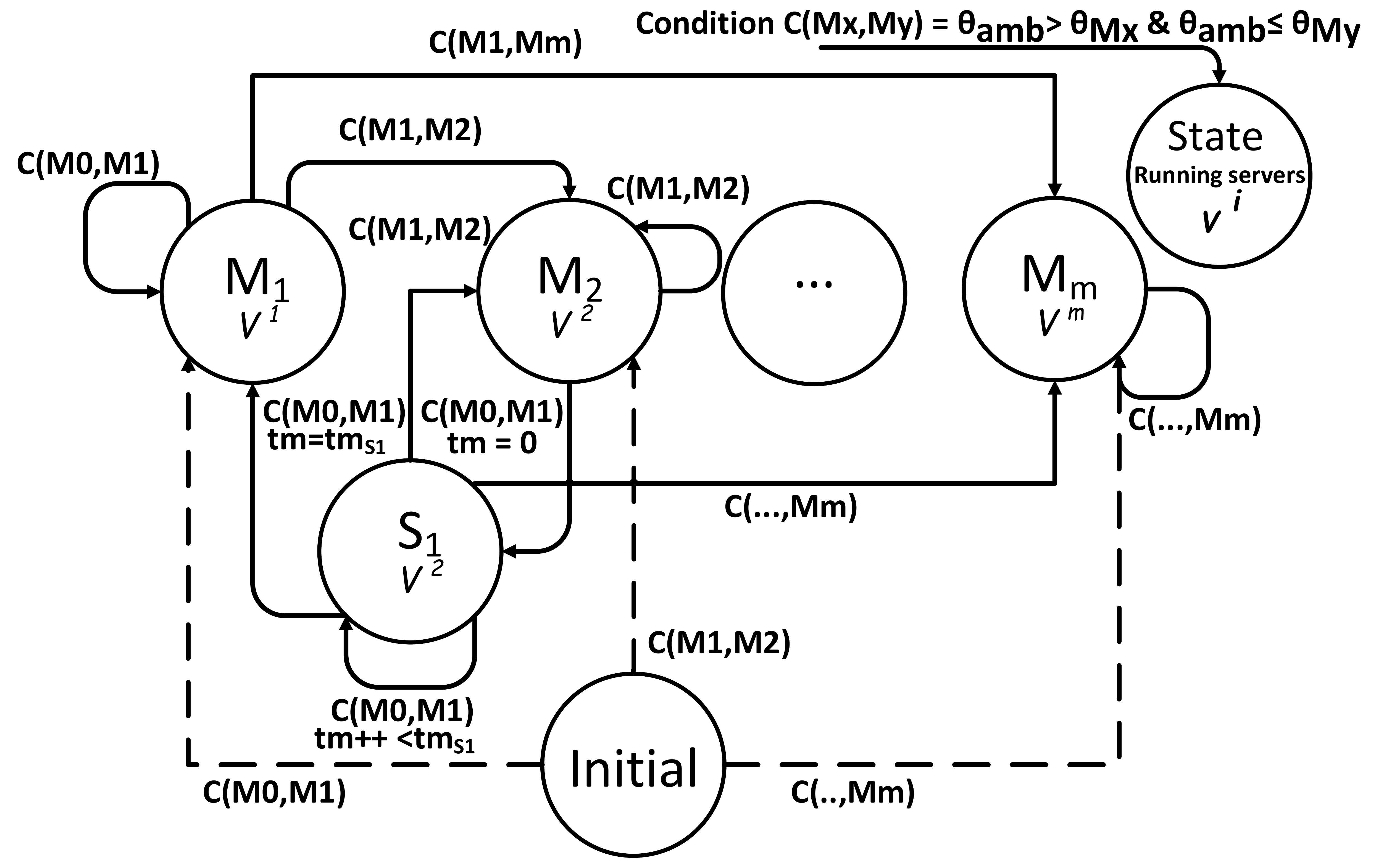}
\caption{Criticality mode change diagram}
\label{fig:chap3_crit_mode}
\end{figure}

The state diagram of criticality mode changes is illustrated in Fig.~\ref{fig:chap3_crit_mode}. The criticality mode of the system is determined by the associated servers of the current state (e.g., $V^l$ for level $l$), and the change of the state is triggered by the monitored ambient temperature. If the ambient temperature goes higher than the critical ambient temperature of the highest criticality mode $l_m$, the system will shutdown. An increase in the ambient temperature leads the system state to transition to a higher criticality mode immediately. However, a transition to a lower criticality mode involves a shifting state. Let $S_i$ denote the shifting state from one criticality mode $l_{i+1}$ to its immediate lower mode $l_i$, and $M_i$ represents the state of the criticality mode $l_i$. After staying in $S_i$ for $tm_{Si}$ time units and the ambient temperature is still under the safe level, the system state will change to $M_i$.

In summary, the design-time analysis and the run-time support of our framework are as follows:
\begin{enumerate}
    \item[\textbf{Design-time:}]
    \item Check the timing schedulability of tasks for each criticality.
    \item Find the parameters of thermal-aware servers that ensure thermal schedulability for each criticality.
    \item Compute the corresponding critical ambient temperature for each criticality.
    \item Compute the shifting time from each criticality level to its immediate lower one.
\end{enumerate}
\begin{enumerate}
    \item[\textbf{Run-time:}]
    \item Monitor ambient temperature.
    \item If ambient temperature exceeds the critical ambient temperature of the current criticality level $l_i$, a) switch to a higher criticality $l_{i+1}$, and b) terminate servers with criticality less than $l_{i+1}$ immediately.
    \item If ambient temperature decreases below the critical ambient temperature of one lower criticality level $l_{i-1}$, a) switch to the lower criticality $l_{i-1}$, b) wait for \textit{shifting time}, and c) resume servers with criticality $l_{i-1}$.
\end{enumerate}

\section{Thermal Analysis}
\label{sec:chap3_thermal_analysis}

In this section, we first develop a generalized thermal model to represent the CPU temperature as a function of a generic input power signal. We show the relation of workload and ambient temperature level under the maximum thermal constraint in a steady state. The shifting time to a lower-criticality level will be discussed. Finally, we prove the worst-case scenario of task arrivals in invariant ambient temperature at a specific criticality level in multi-core platforms.

\subsection{General Thermal Model for Periodic Power Signal}
The thermal behavior of the CPU is modeled when a generic periodic power signal generates heat dissipation, which results in temperature oscillations. The temperature function of the CPU is analytically extracted based on the ambient temperature, workload, physical and geometrical properties, and the thermal resistances between the CPU and surroundings.

A generic power signal is a periodic step function:
\[P(t) = \left\{ {\begin{array}{*{20}{c}}
{{P_S}}&{Sleeping\,time}\\
{{P_S} + {P_D}}&{O.W.}
\end{array}} \right.\]

\begin{figure}[h]
\centering
\vspace{-5pt}
\includegraphics[width=0.80\textwidth]{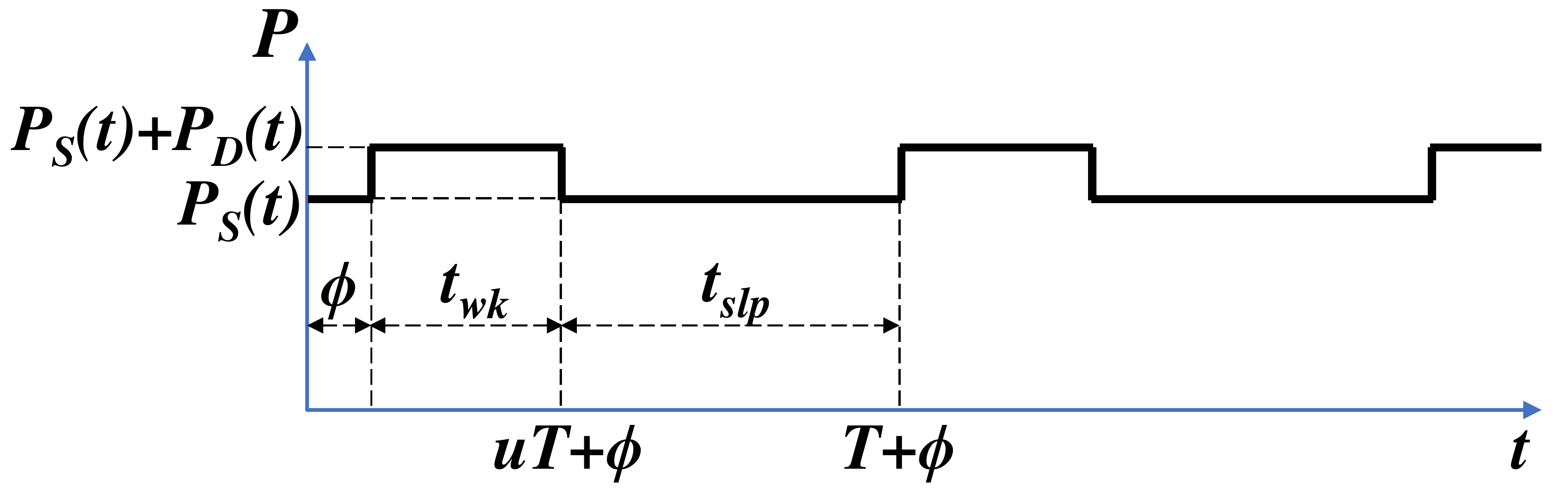}
\caption{A generic periodic power signal.}
\vspace{-5pt}
\label{fig:chap3_power}
\end{figure}

Fig.~\ref{fig:chap3_power} illustrates this power signal function where $t_{wk}$ and $t_{slp}$ denote the execution time and \new{the} sleeping time of \new{a} periodic workload, $u$ is the CPU utilization (i.e., ${u = \frac{t_{wk}}{T}}$), and $\phi$ is the release offset. %Since the heat dissipation is changing with time 

In order to derive a continuous temperature function, here we represent the periodic step function of the power signal as the following Fourier series: 
\begin{equation}\label{eq:chap3_1001}
    P(t) = {P_S} + {P_D}u + {P_O}
\end{equation}
where
\[{P_O} = \sum\limits_{k = 1}^\infty  {\frac{{2{P_D}\sin \left( {uk\pi } \right)}}{{k\pi }}\cos \left( {\frac{{2k\pi }}{T}\left( {t - \frac{{uT}}{2} - \phi } \right)} \right)} \]

Note that using this continuous representation of the power signal is advantageous in deriving
a straightforward formula for a temperature equation, compared to iteratively applying the recursion relation with new initial conditions at every period. First, we define ${\theta(t)  = \Theta_{CPU}(t) - {\Theta_{amb}(t)}}$ as the temperature difference of the CPU core and the ambient. Then, the rate of temperature change can be captured by the following differential equation~\cite{Ahmed2017,DSouza2017ThermalIO}:
\begin{equation}\label{eq:chap3_1002}
    \frac{{d\theta(t) }}{{dt}} = A\theta(t)  + BP(t)
\end{equation}
where $A$ and $B$ are scalar values determined based on the system inner thermal resistance and capacitance. Substituting $P(t)$ from Eq.~\ref{eq:chap3_1001}, we solve Eq.~\ref{eq:chap3_1002} for the CPU temperature. Assuming the temperature difference at the initial time ${t_0}$ is ${\theta _0}$, the temperature of the CPU can be written as:
\begin{equation}\label{eq:chap3_1003}
\begin{split}
    \theta (t) = {\theta _0}{e^{A(t - {t_0})}} - \frac{B}{A}\left( {{P_S} + {P_D}u} \right)\left( {1 - {e^{A(t - {t_0})}}} \right) + \\
    B\left( {S(A,{P_D},u,T,\phi ,t) - S(A,{P_D},u,T,\phi ,{t_0}){e^{A(t - {t_0})}}} \right)
\end{split}
\end{equation}
where $S$ is a function defined as:
\begin{multline*}
    S(\beta ,P,u,T,\phi ,t) =  - \sum\limits_{k = 1}^\infty  \frac{{2PT\sin \left( {uk\pi } \right)}}{{k\pi \left( {{T^2}{\beta ^2} + 4{k^2}{\pi ^2}} \right)}} \times \\
    \left( { - T\beta \cos \left( {\frac{{2k\pi }}{T}\left( {t - \psi } \right)} \right) + 2n\pi \sin \left( {\frac{{2k\pi }}{T}\left( {t - \psi } \right)} \right)} \right)
\end{multline*}
with $\psi  = \frac{{uT}}{2} + \phi$.

The parameters $A$ and $B$ in Eq.~\ref{eq:chap3_1003} are the system characteristics which depend on the thermal resistances, heat capacity, CPU mass, and thermal convection of the ambient. It is worth mentioning that the thermal response of a system with constant power dissipation can be derived as a special case of Eq.~\ref{eq:chap3_1003} by considering ${u = 1}$:
\begin{equation}\label{eq:chap3_1004}
\begin{split}
    \theta (t) = {\theta _0}{e^{A(t - {t_0})}} - \frac{B}{A}\left( {{P_S} + {P_D}} \right)\left( {1 - {e^{A(t - {t_0})}}} \right)
\end{split}
\end{equation}

If ${t_0 = 0}$, the well-known expression for the constant power dissipation case can be obtained from Eq.~\ref{eq:chap3_1004}:
\begin{equation}\label{eq:chap3_1005}
    \theta (t) = \alpha  + \left( {{\theta _0} - \alpha } \right){e^{\beta t}}
\end{equation}
where $\beta  = A$, and $\alpha  =  - \frac{B}{A}\left( {{P_S} + {P_D}} \right)$.

For any ${u}$, the thermal response of the system with a constant power signal reaches the steady state. 
%For the system with a constant power signal, the temperature reaches plateau and stays constant in the steady state conditions. 
From Eq.~\ref{eq:chap3_1004}:
\begin{equation}\label{eq:chap3_1006}
    {\theta _s}(t) = \alpha  =  - \frac{B}{A}\left( {{P_S} + {P_D}} \right).
\end{equation}

For the case where the CPU utilization is not $100\%$, the temperature still oscillates in the steady state, but the oscillation stays within a certain range given by the minimum and the maximum steady-state temperatures (${\theta _L}$ and ${\theta _M}$). We can derive the steady state thermal response from Eq.~\ref{eq:chap3_1003}:
\begin{equation}\label{eq:chap3_1007}
    {\theta _s}(t) =  - \frac{B}{A}\left( {{P_S} + {P_D}u} \right) + BS(A,{P_D},u,T,\phi ,t)
\end{equation}
% where $S(A,{P_D},a,T,\phi,t)$ represents the oscillations. An oscillating steady state temperature is plotted in Fig.~\ref{fig:chap3_steady_sch}. The thermal behavior of the CPUs in the transient and steady conditions can be studied in detail according to the developed general expressions.
where $S(A,{P_D},u,T,\phi,t)$ represents the oscillation. Based on this general expressions, we will give details on the thermal behavior of multiple CPU cores under transient and steady conditions in Sec.~\ref{sec:chap3_thermal_model_multicore}. 

% \begin{figure}[h]
% \centering
% \includegraphics[width=0.35\textwidth]{figs/thermal/Steady_State_Schematic.pdf}
% \caption{Steady state oscillating temperature.}
% \label{fig:chap3_steady_sch}
% \end{figure}

\subsubsection{Workload, ambient, and maximum temperature relations}

In the steady state condition, the relation between workload, ambient temperature ${\Theta_{amb}}$, and the maximum operating temperature ${\Theta_{M}}$ can be determined based on the model developed above. The temperature difference  ${\theta _M}$ can be written as:
\begin{equation}\label{eq:chap3_1008}
    {\theta _M} = {\Theta _M} - {\Theta _{amb}} =  - \frac{B}{A}\left( {{P_S} + {P_D}\frac{{1 - {e^{AuT}}}}{{1 - {e^{AT}}}}} \right)
\end{equation}

Therefore, if ${\Theta_{M}}$ is used as a thermal constraint, the maximum ambient temperature for a given utilization $u$ is expressed as follows:
\begin{equation}\label{eq:chap3_1009}
    {\Theta _{amb}} = {\Theta _M} + \frac{B}{A}\left( {{P_S} + {P_D}\frac{{1 - {e^{AuT}}}}{{1 - {e^{AT}}}}} \right)
\end{equation}

Also, another useful expression can be derived for the workload based on the constraint maximum temperature, ambient temperature, and the power signal:
\begin{equation}\label{eq:chap3_1010}
u = \frac{1}{{AT}}\ln \left( {1 + \frac{A}{B}\frac{{{\Theta _M} - {\Theta _{amb}} + \left( {{B \mathord{\left/{\vphantom {B A}} \right. \kern-\nulldelimiterspace} A}} \right){P_S}}}{{{P_D}}}\left( {1 - {e^{AT}}} \right)} \right)
\end{equation}

\subsubsection{Time shifting and transient analysis}

We derive average steady-state temperature by removing the oscillations from the above equations.  
%Another expression can be developed for the average temperature which removes the oscillations and can be a representative of the average thermal response of the CPUs. 
This is especially useful for transient phase analysis. From Eq.~\ref{eq:chap3_1003}, the following can be derived:
\begin{equation}\label{eq:chap3_1011}
\begin{split}
    {\theta _{ave}}(t) = {\theta _0}{e^{A(t - {t_0})}} - \frac{B}{A}\left( {{P_S} + {P_D}u} \right)\left( {1 - {e^{A(t - {t_0})}}} \right)
    - BS(A,{P_D},u,T,\phi ,{t_0}){e^{A(t - {t_0})}}
\end{split}
\end{equation}
${\theta _{ave}}(t)$ can be approximated by taking the first few terms of the series for $S(A,{P_D},u,T,\phi ,{t_0})$. It can be seen from Eq.~\ref{eq:chap3_1011} that the plateau of ${\theta _{ave}}$ in the steady state is ${ - \frac{B}{A}({P_S} + {P_D}u)}$.

% \begin{equation}
% \begin{split}
% {S_1} & =  - \frac{{2PT\sin \left( {a\pi } \right)}}{{\pi \left( {{T^2}{\beta _1}^2 + 4{\pi ^2}} \right)}} \times\\
% & \left( { - T{\beta _1}\cos \left( {\frac{{2\pi {t_i}}}{T} - a\pi } \right) + 2\pi \sin \left( {\frac{{2\pi {t_i}}}{T} - a\pi } \right)} \right)
% \end{split}
% \end{equation}

% \begin{equation}\label{eq:chap3_10012}
% \begin{split}
% {\theta _{ave}}(t)  & \simeq  - \frac{{{\beta _2}}}{{{\beta _1}}}\left( {{P_S} + {P_D}a} \right) +\\
% & \left( {{\theta _i} + \frac{{{\beta _2}}}{{{\beta _1}}}\left( {{P_S} + {P_D}a} \right) + {\beta _2}{S_1}} \right){e^{{\beta _1}(t - {t_i})}}
% \end{split}
% \end{equation}

Based on the expression of the  ${\theta _{ave}}(t)$, shifting time can be calculated which is defined as the time it takes for the system to reach a new steady\new{-}state condition when system parameters are changed. We calculate the shifting time that the CPUs take to reach to  99\% of the new steady\new{-}state condition from an initial temperature difference of ${\theta _0}$. Assuming that $S_k$ is the value of $S(t_0)$ taking the first $k$ terms, we have:
\begin{equation}\label{eq:chap3_1012}
    {t_{shift}} = \frac{1}{A}\ln \left( {\left| {\frac{{0.01\frac{B}{A}({P_S} + {P_D}u)}}{{{\theta _0} + \frac{B}{A}({P_S} + {P_D}u) + B{S_k}}}} \right|} \right)
\end{equation}

\begin{figure}[h]
\centering
% \vspace{-10pt}
\includegraphics[width=0.64\textwidth]{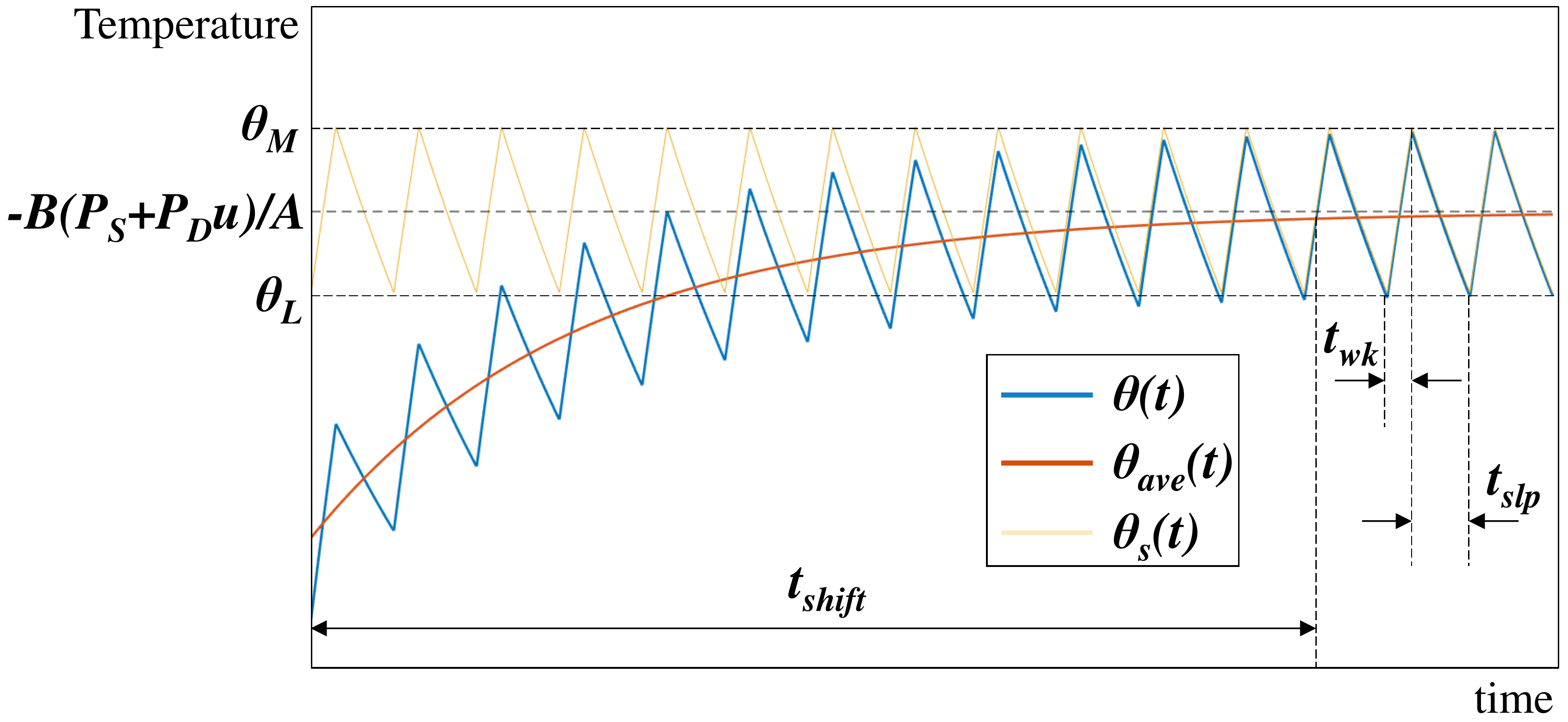}
\caption{Current, steady state, and average temperature profiles.}
% \vspace{-7pt}
\label{fig:chap3_theta}
\end{figure}
Fig.~\ref{fig:chap3_theta} illustrates temperature profile ${\theta}(t)$, oscillating steady state temperature ${\theta}_{s}(t)$, and average temperature ${\theta}_{ave}(t)$.
\subsection{General Thermal Model for Multi-Core Platforms}\label{sec:chap3_thermal_model_multicore}
Similar to the single-core case, a general thermal model can be developed for a multi-core platform with a periodic input power signal. In this subsection, we represent the power signal as a continuous function and show its benefit in simplifying the final solution. For a multi-core platform with $n$ cores, we can have $n$ eigenvalues that define the thermal response of the system. Therefore, we use matrix notations to solve the differential equations. After performing a thermal analysis, the rate of CPUs' temperature change can be denoted as:
\begin{equation}\label{eq:chap3_1013}
    \frac{{d\boldsymbol{\theta}}(t)}{{dt}} = {\bf{A}}\boldsymbol{\theta}(t)  + {\bf{BP}}(t)
\end{equation} 
where {\bf{A}}, an $n \times n$ matrix, and {\bf{B}}, a diagonal $n \times n$ matrix, are determined by the inner thermal resistance and capacitance of the system. ${\boldsymbol{\theta}}$ is an $n \times 1$ matrix of the CPU cores' temperature difference, and {\bf{P}}$(t)$ is the $n \times 1$ matrix of CPU cores' power signal functions. If ${\boldsymbol{\theta_{0}}}$ is the initial temperature difference matrix at $t_{0}$, the solution of Eq. \ref{eq:chap3_1013} can be written as:
\begin{equation}\label{eq:chap3_1014}
    \boldsymbol{\theta} (t) = {e^{(t - {t_0}){\bf{A}}}}{\boldsymbol{\theta} _0} + \int_{{t_0}}^t {{e^{(t - s){\bf{A}}}}{\bf{BP}}(s)ds}
\end{equation}
The first part is the homogeneous solution which is the thermal response due to the initial temperature difference from the ambient. The second part is the non-homogeneous solution caused by the input power signal. The temperature rise due to a power signal is independent of the initial condition. ${e^{(t - {t_0}){\bf{A}}}}$ is the matrix exponential of ${\bf{A}}$. For a platform with $n$ CPU cores, we denote the eigenvalues as ${\beta_{1}}$, ${\beta_{2}}$, ..., and ${\beta_{n}}$ \ali{with ${\beta_1} \ge {\beta_2} \ge ... \ge {\beta_n}$}. To solve Eq.~\ref{eq:chap3_1014}, we first show that the solution of power signals can be obtained as the sum of the solutions of each signal.

\begin{thm}\label{thm:superposition}
(Superposition) Thermal response due to any combinations of power signals is equivalent to the sum of thermal responses caused by each of those power signals. 
\end{thm}
\begin{proof}
% Please refer to \cite{techreport}.

Assume that ${\boldsymbol{\theta}}_1$ and ${\boldsymbol{\theta}_2}$ are the thermal responses caused by executions ${\bf{P}}_1(t)$ and ${\bf{P}}_2(t)$. Also, ${\boldsymbol{\theta}_3}$ is the thermal response caused by  ${\bf{P}}_3(t) = {{\bf{P}}_1}(t) + {{\bf{P}}_2}(t)$. From Eq.~13:
\[\frac{{d{\boldsymbol{\theta} _1}}(t)}{{dt}} = {\bf{A}}{\boldsymbol{\theta}_1(t)} + {\bf{B}}{{\bf{P}}_1}(t)\text{, and  }\frac{{d{\boldsymbol{\theta}_2(t)}}}{{dt}} = {\bf{A}}{\boldsymbol{\theta}_2(t)} + {\bf{B}}{{\bf{P}}_2}(t).\]

By adding these two differential equations, we have:
\[\frac{d}{{dt}}\left( {{\boldsymbol{\theta}_1(t)} + {\boldsymbol{\theta}_2(t)}} \right) = {\bf{A}}\left( {{\boldsymbol{\theta}_1(t)} + {\boldsymbol{\theta}_2(t)}} \right) + {\bf{B}}\left( {{{\bf{P}}_1}(t) + {{\bf{P}}_2}(t)} \right)\]

This is equivalent to the differential equation of ${\boldsymbol{\theta}_3}$:
\[\frac{{d{\boldsymbol{\theta}_3(t)}}}{{dt}} = {\bf{A}}{\boldsymbol{\theta}_3(t)} + {\bf{B}}{{\bf{P}}_3}(t) = {\bf{A}}{\boldsymbol{\theta}_3(t)} + {\bf{B}}\left( {{{\bf{P}}_1}(t) + {{\bf{P}}_2}(t)} \right)\]

Therefore, $\boldsymbol{\theta}_3(t) = \boldsymbol{\theta}_1 (t) + \boldsymbol{\theta}_2 (t)$.
\end{proof}
Let {\bf{V}} denote an $n \times n$ matrix of eigenvectors of {\bf{A}} where column $i$ is the $i^{th}$ eigenvector. Also, {\bf{D}} is a diagonal $n \times n$ matrix in which the diagonal entries are the eigenvalues of {\bf{A}}. The solutions of the non-homogeneous part in Eq.~\ref{eq:chap3_1014} can be written as:
\begin{equation}\label{eq:chap3_1015}
\begin{split}
    \int_{{t_0}}^t {{e^{(t - s){\bf{A}}}}{\bf{BP}}(s)ds}  = \int_{{t_0}}^t {{e^{(t - s){\bf{A}}}}{\bf{B}}\left( {{{\bf{P}}^\infty } + {{\bf{P}}_o}(s)} \right)ds} \\
    = \int_{{t_0}}^t {{e^{(t - s){\bf{A}}}}{\bf{B}}{{\bf{P}}^\infty }ds}  + \int_{{t_0}}^t {{e^{(t - s){\bf{A}}}}{\bf{B}}{{\bf{P}}_o}(s)ds}
\end{split}
\end{equation}
where ${{\bf{P}}^\infty}$ and ${{{\bf{P}}_o}(s)}$ are the constant and oscillating part of the power signal matrix, respectively. 
\[{{\bf{P}}^\infty } = [{P_j}^\infty]_{n\times 1}  = [{P_S}_{_j} + {P_D}_{_j}{u_j}]_{n\times 1}\]
\[{{\bf{P}}_o} = [{P_o}_{_j}] = \left[\sum\limits_{k = 1}^\infty  {\frac{{2{P_D}_{_j}}}{{k\pi }}\sin \left( {{u_j}k\pi } \right)\cos \left( {\frac{{2k\pi }}{{{T_j}}}\left( {s - {\psi _j}} \right)} \right)}\right]_{n \times 1} \]
with ${\psi _j} = \frac{{{u_j}{T_j}}}{2} + {\phi _j}$. Since ${{\bf{P}}^\infty}$ is constant,
\begin{equation}\label{eq:chap3_1016}
    \int_{{t_0}}^t {{e^{(t - s){\bf{A}}}}{\bf{B}}{{\bf{P}}^\infty }ds}  =  - {{\bf{A}}^{ - 1}}\left( {{\bf{I}} - {e^{(t - {t_0}){\bf{A}}}}} \right){\bf{B}}{{\bf{P}}^\infty }
\end{equation}
For the second integral we have:
\begin{equation}\label{eq:chap3_1017}
\begin{aligned}
   &\int_{{t_0}}^t {{e^{(t - s){\bf{A}}}}{\bf{B}}{{\bf{P}}_o}(s)ds}  = {\bf{V}}{e^{t{\bf{D}}}}\int_{{t_0}}^t {{e^{ - s{\bf{D}}}}{{\bf{V}}^{ - 1}}{\bf{B}}{{\bf{P}}_o}(s)ds} \\ &
   = {\bf{V}}{e^{t{\bf{D}}}}.{\left[ {\sum\limits_{i = 1}^n {\int_{{t_0}}^t {{e^{ - {\beta _j}s}}V_{ji}^{ - 1}{B_{ii}}{P_{{o_i}}}(s)} } } \right]_{n \times 1}}
\end{aligned}
\end{equation}

Similar to the single core case, we can use $S$ in Eq.~\ref{eq:chap3_1003} but in a matrix form. We define the matrix ${\bf{S}}$ as:
\[{\bf{S}}(t) = {\left[ {{S_{ij}}(t)} \right]_{n \times n}} = {\left[ {S({\beta _i},{P_{{D_j}}},{u_j},{T_j},{\phi _j},t)} \right]_{n \times n}}\]

We have $\int {{e^{ - {\beta _j}s}}{P_{{o_i}}}(s)ds}  = {e^{ - {\beta _j}s}}{S_{ji}}(s)$. Therefore:
\begin{multline*}
\begin{aligned}
   &{\left[ {\sum\limits_{i = 1}^n {\int_{{t_0}}^t {{e^{ - {\beta _j}s}}V_{ji}^{ - 1}{B_{ii}}{P_{{o_i}}}(s)ds} } } \right]_{n \times 1}} 
   &= {\left[ {\sum\limits_{i = 1}^n {{e^{ - {\beta _j}t}}V_{ji}^{ - 1}{B_{ii}}{S_{ji}}(t) - {e^{ - {\beta _j}{t_0}}}V_{ji}^{ - 1}{B_{ii}}{S_{ji}}({t_0})} } \right]_{n \times 1}}
\end{aligned}
\end{multline*}

To write this in a matrix notation, we define ${\bf{B}}' = diag({\bf{B}})$ an $n \times 1$ matrix containing the diagonal entries of ${\bf{B}}$ such that $B{'_j} = {B_{jj}}$. Then,
\begin{multline*}
\begin{aligned}
   &{\left[ {\sum\limits_{i = 1}^n {{e^{ - {\beta _j}t}}V_{ji}^{ - 1}{B_{ii}}{S_{ji}}(t) - {e^{ - {\beta _j}{t_0}}}V_{ji}^{ - 1}{B_{ii}}{S_{ji}}({t_0})} } \right]_{n \times 1}} \\
   &= {e^{ - t{\bf{D}}}}\left( {{{\bf{V}}^{ - 1}} \circ {\bf{S}}(t)} \right){\bf{B}}' - {e^{ - {t_0}{\bf{D}}}}\left( {{{\bf{V}}^{ - 1}} \circ {\bf{S}}({t_0})} \right){\bf{B}}'
\end{aligned}
\end{multline*}
where ${{{\bf{V}}^{ - 1}} \circ {\bf{S}}(t)}$ is the Hadamard product of matrices ${{{\bf{V}}^{ - 1}}}$ and ${{\bf{S}}(t)}$, which is a matrix of the same dimension of $n \times n$ where each element $ij$ is the product of counterpart elements $ij$ of  ${{{\bf{V}}^{ - 1}}}$ and ${{\bf{S}}(t)}$: ${\left( {{{\bf{V}}^{ - 1}} \circ {\bf{S}}(t)} \right)_{ij}} = V_{ij}^{ - 1}{S_{ij}}(t)$. Substituting this in Eq.~\ref{eq:chap3_1017} gives:
\begin{equation}\label{eq:chap3_1018}
\begin{aligned}
   &{\bf{V}}{e^{tD}}.{\left[ {\sum\limits_{i = 1}^n {\int_{{t_0}}^t {{e^{ - {\beta _j}s}}V_{ji}^{ - 1}{B_{ii}}{P_{{o_i}}}(s)} } } \right]_{n \times 1}}\\
   &= {\bf{V}}\left( {{{\bf{V}}^{ - 1}} \circ {\bf{S}}(t) - {e^{(t - {t_0}){\bf{D}}}}\left( {{{\bf{V}}^{ - 1}} \circ {\bf{S}}({t_0})} \right)} \right){\bf{B}}'
\end{aligned}
\end{equation}
Substituting Eq.~\ref{eq:chap3_1016} and Eq.~\ref{eq:chap3_1018} in Eq.~\ref{eq:chap3_1015} results in a general solution for the thermal response of a $n$-core CPU platform with distinct input power signals to each core:
\begin{equation}\label{eq:chap3_1019}
\begin{aligned}
   \boldsymbol{\theta} (t) = &{e^{(t - {t_0}){\bf{A}}}}{\boldsymbol{\theta} _0} - {{\bf{A}}^{ - 1}}\left( {{\bf{I}} - {e^{(t - {t_0}){\bf{A}}}}} \right){\bf{B}}{{\bf{P}}^\infty } + \\
   &{\bf{V}}\left( {{{\bf{V}}^{ - 1}} \circ {\bf{S}}(t) - {e^{(t - {t_0}){\bf{D}}}}\left( {{{\bf{V}}^{ - 1}} \circ {\bf{S}}({t_0})} \right)} \right){\bf{B}}'
\end{aligned}
\end{equation}

This expression of ${\boldsymbol{\theta}(t)}$, which contains simple matrix operations, can be easily used to compute the general solution for the  transient temperature profile of multi-core CPUs. 
% A similar approach can be implemented for other type of periodic power signals. 

%  age mikhai ye khat ro comment koni ya uncomnt ctrl+ / hast. /Eyval.. tu dore amuzeshiam :D
% \begin{equation}\label{eq:chap3_1000}
% \begin{split}
% %   template for new line equation (copy/paste below)

% \end{split}
% \end{equation}
\subsection{Worst-Case Execution Scenarios}
% \subsection{steady state condition}
% \subsubsection{single core platform}
% In this section, we calculate the thermal steady state condition of the single core CPU on which  periodically a specific amount of workloads execute. First, we prove that the worst-case execution time happens where all workloads execute consecutively. 

It is important to know the workload execution patterns which cause the peak heat dissipation. In this section, based on the model developed in the previous sections, we explore and discuss the worst-case scenarios for workload execution.

\subsubsection{Consecutive workload execution}

First, we prove that the time to reach the maximum temperature is minimized if all workloads execute consecutively. Suppose that for any arbitrary execution of workloads in a period of  $T$, the CPU executes some workload for $t_1$, sleeps for $t_2$, and then wakes up and executes for $t_3$ time units, and these working-sleeping switches occur $n$ times. For the CPU idling time, Eq.~\ref{eq:chap3_1005} changes to ${\theta (t) = {\theta _0}\ {e^{\beta t}}}$ if the static power  is ignored (i.e., $\alpha = 0$). Hence, the temperature change during a period is:
\[ \theta_{t_0}   = \theta_L \]
\[ \theta_{t_1}   =\alpha + (\theta_{t_0} - \alpha) * e^{\beta t_1}\]
\[ \theta_{t_2}   = \theta_{t_1}  * e^{\beta t_2}\]
% \[ \theta_{t_3}   =\alpha + (\theta_{t_2} - \alpha) * e^{\beta t_3} \]
% \[ \theta_{t_4}   = \theta_{t_3}  * e^{\beta t_4}\]
% \[ \theta_{t_5}   =\alpha + (\theta_{t_4} - \alpha) * e^{\beta t_5} \]
\[\vdots\]
\[ \theta_{t_n}   = \theta_{t_{n-1}}  * e^{\beta t_n}\]
where $\sum t_i = T $ for all $n > 1$ and $t_i > 0$.
% Solving the recursion, $t_n$ is obtained by
% \[  \theta_{t_n} = e^{\beta t_{n-1}} (\alpha-e^{\beta t_{n-2}}\cdots  (\alpha-e^{\beta t_{2}} (\alpha-e^{\beta t_{1}} (\alpha-\theta_L ) ) ) ) ) ).\]

In the steady state, the temperature of the beginning and the end of each period is equal.  Therefore, considering $\theta_{t_n} = \theta_{L}$,  
%  \[ \theta_{L} = \frac{e^{\beta t_{n}}(\alpha-e^{\beta t_{n-1}}\cdots (\alpha-e^{\beta t_{2}}(\alpha-\alpha e^{\beta t_{1}})))}{ e^{\beta t_{1}}e^{\beta t_{2}}\cdots e^{\beta t_{n-1}}e^{\beta t_{n}}-1}%
% \]
% so,
\[
    \theta_{L} =\alpha \frac{e^{\beta(t_{1}+t_{2}+\cdots+t_{n-1}+t_{n})}-e^{\beta(t_{2}+\cdots+t_{n-1}+t_{n})}+\cdots\cmnt{+e^{\beta(t_{n-1}+t_{n})}}-e^{\beta t_{n}}}{e^{\beta T}-1}
    \]
where $ t_{wk} = \sum_{i =1 ,\, i= i+2 }t_i$ is the execution time and  ${t_{slp} = \sum_{i =2 ,\, i= i+2 }t_i }$ is the idle time, respectively. We prove that the amount of total execution time is minimized when there is only one execution time chunk. In other words, the worst-case thermal scenario happens when all workloads execute consecutively.

\begin{figure}[h]
\centering
% \vspace{-15pt}
\subfloat [][]{\includegraphics[width=0.40\textwidth]{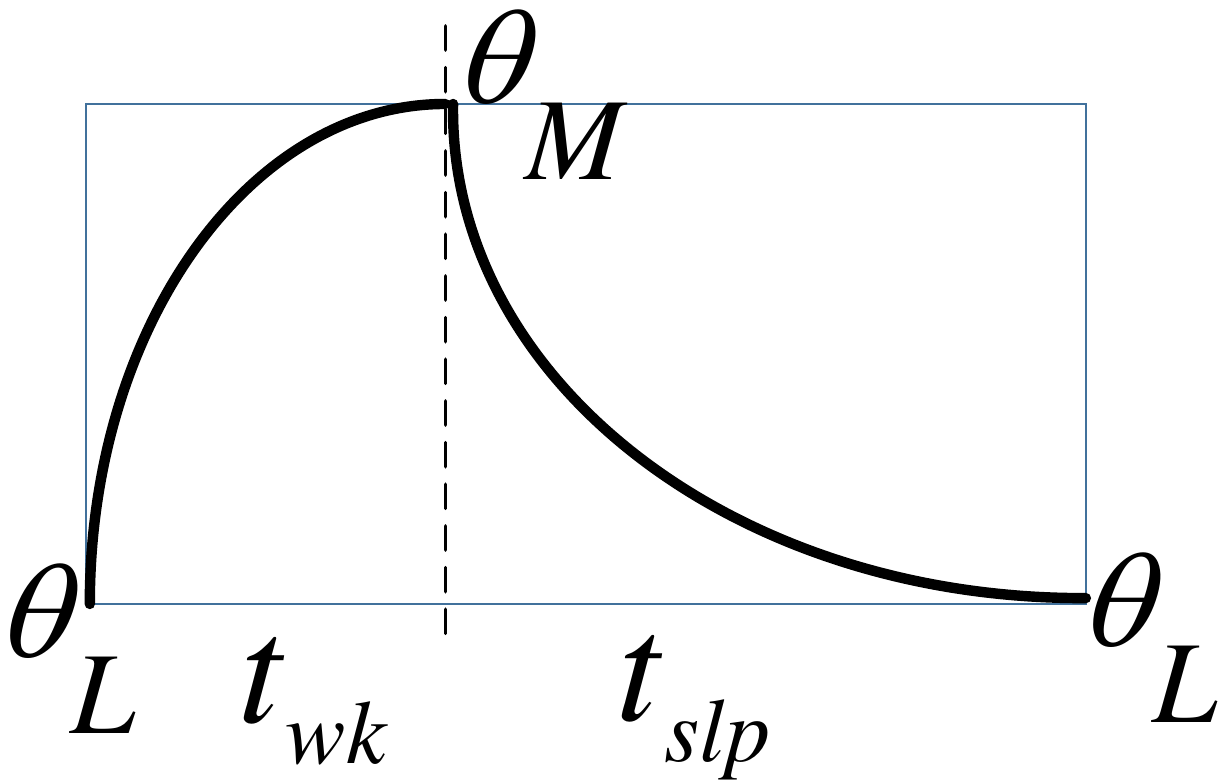}}
 \hspace{10pt}
\subfloat [][]{\includegraphics[width=0.40\textwidth]{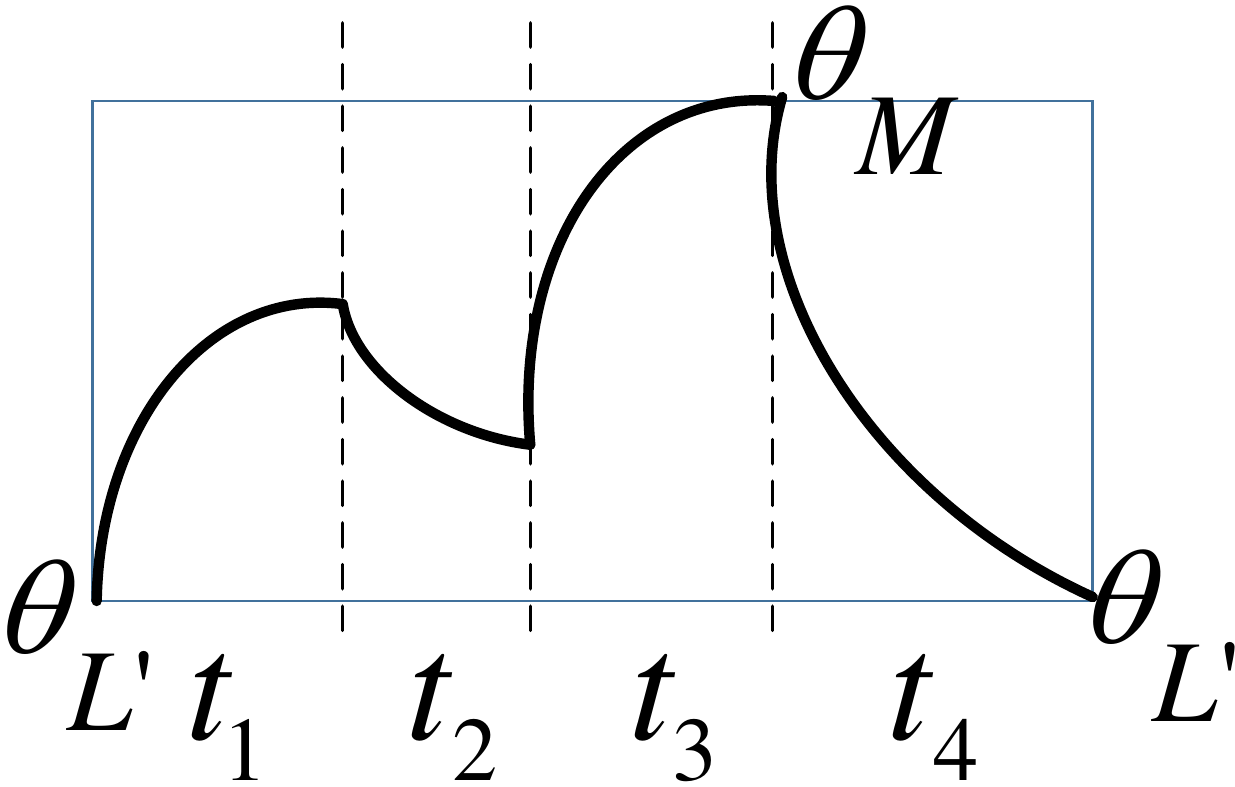}}
\caption{Temperature change in one period when the CPU operates a)  for $t_{wk}$ time units b)  for $t_1$ and \new{$t_3$} time units.}
\label{fig:chap3_budget}
\end{figure}

For clarification, suppose the following two scenarios for a given period under the maximum temperature constraint:
\begin{itemize}
    \item All workloads execute consecutively to reach the maximum temperature for $t_{wk}$ time units; then, the CPU sleeps until the beginning of the next period for $t_{slp}$ time units (Fig.~\ref{fig:chap3_budget}a).
    \item A portion of workloads executes for $t_1$ time units, then the CPU sleeps for $t_2$ \new{time} units, and the rest of the workloads run until the CPU reaches the maximum temperature at \new{$t_1+t_2+t_3$}; afterwards the CPU sleeps for $t_4$ \new{time} units (Fig.~\ref{fig:chap3_budget}b).
\end{itemize}

The following shows that the budget of a thermal server should be bounded based on the first scenario.
\begin{thm}\label{thm:min_waking_time}
The amount of waking time to reach the maximum temperature constraint is minimized when all workloads execute consecutively. 
\end{thm}
\begin{proof}
We are interested to prove that ${t_w \leq t_1+ t_3}$. To prove the statement, we find the relation of the minimum steady state temperature in each scenario, individually. 
 To calculate the budget for the first scenario, we first calculate the minimum steady\new{-}state temperature which is
 \[
 \theta_L = \frac{{\mathrm{e}}^{\beta\,t_{slp}}\,\left(\alpha-\alpha\,{\mathrm{e}}^{\beta\,t_{wk}}\right)}{1-{\mathrm{e}}^{\beta\,t_{wk}}\,{\mathrm{e}}^{\beta\,t_{slp}}}.    
 \]
Accordingly, the maximum temperature in the steady state is 

 \[ 
 \theta_{M} = \alpha-{\mathrm{e}}^{\beta\,t_{wk}}\,\left(\alpha+\frac{{\mathrm{e}}^{\beta\,t_{slp}}\,\left(\alpha-\alpha\,{\mathrm{e}}^{\beta\,t_{wk}}\right)}{{\mathrm{e}}^{\beta\,t_{wk}}\,{\mathrm{e}}^{\beta\,t_{slp}}-1}\right) = \alpha \frac{{\mathrm{e}}^{\beta\,t_{wk}}-1}{{\mathrm{e}}^{\beta\,T}-1}.
 \]

For the second scenario,
%  The minimum temperature in the steady state is 
%  \[
%  \theta_{L'} = \frac{{\mathrm{e}}^{\beta\,t_{4}}\,\left(\alpha-{\mathrm{e}}^{\beta\,t_{3}}\,\left(\alpha-{\mathrm{e}}^{\beta\,t_{2}}\,\left(\alpha-\alpha\,{\mathrm{e}}^{\beta\,t_{1}}\right)\right)\right)}{1-{\mathrm{e}}^{\beta\,t_{1}}\,{\mathrm{e}}^{\beta\,t_{2}}\,{\mathrm{e}}^{\beta\,t_{3}}\,{\mathrm{e}}^{\beta\,t_{4}}}.    
%  \]
 the maximum temperature is reached after ${t_1+t_2+t_3}$ time units. Therefore, 
% \begin{equation}
%      \theta_{t_{1}} = \alpha-{\mathrm{e}}^{\beta\,t_{1}}\,\left(\alpha+\frac{{\mathrm{e}}^{\beta\,t_{4}}\,\left(\alpha-{\mathrm{e}}^{\beta\,t_{3}}\,\left(\alpha-{\mathrm{e}}^{\beta\,t_{2}}\,\left(\alpha-a\,{\mathrm{e}}^{\beta\,t_{1}}\right)\right)\right)}{{\mathrm{e}}^{\beta\,t_{1}}\,{\mathrm{e}}^{\beta\,t_{2}}\,{\mathrm{e}}^{\beta\,t_{3}}\,{\mathrm{e}}^{\beta\,t_{4}}-1}\right)
%  \end{equation}

%  \[
%  \scriptstyle
%     \theta_{M} = \alpha-{\mathrm{e}}^{\beta\,t_{3}}\,\left(\cdots\,\left(\alpha-{\mathrm{e}}^{\beta\,t_{1}}\,\left(\alpha+\frac{{\mathrm{e}}^{\beta\,t_{4}}\,\left(\alpha-{\mathrm{e}}^{\beta\,t_{3}}\,\left(\cdots\,\left(\alpha-\alpha\,{\mathrm{e}}^{\beta\,t_{1}}\right)\right)\right)}{{\mathrm{e}}^{\beta\,t_{1}}\,{\mathrm{e}}^{\beta\,t_{2}}\,{\mathrm{e}}^{\beta\,t_{3}}\,{\mathrm{e}}^{\beta\,t_{4}}-1}\right)\right)\right)
%  \]
%   which is 
\[
\theta_{M} = \alpha \frac{{\mathrm{e}}^{\beta(t_1+t_2+t_3)}-{\mathrm{e}}^{\beta(t_2+t_3)}+{\mathrm{e}}^{\beta(t_3)}-1}{{\mathrm{e}}^{\beta T}-1}
\]

It is worth noting that although the minimum steady-state temperature can be different ($\theta_L$ vs. $\theta_{L'}$), in both scenarios the given maximum temperature is the same. Hence,
\[
\theta_{M} =\alpha \frac{{\mathrm{e}}^{\beta\,t_{wk}}-1}{{\mathrm{e}}^{\beta\,T}-1}= \alpha \frac{{\mathrm{e}}^{\beta(t_1+t_2+t_3)}-{\mathrm{e}}^{\beta(t_2+t_3)}+{\mathrm{e}}^{\beta(t_3)}-1}{{\mathrm{e}}^{\beta T}-1}.
\]
Therefore, we have ${\mathrm{e}}^{\beta\,t_{wk}} = {\mathrm{e}}^{\beta(t_1+t_2+t_3)}-{\mathrm{e}}^{\beta(t_2+t_3)}+{\mathrm{e}}^{\beta(t_3)}$. Now we want to prove that for any value of $t_1$, $t_2$ and $t_3$, $t_1 + t_3 \geq t_w$.\\
\textit{Contradiction:} Suppose the hypothesis is not correct. Hence,
\[
  t_1 + t_3 < t_w \longrightarrow \beta\,(t_{1}+ t_{3}) > \beta\, t_w \longrightarrow {\mathrm{e}}^{\beta\,(t_{1}+ t_{3})} \geq {\mathrm{e}}^{\beta\,t_{wk}}
\]
\[
{\mathrm{e}}^{\beta\,(t_{1}+ t_{3})} \geq {\mathrm{e}}^{\beta\,t_{wk}}= {\mathrm{e}}^{\beta(t_1+t_2+t_3)}-{\mathrm{e}}^{\beta(t_2+t_3)}+{\mathrm{e}}^{\beta(t_3)} \Longleftrightarrow
\]
\[
{\mathrm{e}}^{\beta\,t_{1}} \geq {\mathrm{e}}^{\beta(t_1+t_2)}-{\mathrm{e}}^{\beta(t_2)}+1 \Longleftrightarrow {\mathrm{e}}^{\beta\,t_{1}} - 1 \geq {\mathrm{e}}^{\beta(t_1+t_2)}-{\mathrm{e}}^{\beta(t_2)}
\]
\[
 \Longleftrightarrow{\mathrm{e}}^{\beta\,t_{1}} - 1 \geq {\mathrm{e}}^{\beta(t_2)}({\mathrm{e}}^{\beta(t_1)}-1) \Longleftrightarrow {\mathrm{e}}^{\beta(t_2)} \geq 1 \contra
\]
since $\beta < 0 $ and $t_2>0$.
% Similarly, the general scenario for multiple activation segments can be easily proved by employing induction. For brevity, we omit the proof.
\end{proof}
\begin{corollar}\label{col:1}
The maximum temperature reduces when the period and waking time are halved, since it is the special condition where $t_1 = t_3$ and $t_2=t_4$.
\end{corollar}
\smallskip\noindent\textbf{Server budget calculation under polling server budget replenishment policy.} 
Now we calculate the ``maximum" budget that a server can have while limiting the operating temperature not  to exceed the given thermal constraint in a single-core platform. The worst case for the  polling server happens when it exhausts all of its replenishment budget at the beginning of its period and then it sleeps until the beginning of the next replenishment period. For a server period $T$,
\begin{equation} 
\label{eq:chap3_1}
T = t_{wk} + t_{slp}.
\end{equation}

In the steady state of the system, we are interested in bounding the server's maximum temperature. According to Eq.~\ref{eq:chap3_1005}, $ \alpha + (\theta_L - \alpha ) e^{\beta t_{wk}} \leq \theta_M$. By calculating the minimum temperature at the end of the period and substituting it in this formula, the maximum budget for a period $T$ is given by:

\begin{equation} \label{eq:chap3_2}
t_{wk} <=  \frac{1}{\beta} \ln{\frac{\theta_M(e^{\beta T}-1)+\alpha}{\alpha}}.
\end{equation}
% Consequently, for a given replenishment period, A server ${v_i = (t_{wk}, T)}$ can bound the maximum temperature to $\theta_M$.
\subsection{Thermal Back-to-Back Execution}
Suppose that the CPU workload runs at the end of a period and the workload of the next period execute at the beginning of the period. It causes burst heat generation by back-to-back execution, which can potentially lead to thermal violation. It is noteworthy that this case has been shown in Corollary~\ref{col:1}. The worst-case occurrence is when this phenomenon happens repetitively in the steady state. 

% \begin{figure}[h]
% \centering
% \includegraphics[width=0.35\textwidth]{figs/anl/b2b.pdf}
% \caption{Thermal back-to-back phenomenon of two consecutive periods in the steady state.}
% \label{fig:chap3_b2b}
% \end{figure}

\smallskip\noindent\textbf{Server budget calculation under deferrable server budget replenishment policy.} 
Hereby, the maximum replenishment budget for the deferrable server is determined considering this phenomenon. One can model it as a periodic workload with doubled waking (${2 t_{wk}}$) and sleeping (${2 t_{slp}}$) times, and determine the maximum waking time.

\begin{thm}\label{thm:util}
The maximum waking time in thermal back-to-back execution is

$t_{wk} =  \frac{1}{2\beta} \ln{\frac{\theta_M(e^{2 \beta T}-1)+\alpha}{\alpha}} $.
\end{thm}
\begin{proof} 
\new{
 Considering the period of workload as twice of the previous example, we have $2 t_{wk} + 2 t_{slp}  = 2 T.$
The maximum temperature is reached after $2 t_{wk}$ time units. So,  
${\theta_M =  \alpha + (\theta_L-\alpha) \mathrm{e}^{2 \beta\, t_{wk}}}$.
  Similarly to Eq.~21, we have $t_{wk} = \frac{1}{2\beta} \ln{\frac{\theta_M-\alpha}{\theta_L-\alpha}}$. Since the minimum  temperature in the steady state is reached after $2 t_{slp}$, $\theta_M \mathrm{e}^{2 \beta\, t_{slp}} = \theta_L$ which leads to $t_{slp}  =  \frac{1}{2\beta} \ln{\frac{\theta_L}{\theta_M}}$. Hence,
 \[
     t_{slp}  +t_{wk} =T \Longrightarrow \frac{1}{2\beta} \ln{\frac{\theta_M-\alpha}{\theta_L-\alpha}}  + \frac{1}{2\beta} \ln{\frac{\theta_L}{\theta_M}} = T.
 \]
 Therefore, the minimum temperature in the steady state is ${\theta_L = \frac{\alpha \theta_M \mathrm{e}^{2 \beta\, T}}{\theta_M(\mathrm{e}^{2 \beta\, T}-1)+\alpha}}$. By substituting the value of $\theta_L$ in $t_{wk}$, the maximum waking time for the period $T$ is obtained. 
}
 
\end{proof}
\begin{corollar}
The maximum achievable utilization of a server under the maximum temperature constraint $\theta_M$ is $\frac{\theta_M}{\alpha}$. 
\end{corollar}
\begin{proof} 
Since the maximum utilization is obtained when the period converges to 0, the maximum utilization is 
\[ \lim_{T \to 0} u = \lim_{T \to 0} \frac{\frac{1}{2\beta} \theta_M 2\beta e^{2\beta T} }{\theta_M (e^{2\beta T} -1 ) + \alpha} = \frac{\theta_M}{\alpha}.  \]
The same approach applies to the polling servers.
\end{proof}

Unlike the claim in~\cite{rai2011worst,yang2013real} which states that the worst-case peak temperature occurs when the system warms up by applying a periodic pattern to be in steady state following by a burst workload, we will show that by employing thermal-aware servers for mixed-criticality tasks, this will never happen. 

\begin{thm}
The maximum temperature of a CPU for any workload execution pattern in the ``steady-state" condition is less than the periodic back-to-back execution pattern.
\end{thm}

\begin{proof}

\begin{figure}[h]
\centering
\includegraphics[width=0.50\textwidth]{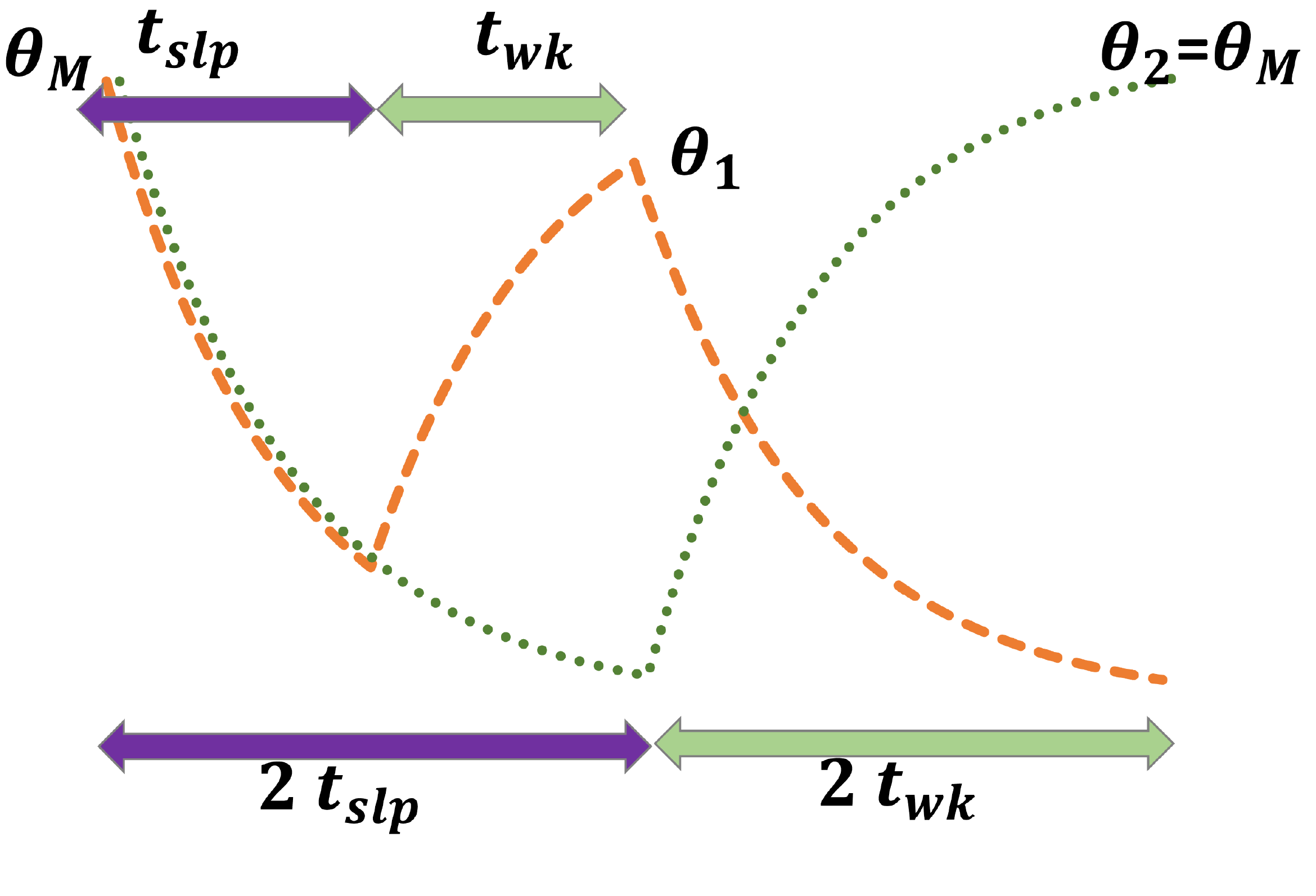}
\caption{Workload execution pattern in the steady state.}
\label{fig:chap3_b2bSteady}
\end{figure}
To prove that, we follow up on the two scenarios illustrated in Fig.~\ref{fig:chap3_b2bSteady}. Without loss of generality, it is assumed that the system has been already warmed up. The green line in Fig.~\ref{fig:chap3_b2bSteady} shows that back-to-back execution pattern continues in the steady-state phase. The orange line depicts the worst-case burst workload that can be executed (according to the Theorem~\ref{thm:min_waking_time} where all of the replenished budget is exhausted at the beginning of the period). Let $\theta_1$ and $\theta_2$ represent the maximum temperature of burst execution and normal back-to-back execution, respectively. Hence, ${\theta_1 = \alpha +\theta _{M}\,{\mathrm{e}}^{T\,\mathrm{\beta}}-\alpha \,{\mathrm{e}}^{\mathrm{\beta}\,t_{w}}}$ and ${\theta_2 = \alpha +\theta _{M}\,{\mathrm{e}}^{2\,T\,\mathrm{\beta}}-\alpha \,{\mathrm{e}}^{2\,\mathrm{\beta}\,t_{w}}}$. Let $\Delta_{\theta}$ denote the temperature difference of these scenarios. Therefore,     
     \[
    \Delta_{\theta} =  \theta_2 - \theta_1= \theta _{M}\,{\mathrm{e}}^{2\,T\,\mathrm{\beta}}-\theta _{M}\,{\mathrm{e}}^{T\,\mathrm{\beta}}+\alpha \,{\mathrm{e}}^{\mathrm{\beta}\,t_{w}}-\alpha \,{\mathrm{e}}^{2\,\mathrm{\beta}\,t_{w}}.
     \]
 By considering $u = \frac{\theta_M}{\alpha}$ and substituting the maximum waking time determined from Theorem~\ref{thm:util}, we show that $\Delta_{\theta}>0$ which means% $u {\mathrm{e}^{2\,\mathrm{\beta}\,T}} - {\mathrm{e}^{2\,\mathrm{\beta}\,t_{wk}}} + {\mathrm{e}^{\,\mathrm{\beta}t_{wk}}} - u {\mathrm{e}^{\mathrm{\beta}\,T}} > 0$.
 \[
 u{\mathrm{e}^{2\,\mathrm{\beta}\,T}} - {\mathrm{e}^{2\,\mathrm{\beta}\,{\frac{1}{2\beta} \ln{\frac{\theta_M(e^{2 \beta T}-1)+\alpha}{\alpha}}}}} + {\mathrm{e}^{\,\mathrm{\beta}{\frac{1}{2\beta} \ln{\frac{\theta_M(e^{2 \beta T}-1)+\alpha}{\alpha}}}}} - u {\mathrm{e}^{\mathrm{\beta}\,T}} > 0.
 \]

For any value of $u\in[0,1]$, since $(u-1) ( 1+ {\mathrm{e}^{\mathrm{\beta}\,T}}) ^2  < 0 $, then we have 
\[u + u {\mathrm{e}^{2\,\mathrm{\beta}\,T}} + 2 {\mathrm{e}^{\mathrm{\beta}\,T}} < 1 + 2 u {\mathrm{e}^{\mathrm{\beta}\,T}} + {\mathrm{e}^{2\,\mathrm{\beta}\,T}}.\]

By multiplying $u$ in the inequality, \new{we have} 
    \[ 
         u ^2  + 1 - 2 u + u^2 {\mathrm{e}^{2\,\mathrm{\beta}\,T}}- 2 (u-1) u {\mathrm{e}^{\mathrm{\beta}\,T}} < u {\mathrm{e}^{2\,\mathrm{\beta}\,T}}  - u  + 1
        \]  
which leads to $
        u - 1 -  u {\mathrm{e}^{\mathrm{\beta}\,T}} >  - {(\frac{\theta_M \mathrm{e}^{2\,\mathrm{\beta}\, T} - \theta_M + \alpha}{\alpha})}^{\frac{1}{2}}$. Therefore,
      
\[
       u {\mathrm{e}^{2\,\mathrm{\beta}\,T}} - \frac{\theta_M \mathrm{e}^{2\,\mathrm{\beta}\, T} - \theta_M + \alpha}{\alpha} +   {(\frac{\theta_M \mathrm{e}^{2\,\mathrm{\beta}\, T} - \theta_M + \alpha}{\alpha})}^{\frac{1}{2}} - u {\mathrm{e}^{\mathrm{\beta}\,T}} > 0.
\]
\end{proof}

\subsection{Peak Temperature Analysis on Multi-Core Platforms}
Now, we extend our analysis to multi-core platforms where each CPU core consists only of one thermal-aware server. We will show that the minimum replenishment budget of polling servers on multi-core CPU happens when all of them exhaust their budget completely at the same time. Afterwards, the maximum available server budget will be determined. 

% It is worth noting that we consider that all servers on CPU cores share the same setting. Moreover, it is assumed that initial temperature of all cores are the same in this chapter. Suppose that $\gamma$ is the conductivity coefficient between CPU core. We investigate an example of a dual-core CPU to determine the worst-case budget. We define two different possible scenarios:

% \begin{itemize}
%     \item \textbf{Synchronized budget depletion:} server budgets are consumed all at once to reach the maximum temperature simultaneously; Accordingly they deactivates until the beginning of the next replenishment period (see Fig.~\ref{fig:chap3_budget}a);
%     \item \textbf{{Asynchronized budget depletion:}} The budget of a server assigned to the second core begins to exhaust after $\phi$ time units of beginning of that of assigned to the first CPU core (see Fig.~\ref{fig:chap3_budget}b).
% \end{itemize}

% \begin{figure}[h]
% \centering
% \subfloat [][synch]{\includegraphics[width=0.23\textwidth]{figs/thermal/phase_synch.pdf}}
% \subfloat [][asynch]{\includegraphics[width=0.23\textwidth]{figs/thermal/phase_asynch.pdf}}
% \caption{An example of single server of a dual-core CPU with heat conduction between CPU cores when a)workloads execute synchronously b) workloads on different cores execute asynchronously with $\phi$ phase difference.}
% \label{fig:chap3_multi-phase}
% \end{figure}

% \new{We are interested to show that ${t_{wk} \geq t_1+t_2}$.} Composability property of the RC model is beneficial in this regard.

% We are interested to show that $t_1+t_2} \geq {t_{wk}$.

\begin{thm}\label{thm:min_waking_time_multi}
The minimum value of waking time for a maximum temperature constraint is when the server on each core exhausts all its budget at once, simultaneously. 
\end{thm}
\begin{proof}
Assume we have a dual-core CPU with the same physical characteristics and surrounding conditions. Both cores have the same periodic step power signal but with a phase difference of $\phi$. For simplicity we assume $P_S$ is zero and $P=P_D$. The input power of the first core $P_1(t)$ starts at $t=t_0$ while the input power of the second core $P_2(t)$ starts with a phase change at $t=t_0+\phi$. We consider two cases: for case I, the phase change is zero ($\phi=0$), and for case II it is positive ($\phi>0$). The schematic of power signal for the two cases is depicted in Fig.\ref{fig:chap3_thrm4_power}. The temperature profiles of in-phase and out-of-phase cases are compared in Fig.\ref{fig:chap3_thrm4_temperature}.

\begin{figure}[ht]
\centering
\includegraphics[width=0.78\textwidth]{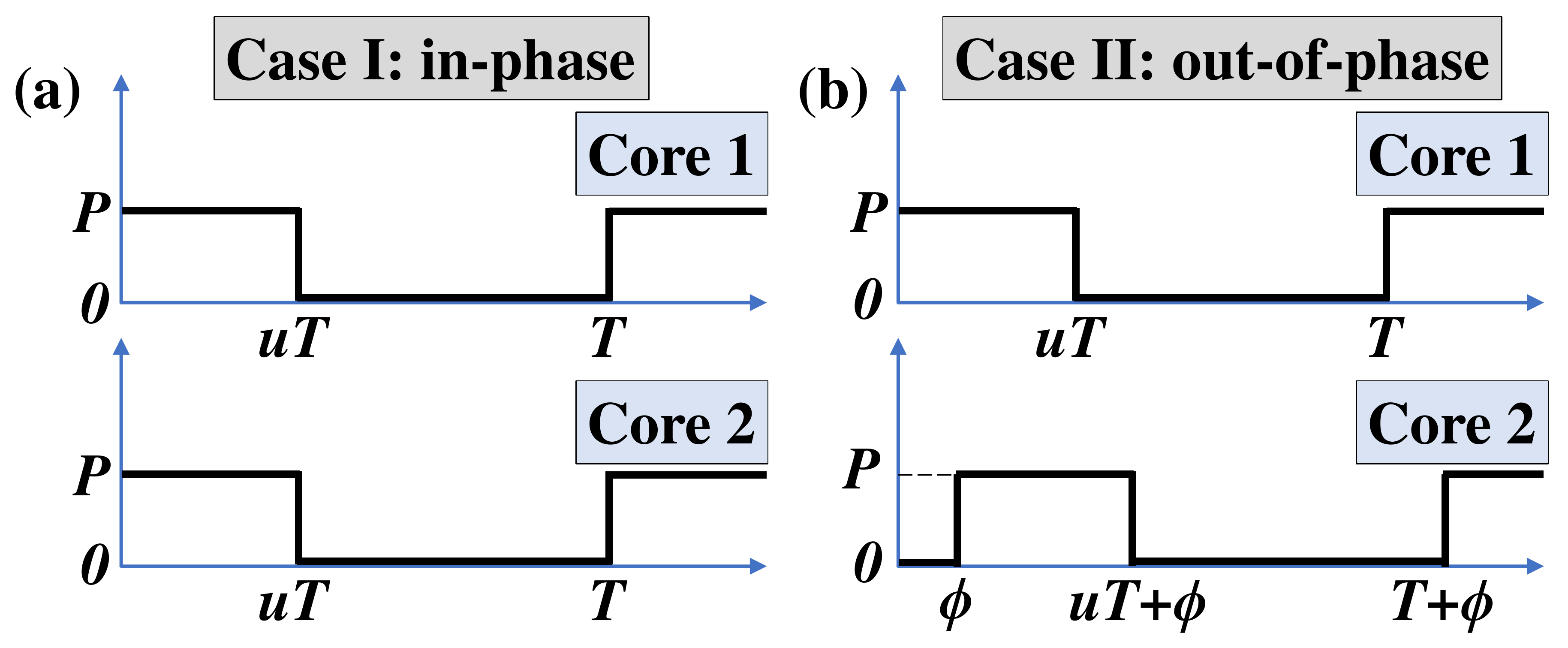}
\caption{Power signal of the cores for (a) case I: in-phase, and (b) case II: out-of-phase.}
\label{fig:chap3_thrm4_power}
\end{figure}

\begin{figure}[ht]
\centering
% \vspace{-10pt}
\includegraphics[width=0.95\textwidth]{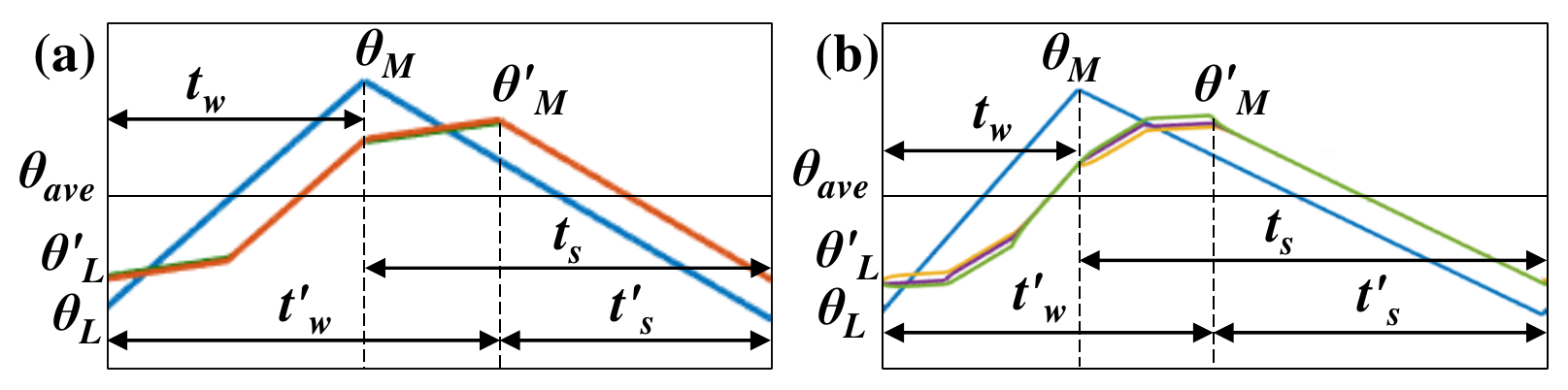}
\caption{Temperature profiles of in-phase and out-of-phase cases for a (a) two-core and (b) three-core system. (Blue lines represent the temperature of the cores in in-phase and other colors represent temperature of the cores in out-of-phase states.) }
\label{fig:chap3_thrm4_temperature}
\end{figure}

\textit{Lemma}. In a multi-core system, if the power signal of each core varies with time in the way described above, the maximum change rate magnitude of temperature is when the powers are in-phase ($\phi=0$). 

\textit{For a \new{dual-core CPU}}. According to the developed model, temperatures of the two cores between $t_0$ and $t$ when the power $P_1$ and $P_2$ are constant are as follows: 
\[
 \scriptstyle
\begin{aligned}
\theta  = \frac{1}{2}{e^{{\ali{\beta_1}}\left( {t - {t_0}} \right)}}\left[ {\begin{array}{*{20}{c}}
{{\theta _0}_1 + {\theta _0}_2}\\
{{\theta _0}_1 + {\theta _0}_2}
\end{array}} \right] + \frac{1}{2}\;{e^{{\ali{\beta _2}}\left( {t - {t_0}} \right)}}\left[ {\begin{array}{*{20}{c}}
{{\theta _0}_1 - {\theta _0}_2}\\
{{\theta _0}_2 - {\theta _0}_1}
\end{array}} \right] + \\
\frac{{{B_1}}}{{2{\ali{\beta_1}}}}\left( {{e^{{\ali{\beta_1}}\left( {t - {t_0}} \right)}} - 1} \right)\left[ {\begin{array}{*{20}{c}}
{{P_1} + {P_2}}\\
{{P_1} + {P_2}}
\end{array}} \right] + 
\frac{{{B_1}}}{{2{\ali{\beta_2}}}}\left( {{e^{{\ali{\beta_2}}\left( {t - {t_0}} \right)}} - 1} \right)\left[ {\begin{array}{*{20}{c}}
{{P_1} - {P_2}}\\
{{P_2} - {P_1}}
\end{array}} \right]
\end{aligned}
\]

\begin{figure}[ht]
\centering
% \vspace{-10pt}
\includegraphics[width=0.95\textwidth]{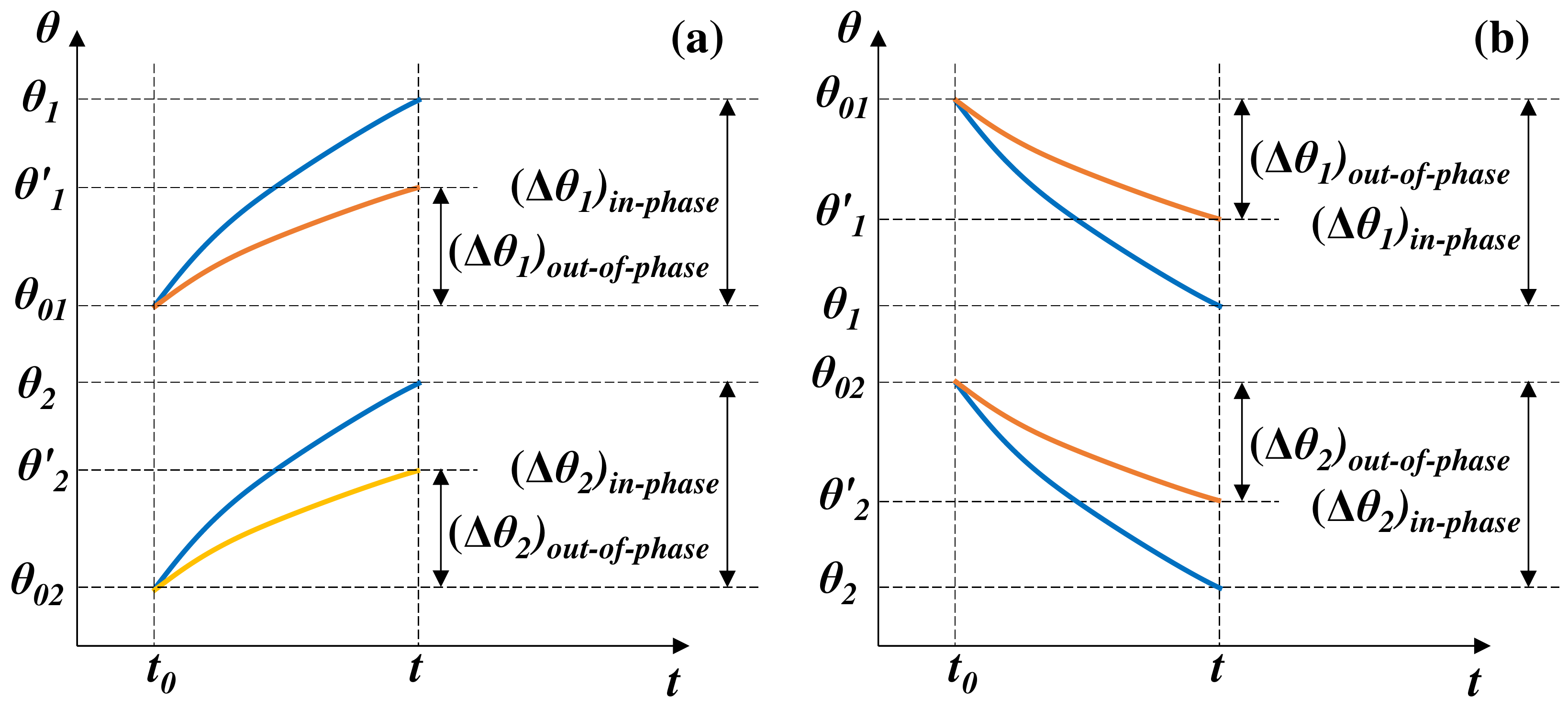}
\caption{Comparison of temperature (a) increase and (b) decrease between in-phase and out-of-phase cases.}
\label{fig:chap3_thrm4_tempderivative}
\end{figure}

Assume that core 1 and 2 start from initial temperatures of $\theta_{01}$ and $\theta_{02}$ at time $t_0$ and get to the final temperatures of $\theta_1$ and $\theta_2$ at time $t$ as shown in Fig.\ref{fig:chap3_thrm4_tempderivative} . We want to show that $\left| {{\rm{\Delta }}{\theta _{in - phase}}} \right| > \left| {{\rm{\Delta }}{\theta _{out - of - phase}}} \right|$  for any time slot between $t_0$ and $t$ where the power remains constant. We calculate the derivative of temperature for the two cases. When there is a power for the in-phase case ($P\neq0$), the temperate change rate is positive. For in-phase case $P_1+P_2=2P$ and $P_1-P_2=0$. For the out-of-phase case, ${P_1+P_2=P}$ and ${P_1-P_2=-P}$ because $P_1=0$ and $P_2=P$ or vice versa. We have: (I represents in-phase and II represents out-of-phase)

\[
\begin{aligned}
{\left( {\frac{{d\theta }}{{dt}}} \right)_{I}} - {\left( {\frac{{d\theta }}{{dt}}} \right)_{II}} =
\frac{{{B_1}P}}{2}\left[ {\begin{array}{*{20}{c}}
{{e^{{\ali{\beta_1}}\left( {t - {t_0}} \right)}} + {e^{{\ali{\beta_2}}\left( {t - {t_0}} \right)}}}\\
{{e^{{\ali{\beta_1}}\left( {t - {t_0}} \right)}} - {e^{{\ali{\beta_2}}\left( {t - {t_0}} \right)}}}
\end{array}} \right]
\end{aligned}
\]

Since \ali{$\beta_1 > \beta_2$}  for both cores and $B_1$ and $P$ are positive, then ${\left( {\frac{{d\theta }}{{dt}}} \right)_{I}} \ge {\left( {\frac{{d\theta }}{{dt}}} \right)_{II}}$. Same conclusion can be drawn if $P_1=P$ and $P_2=0$. We can conclude that ${\rm{\Delta }}{\theta _{in-phase}} \ge {\rm{\Delta }}{\theta _{out-of-phase}}$ based on the fact that if $f\ge g$ in $\left[{a,b}\right]$, and $f$ and $g$ are integrable in $\left[{a,b}\right]$, then $\mathop \smallint \limits_a^b fdx\ge\mathop\smallint\limits_a^b gdx$. If the power in the in-phase case is zero ($P_1+P_2=0$ and $P_1-P_2=0$), the temperature change rate is negative, and for $P_1=0$ and $P_2=P$ we have: 
\[\begin{aligned}
{\left( {\frac{{d\theta }}{{dt}}} \right)_I} - {\left( {\frac{{d\theta }}{{dt}}} \right)_{II}} = \frac{{{B_1}P}}{2}\left[ {\begin{array}{*{20}{c}}
{ - {e^{{\ali{\beta_1}}\left( {t - {t_0}} \right)}} + {e^{{\ali{\beta_2}}\left( {t - {t_0}} \right)}}}\\
{ - {e^{{\ali{\beta_1}}\left( {t - {t_0}} \right)}} - {e^{{\ali{\beta_2}}\left( {t - {t_0}} \right)}}}
\end{array}} \right]
\end{aligned}\]

Since \ali{$\beta_1 > \beta_2$}  for both cores ${\left( {\frac{{d\theta }}{{dt}}} \right)_{I}} \le {\left( {\frac{{d\theta }}{{dt}}} \right)_{II}}$. Same conclusion can be drawn if $P_1=P$ and $P_2=0$. Therefore, in case of an increase in temperature, the in-phase state has the highest temperature change rate, and in case of a decrease in temperature, the in-phase state has the lowest temperature change rate. In conclusion, $\left| {{\rm{\Delta }}{\theta _{in - phase}}} \right| \ge \left| {{\rm{\Delta }}{\theta _{out - of - phase}}} \right|$.

\textit{For \new{a multi-core CPU}}. For a multi-core system when temperature is increasing, the difference between the temperature derivatives of the in-phase (I) and out-of-phase (II) cases between $t_0$ and $t$ when the power signal does not change is as follows:

\[%\resizebox{1\linewidth}{!}{$
\begin{aligned}
{\left( {\frac{{d\theta }}{{dt}}} \right)_I} - {\left( {\frac{{d\theta }}{{dt}}} \right)_{II}} = \frac{{{B_1}}}{n}{e^{{\ali{\beta_1}}\left( {t - {t_0}} \right)}}\left( {n - {k_1}} \right)P{\left[ {\begin{array}{*{20}{c}}
1\\
1\\
 \vdots \\
1
\end{array}} \right]_{n \times 1}} \\
- \frac{{{B_1}}}{{{k_2}}}{e^{{\ali{\beta_2}}\left( {t - {t_0}} \right)}}{\left[ {\begin{array}{*{20}{c}}
{{m_{21}}P}\\
{{m_{22}}P}\\
 \vdots \\
{{m_{2n}}P}
\end{array}} \right]_{n \times 1}} -  \ldots  - \frac{{{B_1}}}{{{k_n}}}{e^{{\ali{\beta_n}}\left( {t - {t_0}} \right)}}{\left[ {\begin{array}{*{20}{c}}
{{m_{n1}}P}\\
{{m_{n2}}P}\\
 \vdots \\
{{m_{nn}}P}
\end{array}} \right]_{n \times 1}}
\end{aligned}
%$}
\]

${k_i}$, $i = 2,..,n$ are integers which satisfy $1\le{k_i}\le n,\,\,i = 2,..,n$. For ${k_1}$ we have $1\le {k_1} < n$ since there is at least one core which is idle when powers are in out-of-phase. $m_{ij}$ can change based on how many cores are active and we have $m_{ij} < k_i$. Since $n > {k_i}$, the first term is positive, and since \ali{${\beta _1} \gg {\beta _2},...,{\beta _n}$} and all \ali{$\beta$s} are negative, the other terms decrease exponentially to zero and can be neglected compared to the first term. Therefore, ${\left( {\frac{{d\theta }}{{dt}}} \right)_I} \ge {\left( {\frac{{d\theta }}{{dt}}} \right)_{II}}$. It can be shown in the same way that when there is no power in the in-phase case, the temperature is decreasing and ${\left( {\frac{{d\theta }}{{dt}}} \right)_I} \le {\left( {\frac{{d\theta }}{{dt}}} \right)_{II}}$. 

\textit{Lemma}. For the described power signals, the maximum temperature for in-phase case is larger than that of the either core for the out-of-phase case (${\theta _M} \ge \theta _M^{'}$). \\
First, let’s point out that ${\theta _{ave}} =  - {{\bf{A}}^{ - 1}}{\bf{B}}{{\bf{P}}^\infty }$ is the same for both cases in the steady state. Therefore, ${\theta _M} + {\theta _L} = \theta _M^{'} + \theta _L^{'}$. Assume that ${\theta _M} < \theta _M^{'}$, then ${\theta _L} > \theta _L^{'}$. Assume that the time it takes for the in-phase case temperature to get from ${\theta _L}$ to ${\theta _M}$ is $t_w$ and the time it takes to get from ${\theta _M}$ to ${\theta _L}$ is $t_s$. Assume these times for the out-of-phase case is $t{'_w}$ to get from $\theta {'_L}$ to $\theta {'_M}$ and $t{'_s}$ to get from $\theta {'_M}$ to $\theta {'_L}$. If ${\theta _M} < \theta _M^{'}$, and ${\theta _L} > \theta _L^{'}$, then $t_w$ would be smaller than $t{'_w}$ because the temperature increase rate is largest for the in-phase case. At the same time, ${t_s} < t_s^{'}$ since temperature decrease rate is largest for the in-phase case. Therefore, ${t_w} + {t_s} < t{'_w} + t{'_s}$. But the periods for the two cases are the same ${t_w} + {t_s} = t{'_w} + t{'_s}$.
\end{proof}

\noindent\textbf{Experimental Example }
To support our claim, we  measure the temperature of big cores on the Exynos 5422 SoC~\cite{exynos} using IR camera FLIR A325sc~\cite{flir} with the sampling rate of 60 frames per second. Four periodic computationally-intensive workloads are ran on four big cores of the board. In all cases, the CPU frequency is set to 1.4 GHz, and each workload executes every 10 seconds with the utilization of $40\%$. Let $\phi(i)$ denote the delay of starting time of a workload execution on core $i$. Table~\ref{tab:chap3_cases} shows the configurations of each case.

\begin{table}[h!]
\centering
 \caption{ Descriptions of workload executions on big cores.}
  \label{tab:chap3_spec}
    \begin{tabular}{ |p{1.1cm}|p{8cm}|p{2.3cm}|}
    \hline
     Name & description & workloads delay settings (seconds)\\
    \hline
   \begin{footnotesize} Case 1 \end{footnotesize}& \begin{footnotesize}all workloads execute at the same time \end{footnotesize}&\begin{tiny}  ${\phi(1)=\phi(2)=0}$ ${\phi(3) = \phi(4) = 0}$ \end{tiny}\\ \hline
   \begin{footnotesize} Case 2 \end{footnotesize}& \begin{footnotesize}workloads on two cores begin after those of other cores \end{footnotesize}&\begin{tiny} ${\phi(1) = \phi(2) = 0 }$ ${\phi(3) = \phi(4)=4}$ \end{tiny}\\
    \hline
    \begin{footnotesize} Case 3 \end{footnotesize}& \begin{footnotesize} workloads on two cores begin 1 second before finishing those of on other cores \end{footnotesize}&\begin{tiny}  ${\phi(1) = \phi(2) = 0}$ ${\phi(3) = \phi(4) = 3} $ \end{tiny}\\ \hline
   \begin{footnotesize} Case 4 \end{footnotesize} & \begin{footnotesize}workloads on each core execute with a 2-second overlap \end{footnotesize}&\begin{tiny}  ${\phi(1) = 0 \ \phi(2) =2}$   ${\phi(3) = 4\  \phi(4) = 6} $ \end{tiny}\\
    \hline
    \end{tabular}
    \label{tab:chap3_cases}
\end{table}

\begin{figure}[h]
\centering
\includegraphics[width=0.62\textwidth]{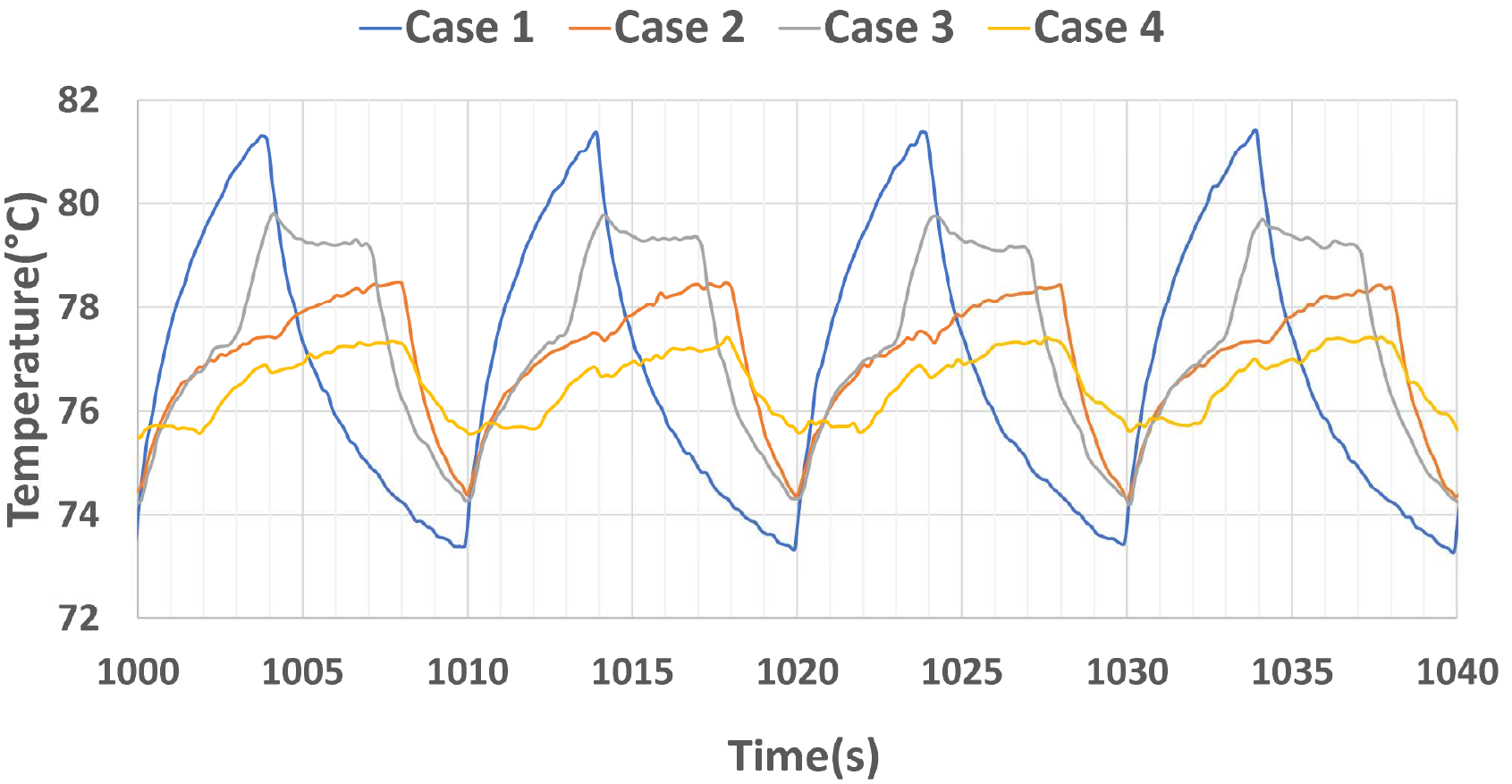}
\caption{The maximum temperature of big cores in the Exynos 5422 captured by FLIR A325sc IR camera in the operating frequency of 1.4 GHz without heat sink.}
\label{fig:chap3_thrm4_observ}
\end{figure}

Fig.~\ref{fig:chap3_thrm4_observ} shows our observations of the maximum temperature of the SoC for the described cases in the steady state. As one can see, in Case 1 where all workloads execute synchronously, the maximum temperature of the chip reaches its highest value and its minimum temperature is the lowest among all cases. On contrary, in Case 4 where workloads of different cores have the least overlap, the maximum temperature of the chip fluctuates the least of other cases and it is close to the average temperature in the steady state. In Case 2 where there is no overlap between executions of two pairs of workloads, the rate of temperature rise is lower than the Case 3 where there is 1 second overlap within all workload executions.

\section{Multiple Server Analysis}

In this section, we extend our analysis for multiple servers running on the multi-core CPU at each criticality level. We are interested to see if there \new{are} enough sleeping slacks between active times of servers to cool down the CPU. The cooling time must be large enough such that the CPU does not exceed the maximum temperature constraint under any circumstance. 
%Since in our framework, the maximum temperature of the CPU is when the system is in the steady state, we focus our analysis on the steady state condition. 
Hence, we should check if the amount of \new{the} sleep\new{ing} time required for cooling is guaranteed for a given period of time. 
%In other words, we propose a mechanism according to the single-server analysis, for the reachability of enough sleeping time at the longest possible period. 

We propose an {\it idle thermal server} technique in this regard.
Unlike regular servers, the idle server does not execute; instead, its budget represents the amount of time that the CPU core needs to be idle in the cooling phase. \new{\chk{The reasoning behind this technique is to simplify the modeling of the resulting idle time from the execution of multiple regular servers as a single budget parameter. \cmnt{Hence, the idle server budget can be determined such that heat dissipation during the idle server's inactive time reaches the same or higher maximum operating temperature than that of running multiple servers under any task execution pattern.}} Hence, the idle server's budget can be determined such that the maximum CPU operating temperature caused by heat dissipation during the idle server's inactive time is the same as or higher than that of running regular servers under any task execution pattern.} If such an idle server is schedulable, one can conclude that the given taskset is thermally schedulable. \new{The thermal effect of multiple servers is analyzed by the complement signal of periodic idle-server execution, the worst-case behavior of which has been proved by Theorems~\ref{thm:superposition}-\ref{thm:min_waking_time_multi}.}

\cmnt{If this idle server is schedulable, the CPU core will never exceed the maximum temperature constraint.} After analytically finding the relation between the budget and period of the idle server, we check its schedulability. 
%In other words, we try to find the feasibility of an idle server equivalent to all servers running of each CPU core. 

Since the idle server does not actually exist on the CPU, it is considered as the lowest-priority server in the schedulability test. In our proposed framework, the CPU is not forced to sleep so that the proposed idle server has no effect on the timing schedulability of regular running servers.

%We define an idle server whose utilization is complimentary utilization of all servers running on the corresponding CPU core. 
The idle server utilization corresponds to the time during which all regular servers are deactivated. Based on this, we investigate the feasibility of the idle server with its minimum possible period within a valid range. In this work, we focus on designing homogeneous idle servers, i.e., all idle servers on different CPU cores share the same parameters at each criticality level. It is worth mentioning that this does not mean that sleeping time in different cores must happen at the same time. Finding the different idle server settings for each CPU core is beyond the scope of this chapter. \cmnt{To this end, the thermal model of the multi-core has to change and the formulae will change according to eigen matrix of the CPU cores and their operating power signal functions.} 

\smallskip\noindent\textbf{Idle server design}
First, we compute the total utilization of the CPU core $c$ at the critical\new{ity} level $l$ by
${u^{l}_c = \sum_{\forall i\ \mathbb{P}(v_i) = c \land v_i \in V^l }\frac{C_i}{T_i}}$, where $\mathbb{P}(v_i)$ is the CPU core assigned to $v_i$.  
The maximum per-core CPU utilization is then $u^{l}_{max} = \max_{\forall c}{u^{l}_c}$. Due to the homogeneity of idle servers on all cores, $u^{l}_{max}$ is considered as the utilization of one core so that each core is supposed to be idle at least $1-u^{l}_{max}$. As a result, we are looking for a server that can be schedulable with the amount of utilization of at least $1-u^{l}_{max}$. Let $u^{l}_{idle}$ denote  this value. According to Theorem~\ref{thm:util}, the period of the idle server $T^{l}_{idle}$ can be modeled as:
\begin{equation}\label{eq:chap3_tuidle}
T^{l}_{idle} \leq \frac{\ln({1+\alpha \frac{ e^{1-u^{l}_{idle}+\frac{1}{2\beta}}-1}{\theta_M}})}{2\beta}.
\end{equation}
 It is worth noting that the period is determined based on the back-to-back execution under in presence of multiple servers for both polling and deferrable budget replenishment policies since servers can preempt each other and sleeping time does not happen in a contiguous manner. 
 $T^{l}_{idle}$ is an increasing function in terms of utilization. As one can figure out from Eq.~\ref{eq:chap3_tuidle}, an increase in the workload on the CPU core (hence resulting in a decrease in $u^{l}_{idle}$) leads to a decrease in $T^{l}_{idle}$. In other words, under heavier workload, the CPU has to sleep more frequently but in a shorter duration, to satisfy the maximum temperature constraint.

\smallskip\noindent\textbf{Thermal schedulability}
Hereby, we present our proposed thermal schedulability test for a specific utilization of idle servers. The  worst-case response time of each idle server of each core at each criticality level can be obtained as follows
\begin{equation}\label{eq:chap3_resp}
R^{n+1,l}_{idle,c} = u^{l}_{idle}\times T^l_{idle} + \sum_{\forall i\ \mathbb{P}(v_i) = c \land v_i \in V^l}\ceil*{\frac{R^{n,l}_{idle,c}}{T_i}}C_i
\end{equation}
where $R^{n+1,l}_{idle,c}$ denotes the worst-case response time of the idle server on the core $c$ at the critical\new{ity} level $l$. As shown in Eq.~\ref{eq:chap3_resp}, the idle server of each core can be preempted by all active servers at $l$. The only unknown parameter in above test is the idle server settings. Next, we discuss the optimal valid range of idle server's setting range to reduce the number of tests for each criticality level.
\begin{figure}[h]
\centering
\includegraphics[width=0.62\textwidth]{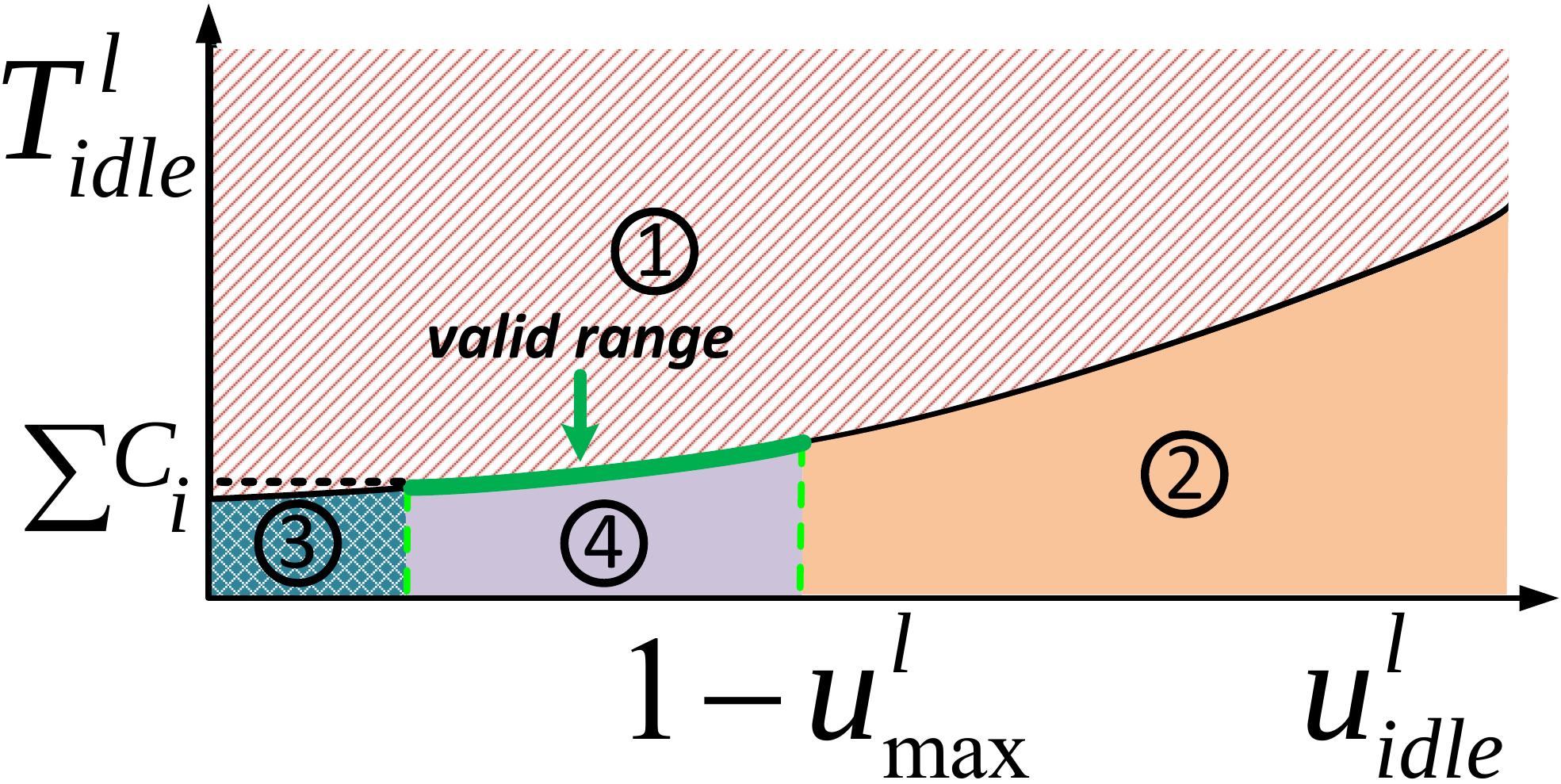}
\caption{ Search space for the idle server settings.}
\label{fig:chap3_multi-phase}
\end{figure}

\smallskip\noindent\textbf{Optimal server setting range}
As discussed in Eq.~\ref{eq:chap3_tuidle}, the period of the idle server is an increasing function in terms of its utilization. Fig.~\ref{fig:chap3_multi-phase} plots the period with respect to utilization of the idle server. The area of the plot is divided to four regions. We discuss that only the highlighted area needs to be searched for finding the valid settings of idle servers.
%the reasons that there is no need to search for the parameters of idle servers in any other regions than the highlighted region.
%and search in only the highlighted line is necessary and sufficient. 

%The reasoning for each region is as follows:
\begin{itemize}
    \item \textbf{region \circled{1}:} In this region, the CPU violates the maximum temperature constraint according to Eq.~\ref{eq:chap3_tuidle}
    \item \textbf{region \circled{2}:} The utilization of idle server has to be less than $1-u^l_{max}$. Choosing a setting in this region causes the system to be unschedulabe at this criticality level.
    \item \textbf{region \circled{3}:} Since idle servers have the lowest priority, they cannot preempt other servers. According Eq.~\ref{eq:chap3_resp}, in the worst case all servers preempt the idle servers.
    \item \textbf{region \circled{4}:} Valid settings can be found in this region, but it is unnecessary to search the entire of this region because there exists a valid setting with period of $T'$, a solution is valid on the highlighted range with the same utilization value.
\end{itemize}

Therefore, the only period of ${T_{idle} = \frac{\ln({1+\alpha \frac{ e^{1-u_{idle}+\frac{1}{2\beta}}-1}{\theta_M}})}{2\beta}}$ within the optimal range of utilization, $u_{idle}$, needs to be checked. 
One may try to insert $T_{idle}$ into Eq.~\ref{eq:chap3_resp} and assign $T_{idle}$ to ${R^{n+1,l}_{idle,c}}$ to find a single optimum point. However, because of the ceiling operation in Eq.~\ref{eq:chap3_resp}, finding the optimum $u_{idle}$ would need an exhaustive search.

\new{After finding the idle server settings, the critical ambient temperature for each criticality level is computed by using  Eq.~\ref{eq:chap3_1009} with  ${u = 1-u^l_{idle}}$. The shifting time from criticality level $l+1$ to $l$ can be determined by applying Eq.~\ref{eq:chap3_1012}.} %Considering $\theta_0$ as the maximum operating temperature of criticality level $l+1$ in Eq.~\ref{eq:chap3_1012}, the shifting time from criticality level $l+1$ to $l$ can be determined.}   
% Next, we will discuss the timing schedulability of servers and tasks at each criticality level.

% \subsection{Timing Schedulability Analysis}

% The thermal analysis in the previous section with the aim of the notion of idle server guaranteed the CPU operates under the maximum temperature constraint of the system at any criticality level. Hereby, we present the schedulability analysis of a task $\tau_i$ in our framework. Our analysis considers each of the polling  and deferrable budget replenishment server policies. It is noteworthy that the response time test is performed for each criticality level.

% Because of separating the thermal schedulability analysis for multiple servers on the multi-core CPU platform, we are able to use the existing response time test for independent tasks with no thermal constraints and under hierarchical scheduling~\cite{saewong2002analysis}, which is 
\noindent\textbf{Timing schedulability}
Due to our separate thermal schedulability analysis,
we are able to use the existing response time test developed for independent tasks with no thermal constraints under hierarchical scheduling~\cite{saewong2002analysis}:
\chk{
\begin{equation}
\label{eq:chap3_hierchical} \begin{split}
 W_i^{n+1,l}   =  E_i + \sum_{\begin{subarray}{h} \tau_h \in \mathbb{V}^l(\tau_i) \\
       h > i \end{subarray} }\ceil*{\frac{W_i^{n,l}  +J_i + (W_{h}^{l} - E_h)}{D_h}}E_h +
       \ceil*{\frac{W^{n,l}+C_j}{T_j}} (T_j - C_j)
\end{split}
\end{equation}
}
% \chk{
% \begin{equation}
% \label{eq:chap3_hierchical}\resizebox{.88\linewidth}{!}{$ \begin{split}
%  W_i^{n+1,l}   =  E_i + \sum_{\begin{subarray}{h} \tau_h \in \mathbb{V}^l(\tau_i) \\
%       h > i \end{subarray} }\ceil*{\frac{W_i^{n,l}  +J_i + (W_{h}^{l} - E_h)}{D_h}}E_h +\\
%       \ceil*{\frac{W^{n,l}+C_j}{T_j}} (T_j - C_j)
% \end{split}
% $}
% \end{equation}
% }
where $W_i^{n,l}$ is the worst-case response time of $\tau_i$ at criticality level $l$, $\mathbb{V}^l(\tau_i)$ is the server of $\tau_i$, $J_j$ is the jitter of a task running in a server $v_j$ (see Sec~\ref{sec:chap3_model}), and $W^{0,l}_i = E_i$. 
% The recursion terminates successfully when $W^{n+1}= W^n$ and  fails when $W^{n+1}> D_i$. This equation considers the budget depletion of a CPU server. The first term is the amount of CPU time used by the task $\tau_i$, the second term captures the back-to-back execution of each higher-priority task $\tau_h$,  and the third term captures the amount of interference that the server can generate due to the periodic budget replenishment. To guarantee that tasks on lower criticality modes have no effect on response time of higher criticality ones, our proposed framework terminates the execution of lower-criticality servers and stops servicing the running lower-criticality task immediately. 

\section{Evaluation}
\label{sec:chap3_eval}
This section gives the experimental evaluation of our framework. First, we show the model validation by measuring the physical system parameters on a real platform. After the analysis of server characteristics in different ambient temperatures, we present our discussion with a case study.

\subsection{Experimental Platform}
The experimental platform is an ODroid-XU4 development board~\cite{ODROIDXU4} equipped with a Samsung Exynos5422 SoC. There exist two different CPU clusters of little Cortex-A7 and  big Cortex-A15 cores, where each cluster consists of four homogeneous cores. Built-in sensors with the sampling rate of 10 Hz with the precision of $1\textdegree$C are on each big CPU core to measure the chip temperature. Note that there are no temperature sensors on little cores since the power consumption and heat generation of the little cluster is considerably low. The DTM throttles the frequency of the big CPU cluster to 900 MHz when one of its cores reaches the pre-defined maximum temperature constraint of $95\textdegree$C. During experiments, the CPU fan is turned off and the CPU is set to run at 1400 MHz. %The temperature threshold when there is no workload on the GPU is 95\textdegree~C and when the CPU fan is forced to off, we observed that the DTM throttles the frequency at 84\textdegree~C. 

%In this section we evaluate the capability and accuracy of the developed model and the stated theorems by performing a series of related experiments. Different experiments can provided various valuable information about the behavior and characteristics of the thermal aware system. Here, we design, perform, and analyze different experiments to extract the thermal characteristics of the CPU cores to calibrate the analytical model and discuss various critical modes. These include setups with different budgets, periods, input power, and ambient temperatures. 

\subsection{Model Validation}
\begin{figure}[ht]
\centering
% \vspace{-10pt}
%\includegraphics[width=0.43\textwidth]{figs/results/Model_Validation.pdf}
\includegraphics[width=0.63\textwidth]{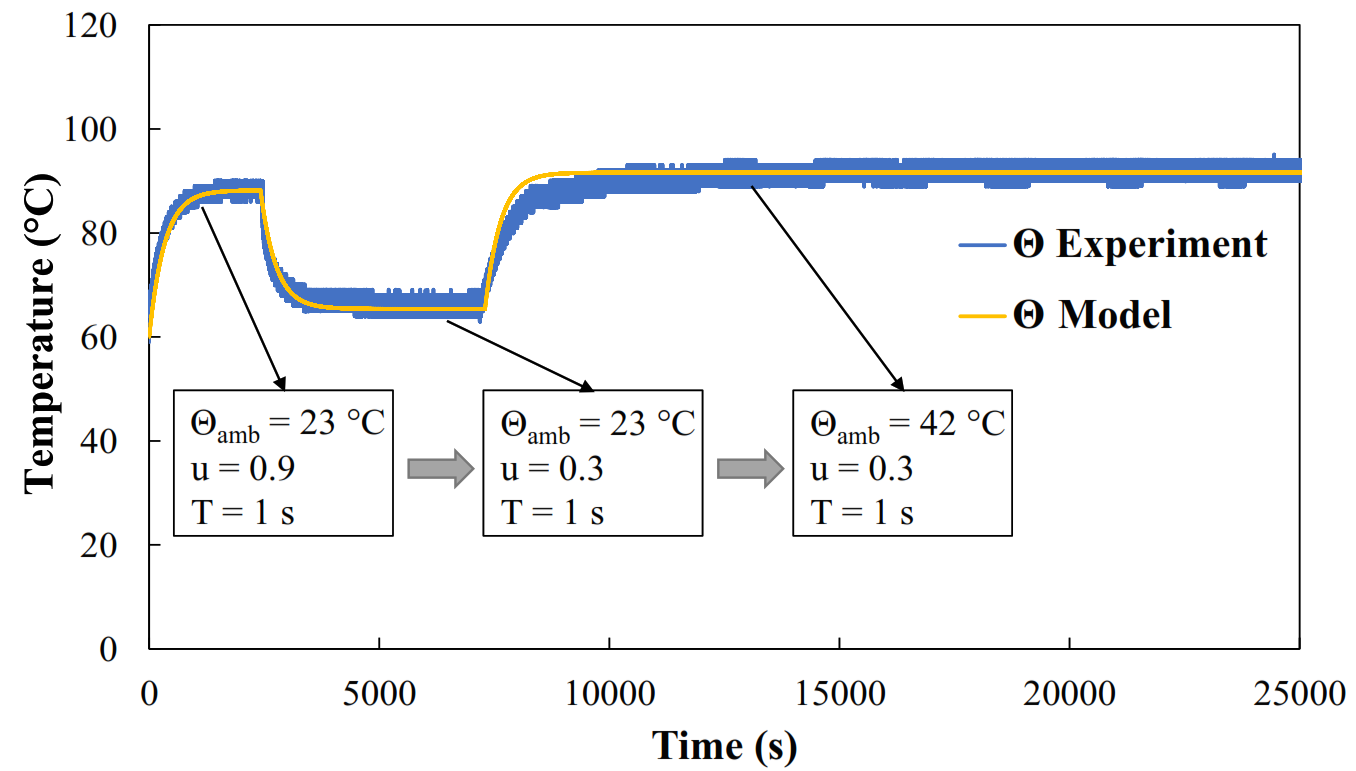}
\caption{Comparison of the experimental results and the model prediction for CPU temperature at different workloads and ambient temperatures.}
\label{fig:chap3_Model_validation}
\end{figure}

According to Eq.~\ref{eq:chap3_1003} and~\ref{eq:chap3_1019}, the characteristic matrices {\bf{A}} and {\bf{B}} can be determined by a utilization test at different CPU frequencies. A zero utilization (idle) at different ambient temperatures can reveal the range of the static and consequently the dynamic dissipation. 
% An example of frequency scaling on the CPU temperature at full utilization is shown in Fig.~\ref{fig:chap3_T_vs_f}.
After finding the thermal parameters of the system and model calibration, the analytical model is validated with the experimental results. For this purpose the CPU temperature is recorded at 90\% utilization with a period of 1 s in $\Theta_{amb}$ of $23\textdegree$C. Then utilization is decreased to 30\% and the CPU is cooled down until reaching a steady state. Afterwards, the CPU working at 30\% is placed in the furnace with $\Theta_{amb}$ of $42\textdegree$C. The CPU temperature is recorded until the steady state. The same conditions are simulated using the developed model. Fig.~\ref{fig:chap3_Model_validation} compares the CPU temperature recorded in the experiment at different stages of workloads and ambient temperature with the predicted values by the model. It can be seen that there is a good agreement between the model and the experimental results. 
%There might be some heat losses and temperature variations during the experiment while the conditions in the prediction model are ideal.

% \begin{figure}[h]
% \centering
% \includegraphics[width=0.3\textwidth]{figs/results/T_vs_f_u1.pdf}
% \caption{An example of frequency scaling on the CPU temperature at full utilization.}
% \label{fig:chap3_T_vs_f}
% \end{figure}

\subsection{Workload, Period, and Ambient Temperature Relations}
\begin{figure}[ht]
\centering
\subfloat [][]{\includegraphics[width=0.45\textwidth]{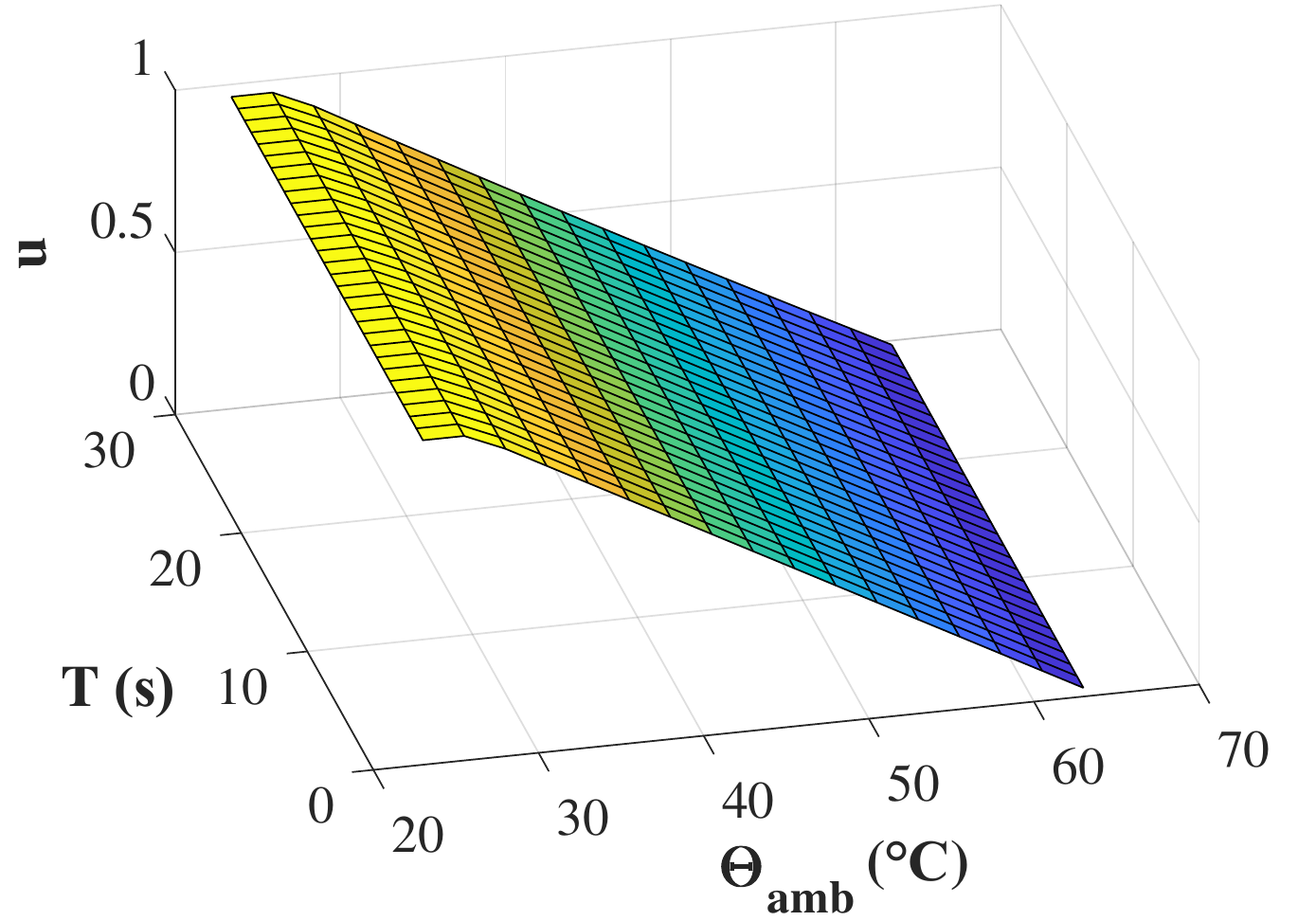}}
\subfloat [][]{\includegraphics[width=0.44\textwidth]{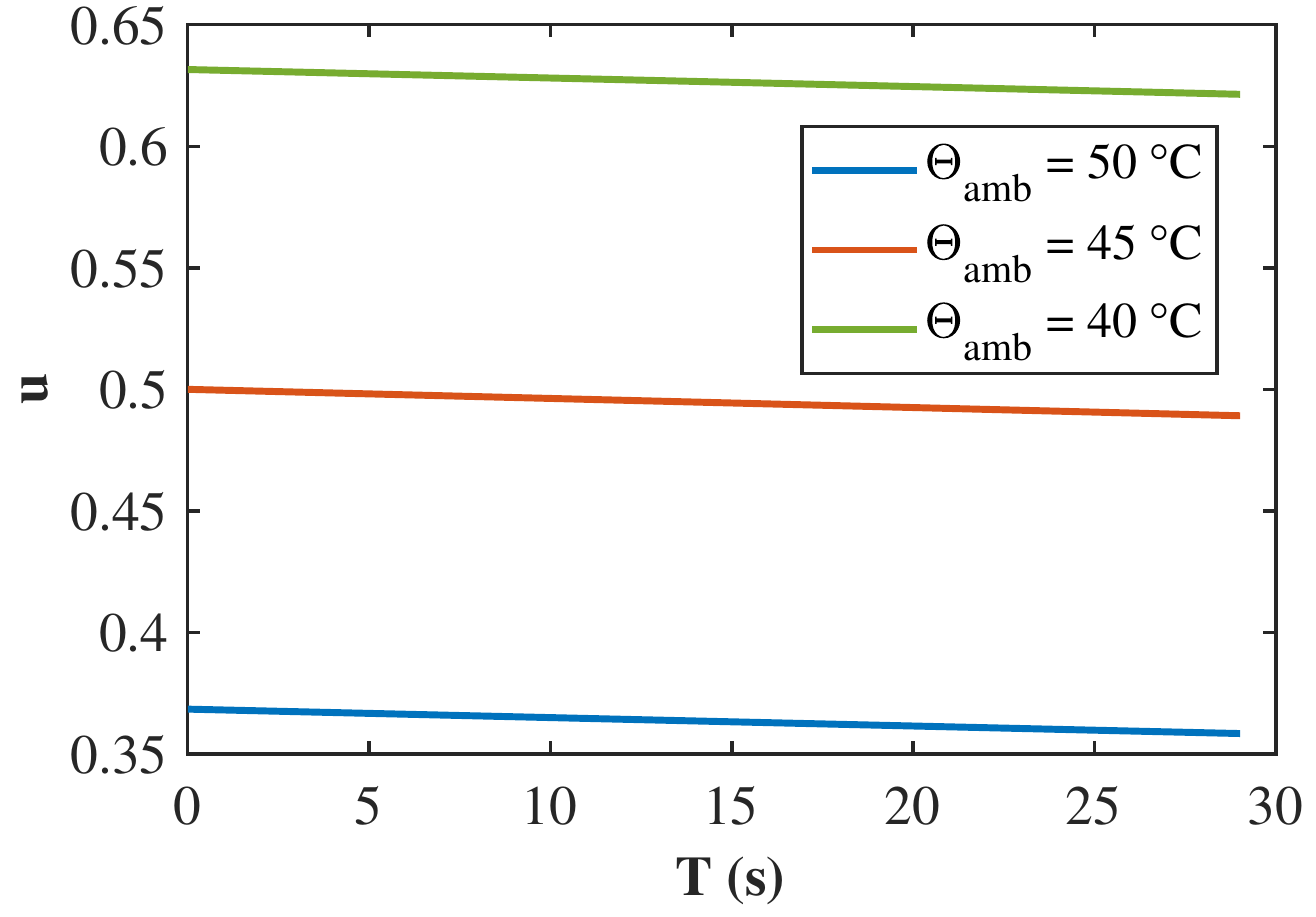}}
\caption{a) Utilization versus period and ambient temperature. b) Utilization versus period at different ambient temperatures.}
\label{fig:chap3_u_vs_T_Tamb}
\end{figure}

We investigate the effect of  ambient temperature, server period, and utilization using Eq.~\ref{eq:chap3_1009} and Eq.~\ref{eq:chap3_1010}. Assuming the temperature threshold of $\Theta_m=95\textdegree$C, the maximum allowable utilization is plotted in Fig.~\ref{fig:chap3_u_vs_T_Tamb}a against period and ambient temperatures. It can be seen that for all considered periods, when the ambient temperature is increased from $23\textdegree$C, the maximum workload decreases almost linearly with the ambient temperature. At higher ambient temperatures lower workloads can be used until the ambient temperature of $69\textdegree$C where even an idle CPU usage will result in a working temperature equal to the threshold temperature. To better see the effect of period, the maximum workload is plotted against period at different ambient temperatures in Fig.~\ref{fig:chap3_u_vs_T_Tamb}b. The period has been changed from 10 ms to 30 s. It can be seen that the maximum workload decreases by increasing the value of the period in an almost linear manner. This has been discussed and confirmed in Theorem~\ref{thm:min_waking_time}. Moreover, it can be concluded that the effect of ambient temperature on the maximum allowable workload is more prominent than the effect of period. 
% \begin{figure}[h]
% \centering
% \includegraphics[width=0.3\textwidth]{figs/results/u_vs_Tamb.pdf}
% \caption{Utilization versus ambient temperature at a period of 1 s.}
% \label{fig:chap3_u_vs_Tamb}
% \end{figure}

% \begin{figure}[h]
% \centering
% \includegraphics[width=0.3\textwidth]{figs/results/u_vs_T.pdf}
% \caption{Utilization versus period at different ambient temperatures.}
% \label{fig:chap3_u_vs_T}
% \end{figure}
\subsection{Shifting Time Analysis}
\begin{figure}[t]
\centering
\subfloat [][]{\includegraphics[width=0.43\textwidth]{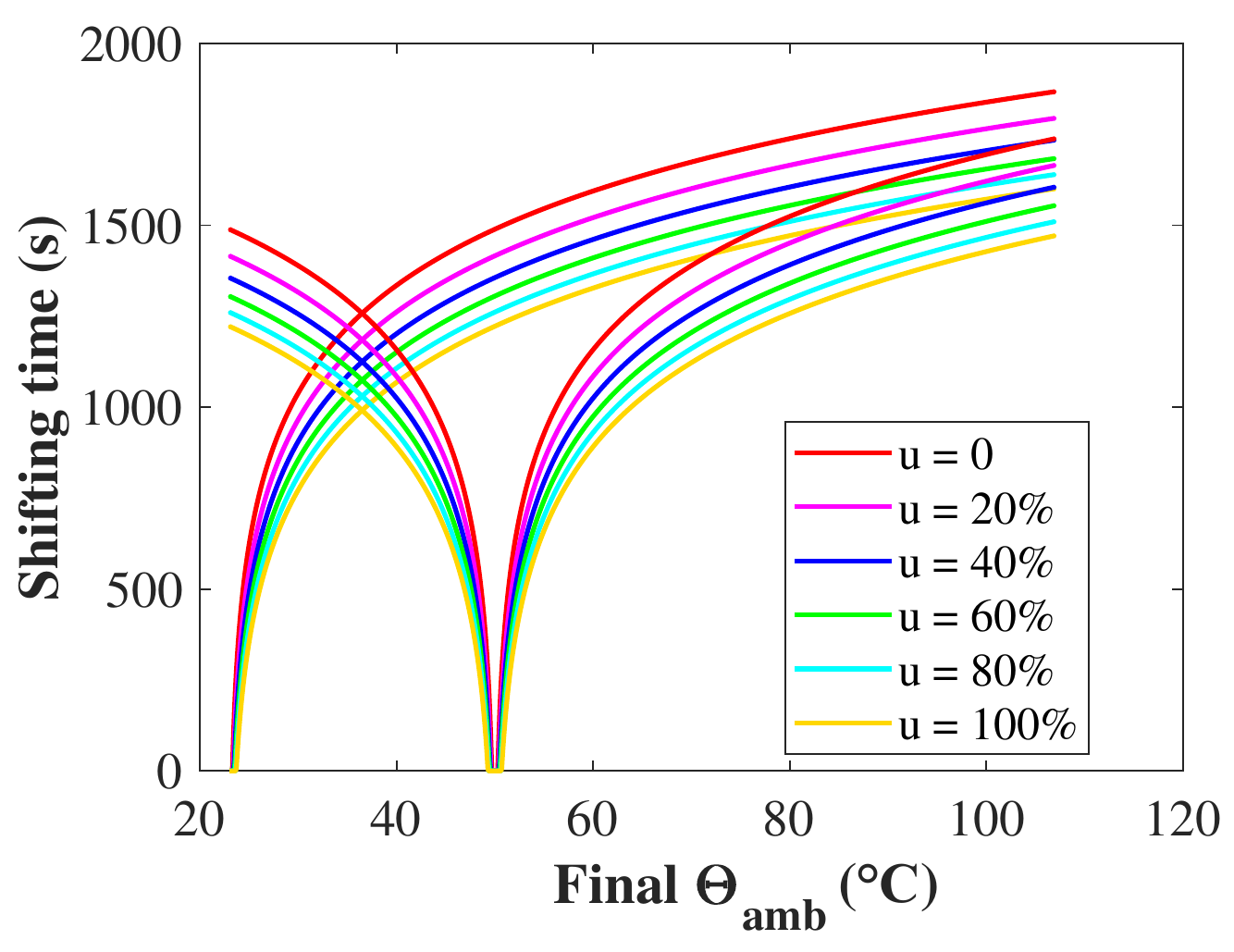}}
\subfloat [][]{\includegraphics[width=0.43\textwidth]{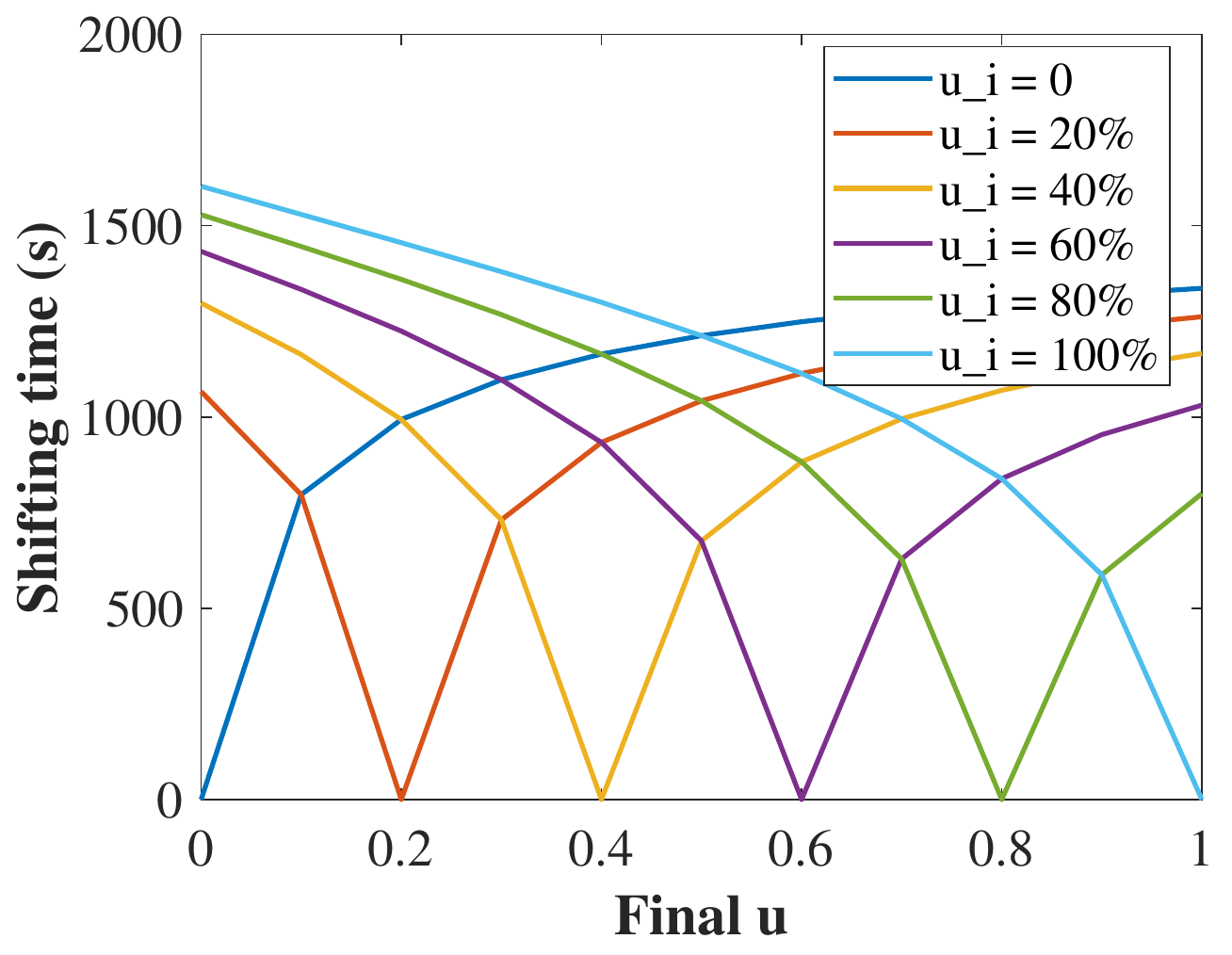}}
% \caption{a) Shifting time from initial ambient temperature of 23\textdegree~C and 50\textdegree~C to different final ambient temperatures at various workloads. b)Shifting time from different initial workloads to different final workloads at an ambient temperature of 23\textdegree~C.}
\caption{Shifting time a) from initial ambient temperature to different final ambient temperatures at various workloads, b) from different initial workloads to different final workloads at an ambient temperature of $23\textdegree$C.}
\label{fig:chap3_Shifting_Tamb}
\end{figure}

As mentioned before, any change in the working parameters of the system may cause a transient thermal behavior and change the steady state conditions. Here, we discuss the effects of changing workload and also ambient temperature on the thermal response of the CPU cores. In Fig.~\ref{fig:chap3_Shifting_Tamb}a. shifting times are plotted against the final ambient temperature at different workloads assuming the initial ambient temperature is $23\textdegree$C or $50\textdegree$C. It can be seen that at higher workloads it takes less time for the CPU to reach the steady state when ambient temperature is changed from $\Theta_{amb_i}$ to $\Theta_{amb_f}$. Also, at all workloads, it takes more time for the CPU to reach the final ambient temperature $\Theta_{amb_f}$ if it starts from a lower initial ambient temperature $\Theta_{amb_i}$. Furthermore, for all utilizations, the time it takes for the CPU to reach the steady state when the ambient temperature goes up by $\Delta \Theta$ is almost equal to the time it takes when it goes down by the same amount. 

In Fig.~\ref{fig:chap3_Shifting_Tamb}b. shifting times are plotted against the final workload $u_f$ for different initial workloads $u_i$. It can be seen that it takes more time to reach the steady state of final $u_f$ when starting from a lower workload $u_i$. Also, opposite to the case of ambient temperature, if ${u_i} < {u_f}$, it takes less time to reach the steady state when shifting from $u_i$ to $u_f$ (heating), compared to when shifting from $u_f$ to $u_i$ (cooling).  

% \begin{figure}[t]
% \centering
% \includegraphics[width=0.22\textwidth]{figs/results/Shifting_Time_u.pdf}
% \caption{}
% \label{fig:chap3_Shifting_u}
% \end{figure}

\subsection{Case Study}
We emulate the mixed-critical\new{ity} Flight Management System (FMS) application~\cite{Ahmed2017, giannopoulou2013scheduling} which comprises  two criticality levels of \textit{H} and \textit{L}. The parameters of the executing real-time tasks are given in ~\cite{Ahmed2017}\cmnt{Table~\ref{tab:spec}}. In our experiment, there exist one high\new{-}criticality and one \new{low-criticality} server on each CPU \new{core}. The budget replenishment period of each thermal-aware server is considered 50 ms under the deferrable budget replenishment policy. The budget for high-criticality and low-criticality servers are 15 ms and 27 ms, respectively. Tasks are assigned to CPU cores by using the worst-fit decreasing (WFD) heuristic for load balancing across cores and are scheduled by the Rate Monotonic (RM) policy. %Task index is chosen for tie-breaking in priority assignments. 
Since the amount of the workload in low-criticality level is insignificant to reach the maximum temperature, non-real-time tasks are also assigned to low-criticality servers with the lowest priority level. 

\begin{figure}[ht]
\centering
\includegraphics[width=0.45\textwidth]{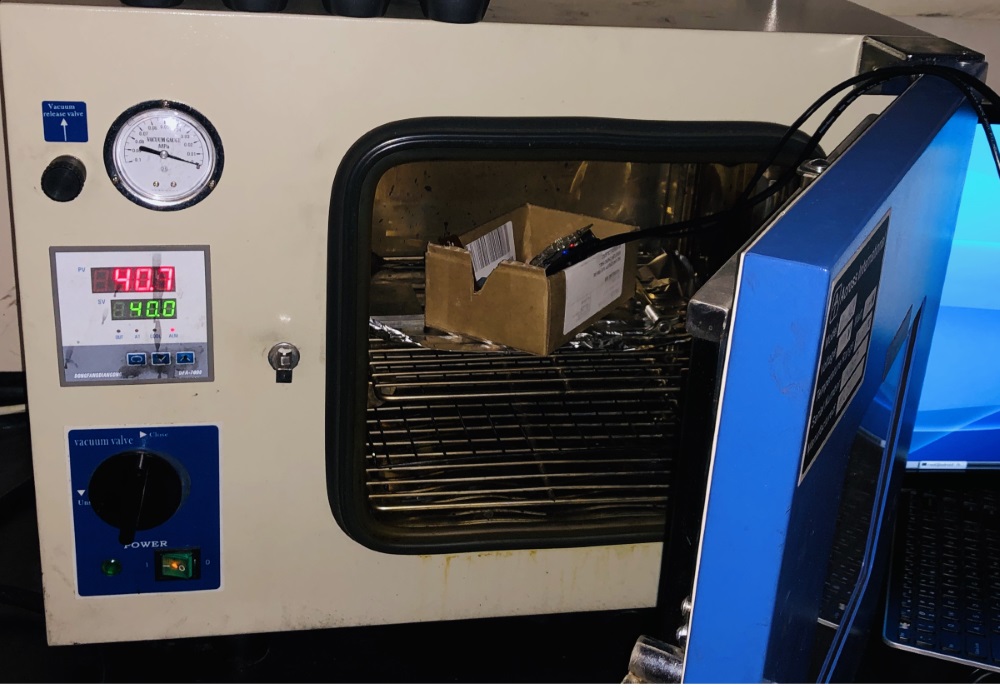}
\caption{Experimental environment using furnace.}
\label{fig:chap3_furnace}
\end{figure}

Critical ambient temperatures have been determined by Eq.~\ref{eq:chap3_1009}: $24\textdegree$C for the low-criticality and $40\textdegree$C for the high-criticality level. %These ambient temperature levels cause the operating system temperature to fall slightly lower than the maximum temperature threshold to deal with the noise of the temperature sensors. 
As shown in Fig.~\ref{fig:chap3_furnace}, the experiment has been performed in the furnace. Nordic Semiconductor Thingy:52™ IoT sensor development kit~\cite{Thingy52} is used to capture the ambient temperature with the sampling rate of 10~Hz. 

\begin{figure}[ht]
\centering
\includegraphics[width=0.55\textwidth]{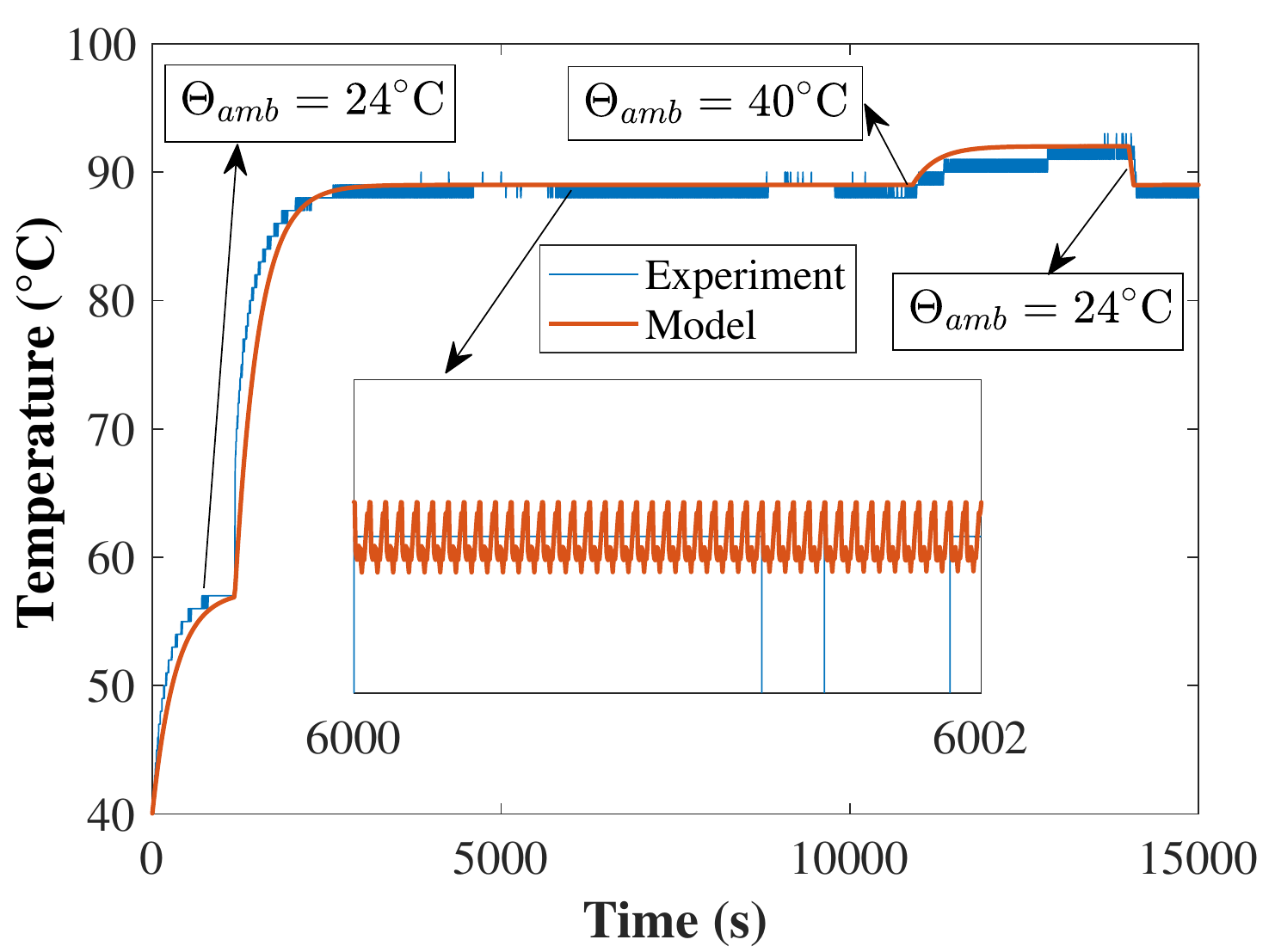}
\caption{Experimental and model results for a case study with a period of 50 ms at two ambient temperatures and workloads.}
\label{fig:chap3_case_study}
\end{figure}

Fig.~\ref{fig:chap3_case_study} shows the experimental and model results for a case study with period of 50 ms. In step I, the CPU is idling for 1800 s, and then in step II, workload with $u=95\%$ is applied at $\Theta_{amb}=24\textdegree$C. The system is left to work with $u=95\%$ until it reaches steady state conditions and keeps working for about 10000 s. The CPU temperature reaches to a value around 89$\textdegree$C. Afterwards, in step III, the CPU is placed in the furnace with $\Theta_{amb}=40\textdegree$C and the workload is changed to 30\% at the same time. The temperature increases to 92$\textdegree$C and remains steady for about 4000 s. The CPU is then taken out of the furnace and left to work with $u = 30\%$ for a fast cooling. Finally, in step IV, the workload is set to 95\% at ambient temperature $\Theta_{amb}=24\textdegree$C. It can be seen that the developed model matches the experimental results with a good accuracy. The time step in the model is more accurate than the actual temperature sensor and temperature variations for each period can be captured by the developed model. Temperature curve from 6000 s to 6002 s is zoomed for a better comparison of the variations.

\section{Summary}

In this chapter, we proposed a novel mixed-criticality thermal-aware server framework to bound the maximum temperature of CPU cores in the presence of dynamic ambient temperature. In this framework, the server schedule is flexible -- fully preemptive and priority-based. We investigated the thermal feasibility by analyzing the amount of slack between execution of preemptive thermal servers with the notion of idle servers. We presented a mechanism to optimally search the maximum ambient temperature for every critical\new{ity} level.  We provided analytical foundations to check thermal safety while temporal safety is guaranteed. Experimental results show that our proposed framework is effective in bounding the maximum temperature at every critical\new{ity} level.

\chapter{Data-driven Thermal Parameter Estimation for COTS-based Mixed-criticality Systems}

Thermal awareness is increasingly important for mixed-criticality systems deployed in harsh environments. As high chip temperature can cause frequency throttling or shutdown of processor cores at an unexpected time, many real-time scheduling techniques have been developed to ensure continuous, fail-safe operation of safety-critical tasks with stringent timing constraints. However, their practical use remains largely limited due to the fact that it is extremely difficult to obtain a precise thermal model of commercial processors without using special measurement instruments or access to proprietary information, such as the power traces of micro-architectural units and detailed floorplan maps. 

In this chapter, we propose a data-driven thermal parameter estimation scheme that is directly applicable to commercial off-the-shelf multi-core processors used in real-time mixed-criticality systems. By using a small number of thermal profiles obtained from on-chip temperature sensors, our scheme can predict and bound the processor operating temperature under dynamic real-time workloads at various CPU frequencies and ambient conditions. The thermal model derived from our scheme is fast to converge and robust against different sources of errors. Our scheme is non-intrusive, meaning that it does not require changes to the software code or the hardware packaging of the target system. Furthermore, our scheme can estimate the relative power consumption of the processor for given workload and clock frequency level. We evaluate the effectiveness of our scheme on a multi-core ARM platform.

% \begin{IEEEkeywords}
% IEEEtrtn, journal, \LaTeX, paper, template.
% \end{IEEEkeywords}

% \IEEEpeerreviewmaketitle
\section{Introduction}
One of the major concerns in recent embedded systems is the high heat dissipation caused by complex applications running on high-performance multi-core systems-on-chips (SoCs). Ambient temperature in the physical environment is another key factor that increases chip operating temperature. The high temperature also increases power consumption~\cite{ahmed2016necessary}, reduces the chip reliability~\cite{srinivasan2004impact, viswanath2000thermal}, and leads to chip burnout. Due to these reasons, many of today's operating systems (OSs) monitor the chip operating temperature using on-chip temperature sensors to check if it is within the safe thermal constraint. 

%The \textit{reactive} thermal management mechanisms implemented to protect the system from burnout rise a challenging problem. 
To protect the processor chip from thermal damage, a set of policies are defined in the thermal governor of the OS for different scenarios.
When the chip temperature crosses a trip point, a predefined cooling scenario is performed such as frequency throttling or shutting down of CPU cores~\cite{choi2004fine, herbert2007analysis, wang2009temperature}. Such thermal countermeasures, however, lead to timing unpredictability in real-time mixed-criticality systems (MCS) since the deadlines of tasks could be unexpectedly violated by reduced processing speed or temporarily unavailable CPU cores.
%task execution, and real-time tasks may miss their deadlines. This is particularly important for safety-critical applications with real-time mixed-criticality components, e.g., automotive, aerospace, manufacturing, and defense systems, where even occasional timing failures of high-criticality components can lead to catastrophic consequences. 
Extensive studies have been have been conducted to prevent the negative impact of such performance disruption, including the techniques based on Dynamic Voltage Frequency Scaling (DVFS)~\cite{fu2010feedback,ma2015improving, 7372657, 7746768, zhang2010thermal} and forced sleeping during the execution of \textit{hot} applications~\cite{chantem2010temperature, Huang2014, Youngmoon2018, kumar2011system, jayaseelan2008temperature, zhang2010thermal}. Moreover, offline analysis has been proposed to guarantee the thermal safety of real-time tasks by bounding the maximum operating temperature in  the steady~\cite{ mehdi2020dynamic, hosseinimotlagh2019thermal} and transient states~\cite{pagani2015matex} of a given system. They all assume to have a priori knowledge of precise thermal models for temperature prediction, resource management, and task scheduling, and need accurate simulation tools or extra equipment for validation. Therefore, obtaining thermal parameters of a given system is the fundamental requirement to substantiate these techniques in practice.
%key point in temperature estimation for those models. 

Despite its importance, the thermal parameter estimation of commercial off-the-shelf (COTS) processors still remains a challenging problem. 
%thermal parameters of a given system rises a crucial challenge in using the current tools and thermal management algorithms for commercial platforms and limits their practical applicability. \cmnt{These tools are designed to capture the spatial and temporal thermal dependencies on the cores using heat map.}
Existing numerical simulation tools like HotSpot~\cite{huang2006hotspot} can construct a compact thermal model for modern VLSI devices by using the resistance-capacitance (RC) thermal network to capture the transient temperature and generate the heatmap at each time instant. However, they are not directly applicable to modern COTS multi-core processors because the information required by these tools, such as power traces of micro-architectural units, detailed floorplan maps, and cooling package, is proprietary and not publicly available. An exhaustive search approach for approximating this information through reverse engineering is time-consuming and prone to unacceptably high inaccuracies, especially for transient temperature estimation. 

The latest work~\cite{rai2015calibration, rai2012power} partially addresses these issues by calibrating thermal parameters without power traces and detailed floorplans. However, it is not applicable to MCS where execution patterns are dynamic due to preemptive, priority-driven, work-conserving task schedulers~\cite{mehdi2020dynamic, hosseinimotlagh2019thermal, pagani2015matex, Ahmed2017, DSouza2017ThermalIO}. Similar to HotSpot, it uses a time-driven prediction model where the temperature is calculated for each time instant under static workloads. This makes the model not only very slow but also inflexible to capture the correct operating temperature when multiple real-time tasks with different periodicity are concurrently scheduled.
%It is also designed for single task per CPU core with invariant frequency and power consumption during execution. It is not tailored for real-time systems where tasks are suspended or preempted with other tasks. Therefore, the proposed model cannot capture the correct operating temperature when different real-time tasks are periodically run in mixed-criticality systems.  

In this chapter, we propose a fast and accurate scheme to estimate the thermal parameters of COTS multi-core processors for real-time MCS. Our scheme requires only a small number of temperature traces from on-chip thermal sensors which are widely available in today's processors. Our scheme also improves the accuracy of thermal parameters through the ensemble of measurements from different frequency levels and execution patterns. 

\smallskip\noindent\textbf{Contributions.} The contributions of this chapter are as follows:
\begin{itemize}
    \item 
    %We propose a data-driven thermal parameter estimation scheme for COTS-based multi-core real-time systems. Our scheme first estimates the relative thermal parameters of a given system using steady-state profiles, and then calibrate these values with transient-state profiles. 
    We present a thermal estimation scheme that has low computational cost by design. Given that steady-state profiles are much compact than transient-state profiles, our scheme first estimates the thermal parameters of a given system using only steady-state profiles, and then uses transient-state data for calibration purpose. 
    
    \item We characterize various sources of errors in thermal parameter estimation, and reduce their negative effects through the multiple refinement stages of our scheme. Our scheme also enables locating errors in the temperature profiles.
    \item Our scheme can identify the relative distance between CPU cores and produce an estimated chip floorplan from temperature profiles. It can also estimate the relative power consumption for a given  workload on each CPU core.
    \item We present techniques to further improve the accuracy of thermal parameters by exploiting the ensemble of measurement data obtained at various frequency and workload settings.
    %We present two improvement mechanisms for reaching more accuracy and generalizing our methods by proposing the ensemble of multi-frequency and multiple-trace.
    %\item We extract the relative power consumption of each CPU core in different frequency levels embedded in the temperature profiles and evaluate it with real power data.
    \item The effectiveness and accuracy of our proposed scheme is demonstrated with extensive experimental results on a real ARM embedded platform.
    %We evaluate our proposed scheme on a ARM and x86 multi-core platforms.
\end{itemize}

\section{Related Work}

There exist well-known numerical tools to estimate the chip operating temperature. For instance, HotSpot solves the system of differential equations using the fourth-order Runge-Kutta numerical method through very fine-granularity iterations. The authors of ~\cite{7904613, 8442110} constructed the thermal model with the measured power and temperature trace of each subsystem on a real mobile platform.  Power Blurring~\cite{ziabari2014power} calculates temperature distributions using a matrix convolution technique, in contrast to the finite-element analysis (FEA) used in other simulation tools like HotSpot. To use these tools in practice, one needs to obtain the detailed information of the target device including power traces and floorplans or to arrange special measurement equipment such as a high-precision IR camera.

The authors of \cite{rai2015calibration, rai2012power} proposed a calibration-based method to predict thermal behavior by using an impulse response model. They assume that the power consumption of each application task remains unchanged during execution. Similar to~\cite{5456979, Ahmed2017}, they employed the Generalized-Pencil-Of-Function (GPOF)~\cite{hua1989generalized} to calculate the impulse response of each application from utilization and temperature traces. In their work, the thermal effects due to conduction between CPU cores are represented as impulse responses.
%, and then constructed for each application~\cite{rai2012power} and an adaptive fitness model proposed~\cite{rai2015calibration}.
However, if an application task is preempted by another task or getting suspended, the impulse function has to switch. This spatio-temporal thermal dependency cannot be captured in their model for preemptively-scheduled tasks on the same CPU core or concurrently-executing tasks on other CPU cores. Due to this reason, despite all the other benefits provided, the applicability of their approach to real-time mixed-criticality systems is limited. In this chapter, we present a thermal model based on matrix exponential, which overcomes the aforementioned limitations of prior work and allows estimating the key thermal parameters that represent the characteristics of the semiconductor technology independent of the specificity of applications. 
\section{System Model}
We consider a homogeneous multi-core processor where each CPU core uses the same microarchitecture. Each core is assumed to have a dedicated temperature sensor that is accessible by the OS at runtime. This assumption can be easily met in many commercial processors such as Intel Core i7 and Samsung Exynos products. Reflecting reality, it is assumed that the following information is not available to use: the chip floorplan, the exact locations of on-chip temperature sensors, and the power traces of the processor.

In the rest of this section, we introduce the power and temperature model used in this chapter.
\subsection{Power Model}

 The total power consumption of CMOS circuits at time $t$ is modeled as the summation of dynamic and static powers~\cite{4484694}, i.e., ${P(t) = P_S(t) + P_D(t)}$. Static power $P_S$ depends on the semiconductor technology and the operating temperature caused by current leakage. 
Hence, it can be modeled as: ${P_S(t) = k_1 \theta(t) + k_2}$, where $k_1$ and $k_2$ are technology-dependent system constants, and $\theta(t)$ is the operating temperature \cite{4212027}. Dynamic power $P_D(t)$ is the amount of power consumption due to the processor operating frequency $f$ at time $t$, modeled as ${P_D(t) = k_0 f(t)^s}$, where $s$ and $k_0$ are the system constants that depend on the semiconductor technology.

% We profile the thermal data of CPU cores when they run at the same fixed operating frequency. This is to ensure that other on-chip components such as caches and buses give a constant thermal effect to the CPU cores.
% The total power is the function of temperature at any time instant $t$ because the other factors such as frequency remain invariant during task execution. For homogeneous multi-core CPUs, the power for each core is $P_S(t)$ when CPU is idle, and $P_S(t) + P_D(t)$ when the CPU executes some workload in which all cores are fully utilized.

\subsection{Temperature Model}
\label{sec:chap4_tempMdl}
We consider the temperature model widely used in real-time mixed-criticality systems~\cite{Ahmed2017,DSouza2017ThermalIO}, which follows the well-known linear time-invariant (LTI) model. Hence, the temperature model for a multi-core CPU with $n$ cores is given by the following equation:
\begin{equation}\label{eq:chap4_tempMdl}
    [ \boldsymbol{\bf{\theta'}}(t)]_{n \times 1} = \boldsymbol{\bf{A}}_{n \times n}\ [\boldsymbol{\bf{\theta}}(t)]_{n \times 1} + \boldsymbol{\bf{B}}_{n \times n}\ [\boldsymbol{\bf{P}}(t)]_{n \times 1}
\end{equation}
 where $\myb[\theta](t)$ is the $n \times 1 $ matrix of the CPU core operating temperatures relative to the ambient temperature, and $\myb[P](t)$ is the power consumption of all cores at time $t$. $\myb[A]$ is an invariant $n \times n$ matrix and it is based on the characteristics of the semiconductor technology. It quantifies the effect of the conduction between adjacent cores, the convection among all cores, and the difference between ambient and operating temperature. 

$\myb[B]$ is the diagonal $n \times n$  matrix and it captures the effect of power consumption on the temperature of each core. For homogeneous multi-core CPUs, since the total power of CPU cores are the same (either $P_S$ or $P_S + P_D$), the matrix $\myb[B]$ can be represented as $b \times \myb[I]$, where $\myb[I]$ is the $n \times n$ identity matrix. Similar to $\myb[A]$, the values of $b$ are invariant to the changes in static or dynamic power consumption.% related to fluctuations in the core frequency and operating temperature.

Hence, the problem will be estimating the values of the matrices $\myb[A]$ and $\myb[B]$ ($=b \times \myb[I]$) without any prior knowledge or direct measurement of CPU power consumption. We will discuss that it is impossible to estimate the value of $b$ without having any knowledge of CPU power consumption; instead, we estimate $\myb[B]\times \myb[P]$ which matters in calculating the temperature in the LTI model.
\subsection{Problem Description}
Given a multi-core CPU equipped with on-chip temperature sensors, construct an accurate and fast thermal RC model by the estimation of $\myb[A]$ and $\myb[B]\times \myb[P]$ of the CPU, exclusively from a limited number of temperature profiles without requiring a priori knowledge of the floorplan, cooling package, and power traces.
\section{Limitations and Errors}
\label{sec:chap4_limitations_and_errors}
Before continuing our discussion, we must address some limitations to thermal parameter estimation on real-life platforms. Identifying these potential sources of error is critical to our data analysis and comprehension. It allows us to address the noise and limitations of our profile data set by preprocessing raw data and improve the accuracy of thermal parameters through the ensemble of measurements from different frequency levels and workload settings.

\subsection{Built-in Sensors}
The largest sources of error in the raw data are the built-in temperature sensors.

\subsubsection{Sensor Locations} The data sampled from the CPU's built-in temperature sensors is sensitive to the physical location of the sensors within CPU cores. For example, the thermal sensor for one CPU core may be located near its primary source of heat dissipation (i.e., hotspot) while the thermal sensor for another CPU core may be located further away from the CPU's hotspot. In addition to the proximity of the thermal sensor to the CPU's hotspot, the physical location of the hotspot may vary depending on the type of workload. Thus, distinct thermal footprints of various applications may not be precisely captured by a single built-in sensor. 

\subsubsection{Sensor Sampling Frequency} In CPUs for embedded systems, the sampling rate of temperature sensors is, generally, very low.  This limitation causes inaccuracies in raw data. For instance, the ODroid XU4 board maintains a 10Hz sampling rate for the thermal sensors. If the utilization pattern of a prospective application changes in less than 100~ms, e.g., real-time control tasks activated every 30~ms, the thermal footprint cannot be captured with the built-in sensors. 

\subsubsection{Sampling Precision} The data from on-chip CMOS temperature sensors is subject to quantization. This generates another source of error because the accuracy of temperature measurements is reduced to the granularity of quantization. Hence, the data needs to be processed before and after the estimating procedure to ensure the correctness of our results. For instance, if the sampling precision of all CPU core temperature sensors is 1\textdegree~C, then according to the superposition law in thermal modeling~\cite{mehdi2020dynamic}, the magnitude of total temperature error for a quad-core processor can reach up to 4\textdegree~C.  

\subsubsection{Sensor Response Function} Due to differences in the construction and architecture of on-chip thermal sensors, their response time may vary. Hence, the sensor data may not represent the actual temperature if the CPU utilization changes in a relatively short time interval. Over time, if there is no fluctuation in CPU utilization, the sensor data will converge to the actual CPU core temperature. Therefore, we can assume that this type of error only affects the transient-state data while the steady-state data is unaffected. 

\subsection{Ambient Temperature} Temperature is a relative value to the ambient temperature in our thermal model. Although it is assumed that the ambient temperature would remain invariant during profiling, in reality, it may change even in a room or a thermal furnace.

\subsubsection{Varying Ambient Temperature} The ambient temperature in room can change slightly, or even several degrees given circumstances that are difficult to regulate such as poorly insulated walls, drastic changes in the outside temperature, and even heat dissipated from idle electrical outlets. This change can affect the operating temperature of the CPU. Therefore, the fluctuation of the ambient temperature while profiling can introduce noise into the raw data. This type of noise impacts the relative operating temperature. 

\subsubsection{Varying Air Convection} Heat convection may also introduce noise into our data sets (thermal profiles). Forced air convection caused by air conditioning, active cooling package of the CPU or even people just moving around the board can all contribute to fluctuations in the CPU core's heat dissipation. This type of noise will directly affect the heat transfer from the CPU cores to the ambient air, hence the values of diagonal elements of the matrix $\myb[A]$ change.

\subsection{Thermal Interference} 
Heat dissipation due to miscellaneous tasks can impact the accuracy of the data set. Even though one expects the operating temperature of CPU cores to approach the ambient temperature when the CPU cores are disabled (e.g., by using the CPU hotplug mechanism in Linux), the operating temperature is much higher than this level.

\subsubsection{OS Tasks} There are some essential OS service tasks that cannot be terminated while profiling thermal data. Some of them even have predefined CPU affinity, thereby affecting the observed thermal behavior of target CPU cores.
%Even isolating the CPU cores with the ongoing applications can be violated by interrupting with operating system modules.

\subsubsection{Cache Coherent Interconnect (CCI)} Under cache consistency policies, CCI which is near to the CPU cores, generates some heat during task execution which causes an increase in the operating temperature of the CPU cores.

\subsubsection{Workload on Intellectual Property (IP) Blocks} Running tasks on IP blocks such as integrated GPUs, video encoders/decoders, and digital signal processors (DSP), can significantly affect the overall heat dissipation. Additionally the static power consumed while IPs are idle can also cause non-trivial heat dissipation. If there is no change in the utilization status of IPs, we can consider the amount of heat dissipation from those IPs as a constant quantity in thermal modeling. Nevertheless, this type of the noise can continuously or temporarily affect the profiling of thermal data. 

%It is worth noting that, in this chapter, we assume that only the data of CPU's temperature exists and there is no temperature profile of IPs. 

\section{Proposed Scheme}
\begin{figure}[t]
\centering
\includegraphics[width=0.9\textwidth]{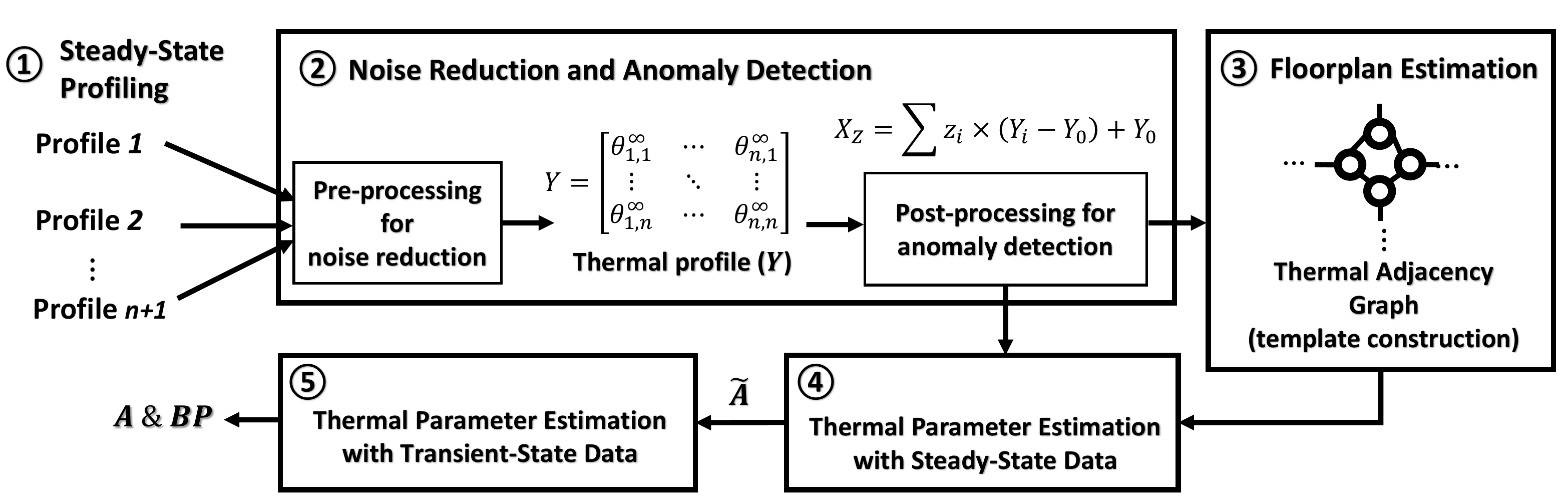}
\caption{Parameter Estimation Scheme}
\label{fig:chap4_scheme}
\end{figure}

In this section, we introduce our scheme for estimating the thermal parameters of a given multi-core processor. 
%We divide the section into several segments.
The entire workflow of the proposed scheme is illustrated in Fig.~\ref{fig:chap4_scheme}. The very first step is profiling steady-state temperature data for a set of designed workloads. Then, the scheme removes noise from the raw data set (\circled{2}), and performs the floorplan estimation (\circled{3}). By using the estimated floorplan template and the collected steady-state data, the value of the matrix $\myb[A]$ is estimated in terms of the power parameters (\circled{4}). The parameters $\myb[B] \times \myb[P]$ are then estimated by analyzing a subset of the transient-state data at the final stage (\circled{5}). One of the reasons that we propose dividing analysis into the steady-state and the transient-state stages (\circled{4} and \circled{5}) is to cope with the various types of errors that may be introduced during temperature profiling. If one tries to tackle this problem by using both data at the same time, errors in the transient-state data can adversely affect the characterization of the system in steady state because the transient-state data has a much larger number of data points than the steady-state data. Moreover, our proposed scheme has a low computational cost as it requires processing only a few data points in the steady-state stage to obtain the thermal characteristics of the system. %Later in this section, we will describe the phases and reasons in details.  

\subsection{Thermal Analysis and Steady-State Profiling}
\label{sec:chap4_SSI}
The proposed scheme primarily uses the state-state data for thermal parameter estimation since it is more robust to measurement errors than the transient-state data but still contains the required information about semiconductor technology and power consumption. Hence, before introducing the detailed stages of our scheme, we analyze the thermal model and provide the reasoning behind the steady-state profiling. 

Our goal is to estimate the thermal parameters related to the steady-state data. After solving the first order equation of Eq.~\ref{eq:chap4_tempMdl}, we have

\begin{equation}\label{eq:chap4_1}
    \boldsymbol{\bf{\theta}} (t) =  \boldsymbol{\bf{\theta}}_{chip} + {e^{(t - {t_0}){\bf{A}}}}{\boldsymbol{\theta} _0} + \int_{{t_0}}^t {{e^{(t - s){\bf{A}}}}{\bf{BP}}(s)ds}
\end{equation}
where $[\boldsymbol{\bf{\theta}}_{chip}]_{n \times 1}$ is the total heat dissipation caused by IP blocks on the chip and idle power of the CPU cores. We assume that all IPs generate a constant amount of heat. This term captures the heat conduction between all parts of the chip and the CPU. When the CPU cores are \new{idle} for a very long time, $\boldsymbol{\bf{\theta}}_{chip}$ can be captured. It is worth noting \new{that  due to the location CPU cores and their sensors on the floorplan, the steady-state temperature of idle CPU cores are different}. \cmnt{value is different for each CPU core because of the the location of CPU cores on the floorplan and also the location of each thermal sensor.} \new{The second term is the homogeneous solution which is the thermal response due to the initial temperature difference from the ambient. The third term is the non-homogeneous solution caused by the input power signal.}

\new{The total power is the function of temperature at any time instant $t$ because the other factors such as frequency remain invariant during task execution. For homogeneous multi-core CPUs, the power for each core is $P_S(t)$ when CPU is idle, and $P_S(t) + P_D(t)$ when the CPU executes some workload in which all cores are fully utilized. Hence,} \cmnt{If the power is considered to be fixed,} Eq.~\ref{eq:chap4_1} can be written as: 
\begin{equation}\label{eq:chap4_2}
\boldsymbol{\bf{\theta}} (t) = \boldsymbol{\bf{\theta}}_{chip} + {e^{(t - {t_0}){\bf{A}}}} \boldsymbol{\theta}_0 -\bf{A}^{-1}\ (I-e^{\bf{A}t}\ )\bf{BP}.
\end{equation}

In the steady state, the second term disappears because the steady-state operating temperature depends only on the power consumption (third term) and it is unaffected by initial temperature $\myb[\theta]_0$. Suppose $\boldsymbol{\bf{\theta}}^\infty$ represents the operating temperature in the steady state, then:
\begin{equation}\label{eq:chap4_3}
\boldsymbol{\bf{\theta}}^\infty  = \lim_{t \rightarrow \infty}   \boldsymbol{\bf{\theta}} (t) =  \boldsymbol{\bf{\theta}}_{chip}  -\bf{A}^{-1}\bf{BP}.
\end{equation}

We are interested in finding the value of matrices $\myb[A]$ and $\myb[BP]$. It is worth noting that the only control parameter is the power signal. It means that for each profiling procedure, it is possible to execute workload on any subset of CPU cores or hot-unplug them but the actual value of the power remains unknown. Let $[\myb[Y_i]]_{n \times 1}$ denote the operating temperature of the CPU cores when the $i$-th core is fully-utilized and $\myb[Y]_0$ represent the temperature of the CPU when all CPU cores are idle. Furthermore, let $[\myb[Y]]_{n \times n} = [\myb[Y_1] \myb[Y_2] \dots \myb[Y_n]]^T$ be the matrix of temperature profiles of the CPU in the steady state. For instance, $y_{i,j}$ in the row $i$ and column $j$ of $\myb[Y]$ is the temperature of the $j$-th CPU core when tasks are executing only on the $i$-th CPU core. Hence, according to Eq.~\ref{eq:chap4_3}, 
\begin{equation}\label{eq:chap4_4}
\myb[Y] - [\myb[Y_0] \myb[Y_0] ... \myb[Y_0]]^T_{n \times n} =  P_D\ (-\bf{A}^{-1}\bf{B}).
\end{equation}

We solve the equation to find the value of $\myb[A]$. Therefore, 
\begin{equation}\label{eq:chap4_5}
  \myb[A]=- P_D\ b\ (\myb[Y] - [\myb[Y_0] \myb[Y_0] ... \myb[Y_0]]^T)^{-1}.
\end{equation}

By denoting ${\Tilde{\myb[A]} = (\myb[Y] - [\myb[Y_0] \myb[Y_0] ... \myb[Y_0]]^T)^{-1}}$ and ${\gamma = b\ P_D}$, we have
\begin{equation}\label{eq:chap4_6}
  \myb[A]=- \gamma \Tilde{\myb[A]}.
\end{equation}
 Therefore, by estimating $\Tilde{\myb[A]}$ and $\gamma$, we can determine the value of $\myb[A]$. $\Tilde{\myb[A]}$ can be determined by  profile without any knowledge of power consumption. The value of $\gamma$ cannot be determined by the information from the steady-state profiles even with temperature profiles encompassing various CPU frequencies. 
 
 As shown in Eq.~\ref{eq:chap4_6}, we only need $n+1$ profiles to estimate the value of $\Tilde{\myb[A]}$, i.e., one profile measured when all cores are off, and $n$ profiles when each core is fully utilized one at a time. However, because of errors and limitations discussed in Section~\ref{sec:chap4_limitations_and_errors}, there may be a considerable \new{amount of noises} in estimating the value of $\Tilde{\myb[A]}$. If the profiles are noiseless, $\Tilde{\myb[A]}$ will be a symmetric matrix with positive values on diagonal and non-positive values on non-diagonal elements.  \new{Zeros at $\Tilde{a}_{i,j}$ represent} that there is no heat conduction between core $i$ and $j$. It is noteworthy that a non-symmetric $\Tilde{\myb[A]}$ caused by noisy data sets can lead to imaginary eigenvectors which is impossible \new{in practice}. We will later propose several methods to address different types of noises. Moreover, we will extend our analysis to reach a more precise $\Tilde{\myb[A]}$ by infusing more data profiles. By using those techniques, it is not necessary to have the exact CPU core combinations of $\myb[Y]_i$ for profiling purposes, but still possible to benefit from multiple auxiliary data obtained from the same core configuration. 
 
 One interesting property of $\myb[A]$ and $\Tilde{\myb[A]}$ is that both have the same eigenvectors. Additionally the eigenvalues of $\myb[A]$ are $-\gamma$ times the eigenvalues of $\Tilde{\myb[A]}$. We will use these properties to find the value of $\gamma$ later and justify why it is infeasible to estimate the value of power consumption from thermal profiling.

\subsection{\new{Noise Reduction and Anomaly Detection}}
Compensating for various sources of errors in the steady-state data is critical to accurately estimate the system thermal parameters. In this section, we will discuss two low-cost noise reduction and detection procedures for the steady-state data.

\subsubsection{Pre-processing for Noise Reduction}
The most challenging sources of errors in the steady-state data are the built-in CPU core temperature sensors and the uncontrollable fluctuations in the ambient temperature. Heat dissipation due to all types of \new{task interference} can be captured in $\myb[\theta]_{chip}$. We will show in the evaluation section that this amount is almost invariant for any given operating frequency. Therefore, we focus our discussion on other types of errors. \new{The accuracy of  the steady-state data} is crucial to calibrating thermal parameters with transient-state data  in the later stage of our scheme. It is also important for estimating \new{the relative locations of CPU cores on} the floorplan since they depend on the accuracy of this stage.

The errors that could be found in transient-state data, such as due to the sampling frequency and response time of built-in sensors, do not occur in steady-state data because the steady-state temperature converges to a certain level and remains invariant. To overcome the limitation of sampling precision, we pre-process the steady-state temperature data by taking a moving average. Although the temperature should remain invariant throughout the steady state, the limitation introduced by sensor data quantization may cause fluctuations. Suppose that the precision of built-in temperature sensor is 1\textdegree~C and the actual steady-state core temperature is 55.1\textdegree~C. The data will usually reflect a sensor reading of 55\textdegree~C. However, on less frequent occasions, the data may reflect a sensor reading of 56\textdegree~C. Aside from the precision concern, some transient noises such as varying air convection, temporary OS tasks, or variant ambient temperature can also be added up in the steady state. To alleviate the effects of the limitation of sensed data precision and also transient noises, one can apply a low-pass filter such as the moving average in an arbitrary time window on the steady-state data before constructing the observation matrix $\myb[Y]$.

\subsubsection{\new{Post-processing for Anomaly Detection}}
\label{sec:chap4_anomaly}
%This method is to detect the \new{error] from the experiment and correct the observed data.  
% The other approach can be ensemble multiple combinations of workload scenarios. It is possible to extend our analysis by changing the representation of $X$.Hereby, we proposed to ensemble multiple temperature profiles when two or more CPU cores are switched on and execute some workload. The reason is that using different combination of running CPU cores leads to overcome the precision limitation.It is noteworthy that more test profiles leads to obtain more accuracy. Therefore, under a user accuracy demands, the number of tests can be changed.
As we discussed in Sec.~\ref{sec:chap4_SSI}, our scheme needs steady-state temperature profiles which are obtained when only one of the CPU cores is fully utilized at a time. If there are more profiled data, it is possible to detect if there is any persistent error in the temperature profiles. For instance, if the ambient temperature in one profile is different from that in other profiles, it can be detected and rectified. Our post-processing error detection tests if the observed data $\myb[Y]_i$ are consistent with other auxiliary profiles that are obtained when more than one CPU core is fully utilized.  If any error is found from the observed data $\myb[Y]_i$, the corresponding column of $\myb[Y]$ will be rectified with the correct value.  

We now present the details. Let $[\myb[X]_Z]$ denote the steady-state temperature of CPU cores and $Z=\overline{z_1 z_2 \dots z_n}$ is the predicate that shows the status of CPU cores in the profile test where $z_i \in \{0,1\}$ represents whether the $i$-th CPU core is busy ($z_i = 1$) or not ($z_i = 0$). We are interested to test primary observed data $Y_i$ by using auxiliary data $\myb[X]$s to detect the prospective error in $Y$. According to the thermal superposition theorem~\cite{mehdi2020dynamic}, 
\begin{equation}\label{eq:chap4_11}
\myb[X]_{Z} = \sum_{i=1}^n z_i \times (\myb[Y]_i -\myb[Y]_0) + \myb[Y]_0
\end{equation}

Suppose that the available tests for a hypothetical 3-core CPU are $\myb[Y]_i$ for $i \in [1,3]$ and the auxiliary test $\myb[X]_{\overline{101}}$. Hence, $\myb[X]_{\overline{101}} = \myb[Y]_1 +  \myb[Y]_3 - \myb[Y]_0$. In the idle condition where there is no errors in the data, the equation much hold. Let's suppose there is an error data in one of them. We are interested to detect or correct it. By adding an error term $\epsilon$ to the equation, we have $\myb[X]_{\overline{101}} = \myb[Y]_1 + \myb[Y]_3 - \myb[Y]_0 + \epsilon_{1,3}$. If there is no error in data, one can assume that ${\epsilon_{1,3} = [0, 0, 0]^T}$. If there is an error in one of the tests, it can be detectable but not correctable. Now suppose that there is another test $\myb[X]_{\overline{011}}$ available, hence $\myb[X]_{\overline{011}} = \myb[Y]_2 + \myb[Y]_3 - \myb[Y]_0 + \epsilon_{2,3}$.

Now in this case, it is possible to detect not only the presence of error but also the location of error in the profiles. In our example, 

\textbf{I)} if $\epsilon_{1,3} = \epsilon_{2,3} = 0 $, there is no single error in the steady state values of any of $\myb[Y_1]$, $\myb[Y]_2$ or $\myb[Y]_3$.

\textbf{II)} if $\epsilon_{1,3} = 0 $ but $\epsilon_{2,3} \neq 0$, there is error in either $\myb[Y]_2$ or $\myb[X]_{\overline{011}}$. 

\textbf{III)} if $\epsilon_{1,3} \neq 0  $ but $\epsilon_{2,3} = 0$, there is error in either $\myb[X]_{\overline{101}}$ or $\myb[Y]_1$.

\textbf{IV)} if $\epsilon_{1,3}\neq 0$ and also $\epsilon_{2,3} \neq 0$, there is error in the steady state values of $\myb[Y]_3$.

% Therefore, It is possible to detect a single error in the test by two auxiliary tests. However, it is not always possible to detect the error. To detect the spot of the error, at least one other test is needed. Suppose that the steady state value of the test  $X_{\overline{111}}$ is also available. In this case, it is possible to check if there is an error in one of the auxiliary tests $X_{\overline{101}}$ or $X_{\overline{011}}$ or there is an error in $Y_2$ or $Y_3$. Therefore, aside from four main observation profiles, two additional tests are needed to verify their correctness and three additional tests are required to detect the spot of one single error. Some interesting questions are as follow

% \begin{itemize}
%     \item How many tests are at least required for detect/correct some given errors?
%     \item What are the maximum number of errors detectable/correctable?
%     \item What is the mechanism to select tests to detect/correct  primary tests?
% \end{itemize}

In order to detect and correct the error, we design a test in an n-hypercube format. Each corner of the hypercube represents one measurement setting $Z$, and there is only a one-bit difference in the $Z$ values of two neighboring corners. In this way, it is possible to examine the correctness of each test with its neighbors. It is also possible to spot the error with a sufficient number of tests. 

\begin{figure}[ht]
\centering
\subfloat [][]{\includegraphics[width=0.32\textwidth]{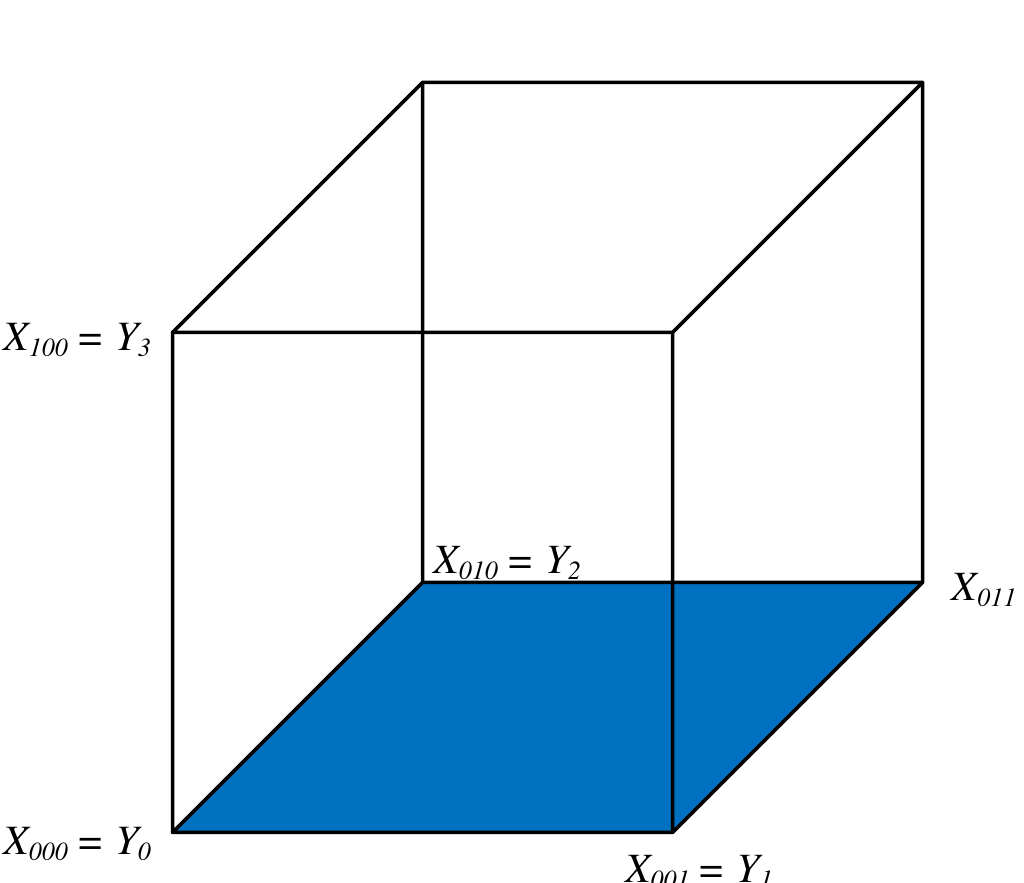}}
\subfloat [][]{\includegraphics[width=0.32\textwidth]{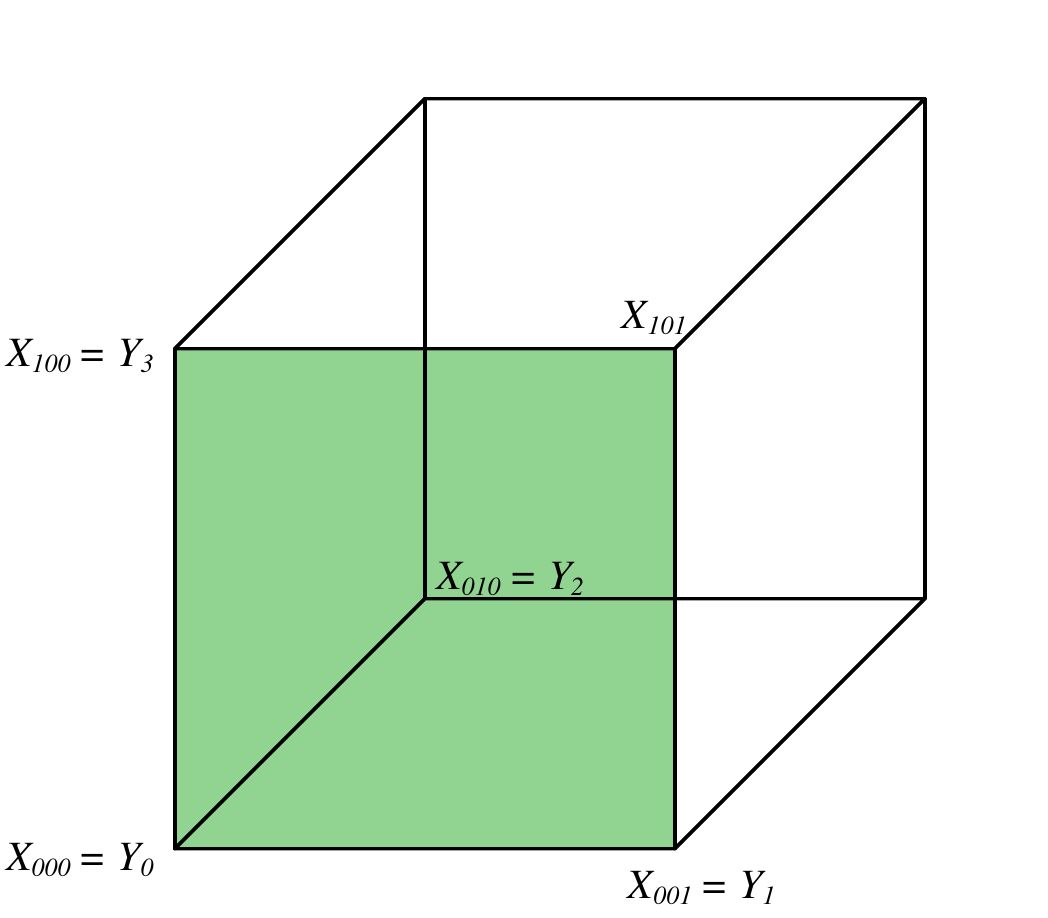}}
\subfloat [][]{\includegraphics[width=0.32\textwidth]{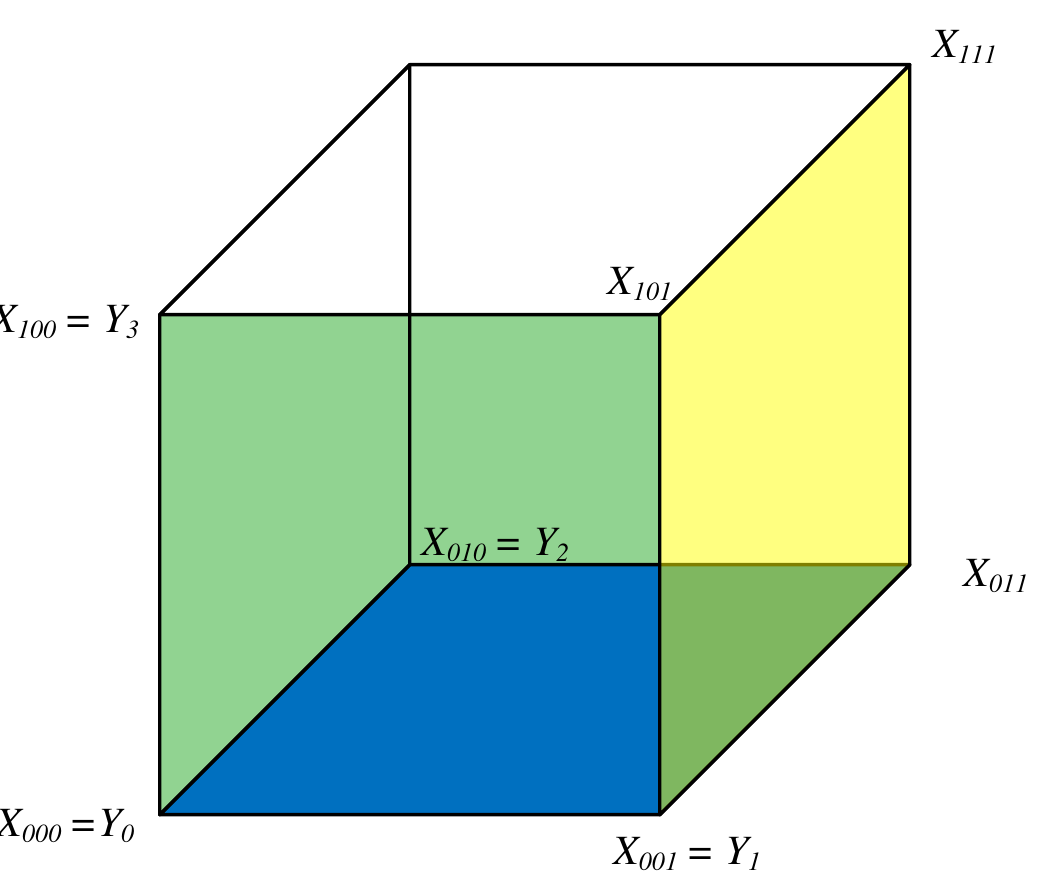}}
% \caption{Absolute error in different frequencies}
\caption{a) Auxiliary test for detect one error in either $\myb[Y]_1$ or $\myb[Y]_2$. b) Auxiliary test for detect one error in either $\myb[Y]_1$ or $\myb[Y]_3$.
c) Second-tier testing of auxiliary tests.}
\label{fig:chap4_err}
\end{figure}

 We explain the procedure by making an example. Fig.~\ref{fig:chap4_err}a illustrates the auxiliary test $\myb[X]_{\overline{011}}$. The blue side shows the verification of the $\myb[Y]_1$ and $\myb[Y]_2$. If all the data are consistent, one can conclude that there is no single error in $\myb[Y]_1$ and $\myb[Y]_2$. If there is an error in one data an additional test is needed that we will explain later.

The green side in Fig.~\ref{fig:chap4_err}b shows the verification for $Y_1$ and $Y_3$. If there is no single error, the data on this side is also consistent. 

% \begin{figure}[ht]
% \centering
% \includegraphics[width=0.25\textwidth]{5-Analysis/figs/error2.pdf}
% \caption{}
% \label{fig:chap4_e101}
% \end{figure}

Now suppose that both sides are inconsistent, then it is easy to say that  the data of $\myb[Y]_1$ is faulty because we assume that there is only one error in observation profiles and the edge of $(\myb[Y]_0,\myb[Y]_1)$ is the intersection of both sides. Since $\myb[Y]_0$ is zero base and not faulty, $\myb[Y]_1$ is noisy. 

We now discuss the case III where  $\epsilon_{1,3} \neq 0  $ and $\epsilon_{2,3} = 0$. To locate the error in the data, we use the auxiliary test that make a side with two of suspicious tests (the yellow side in Fig.~\ref{fig:chap4_err}c). In this case, we consider $\myb[X]_{\overline{111}}$ that can make a side with $\myb[X]_{\overline{001}}$, $\myb[X]_{\overline{011}}$ and $\myb[X]_{\overline{101}}$. If the data of this side is consistent, the error will belong to $\myb[Y]_1$, otherwise to the test $\myb[X]_{\overline{101}}$.

% \begin{figure}[ht]
% \centering
% \includegraphics[width=0.25\textwidth]{5-Analysis/figs/error3.pdf}
% \caption{}
% \label{fig:chap4_e111}
% \end{figure}

By employing the same technique, it is easily possible to extend the proposed method for two errors in the tests by adding another tier to test the auxiliary tests with each other. Additionally, it is possible to conclude that there exists a single error or two errors in the data. It is important to mention that in this chapter, it is assumed that errors in the two tests cannot conceal each other.

\subsection{Floorplan Estimation}
Hereby, we introduce a breadth-first-search (BFS) algorithm to estimate the topography of the CPU cores in the chip. Later, we use this information to calibrate the thermal term $b\ P_D$ in the temperature model by using the transient-state data. One of the steps to refine the thermal parameters of a chip is to estimate the parameters complying with the floorplan. We present a mechanism to estimate the relative location of the CPU cores on a given chip. 

\begin{figure}[ht]
\centering
\includegraphics[width=0.22\textwidth]{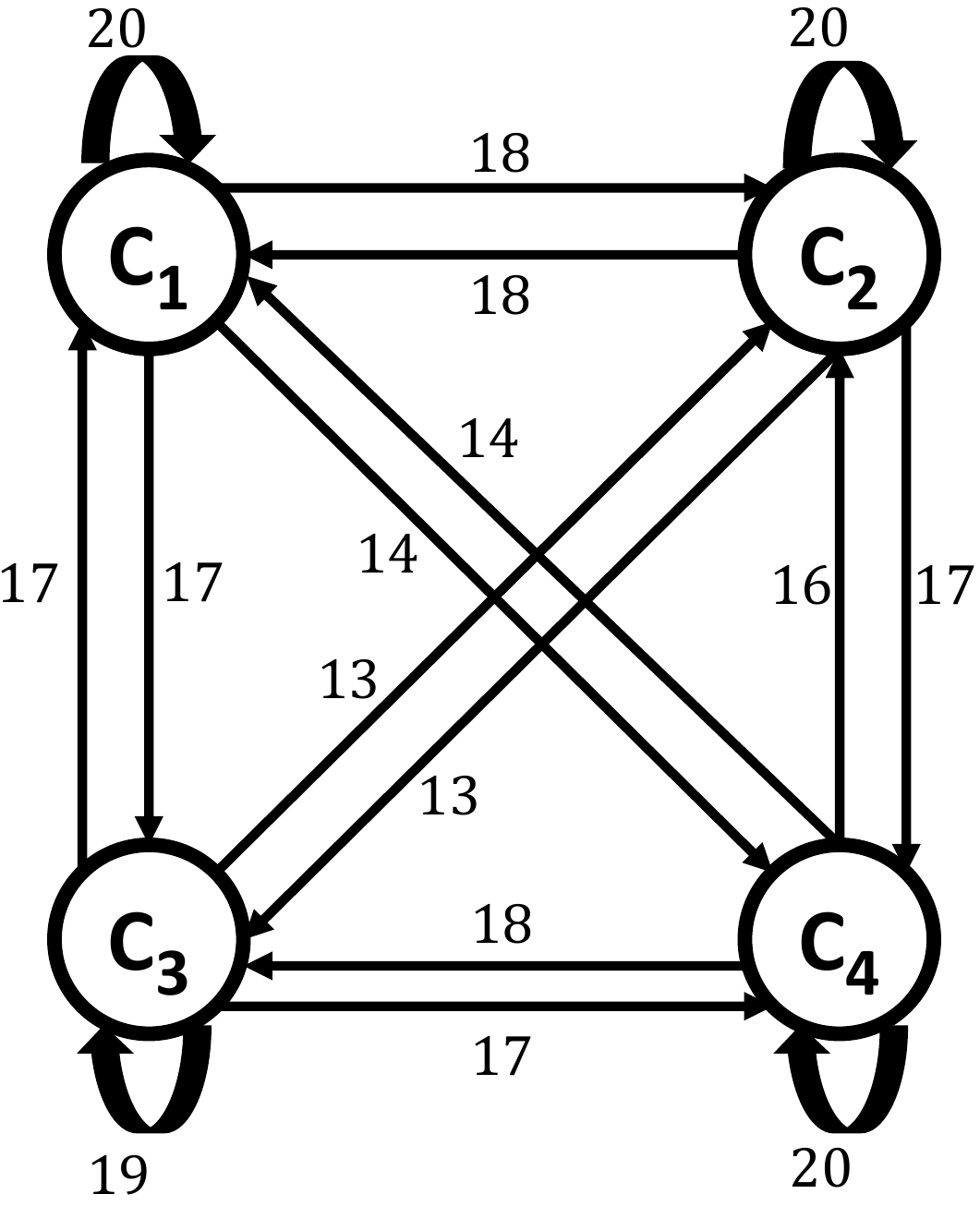}
\caption{Example of adjacency graph of a quad-core CPU.}
\label{fig:chap4_001}
\end{figure}

The intuition behind our proposed algorithm is that the amount of heat dissipation from a heat source to a closer object is more than that to a farther object. Therefore, we expect that the temperature increase due to heat conduction of adjacent CPU cores is more than that of non-adjacent CPU cores. We introduce a fully-connected weighted graph where the edge weight represents the temperature increase of a CPU core when its neighboring cores are fully utilized. For instance, suppose there is a quad-core CPU where each core is labeled as $C_i$ with $i \in [1,4]$. The fully-connected graph of this CPU is illustrated in Fig.~\ref{fig:chap4_001}. The weight of each edge, $w_{i,j}$, represents the temperature increase of a core $C_j$ when another core $C_i$ is fully utilized (i.e., $y_{i,j}-y^0_i$). 
%Hence, the weight of the edge, $w_{i,j}$ is the temperature increase of core $C_j$ due to heat dissipation of core $C_i$. 
It is worth noting that $w_{i,j} \neq w_{j,i}$ not only because of noise but also due to the location of the built-in temperature sensor relative to the hotspot of each CPU core, although the amount of heat dissipation from each homogeneous core is ideally the same.

\begin{figure}[ht]
\centering
\includegraphics[width=0.22\textwidth]{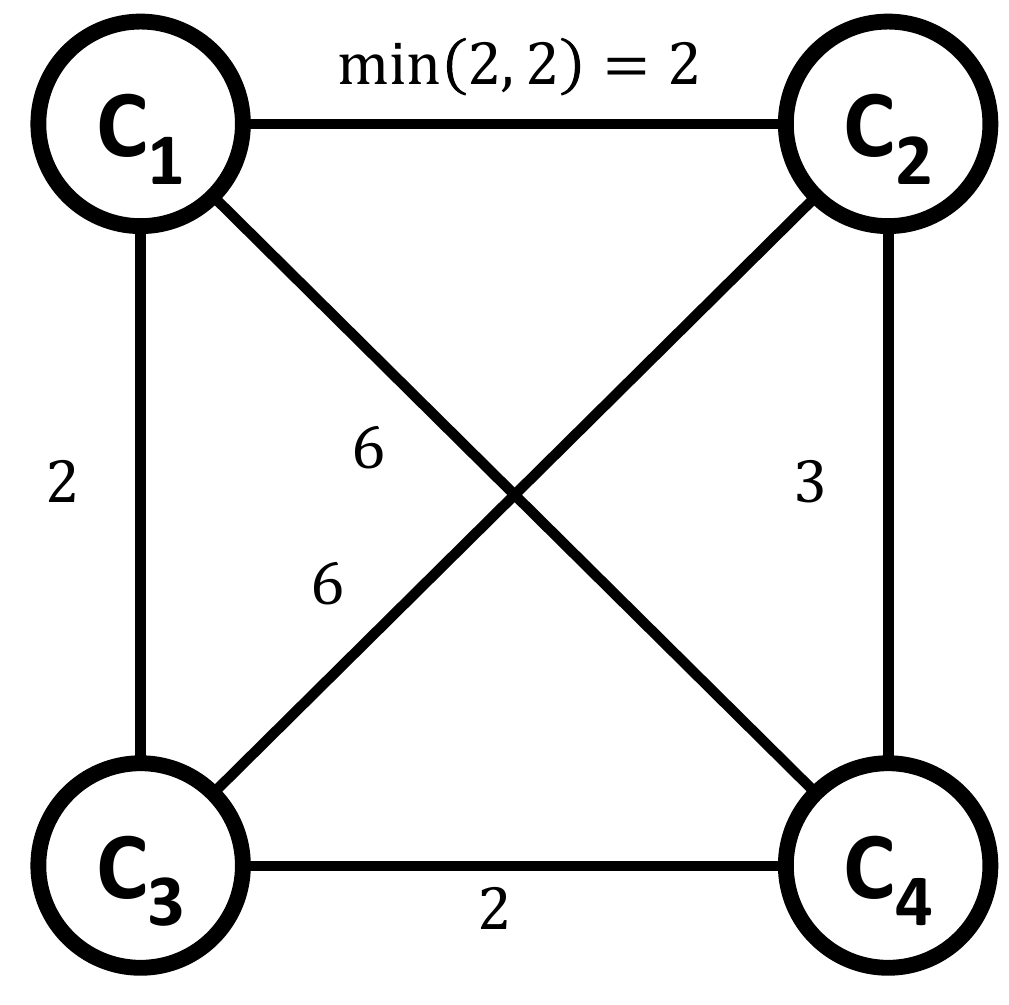}
\caption{Example of adjacency graph of a quad-core CPU.}
\label{fig:chap4_002}
\end{figure}

We are interested in the relative temperature increase of core $C_i$ when it is fully utilized compared to when another core $C_j$ is fully-utilized. Hence, we have ${w'_{i,j} = y_{i,i} - y_{i,j}}$. We also remove self-loops from the graph for simplicity. After this, the graph of Fig.~\ref{fig:chap4_001} is reduced to a graph where all weights are positive. Next, we convert the bi-directed graph to a uni-directed graph by taking maximum of the edges in the different direction (i,e., ${\min\{w'_{i,j},w'_{j,i}\}}$) as shown in Fig.~\ref{fig:chap4_002}. It is noteworthy that one can use average, minimum or any reduction operation for this stage. As shown in Fig.~\ref{fig:chap4_002}, the temperature increase of CPU core $C_4$ when the CPU core $C_1$ is fully utilized is because of heat conduction of CPU cores $C_2$ and $C_3$. In other words, the CPU core $C_4$ is heated up because of transitive heat transfer from $C_1$ to both $C_2$ and $C_3$ and then heat transfer occurs from these CPU cores to the CPU core $C_4$. Therefore, there is only transitive heat conductivity between the CPU core $C_1$ and the core $C_4$. 

% \begin{figure}[ht]
% \centering
% %\includegraphics[width=0.22\textwidth]{figs/results/furnace.jpg}
% \includegraphics[width=0.22\textwidth]{6-floorplan/figs/graph2.pdf}
% \caption{Example of adjacency graph of a quad-core CPU.}
% \label{fig:chap4_002}
% \end{figure}

\begin{figure}[ht]
\centering
\includegraphics[width=0.22\textwidth]{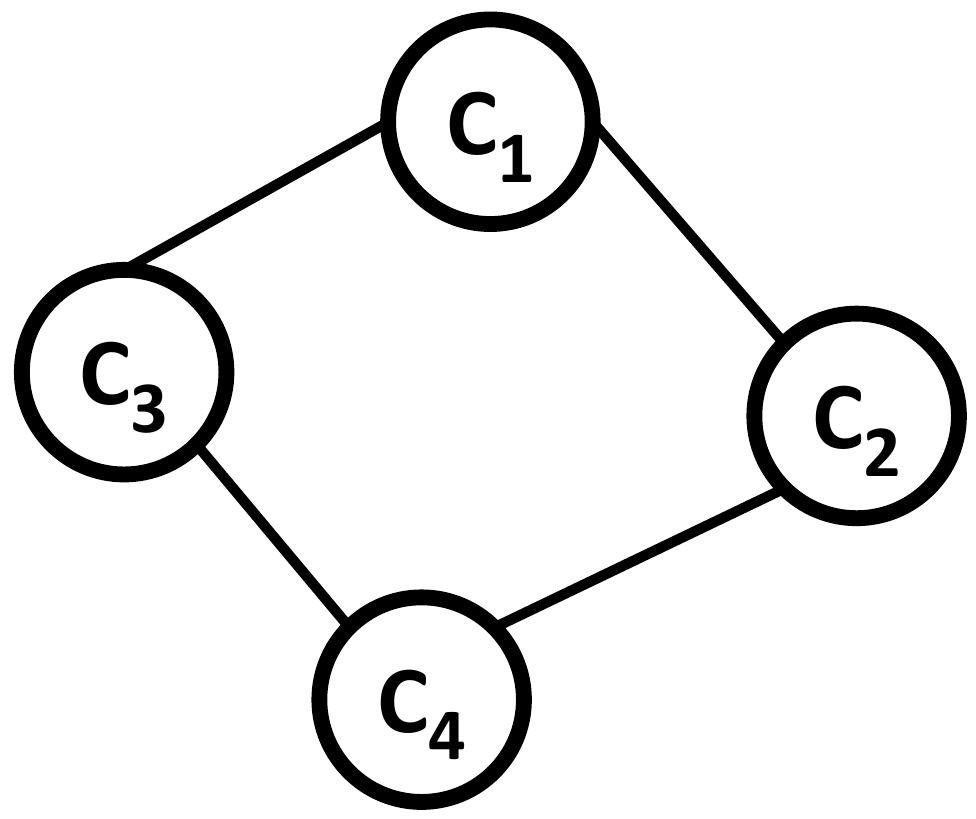}
\caption{Final adjacency graph of a quad-core CPU.}
\label{fig:chap4_004}
\end{figure}

By using this graph and also the aforementioned intuition, we estimate the location of the CPU cores on the floorplan. To do this end, we begin from one arbitrary core (let's say $C_1$) and find the minimum value of all weights connected to its corresponding node in the graph. In this example, the maximum weight is 2 for both CPU cores $C_2$ and $C_3$. Next, we find the maximum weight of CPU cores $C_2$ and $C_3$ from the unvisited node list, one at a time. We continue this step until all nodes are visited. If some of the visited CPU cores have the same value (within some error margin), they are connected. In this example, the error margin is 1 so the cores are connected as illustrated in Fig.~\ref{fig:chap4_004}. It is shown that the core $C_4$ (or at least its built-in temperature sensor) is closer to the core $C_3$ than $C_2$.
%The final adjacency graph of the CPU cores is illustrated in Fig.~\ref{fig:chap4_004}. 

Furthermore, if the temperature profiles of on-chip IPs such as an integrated GPU are available, the same approach can be applied to  locate the corresponding IP \new{by using the temperature increase when each CPU core is fully-utilized.} In this way, there is only one edge from each CPU core $C_i$ to the IP.

% Additionally, one can infer the location of the temperature sensor for each CPU core by considering the value of directed weights of the corresponding edge in the final adjacency graph. For instance, it is likely that the location of the thermal sensor of $C_3$ is on the bottom right of the CPU core since this is the floorplan of the homogeneous quad-core CPU and it shows the lower temperature corresponding to other cores. However, since this information is not usually released, it is not possible to evaluate the estimation.  
\subsection{Thermal Parameter Estimation with Steady-State Data}
\label{sec:chap4_estimate}

We propose a method to calibrate the thermal parameters based on the floorplan estimation discussed in the previous section. The thermal parameter estimation without considering the floorplan may not be applicable in practice.\footnote{This is because $\myb[Y]$ can be inaccurate and the properties of $\myb[A]$ may be violated if the floorplan is not considered.} Some important properties of $\myb[A]$ according to heat transfer are as follows:
\begin{itemize}
    \item $\myb[A]$ is a symmetric matrix. 
    \item The elements on diagonal of $\myb[A]$ are positive.
    \item The zero values of non-diagonal elements represent a pair of corresponding CPU cores are not adjacent.
    \item All non-zero values of non-diagonal elements are negative.
\end{itemize}
 These properties of $\myb[A]$ guarantees that there exist $n$ real eigenvalues and a set of $n$ eigenvectors, one for each eigenvalue, that are mutually orthogonal. This stage estimates the values of thermal parameters in way that it not only complies with the floorplan but also reduces the effect of noisy data observed data in $\myb[A]$. 

The matrix $\Tilde{\myb[A]}$ is constructed according to the estimated floorplan. If there is no thermal interference between two cores $C_i$ and $C_j$ in the estimated floorplan, the corresponding value in the matrix, i.e., $\Tilde{a}_{i,j}$, is zero. A single value can be used for all adjacent CPU cores in the matrix $\Tilde{\myb[A]}$ as they share the same heat dissipation increase value. 
% However, if different variables considered for $\Tilde{A}$ as long as they preserve the symmetrical and adjacency relations, it is more than likely that the values lead be close to each other.
 The gradient descent algorithm will then estimate the non-zero values. For the quad-core CPU exemplified in the previous section, there exist 4 zeros in the matrix $\Tilde{\myb[A]}$ because each CPU core has only 2 adjacent CPU cores. The number of distinct non-zero values in $\Tilde{[A]}$ are six in total because four different values are on the diagonal of $\Tilde{\myb[A]}$ and two distinct values are used for the other non-zero elements. It is worth noting that,  depending on the user's accuracy demand, more distinct non-zero values may be used. For the explained example, the matrix $\Tilde{\myb[A]}$ will be 
\[
\Tilde{A} = \begin{bmatrix}
a_1 & a_5 & a_5 & 0\\ 
a_5 & a_2 & 0 & a_6\\ 
a_5 & 0 & a_3 & a_5\\ 
0 & a_6 & a_5 & a_4\\ 
\end{bmatrix}.
\]
We propose a gradient descent algorithm to \new{find the parameters according the estimated floorplan}. For the calibration of the parameters, $\Tilde{\myb[A]}$ is compared with the inverse of $\myb[Y]$, which means that $\Tilde{\myb[A]}$ follows the estimated floorplan template and its inverse must be close to the temperature profiles. We propose the cost function as follows
\begin{equation}
\operatorname*{argmin}_{\substack{\text{$a_{i}$} \\ \text{$i \in [1,r]$}}}  ||\Tilde{\myb[A]}^{-1} - Y ||^2_F
\label{eq:chap4_cost}
\end{equation}
where $r$ is the number of unknown parameters. The sign of each parameter has to be carefully considered in the implementation, taking into account the properties of $\myb[A]$ discussed before (e.g., diagonal elements should be positive and the rest be non-positive).

% \subsection{floorplan-agnostic parameter estimation by transient information}
\subsection{Thermal Parameter Estimation with Transient-State Data}
\label{sec:chap4_trans}

In this section, we discuss how to estimate the matrix $\myb[B]$ for homogeneous multi-core platforms. As we mentioned in Sec. \ref{sec:chap4_tempMdl}, it is impossible to determine $\gamma$ from the steady-state observations, hence, we must process the transient-state data. 

As discussed in the steady-state section, if the power remains the same, the temperature equation will become Eq.~\ref{eq:chap4_2}.  To estimate the value of $\myb[A]$, we only need $\gamma$  which is commensurate to the power consumption. Substituting Eq.\ref{eq:chap4_6} in Eq.~\ref{eq:chap4_2}, we have:

\begin{equation}\label{eq:chap4_7}
\boldsymbol{\bf{\theta}} (t) = \boldsymbol{\bf{\theta}}_{chip} + {e^{(t - {t_0}){\bf{- \gamma \Tilde{A}}}}} \boldsymbol{\theta}_0 +\bf{(\gamma \Tilde{A})}^{-1}\ (I-e^{\bf{A}t}\ )\bf{b I P}.
\end{equation}

Suppose that $\myb[V]$ and $\myb[\Lambda]$ are the eigenvectors and eigenvalues of $\Tilde{\myb[A]}$, respectively. Therefore, ${e^{(t - {t_0}){\bf{- \gamma \Tilde{\myb[A]}}}}}$ can be represented as ${\myb[V]\ e^{(t - {t_0}){\bf{- \gamma \Lambda}}}\ \myb[V]^{-1}}$. From our proposed steady-state scheme, $\myb[V]$ and $\lambda$ are determined. We can estimate the value of $\gamma$ from the transient-state data of a singular observation.  To better calibrate  $\gamma$, multiple profiles may also be considered. Hence,
\begin{equation}\label{eq:chap4_8}
\boldsymbol{\bf{\theta}} (t) = \boldsymbol{\bf{\theta}}_{chip} + {e^{-(t - {t_0}){\bf{\gamma \Tilde{A}}}}} \boldsymbol{\theta}_0 +\bf{\Tilde{A}}^{-1}\ (I-e^{\bf{A}(t-t_0)}\ )\bf{H}
\end{equation}
where $\myb[H]$ is an $n\times 1$ input signal whose $i$-th element is $h_i \in \{0,1\}$. It is noteworthy that $h_i = 1$ when $i$-th core is fully utilized. By substituting the eigenvalues and eigenvectors, the equation can be represented as 

\begin{equation}\label{eq:chap4_9}
\begin{aligned}
\boldsymbol{\bf{\theta}} (t) = \boldsymbol{\bf{\theta}}_{chip} + V\ e^{-(t - {t_0}){\bf{ \gamma \Lambda}}}\ V^{-1} \boldsymbol{\theta}_0 + \\
\bf{\Tilde{A}}^{-1}\ (I-V\ e^{-(t - {t_0}){\bf{\gamma \Lambda}}}\ V^{-1}\ )\bf{H}.
\end{aligned}
\end{equation}

The only missing part in the temperature model is $\gamma$ which can be estimated by curve fitting on transient-state temperature. The values of $P_d(t)$, $\boldsymbol{\bf{\theta}}_{chip}$ and $\bf{\Tilde{A}}$ change at different frequency levels but the values of $\myb[\Lambda]$, $\myb[V]$ and $b$ remain invariant against frequency change. Although the value of $b\gamma$ can be estimated and is embedded in the temperature data, it is not possible to estimate the value of power even with temperature information at different frequency levels.

Since $\gamma$ is a scalar value, we observed that curve fitting on only a few cores can provide good results. The simplest test for estimating the value of $\gamma$ is when all cores are cooling down. In this case the third term is eliminated and \new{the temperature equation} for the $i$th core can be reduced to 

\begin{equation}\label{eq:chap4_10}
\boldsymbol{\bf{\theta}}^i (t) = \boldsymbol{\bf{\theta}}_{chip}^i + \sum_{ \lambda_j \in \Lambda} v_{i,j} y_{i,j} e^{-(t - {t_0}){\bf{ \gamma \lambda_j}}}  \boldsymbol{\theta}_0^i 
\end{equation}
where $v_{i,j}$ and $y_{i,j}$ are the elements of the $i$th row and the $j$th column in the matrices $V$ and $V^{-1}$, respectively. 

\section{Accuracy Enhancement}
In this section, we extend our proposed scheme to improve accuracy using additional steady-state data from different core settings and extra data observed at different frequencies.
 \subsection{Use of Steady-State Data Ensembles}
We discuss how to use the ensemble of extra observations at one frequency level to obtain a more accurate thermal observation profile. As discussed in the previous section, our method requires $n+1$ observed steady-state temperature profiles for a $n$-core system to construct the matrix $\myb[Y]$: one profile when all cores are idle and $n$ profiles when each of CPU core is only fully-utilized. 
Now, we extend our analysis to answer the following questions:
\begin{itemize}
    \item Is it possible to collect different CPU core usage settings (other than the aforementioned $n+1$ observed data) to construct $\myb[Y]$?
    \item Is it possible to have a more precise $\myb[Y]$ if there exist multiple instantiations from an identical setting and use all of them?
    \item Is it possible to construct $\myb[Y]$ with more than $n+1$ steady-state data?
\end{itemize}

We are interested in generalizing the construction of $\myb[Y]$ to include any possible thermal traces of fully-utilized CPU core combinations and the ensemble of more thermal traces.
It is noteworthy that if there is no noise in the observed data and also data is not quantized, no extra data or multiple instantiations are needed due to superposition law in Eq.~\ref{eq:chap4_11}. %Besides, constructing from different CPU core settings would be trivial. 

Now, suppose we have $m$ profiles such that $n \leq m \leq 2^n-1$ and each profiling was done under a different CPU core setting $Z_i$. Based on that, we can construct the predicate matrix $[\myb[D]]_{m \times n} = [\myb[Z]_1 \myb[Z]_2 \dots \myb[Z]_m]^T$ from the auxiliary profiles in $\myb[U] = [\myb[X]_{z_1} X_{z_2} \dots X_{z_m}]^T$. The matrix $\myb[Y]$ is then generated as follows

\begin{equation}
    \myb[Y] = ((\myb[D] \times \myb[D]^{T})^{-1} \times \myb[D] \times \myb[U]^T)^{-1}.
\label{eq:chap4_ensemble}
\end{equation}

Because of the inversion in the equation, the equation works when the number of the profiles is more than the number of CPU cores. It also works when there exist enough orthogonal profiles, meaning that there is at least one profile for each CPU core where the core is idle. The matrix $\Tilde{\myb[A]}$ is then calculated as explained in Sec.~\ref{sec:chap4_estimate}.
\subsection{Use of Multi-Frequency Data Ensembles}
\label{sec:chap4_multi_frequency_ensembles}
We now discuss how to obtain a more accurate $\myb[A]$ by using additional data collected at multiple frequency levels. As explained in Sec.~\ref{sec:chap4_tempMdl}, the thermal parameter $\myb[A]$ is dependent on the semiconductor technology and remains invariant against frequency changes. Therefore, it would be logical to assume that having observed temperature profiles from different frequency levels would lead to an identical $\myb[A]$. However, due to the term $\gamma$ in Eq.~\ref{eq:chap4_6}, $\Tilde{\myb[A]}$ is proportional to the power consumption and the clock frequency of CPU cores. Based on this, we extend our scheme to answer the following questions:

\begin{itemize}
    \item Is it possible to have an identical $\myb[A]$ from different $Y$s which are collected at different frequency levels?
    \item Is it possible to estimate the power consumption with respect to different frequency levels?
\end{itemize}

To answer these questions, we propose a thermal parameter estimation approach based on multi-frequency data ensembles. 
%It is worth mentioning that for this stage, the steady-state data from different frequency levels is still required. 
Let $\myb[Y]^i$ denote the temperature profile matrix constructed from the data when the CPU frequency is $f_i$, and $\gamma_i$ denote its power effect on temperature at the frequency $f_i$. Unlike the method presented in Sec.~\ref{sec:chap4_estimate} which estimates the parameters in $\Tilde{\myb[A]}$, we will calculate a $\Tilde{\myb[A]_1}$, which is the base at $f_1$. Additionally, we consider $\gamma_1 = 1$  and estimate $\frac{\gamma_i}{\gamma_1}, \forall i >1$. In such case, tracking the values of $\gamma$ over different frequency levels gives the power consumption of CPU cores proportional to the $\gamma_1$ of the base frequency level $f_1$, and all $\myb[Y]$s share the same thermal parameter $\myb[A]$. Using the same procedure as in Sec.~\ref{sec:chap4_trans} allows estimating the value of $\gamma_1$;  hence, all $\gamma$s can be estimated. It is noteworthy that, even in this case, the actual value of the power consumption cannot be computed, but the relative power consumption at different frequency levels can be obtained. Therefore, one can see the effect of power consumption embedded in the temperature profiles.

We change the cost model of Eq.~\ref{eq:chap4_cost} to

\begin{equation}
\operatorname*{argmin}_{\substack{\text{$a_{j}$} \\ \text{$ j \in [1,r]$} \\ \text{$\frac{\gamma_i}{\gamma_1}$} \\  \text{$i \in [1,|f|]$}}}  \sum_{1}^{|f|} ||\Tilde{\myb[A]_1}^{-1} - \frac{\gamma_i}{\gamma_1}\myb[Y]^i ||^2_F
\label{eq:chap4_mfreqs}
\end{equation}

The problem is estimating the values of $\Tilde{\myb[A]}_1$ and also $\frac{\gamma_i}{\gamma_1}$. One can expect that $f_i > f_j$ leads to $\gamma_i < \gamma_j$ due to dynamic power increase and this can be considered as a constraint in the implementation. 
\section{Evaluation}
 This section describes the experimental evaluation of our scheme. We explain our implementation on a real embedded platform. We show the results of our proposed method to find the thermal parameters from steady-state temperature data. We also explore the effect of our approaches in the error correction of estimated thermal parameters. Furthermore, we validate our method of finding the location of CPU core on the floorplan. We also show the results of our parameter-based power estimation model and compare it with data collected from on-chip power sensors. 
 %Finally, we construct our model by finding the absolute values of thermal parameters with transient-state data and evaluate our proposed model in different scenarios.

\subsection{Platform} 
We performed our experiments on an ODroid-XU4 development board \cite{ODROIDXU4} equipped with a Samsung Exynos5422 SoC based on the ARM big.LITTLE heterogeneous computing architecture. The Exynos CPU package contains two different quad-core CPU clusters of little Cortex-A7 and big Cortex-A15 cores. Built-in temperature sensors with a sampling rate of 10 Hz and precision of 1\textdegree~C are available for each big CPU core as well as the GPU to measure the operating temperature\footnote{There is no temperature sensor for little cores since the power consumption and heat dissipation of the little cluster is substantially lower than that of the big cluster.}. The DTM throttles the frequency of the entire big CPU cluster to 900 MHz when one of its cores exceeds the hardware-defined maximum temperature threshold of 95\textdegree~C. There is no active or passive cooling mechanism enabled on the CPU. The big CPU cluster frequency can be dynamically adjusted within the range of $[0.2, 2.0]$ GHz. However, for each experiment, it was pinned at a fixed frequency in the range of $[0.7,1.4]$ GHz to avoid thermal violations that occur when all CPU cores run fully-utilized beyond 1.4 GHz. Moreover, as the frequency increases, the data incurs a higher magnitude noise compared to the noise observed when operating at lower frequencies. It is worth noting that frequency changes between tests affect all homogeneous cores on the CPU, ensuring that each core in the cluster maintains the same frequency. Also, during each experiment, ambient temperature is regulated at 21\textdegree~C.
\subsection{Thermal Parameter Estimation with Steady-State Data}

% We divide the results into two parts of idle and fully-utilized. The data then used to determine the thermal parameters in steady state pf the SoC.
\subsubsection{Idle Steady-State Temperature}
First, we measure the temperature of CPU cores at different frequencies and determine the steady state temperature when all big CPU cores are idle. %Fig.~\ref{fig:chap4_off} shows the temperatures of CPU cores operating at different CPU frequencies. 
As we show in Fig.~\ref{fig:chap4_off}, the temperatures of the big CPU cores during the idle state changes with CPU frequency for two possible reasons: 
i) some parts of big CPU package such as CCI, big CPU controlling unit, cache memory and its peripherals still operate at their designated frequencies and cause heat dissipation, and ii) other IPs on the SoC may operate with big CPU frequency. This temperature increase can also be caused by chip leakage power-induced heat dissipation.

\subsubsection{Dynamic Temperature} 
\begin{figure}[t]
\centering
\subfloat [][]{\includegraphics[width=0.49\textwidth]{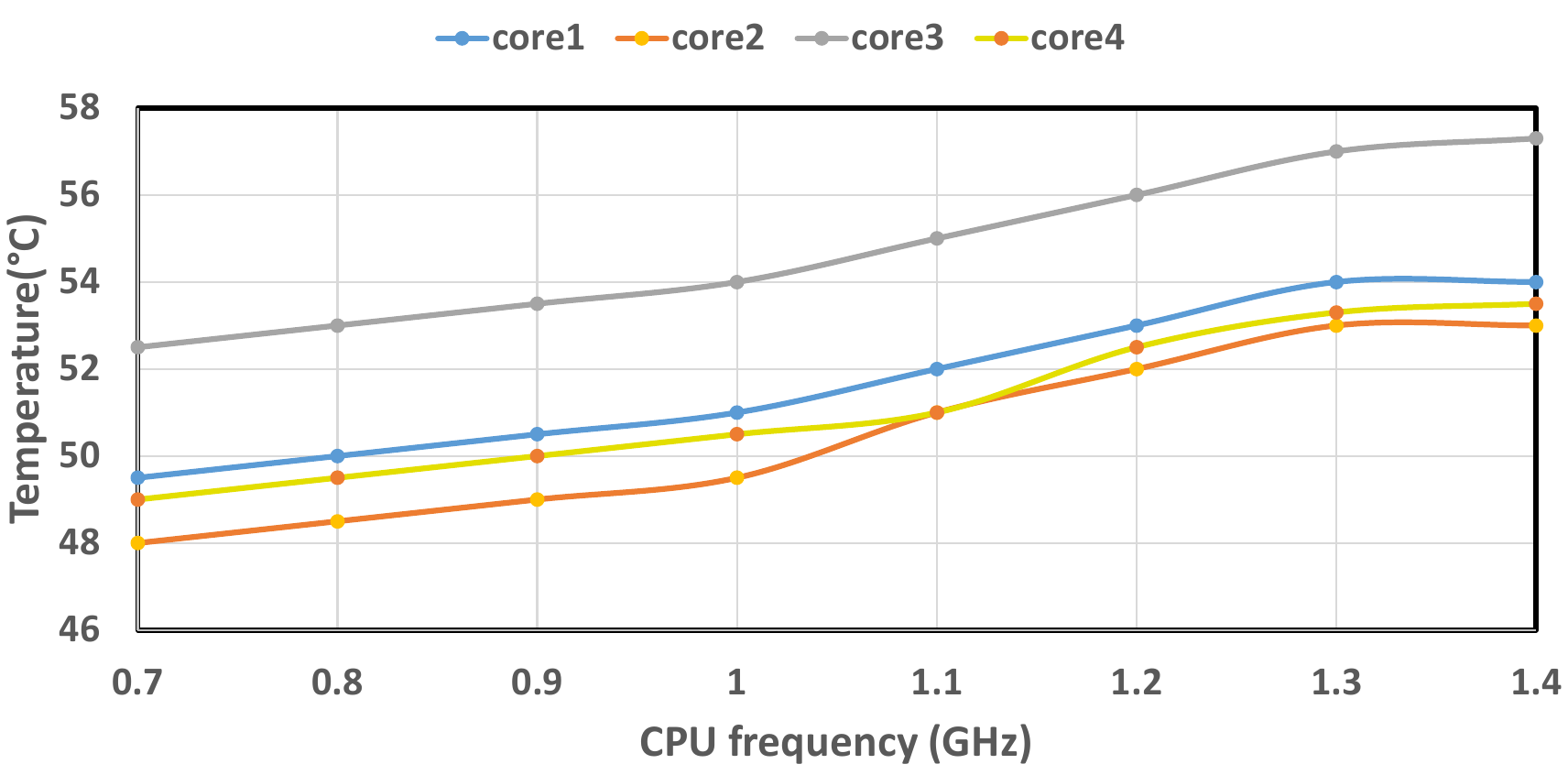}\label{fig:chap4_off}}
\subfloat [][]{\includegraphics[width=0.49\textwidth]{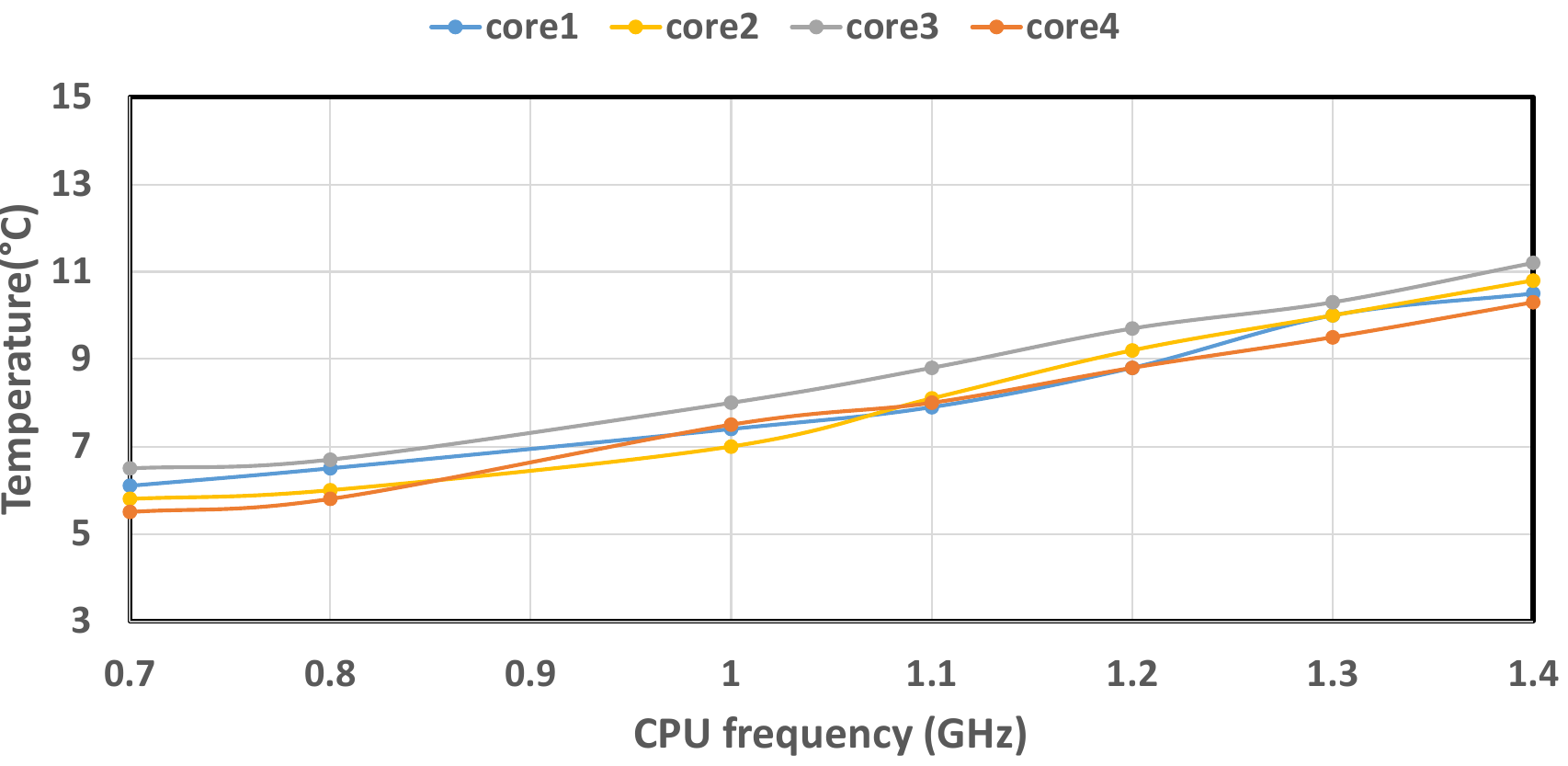}\label{fig:chap4_temp_increase}}
\caption{(a) Idle steady-state temperature of each core at different frequency levels. (b) Temperature increase of CPU cores due to computation.}
\label{fig:chap4_obsrv_temp}
\end{figure}

Fig.~\ref{fig:chap4_temp_increase} depicts the relative temperature increase of each CPU core at different frequency levels when that core is fully-utilized. There is a slight difference in the temperature increase between the CPU cores despite identical workloads and micro-architectures. Although there is noise related to the precision of on-chip thermal sensors, this difference is mainly due to a variation in the location of those sensors on the CPU cores.
% \begin{figure}[ht]
% \centering
% \includegraphics[width=0.50\textwidth]{8-Evaluation/figs/dynamic.pdf}
% \caption{Temperature increase of fully-utilized of all cores.}
% \label{fig:chap4_full}
% \end{figure}

\subsection{Noise Detection and Steady-State Data Ensembles}

\begin{table}[t]
\centering
 \caption{Descriptions of steady-state traces on big cores.}
  \label{tab:chap4_spec}
    \begin{tabular}{ |p{2.1cm}|p{4.2cm}|p{4.5cm}|p{2.0cm}|}
    \hline
     Case name & Descriptions & Traces (decimal) & \# of traces\\
    \hline
   \begin{footnotesize} $CA1$ \end{footnotesize}& \begin{footnotesize}one-core experiments \end{footnotesize}&\begin{tiny}  ${Z_1, Z_2, Z_4, Z_8}$ \end{tiny} & 4\\ \hline

   \begin{footnotesize} $CA3$ \end{footnotesize}& \begin{footnotesize}three-core experiments \end{footnotesize}&\begin{tiny}  ${Z_7, Z_{11}, Z_{13}, Z_{14}}$ \end{tiny} & 4\\ \hline
   
   \begin{footnotesize} $CA2$ \end{footnotesize}& \begin{footnotesize}two-core experiments \end{footnotesize}&\begin{tiny}  ${Z_3, Z_5, Z_6, Z_9, Z_{10}, Z_{12}}$ \end{tiny} & 6\\ \hline

   \begin{footnotesize} $CA1,3$ \end{footnotesize}& \begin{footnotesize}one-core \& three-core experiments \end{footnotesize}&\begin{tiny}  ${Z_1, Z_2, Z_4, Z_7, Z_{8}, Z_{11}, Z_{13}, Z_{14}}$ \end{tiny} & 8\\ \hline
   
   \begin{footnotesize} $CA1,2$ \end{footnotesize}& \begin{footnotesize}one-core \& two-core experiments \end{footnotesize}&\begin{tiny}  $Z_1, Z_2, Z_3, Z_4, Z_5, Z_6, Z_8, Z_9, Z_{10}, Z_{12}$ \end{tiny} & 10\\ \hline
   
   \begin{footnotesize} $CA1,2,3,4$ \end{footnotesize}& \begin{footnotesize}all experiments \end{footnotesize}&\begin{tiny}  ${Z_1, Z_2, \dots, Z_{15}}$ \end{tiny} & 15\\ \hline
  
    \hline
    \end{tabular}
    \label{tab:chap4_cases}
\end{table}

\begin{figure}[t]
\centering
\subfloat [][Core 1]{\includegraphics[width=0.45\textwidth]{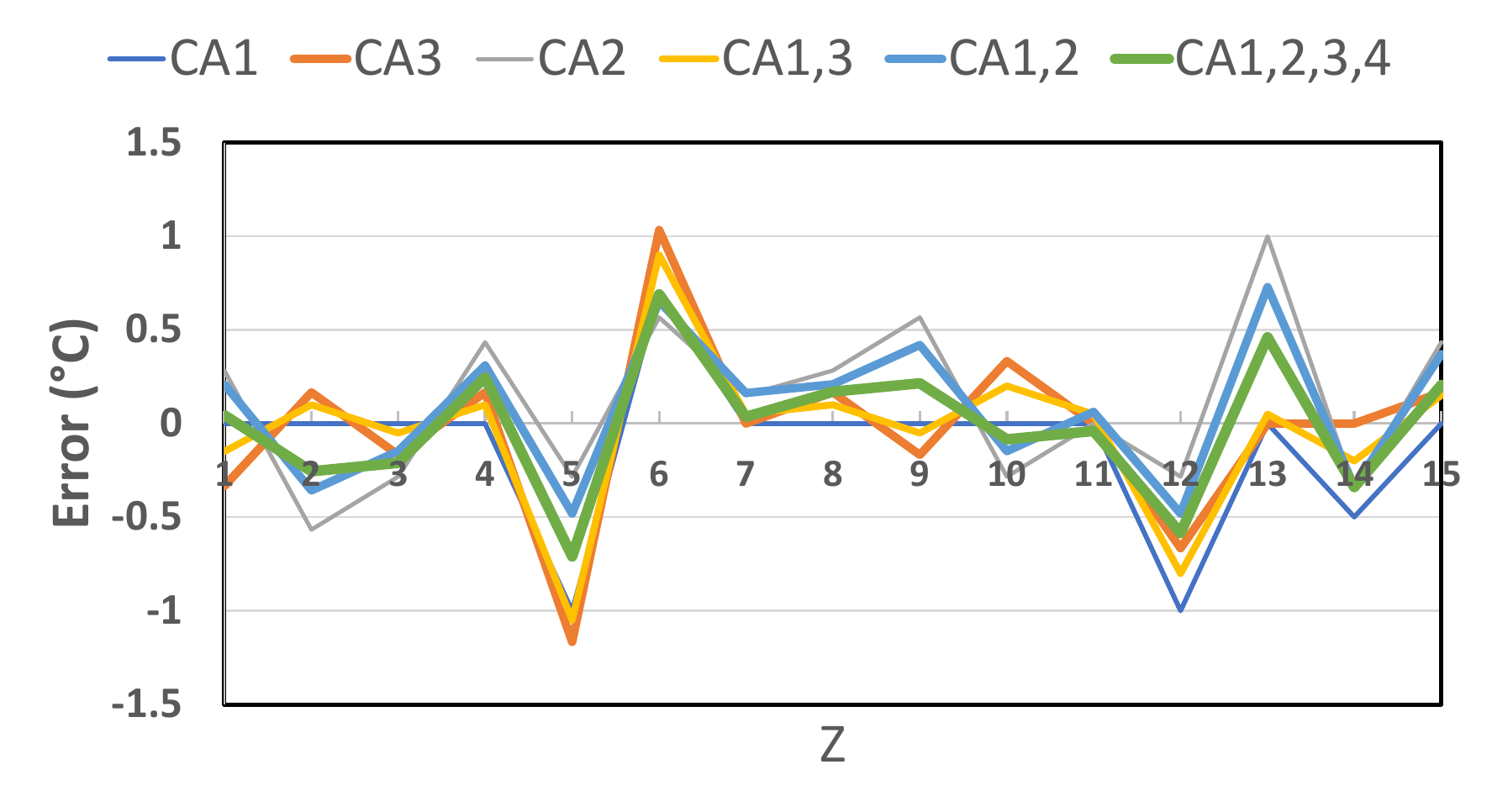}}
\subfloat [][Core 2]{\includegraphics[width=0.45\textwidth]{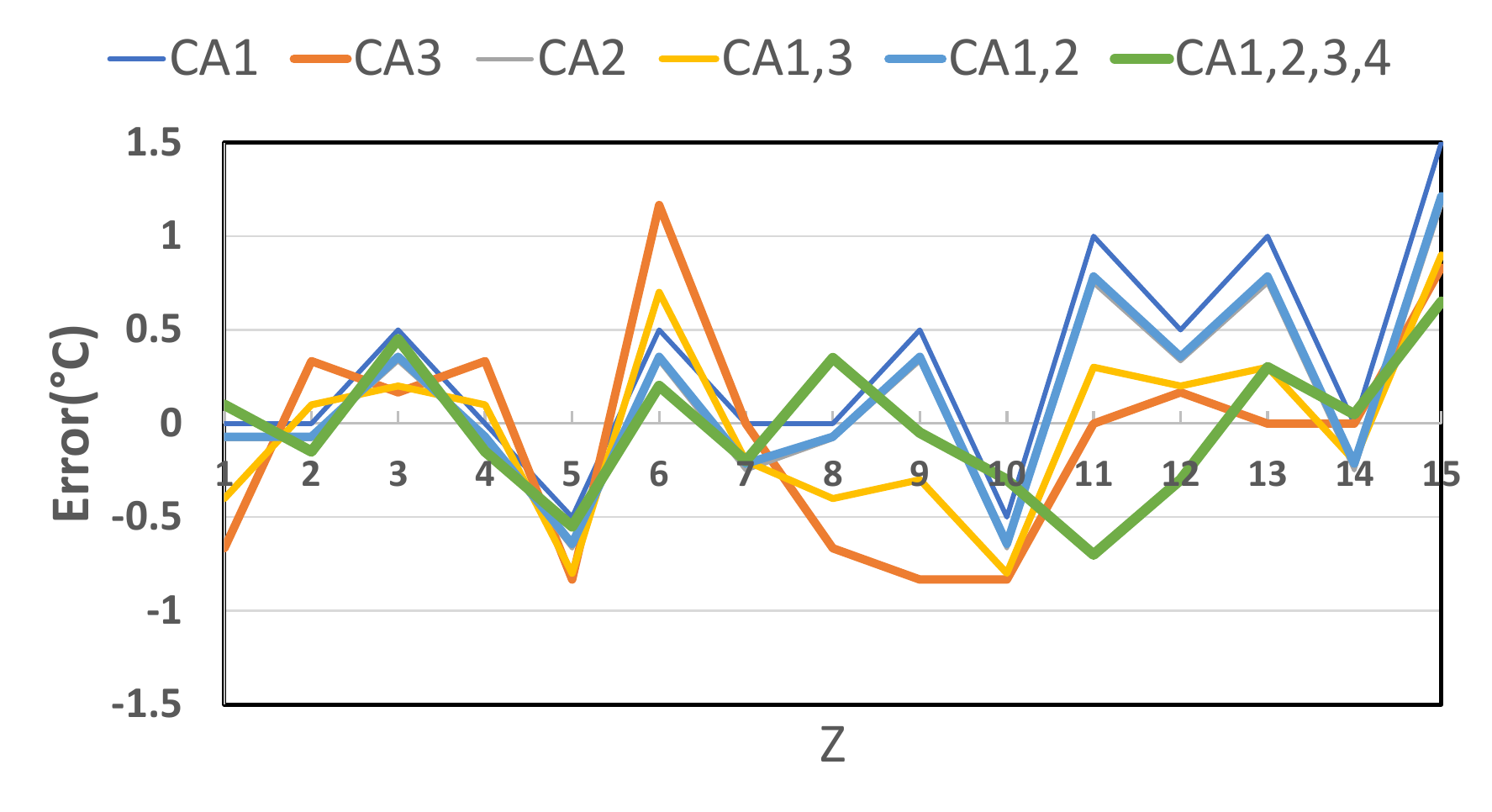}}\\
\subfloat [][Core 3]{\includegraphics[width=0.45\textwidth]{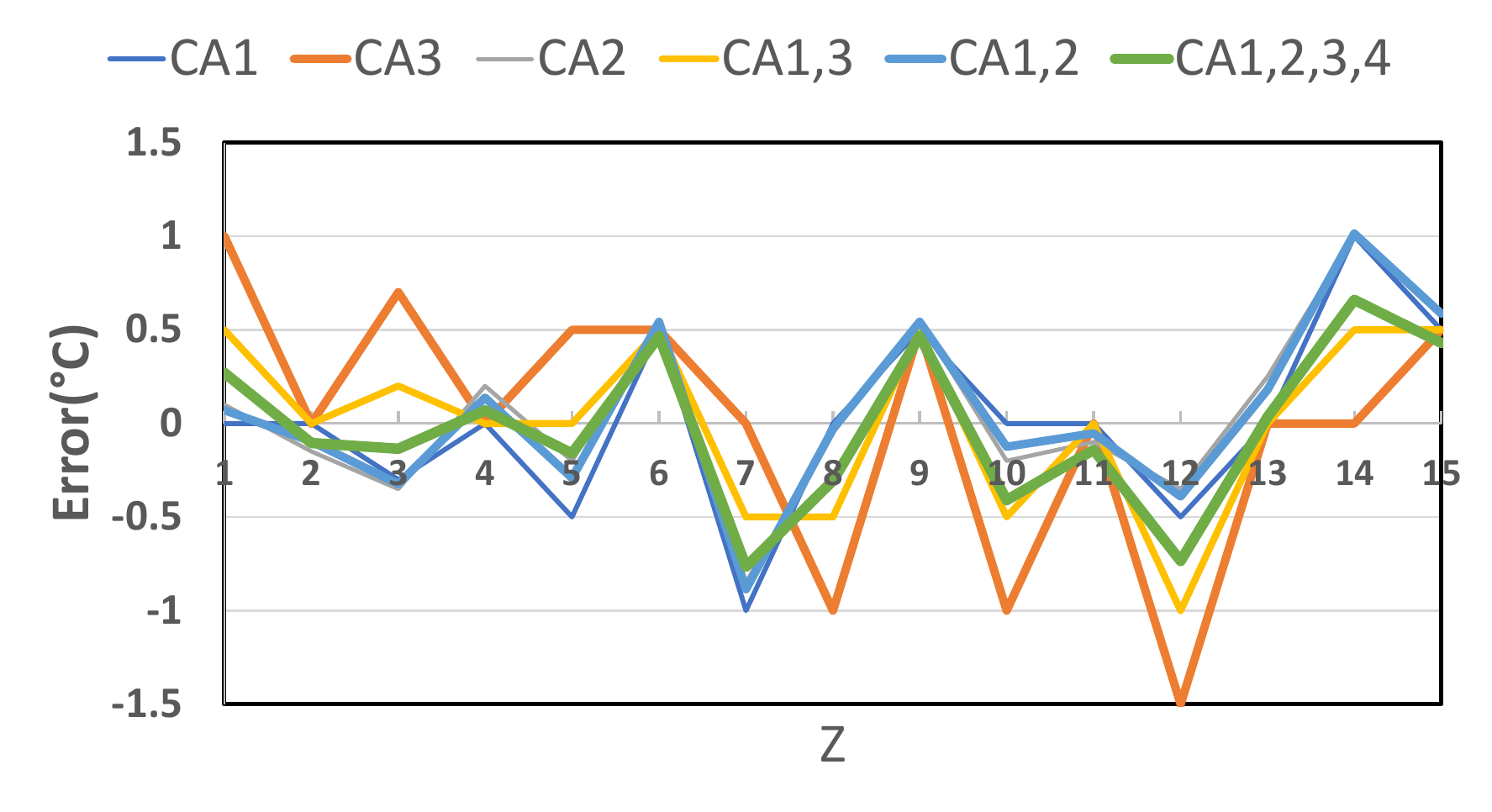}}
\subfloat [][Core 4]{\includegraphics[width=0.45\textwidth]{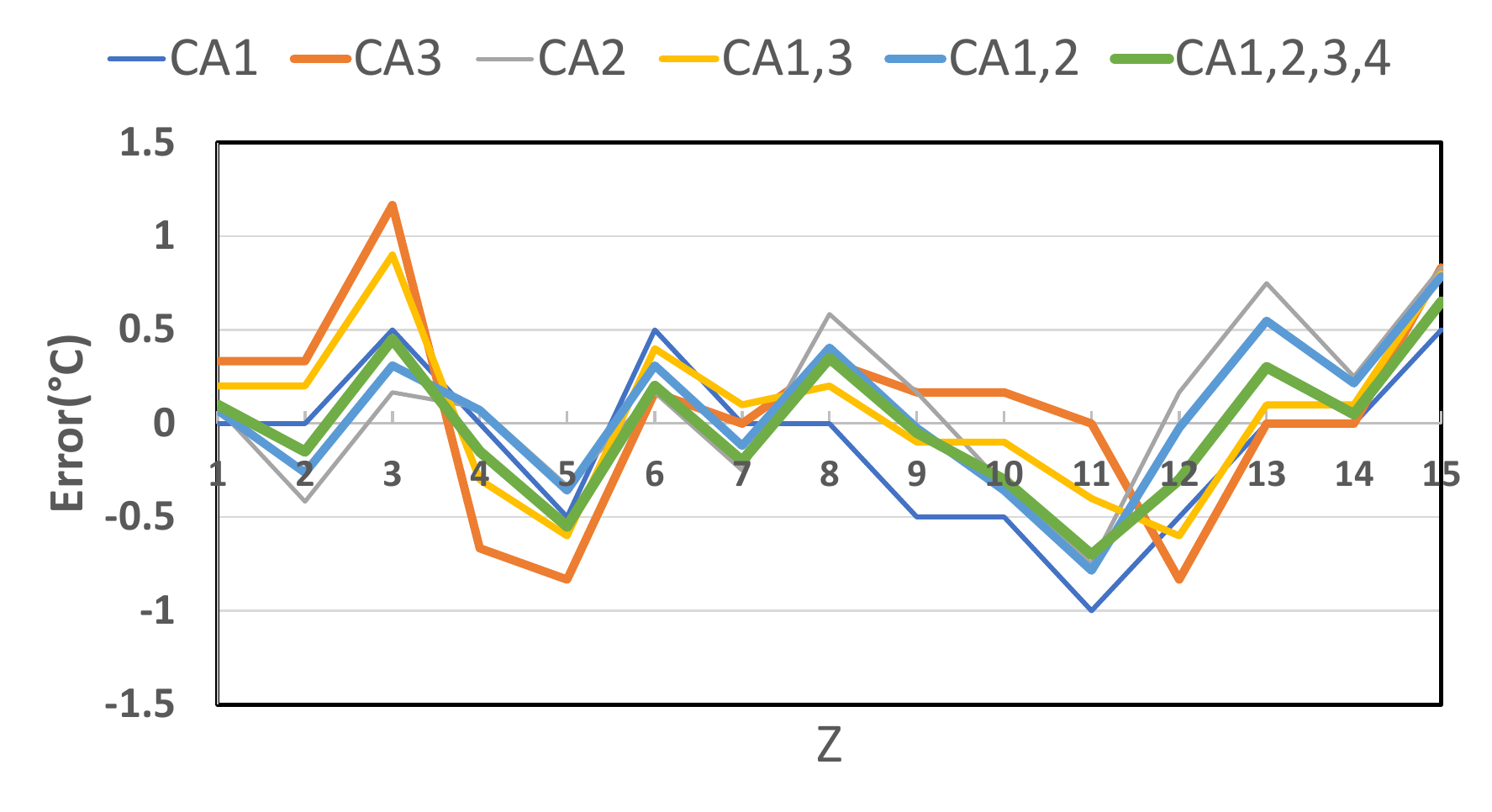}}
\caption{The error of steady-state temperature of CPU cores by using different cases in construction of $Y$.}
\label{fig:chap4_error}
\end{figure}
We evaluate our scheme in improving the accuracy of thermal parameter estimation by using the ensemble of steady-state profiles from multiple frequency levels. We apply the superposition theorem as in Eq.~\ref{eq:chap4_11} to estimate the steady-state temperature of CPU cores when different subsets of them are fully-utilized.
%% ..
We use the data collected from a subset of  all possible combinations of CPU cores at the fixed clock frequency of 1.4~GHz.\footnote{Note that the total number of possible core combinations is $2^4$ in a quad-core system and the total number of selecting a subset of the combinations is $2^{2^4}$.} 
The details on the subsets used to construct the matrix $\myb[Y]$ are given in Table~\ref{tab:chap4_cases}. Some cases contain the least number of orthogonal profiles (i.e., 4 profiles) while others contain more profiles. The case $CA1,2,3,4$ includes all profiles to construct the matrix $\myb[Y]$. The name of each case indicates the number of fully-utilized CPU cores. For instance, $CA1,2$ includes four profiles of one CPU core and six profiles of two cores. 
%% ^

Fig.~\ref{fig:chap4_error} depicts the error in steady-state temperature of each CPU core under different utilization scenarios ($\myb[Y]$ from Eq.~\ref{eq:chap4_ensemble}).
%of fully-utilized CPU cores by constructing $\myb[Y]$ from Eq.~\ref{eq:chap4_ensemble}. 
As shown in the figure, using more profiles helps reduce the effect of noise in  constructing $\myb[Y]$. The proposed anomaly detection mechanism presented in Sec.~\ref{sec:chap4_anomaly} was applied to the profiles and excluded the outliers from being considered for $\myb[Y]$. For instance, applying this mechanism detected an anomaly in the profile $Z_6$ by comparing it with the other profiles. Hence, while constructing $\myb[Y]$, we were able to find out that the anomaly was not because of an error in the other profiles but rather due to the error in $Z_6$. 

\begin{figure}[t]
\centering
\includegraphics[width=0.64\textwidth]{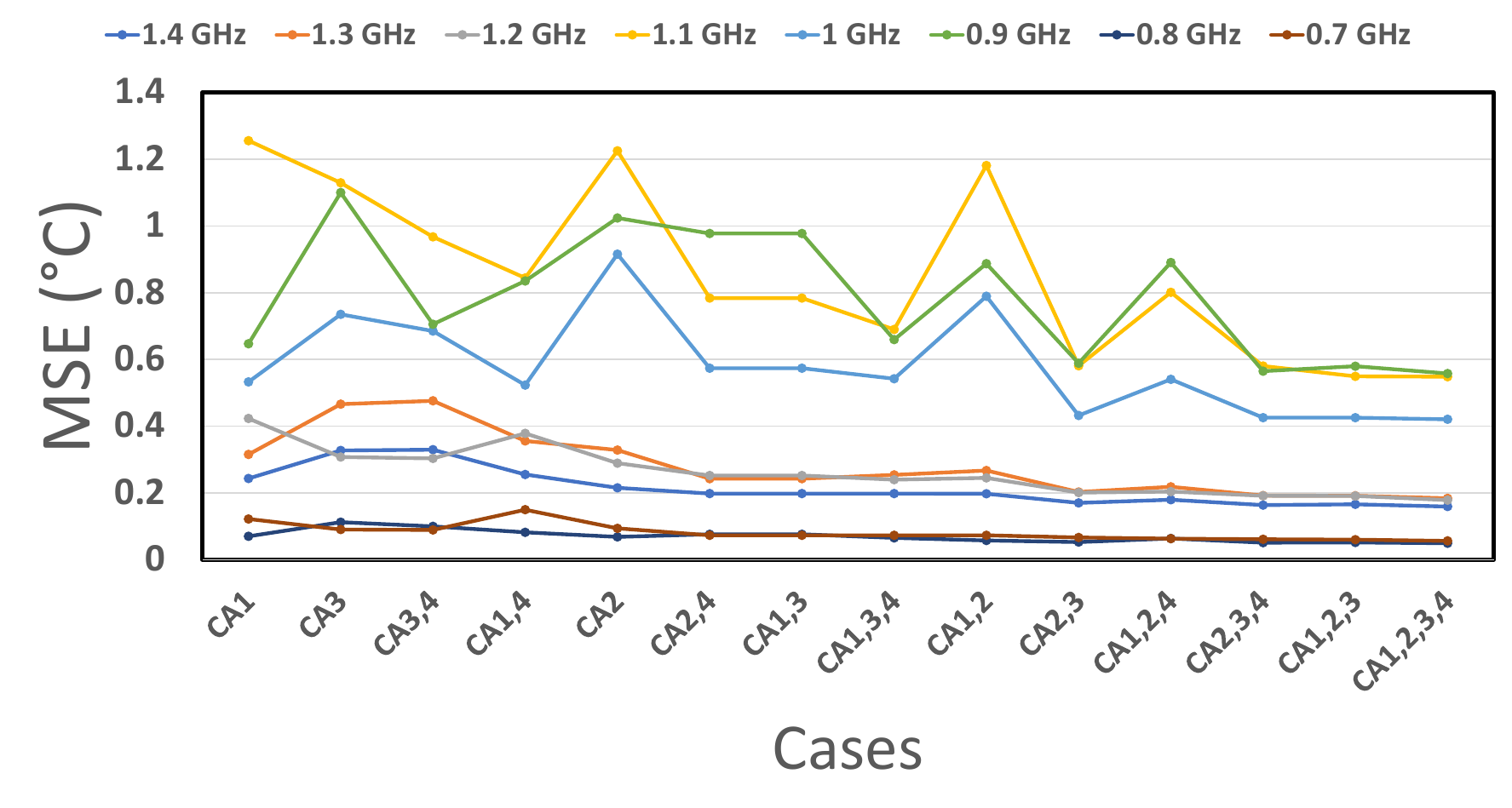}
\caption{MSE of CPU cores from all setting in different cases.}
\label{fig:chap4_mse}
\end{figure}

The mean square errors (MSE) of the temperature model for all CPU cores under all different settings are shown in Fig.~\ref{fig:chap4_mse}. The $x$ axis is sorted in ascending order in terms of the number of profiles used during the construction of $\myb[Y]$. Some spikes in the results, e.g., the yellow line at $CA1,2$, are due to that  additional traces contained noisy data. However, the error generally decreases as more profiles are used for the construction of $\myb[Y]$. This trend indicates that our proposed approach to considering data ensembles can reduce the negative impact of noisy data and improve the accuracy of thermal parameter estimation. 
%The results from different frequencies reveal the necessity of considering the proposed multi-frequency approach in estimating the thermal parameters to reduce the impact of noise on data of one frequency level. 

\subsection{Floorplan Estimation}
\begin{figure}[t]
\centering
\subfloat [][core 1]{\includegraphics[width=0.49\textwidth]{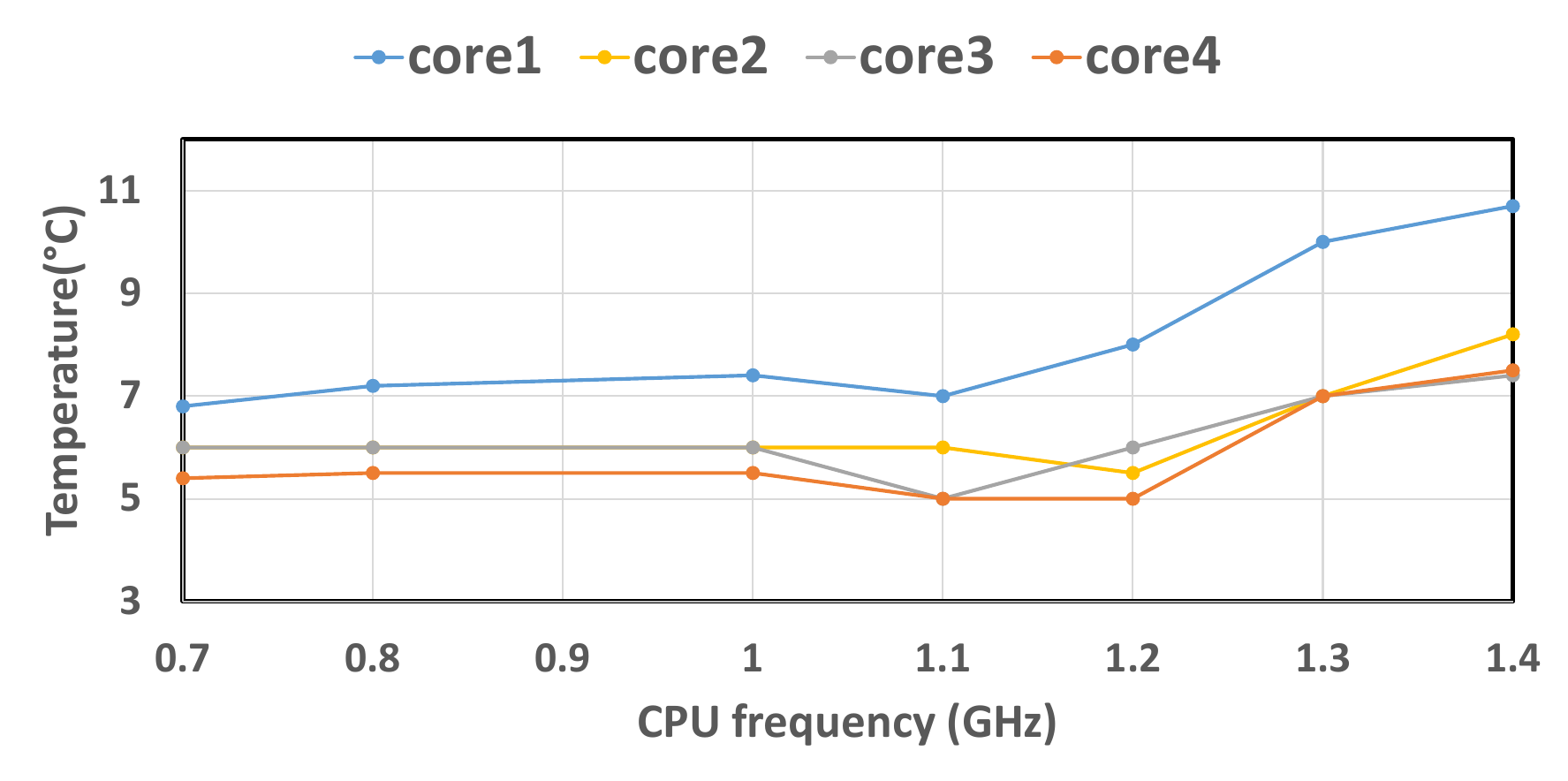}}
\subfloat [][core 2]{\includegraphics[width=0.49\textwidth]{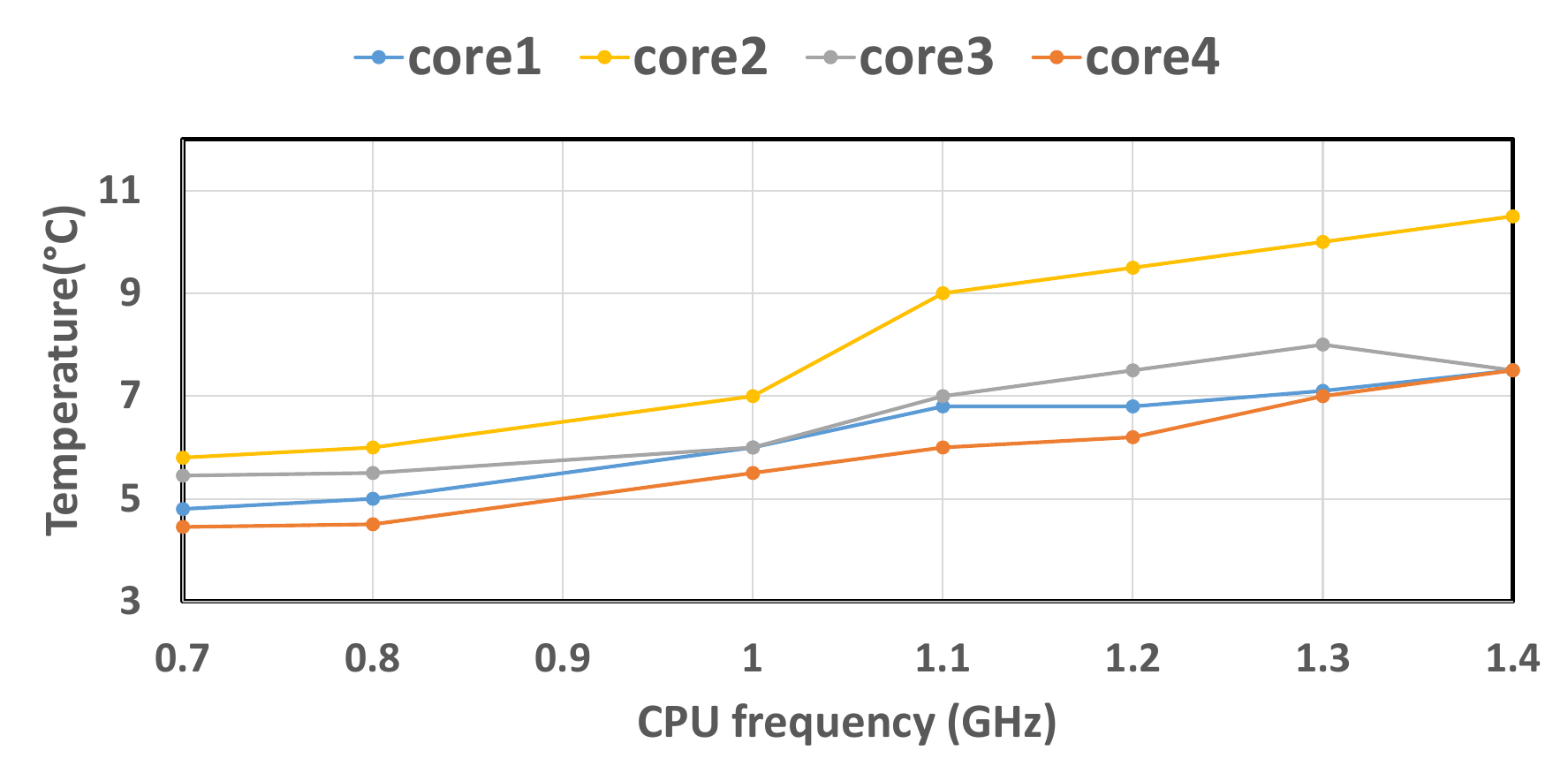}}\\
\subfloat [][core 3]{\includegraphics[width=0.49\textwidth]{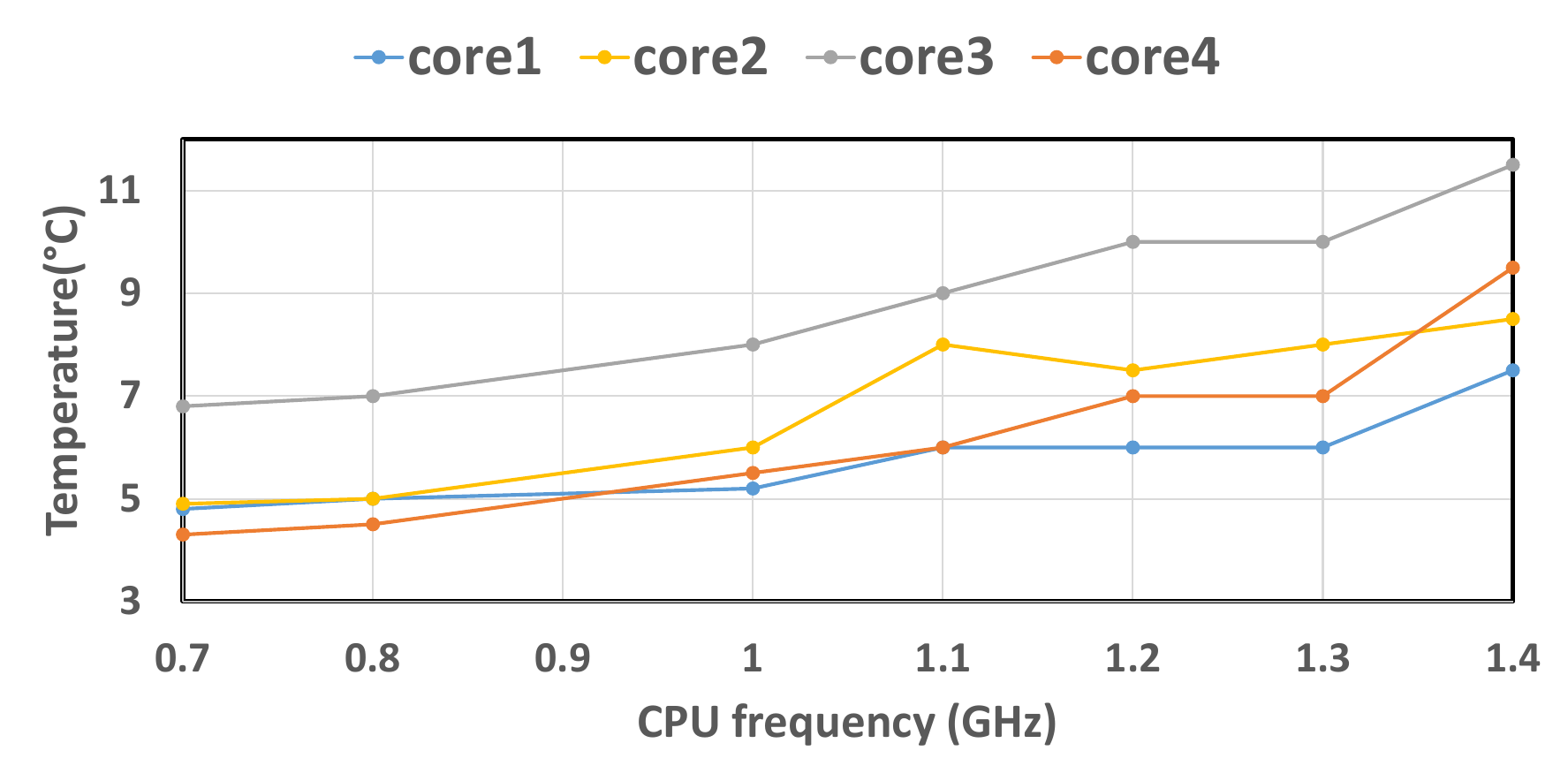}}
\subfloat [][core 4]{\includegraphics[width=0.49\textwidth]{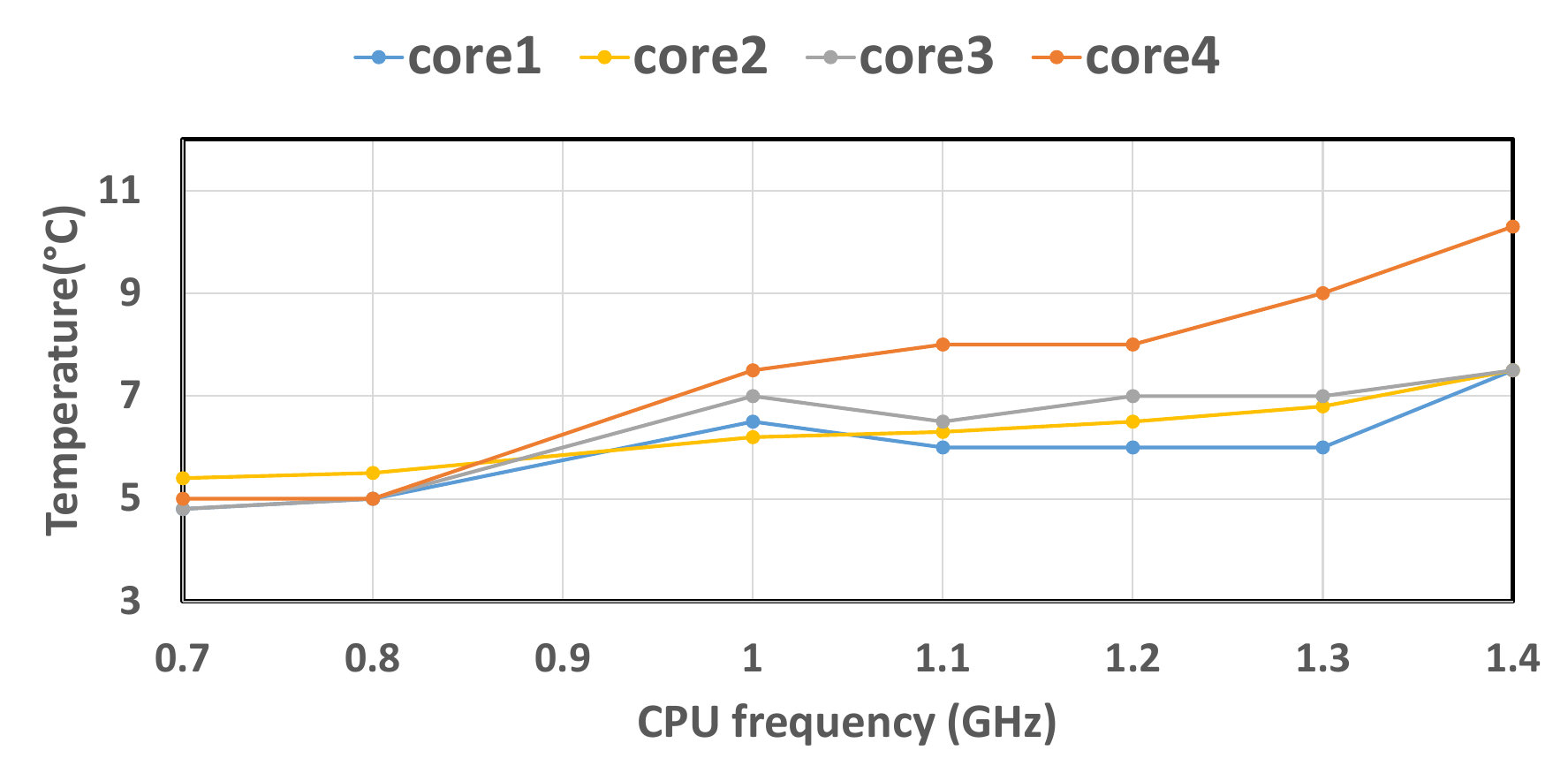}}
\caption{Temperature increase when a)core 1 2)core 2 3)core 3 4)core 4 operates fully-utilized }
\label{fig:chap4_full_oneByone}
\end{figure}

We validate our proposed floorplan estimation method on the Exynos5422 SoC. The temperature increase of each CPU core when only one CPU core is fully-utilized at a time is depicted in Fig.~\ref{fig:chap4_full_oneByone}.

\begin{figure}[t]
\centering
\subfloat [][]{\includegraphics[width=0.18\textwidth]{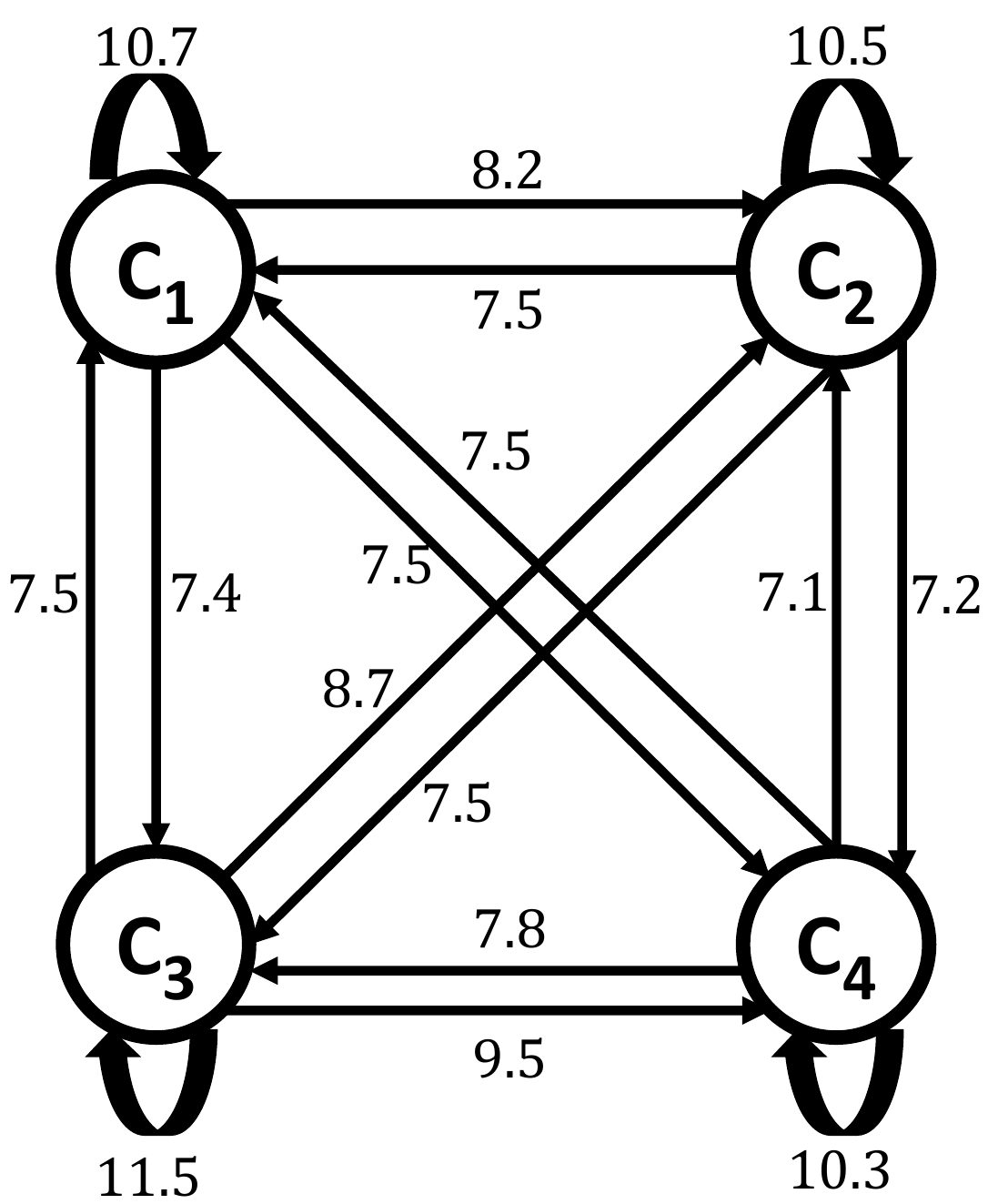}}
\subfloat [][]{\includegraphics[width=0.2\textwidth]{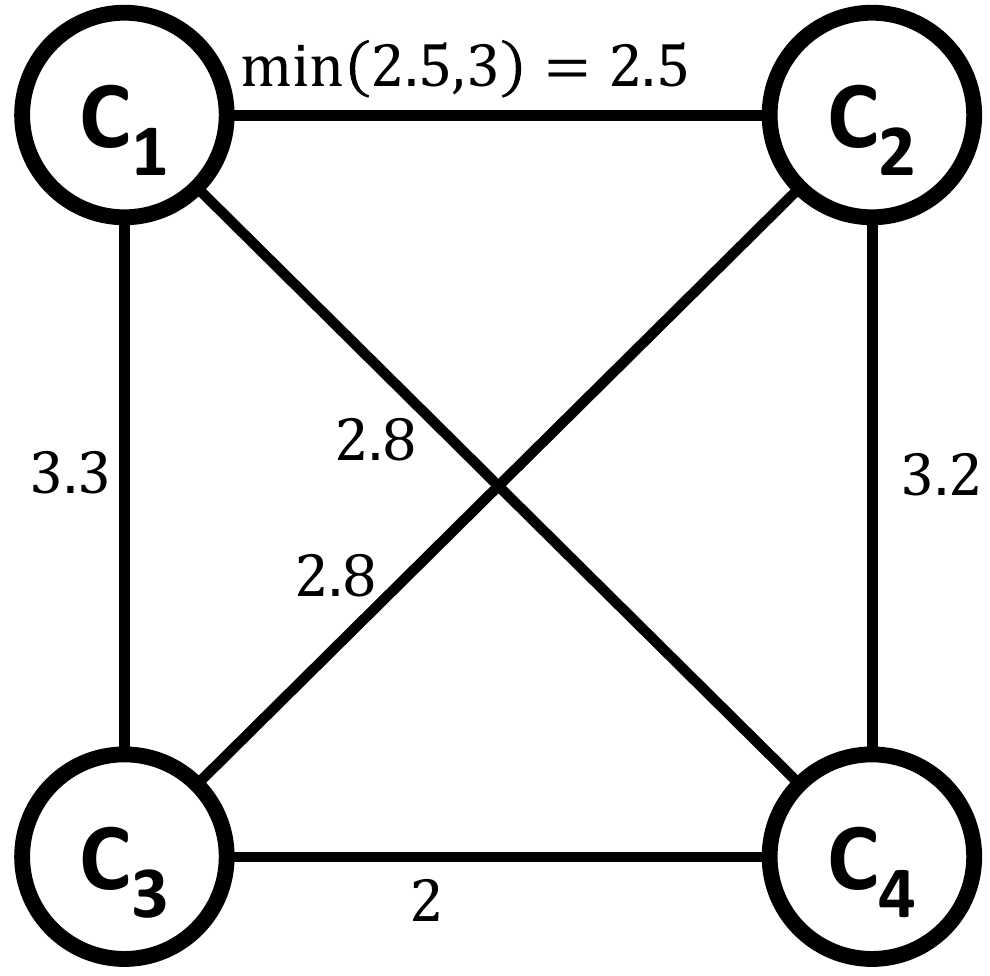}}
\subfloat [][]{\includegraphics[width=0.2\textwidth]{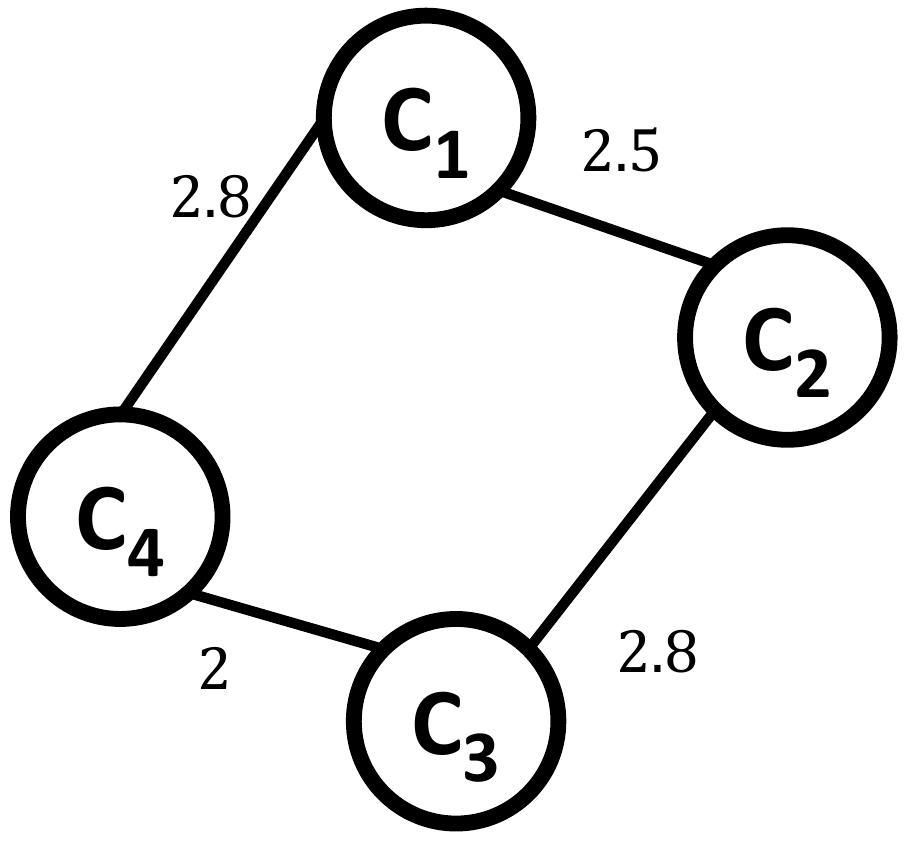}}
\subfloat [][core 1]{\includegraphics[width=0.2\textwidth]{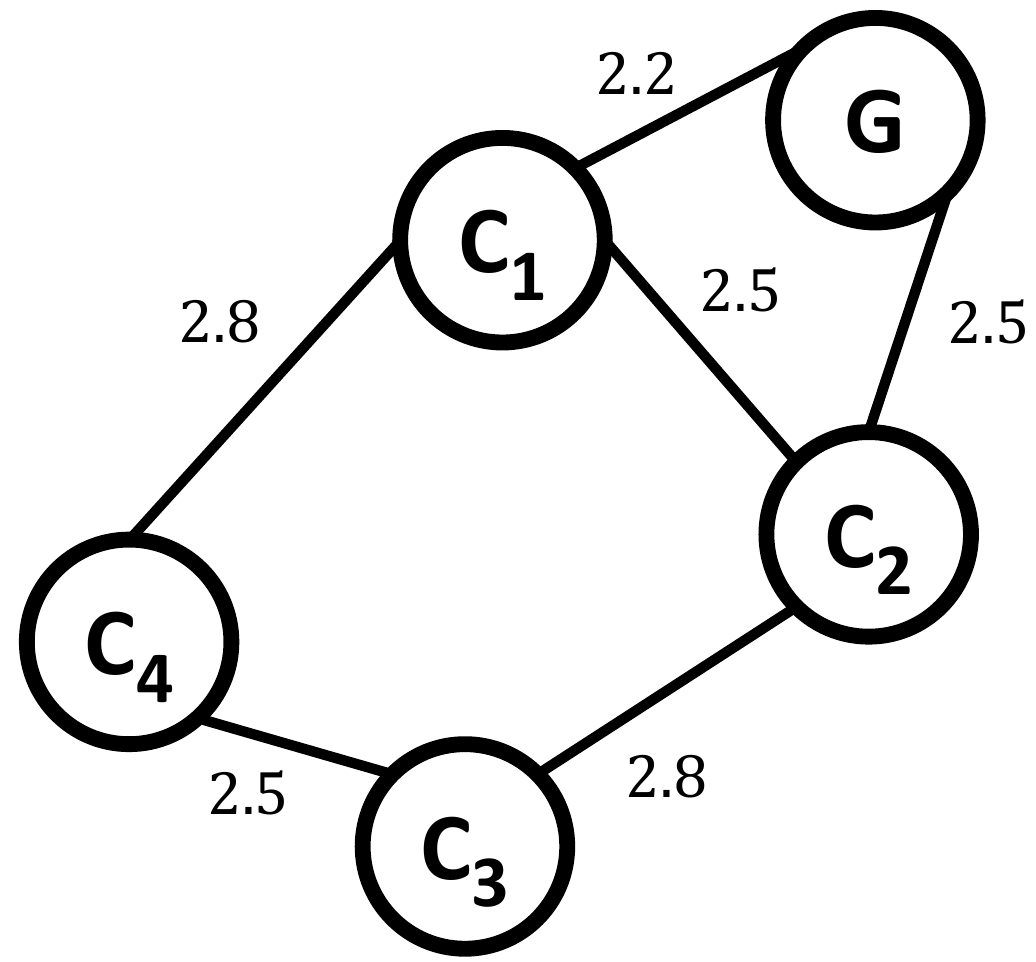}}
\subfloat [][]{\includegraphics[width=0.22\textwidth]{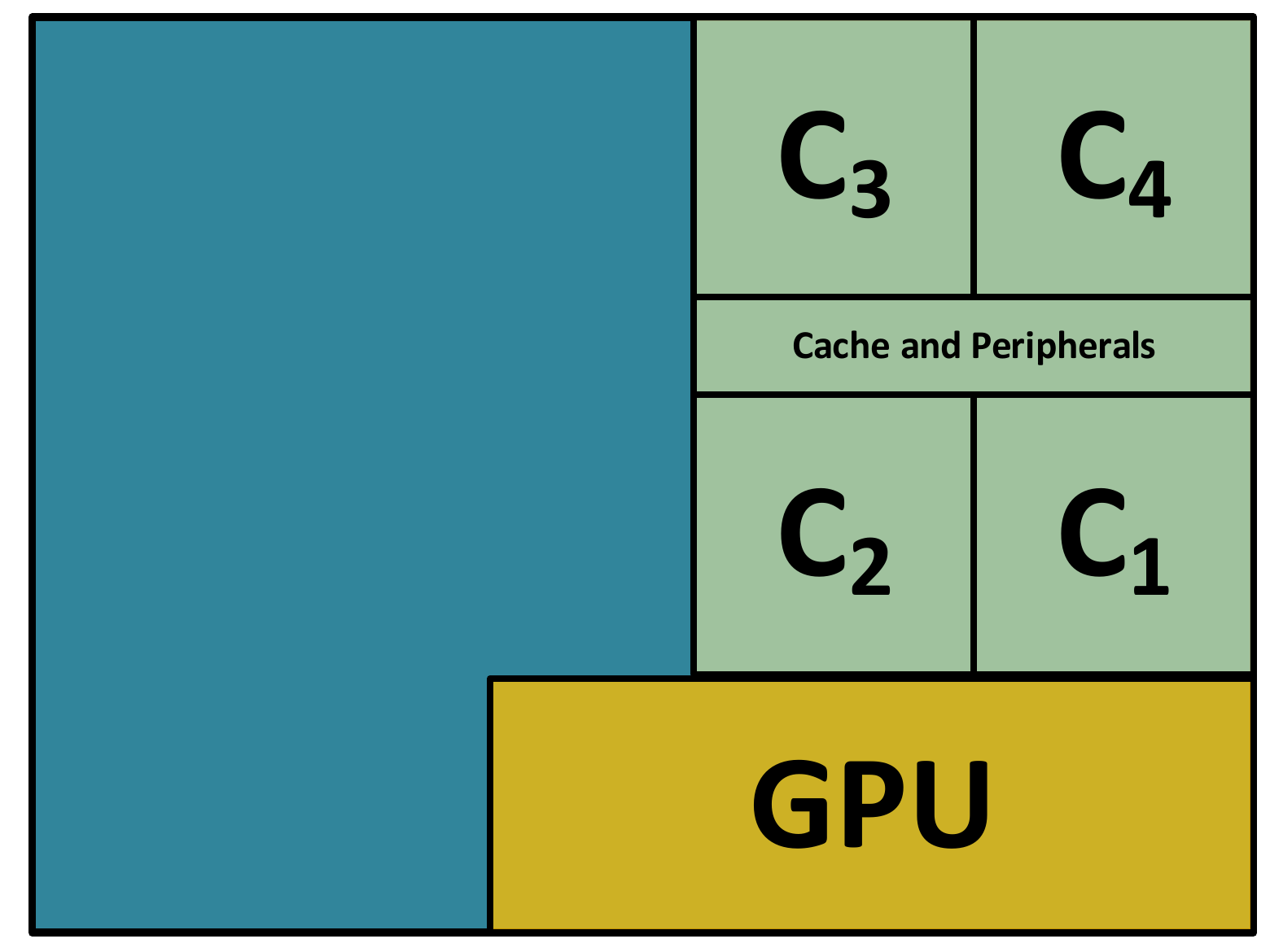}}
\caption{Floorplan estimation of Exynos5422 based on data of 1.4 GHz. (a) The fully-connected graph from the temperature increase data, (b) Graph reduction stage (c) The CPU affinity graph, (d) Estimation of GPU location relative to CPU location, and (e) The actual Exynos5422 floorplan~\cite{7904613}. }
\label{fig:chap4_exp_graphs}
\end{figure}

We draw the proposed adjacency graph according to data of $1.4$ GHz. Figures~\ref{fig:chap4_exp_graphs}a-c show the steps to construct the final adjacency graph for the Exynos5422. As we observe in Fig.~\ref{fig:chap4_exp_graphs}d, the cores $C_3$ and $C_4$ have greater spatial proximity than cores $C_3$ and $C_2$. Although the physical layout of the cores on the CPU package is bi-symmetrical, the primary reason for the asymmetrical layout shown in Fig.~\ref{fig:chap4_exp_graphs}d is that the location of the built-in temperature sensors on each processor may vary between cores. Additionally the L2 cache memory and peripheral controllers may have an effect on the modeled spatial proximity between the core pairs of ${C_1, C_2}$ and ${C_3,C_4}$ .
 % As shown, CPU cores $C_3$ and $C_4$ are closer to each other than CPU cores $C_3$ to $C_2$ are because aside from the location of built-in sensors, it is possible that the L2 cache memory and peripheral are between the pair of ${C_1, C_2}$ and ${C_3,C_4}$. 
Using the same approach and GPU temperature data, we are also able to locate the embedded GPU by profiling the heat conduction between each CPU cores and the GPU. Fig.~\ref{fig:chap4_exp_graphs}d depicts the estimated location of the embedded GPU in the Exynos5422. The floorplan estimation is validated with the data reported in~\cite{7904613}\footnote{The CPU core labeling in ~\cite{7904613} is different from the labels in the driver and it is verified with infra-red imaging  captured  by an FLIR A325sc  IR  camera~\cite{flir}.}.

We construct the template of the matrix $\myb[A]$ to be compatible with the estimated floorplan. The matrix $\Tilde{\myb[A]}$ at 1.4 GHz is then computed by using the steady-state data of $CA1,2,3,4$ at different frequency levels:

\[
\Tilde{\myb[A]} = \begin{bmatrix}
0.2961 & -0.1324 & 0 & -0.1194\\ 
-0.1324 &  0.3017 &-0.1579 & 0\\ 
0 & -0.1579 & 0.3088 & -0.1269\\ 
-0.1194 & 0 & -0.1269 & 0.2798\\ 
\end{bmatrix}.
\]

% \[
% \Tilde{Y}^{-1}_{1.2GHz} = \begin{bmatrix}
% 0.2996 &  -0.1600 & 0.0085 & -0.1121\\ 
% -0.0474 &   0.3113 &-0.1355 & -0.0793\\ 
% -0.0962& -0.0546 & 0.2939  & -0.1441\\ 
% -0.0713 & -0.1007 & -0.1468 & 0.3824\\ 
% \end{bmatrix}.
% \]
% As one can see the parameter on the diagonal of the matrix are very close to each other because of the homogeneity of the CPU cores in XU4 and the difference between is because heat dissipation of IPs around each of them and their location of built-in sensors 
\subsection{Relative Power Estimation}
\begin{figure}[t]
\centering
\subfloat [][]{\includegraphics[width=0.49\textwidth]{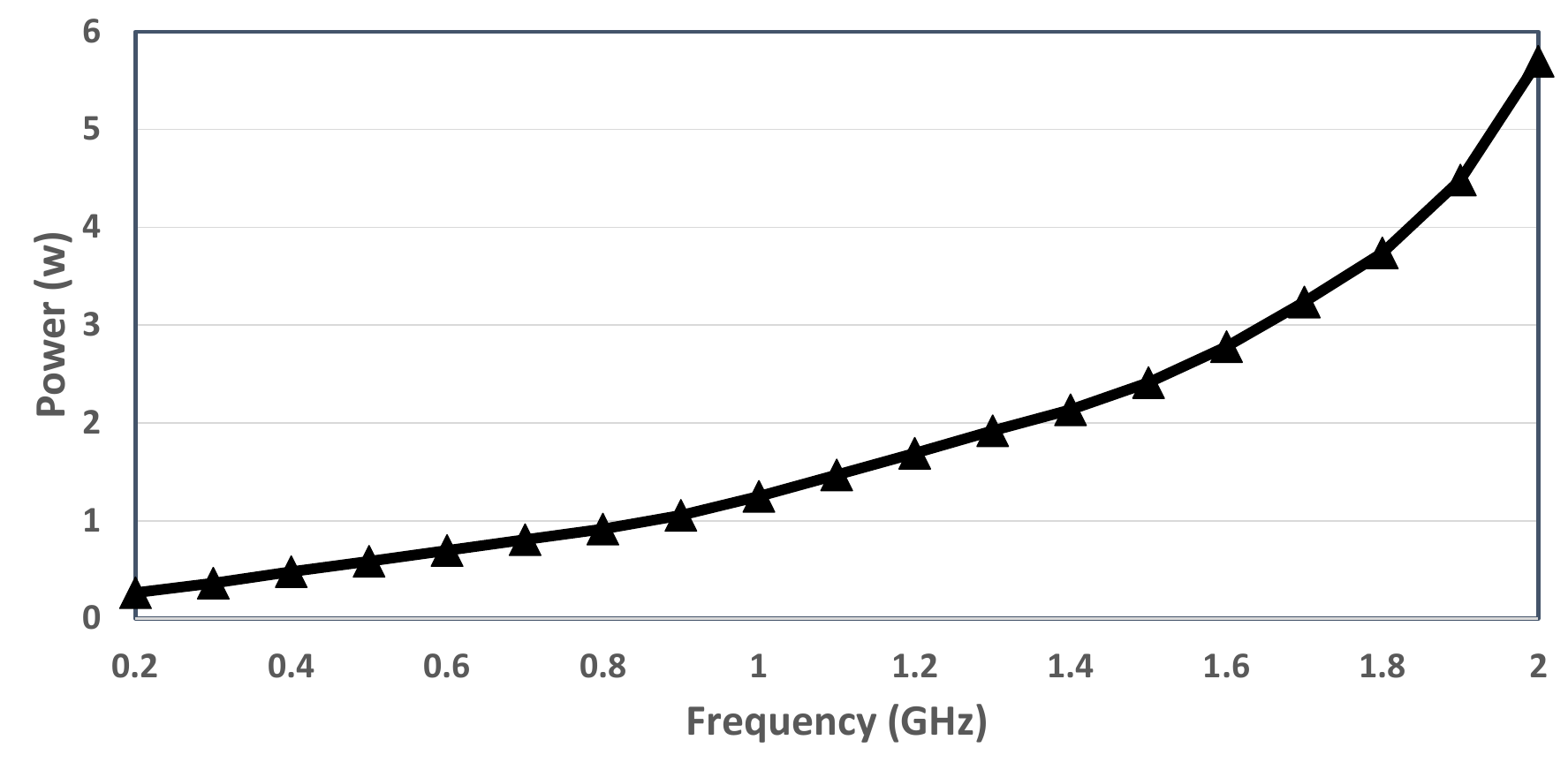}}
\subfloat [][]{\includegraphics[width=0.49\textwidth]{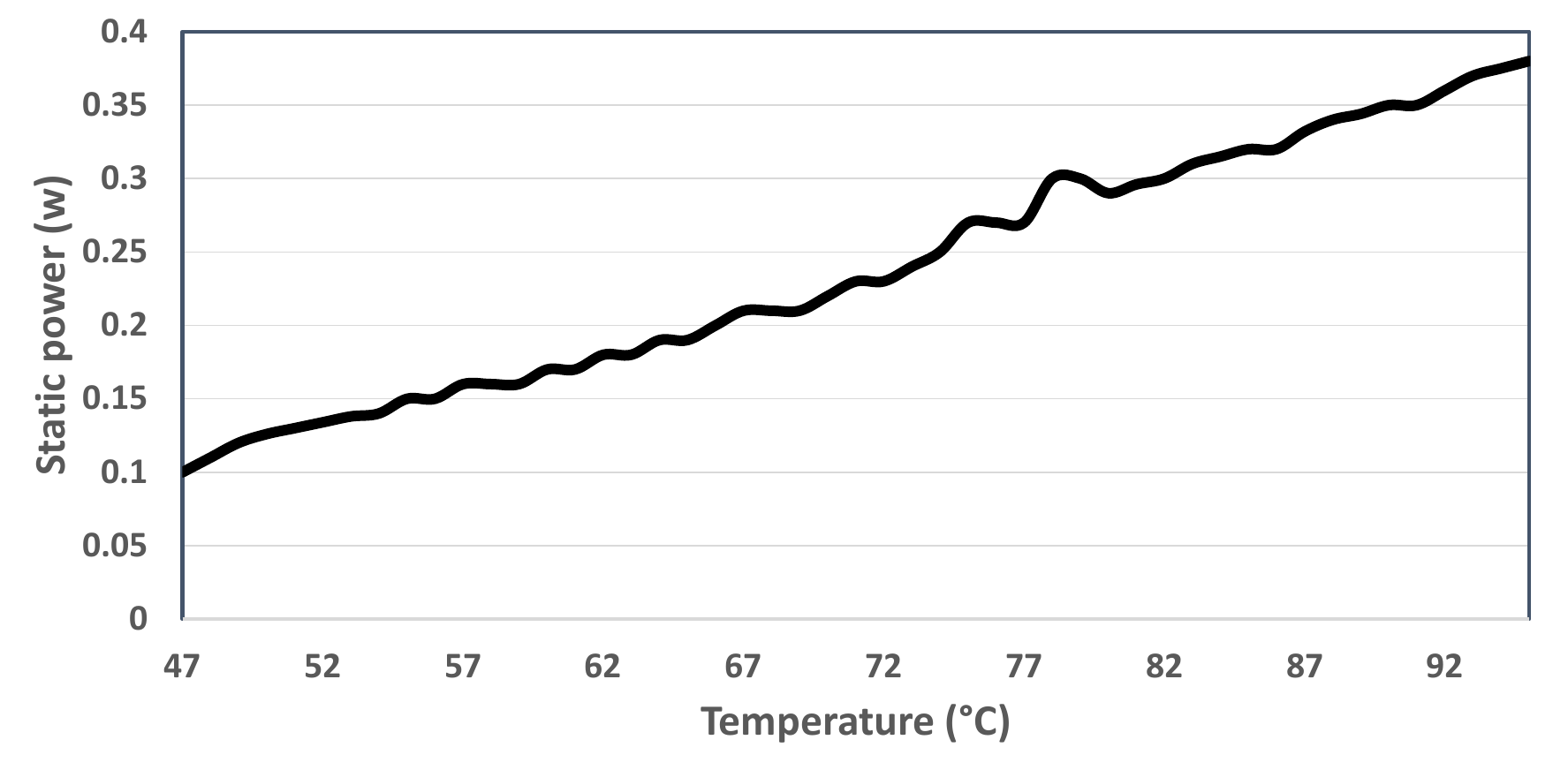}}\\
\subfloat [][]{\includegraphics[width=0.49\textwidth]{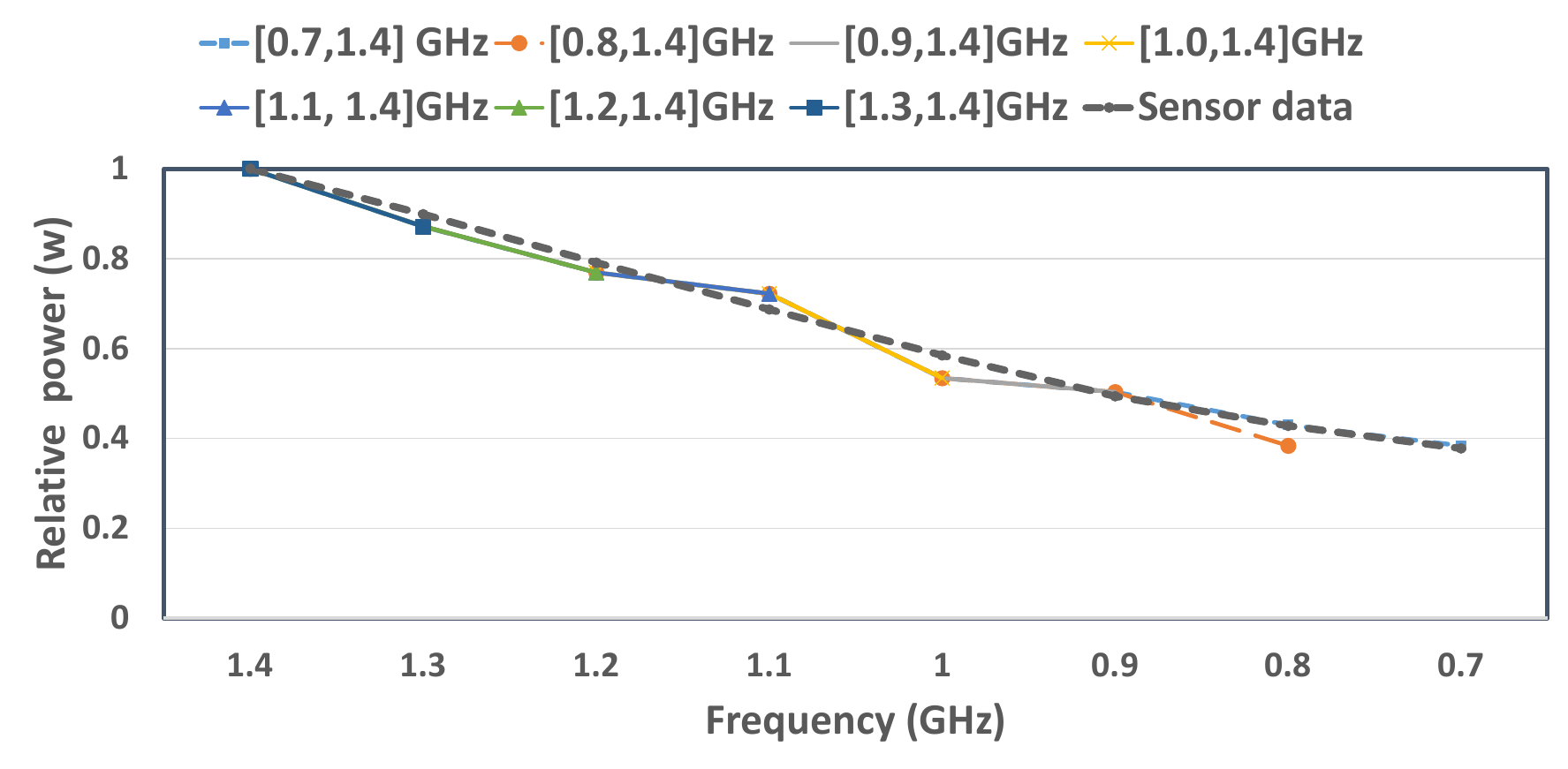}}
\subfloat [][]{\includegraphics[width=0.49\textwidth]{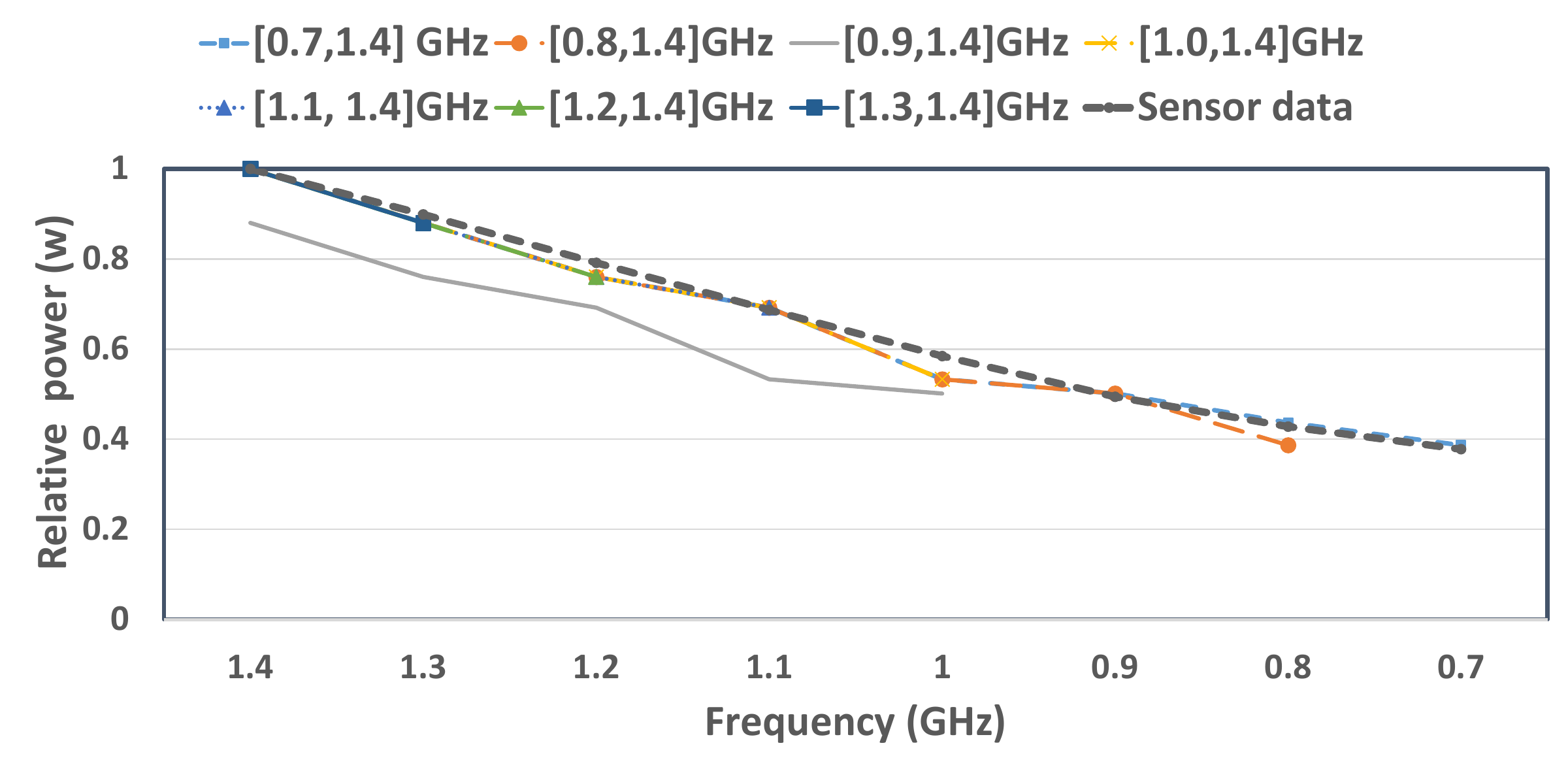}}
\caption{Power data of CPU cores in Exynos5422. (a) Total power of the big cluster with built-in sensors, (b) The leakage power data; the relative power estimates with different frequency ranges used, (c) Comparison between the estimated relative power consumption and the normalized actual power data from built-in power sensors for $CA1$, and (d) Comparison for $CA1,2,3,4$.}
\label{fig:chap4_power_est}
\end{figure}

The relative power consumption of each CPU core can be estimated while estimating $\Tilde{\myb[A]}$, as explained in Sec.~\ref{sec:chap4_multi_frequency_ensembles}. 
Fig.~\ref{fig:chap4_power_est} illustrates the estimated power consumption. The results are obtained using profiles from different frequency ranges. As depicted in the figure, the estimated relative power closely follows the actual data collected from the built-in power sensors of the XU3 board that is equipped with the same Exynos5422 SoC.

\subsection{Temperature Prediction}
We estimate the parameter $\gamma_1$ as the base parameter for the clock frequency 1.4 GHz in our experiments. Based on this, the absolute value of other $\gamma$s can be determined. As discussed in Sec.~\ref{sec:chap4_trans}, we use the transient-state trace when all CPU cores are cooling down. 

\begin{figure}[t]
\centering
\subfloat [][core 1]{\includegraphics[width=0.4\textwidth]{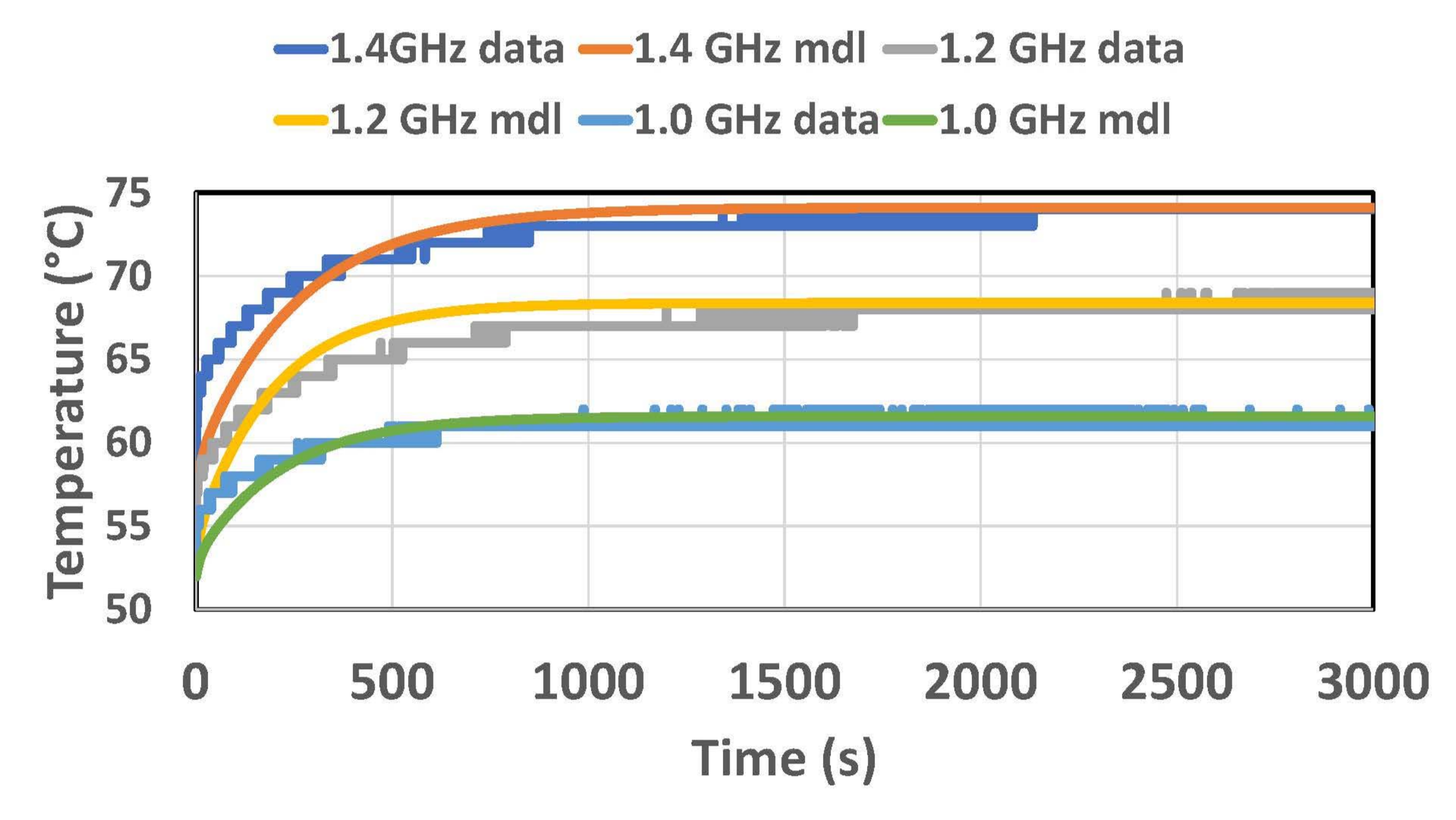}}
\subfloat [][core 2]{\includegraphics[width=0.4\textwidth]{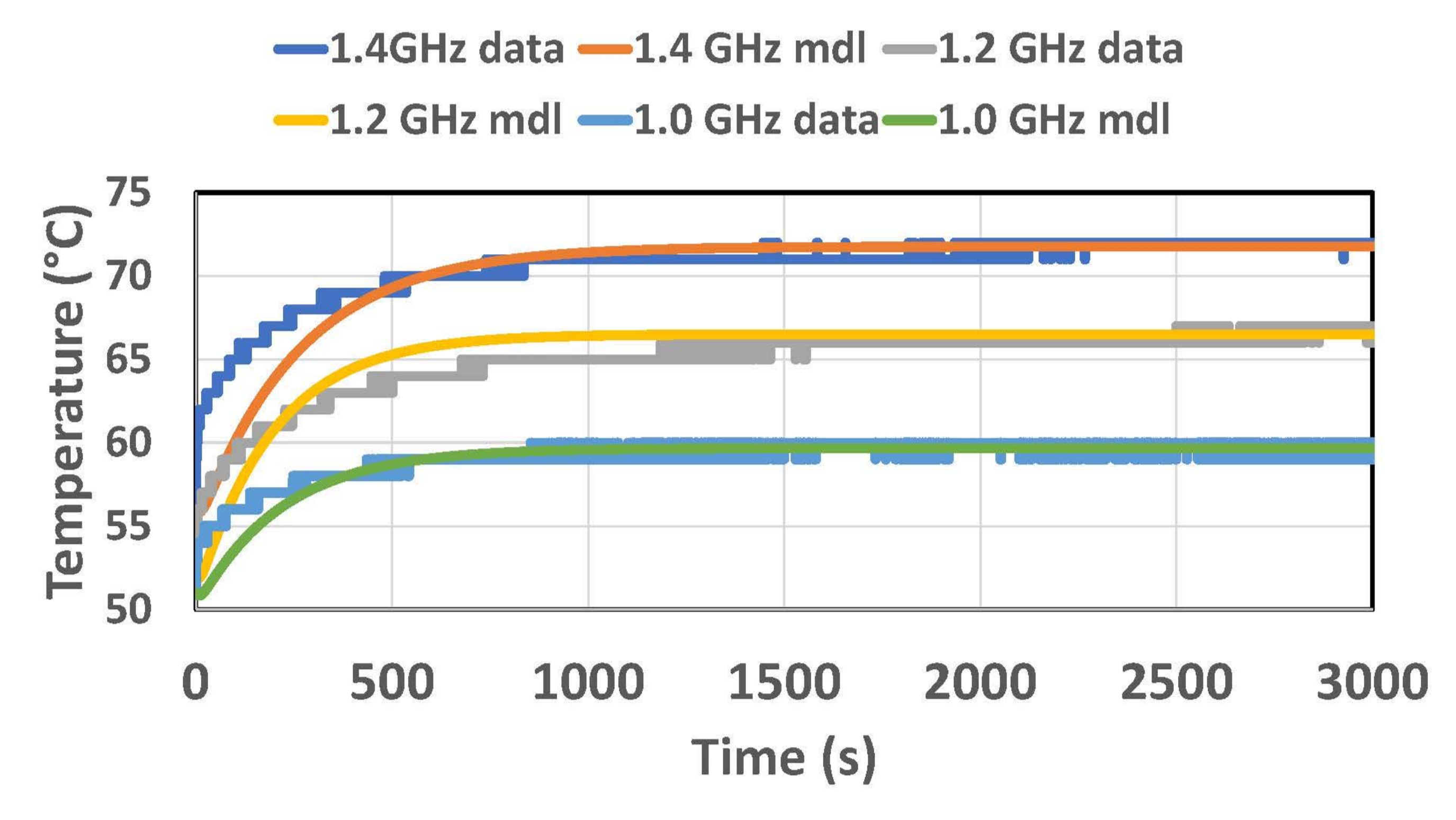}}\\
\subfloat [][core 3]{\includegraphics[width=0.4\textwidth]{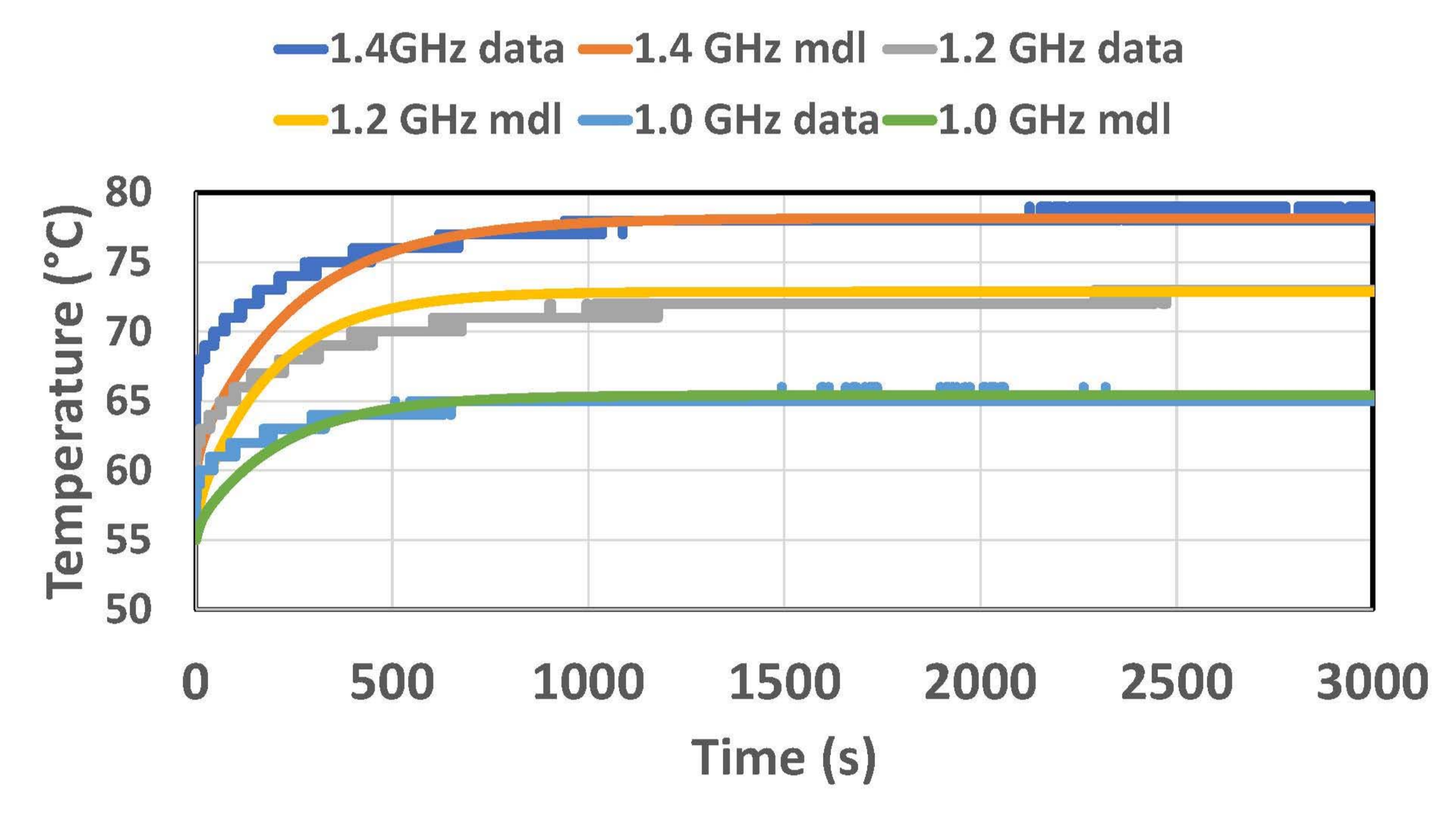}}
\subfloat [][core 4]{\includegraphics[width=0.4\textwidth]{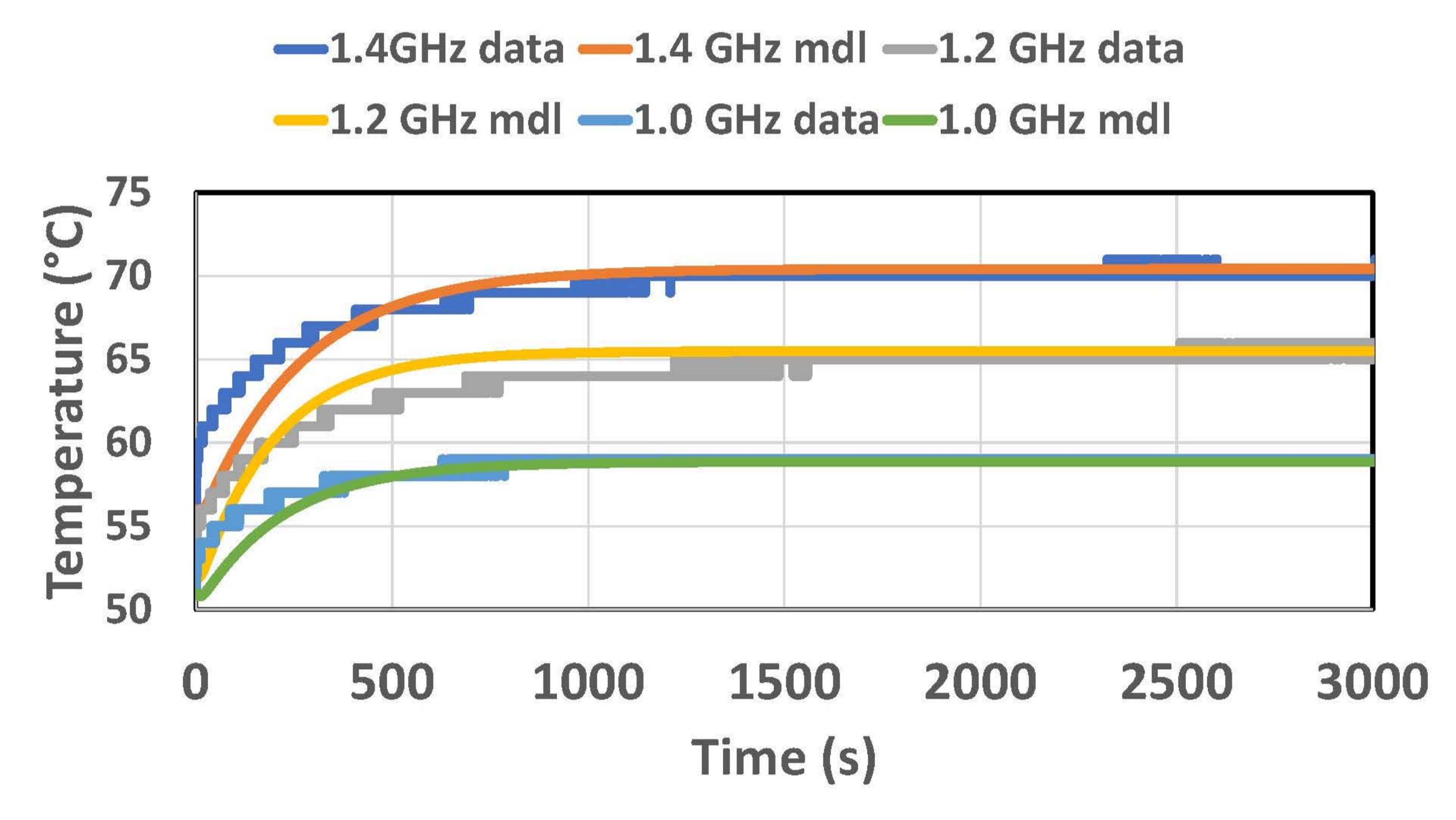}}\\
\subfloat [][core 1]{\includegraphics[width=0.4\textwidth]{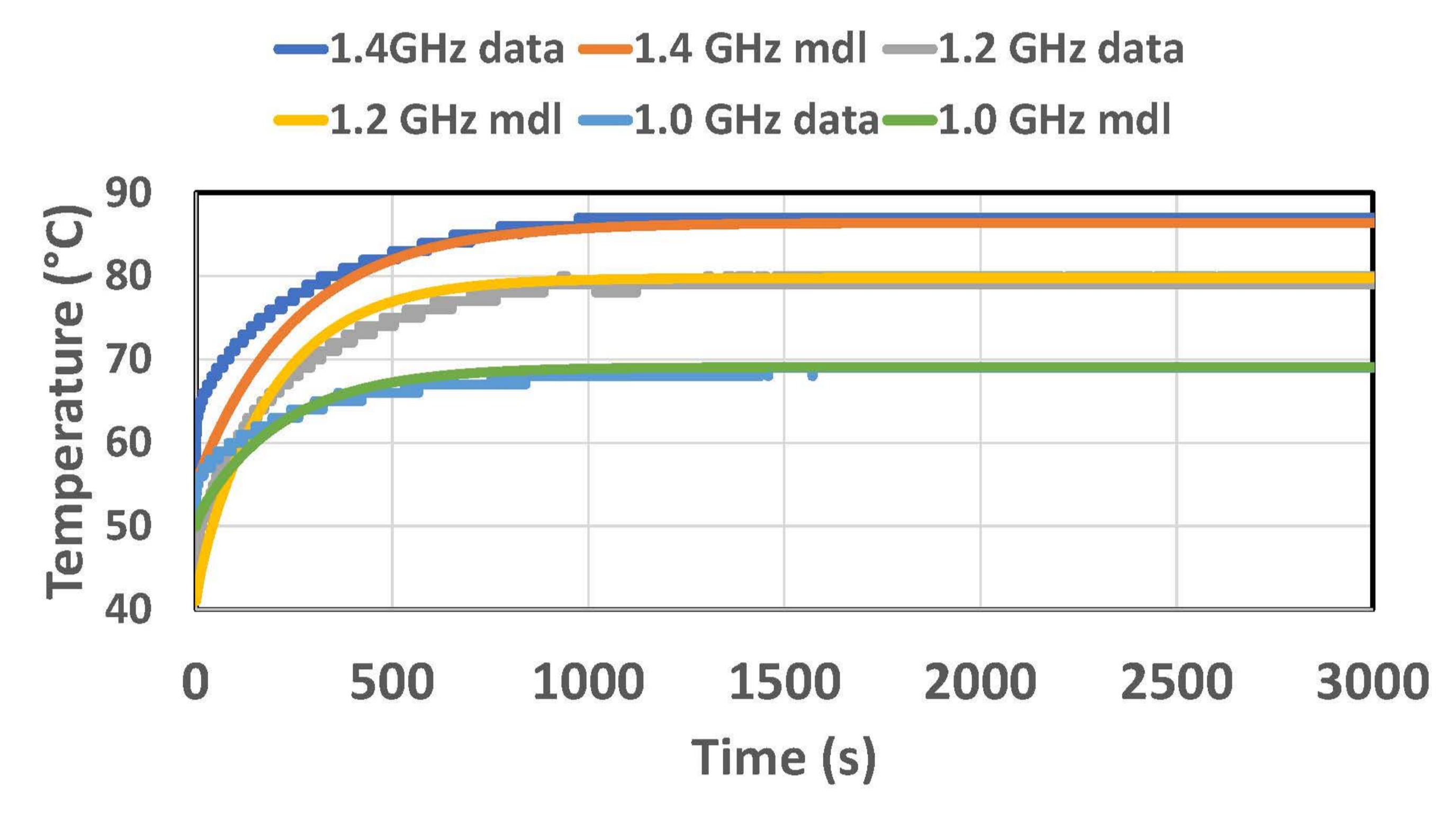}}
\subfloat [][core 2]{\includegraphics[width=0.4\textwidth]{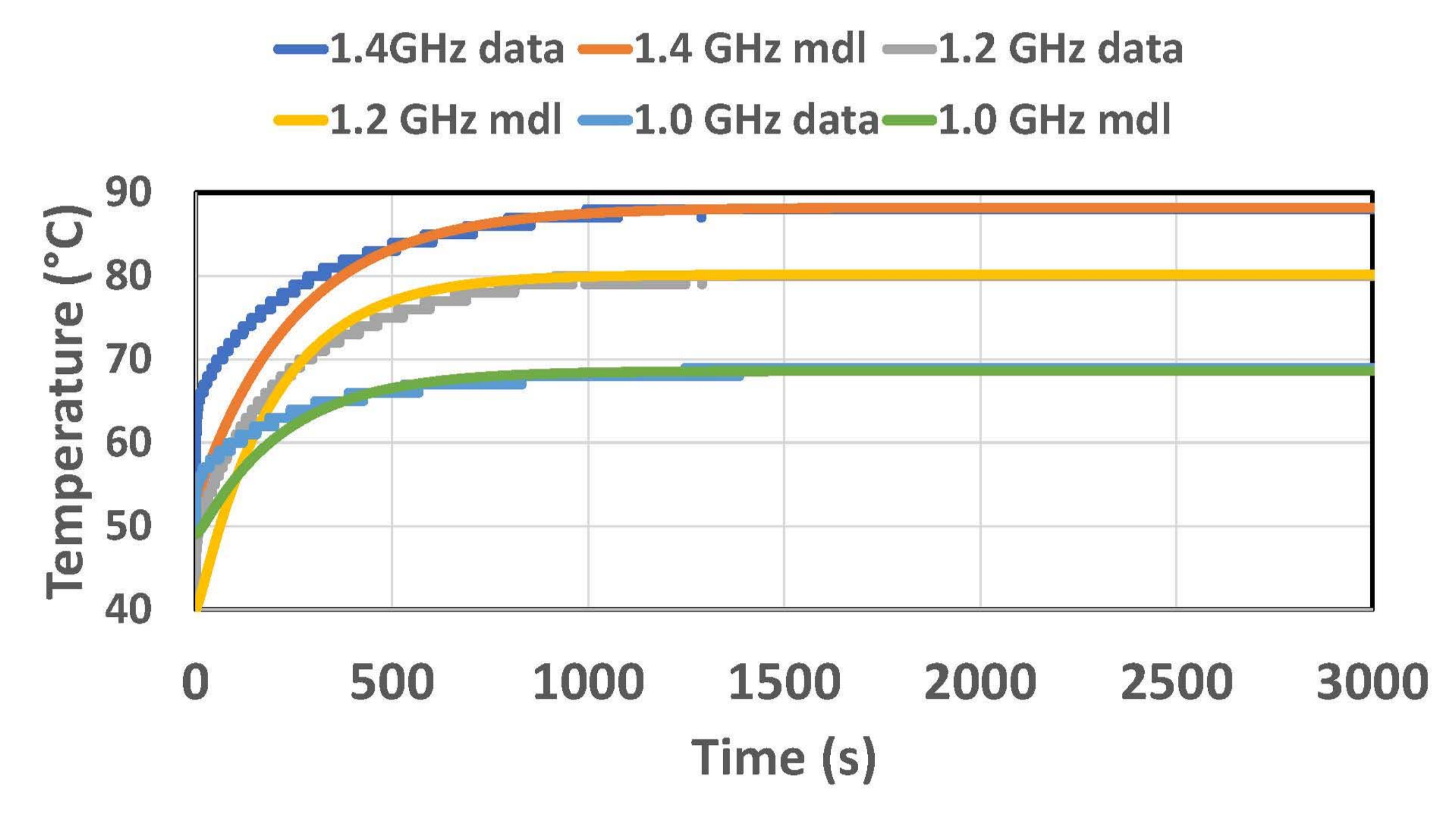}}\\
\subfloat [][core 3]{\includegraphics[width=0.4\textwidth]{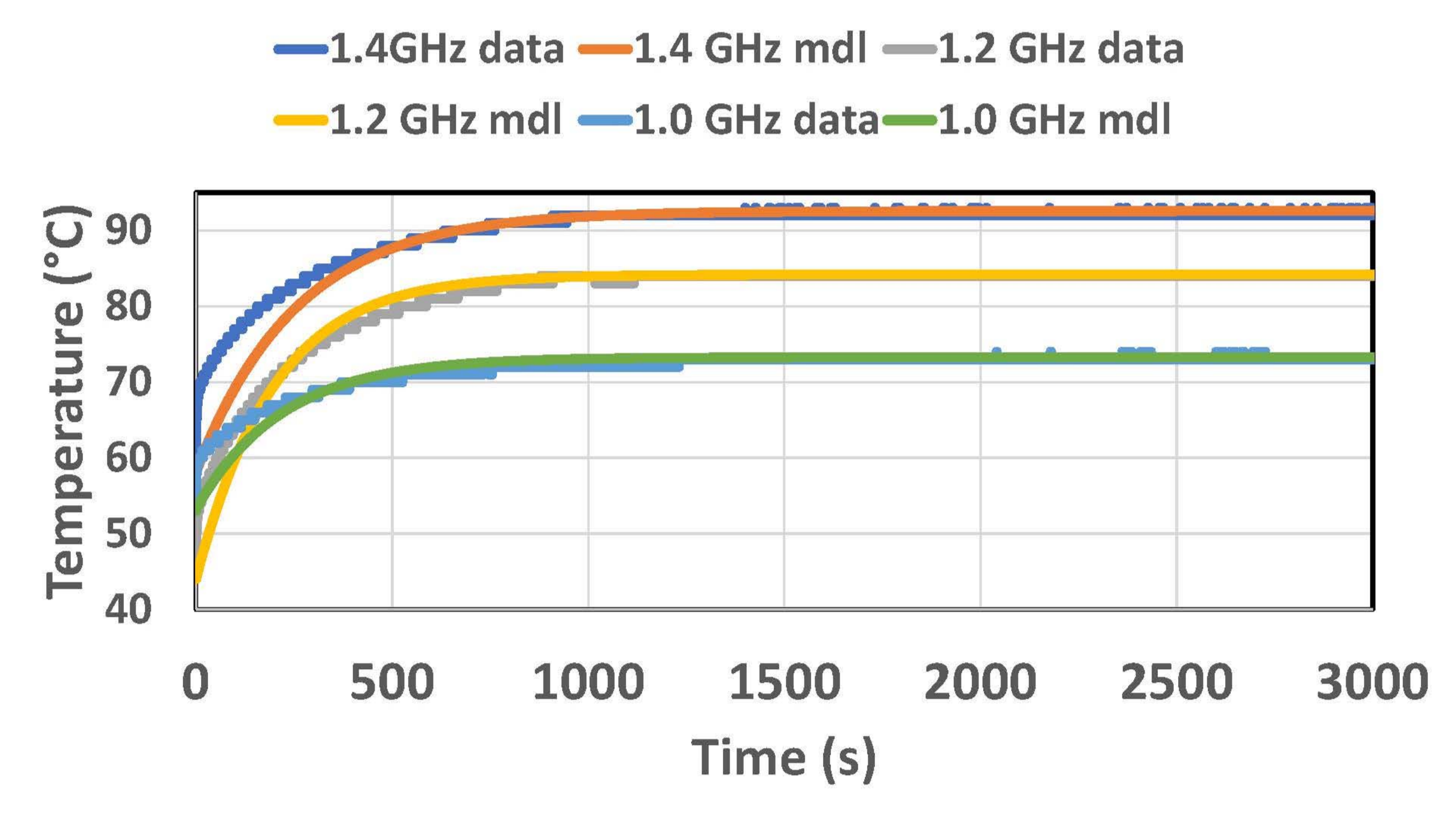}}
\subfloat [][core 4]{\includegraphics[width=0.4\textwidth]{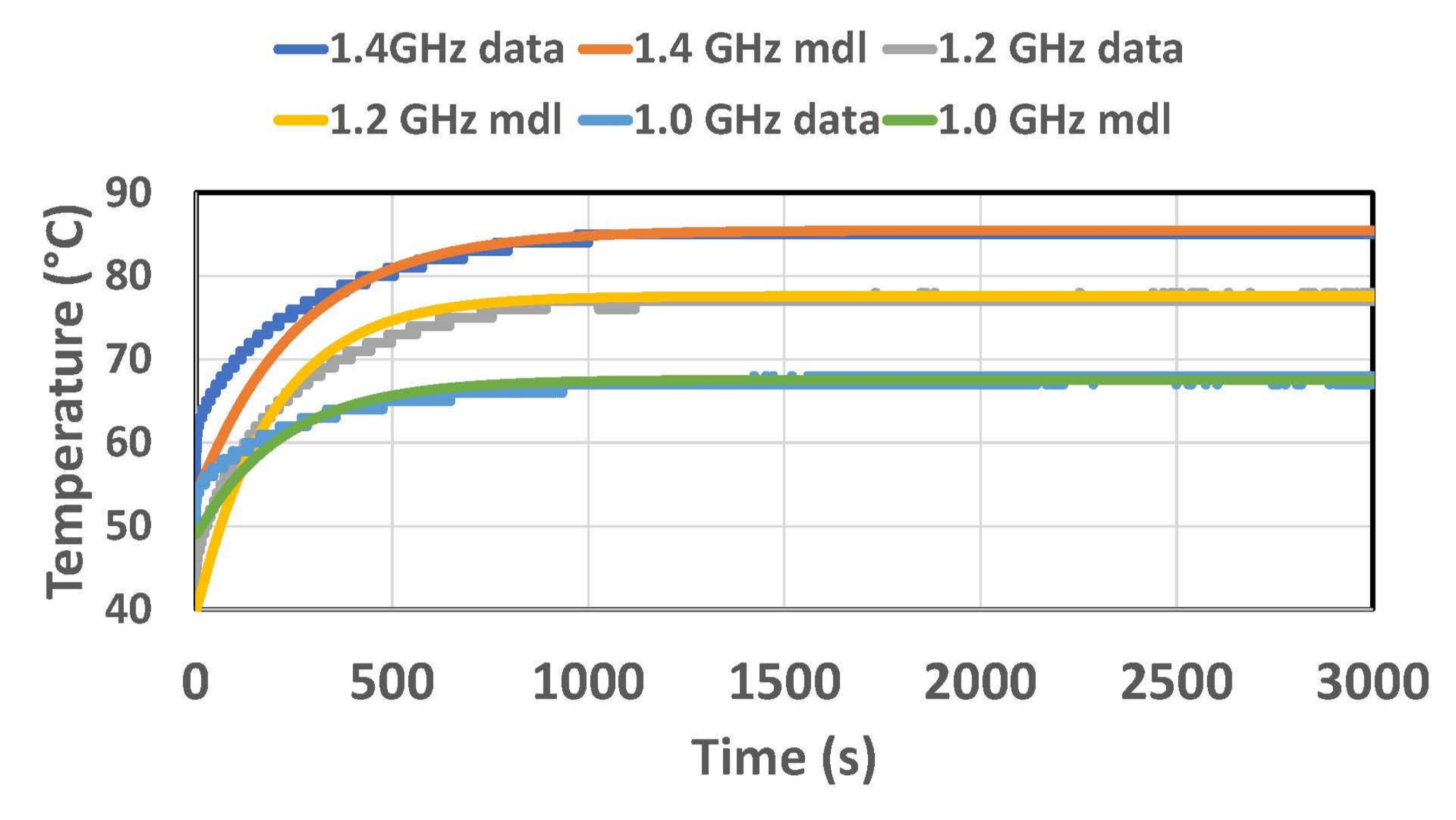}}
\caption{The CPU temperature from sensor data and the model when a-d) two cores are fully-utilized $Z_{10}$ e-h) all cores are fully-utilized $Z_{15}$. }
\label{fig:chap4_transData}
\end{figure}

After estimating the values of $\gamma$s, we apply our model to predict the operating temperature of CPU cores in two different scenarios. Figures~\ref{fig:chap4_transData}a-d  show the CPU temperatures when only two CPU cores are fully utilized and Figures~\ref{fig:chap4_transData}e-h show that when all CPU cores are fully utilized. In each sub-figure, ``data'' means the CPU operating temperature measured from a real platform and ``mdl'' means the temperature predicted by our model. 
As shown, there are some differences in transient-state temperature values, especially at the earlier part of the experiments. The difference is relatively small when the frequency is low. 
%However, because of high impulse response of built-in sensors, there is a slight difference at the beginning of each experiments. This difference is considerably low in lower frequency levels.
On the other hand, in steady state, the temperature predicted by the model is very close to the real one and the difference is less than 1.25\textdegree~C in all cases. Since the steady-state temperature is the metric to test the thermal safety of the system, we expect that our proposed scheme can be effectively used in the thermal-aware design of COTS-based mixed-criticality systems.

\section{Summary} 

In this chapter, we proposed a novel, rapid, and accurate scheme to estimate the thermal parameters of COTS multi-core processors for real-time mixed-criticality systems. By decomposing our estimation scheme into steady-state and transient-state stages, we substantially reduce the number of transient-state profiles needed to estimate the system's thermal parameters. We presented methods to improve the accuracy of our scheme by utilizing additional temperature profiles in the parameter estimation. Our proposed scheme is fast to converge and has a low computational cost for the prediction of chip operating temperature. Hence, it can be even used in an event-driven manner, e.g., at the arrival and the departure of each job of periodic real-time tasks, with negligible memory and computational overhead. We derived the thermal characteristics of a multi-core processor which remain unchanged across different frequency levels. We also showed the effectiveness of our scheme in extracting the relative power consumption information from  temperature profiles.

% In this chapter, we proposed a novel fast and accurate scheme to estimate the thermal parameters of COTS multi-core processors for real-time mixed-criticality systems. By decomposing our estimation scheme into steady-state and transient-state stages, the number of transient-state profiles needed for estimation of thermal parameters reduces substantially. We presented an improvement to our proposed scheme by using more profiles in the parameter estimations for the sake of accuracy improvement. Our proposed model is fast and has low computational overhead since it is event-driven, hence the operating temperature only needs to be calculated at the arrival and departure of a real-time task with incurring negligible memory and computational overhead. We derived the thermal characteristics of mixed-criticality systems which remains unchanged during different frequency levels. We also showed the effectiveness of our scheme in extracting the relative power consumption information embedded in temperature profiles. 

% For future work, we plan to extend our proposed scheme to capture the thermal footprint of different tasks on heterogeneous SoCs by using a minimum number of profiles. We will develop our analysis to capture the effect of cooling packages and forced heat convection so that operating temperature of SoCs can be estimated with various CPU cooling conditions during run-time. Additionally, we will study different statistical approaches to construct the adjacency matrix of the location of CPU cores and IPs.    

\chapter{Conclusions}
This dissertation shows that the timing predictability of real-time tasks in dynamic thermal conditions is achievable on multi-core GPU-integrated SoC platforms by designing a novel thermal-aware system framework with the support of analytical timing and thermal models.

We discussed challenges arising from the thermal interference on heterogeneous multi-core SoC platforms. In Chapter 2, we focused on bounding the temperature increase due to heat conduction among CPU cores as well as integrated computational accelerators like GPUs. We proposed a thermal-aware CPU-GPU framework to handle both real-time CPU-only and GPU-using tasks by designing thermal-aware servers that bound the maximum operating temperature of SoCs with the assurance of performance predictability. In Chapter 3, we introduced a novel multi-core mixed-criticality scheduling framework with ambient temperature awareness to bound heat generation at each criticality level and to trigger the criticality mode change by ambient temperature changes. We tackled the thermal-aware real-time scheduling challenges in dynamic environment conditions by decomposing the analysis into thermal schedulability and timing schedulability. We introduced the notion of \textit{idle thermal servers} that allow bounding the maximum operating temperature in MCS in way that the temperature increase due to heat dissipation during an inactive time of idle servers is not less than that of running preemptive servers under any task execution pattern. In Chapter 4, we presented our work on the estimation of processor thermal parameters to obtain a precise thermal model without using special measurement instruments or access to proprietary information. We showed that even a small number of temperature traces from on-chip thermal sensors is sufficient to achieve an accurate thermal parameter estimation. Not only can our framework detect anomalies in thermal profiles, but increases accuracy using  multi-frequency data ensembles and takes into account extra observations made at one or a subset of frequency levels. 

\section{Future Work}
There are possible extensions and improvements to the work presented in this dissertation. The following suggests several future research directions that can be built upon our work.
\subsection{Timing Assurance and Thermal Safety on SoC Platforms}
%Task allocation to CPU and GPU cores can further increase the schedulability for a given temperature constraint.
By considering task allocation to CPU and GPU cores, the schedulability of real-time tasks with a given temperature constraint can be further increased.
We believe that it is possible to develop an offline mapping algorithm to increase the schedulability of a taskset by determining whether a task executes on the CPU or the GPU according to the system parameters and its execution timing model. 

In our task model, tasks may or may not include concurrent segment which executes on GPU. When there is a GPU segment request, the corresponding task issues a GPU access request. However, it is possible to design an online algorithm to determine if concurrent task segment executes on GPU or CPU. In this way, some jobs of a task can execute only on CPU while the rest executing on the integrated GPU. We believe this technique can reduce the maximum peak temperature of embedded systems and also increase the timing schedulability of tasksets. 

The miscellaneous operations contains a single data transfer to/from GPUs. However, in practice, for improving performance or overcoming the resource limit of GPUs, data can be divided into several chunks and transmitted to GPUs so that data copy and kernel execution can be overlapped. We believe that updating the GPU execution model and revisiting some of our proposed protocols will enable this feature and the practical applicability of our proposed framework will also increase.

The thermal behavior of GPU-using tasks may depend significantly on the type of resources used by their kernels. For instance, a GPU kernel frequently accessing local memory may generate much less heat than those using global memory or being computationally-intensive. A potential direction to this issue would be developing an analytical model based on the thermal footprint of real-time applications to bound the maximum peak temperature with acceptable margins. 

\subsection{Dynamic Ambient Temperature-Aware Framework}
In our proposed thermal-schedulability analysis for multiple real-time tasks on multi-core SoCs, identical idle servers are assumed for all CPU cores at each criticality level. This assumption can affect the thermal schedulability of a real-time taskset. We believe that extending our analysis to allow different idle server settings per CPU core can improve the range of ambient temperature supported by mixed-criticality levels. Besides, developing our analysis to capture the effect of cooling packages and forced heat convection can be a challenging issue.

In our proposed framework, a  criticality  mode  change  is  triggered  by  ambient  temperature which is different from the well-known Vestal model~\cite{vestal2007preemptive}, which focuses on varying assurance of execution timing. It is intriguing to extend our framework such that changes in both ambient temperature and execution time of tasks trigger the criticality mode of MCS.

\subsection{Thermal Parameter Estimation Scheme}
We believe that our proposed scheme can be extended to capture the thermal footprint of different tasks on heterogeneous SoCs by using a minimum number of profiles. We also believe that the idea of  further development of the analysis to capture the effect of cooling packages and forced heat convection lead to the  estimation of operating temperature of SoCs in various CPU cooling conditions at runtime. Additionally, statistical approaches to construct the adjacency matrix of the location of CPU cores and IPs can be a potential intriguing extension.    

The robustness of our proposed estimation scheme against noise in the thermal profiles has not been discussed through mathematical analysis. To be more precise, although our proposed method estimates the thermal parameters of SoCs with acceptable margins, we did not quantify the degree of noise or errors in the estimation with respect to the amount of data used. Given that it is essential to assess the trustworthy of each thermal profile, we consider this as interesting future work.

 \nocite{*}
%  \singlespacing
% \bibliographystyle{alpha}
%  \newcommand{\newblock}{}
\bibliographystyle{plain}
\bibliography{thesis}

\begin{thebibliography}{100}

\bibitem{techreport}
{Technical Report: On Dynamic Thermal Conditions in Mixed-Criticality Systems
  (available upon request from the track chair)}.
\newblock Technical report, 2019.

\bibitem{ahmed2011minimizing}
Masud Ahmed, Nathan Fisher, Shengquan Wang, and Pradeep Hettiarachchi.
\newblock Minimizing peak temperature in embedded real-time systems via
  thermal-aware periodic resources.
\newblock {\em Sustainable Computing: Informatics and Systems}, 1(3):226--240,
  2011.

\bibitem{Ahmed2017}
Rehan Ahmed, Pengcheng Huang, Max Millen, and Lothar Thiele.
\newblock On the design and application of thermal isolation servers.
\newblock {\em ACM Trans. Embed. Comput. Syst.}, 16(5s):165:1--165:19,
  September 2017.

\bibitem{ahmed2013thermal}
Rehan Ahmed, Parameswaran Ramanathan, and Kewal~K Saluja.
\newblock On thermal utilization of periodic task sets in uni-processor
  systems.
\newblock In {\em 2013 IEEE 19th International Conference on Embedded and
  Real-Time Computing Systems and Applications}, pages 267--276. IEEE, 2013.

\bibitem{ahmed2014temperature}
Rehan Ahmed, Parameswaran Ramanathan, and Kewal~K Saluja.
\newblock Temperature minimization using power redistribution in embedded
  systems.
\newblock In {\em 2014 27th International Conference on VLSI Design and 2014
  13th International Conference on Embedded Systems}, pages 264--269. IEEE,
  2014.

\bibitem{ahmed2016necessary}
Rehan Ahmed, Parameswaran Ramanathan, and Kewal~K Saluja.
\newblock Necessary and sufficient conditions for thermal schedulability of
  periodic real-time tasks under fluid scheduling model.
\newblock {\em ACM Transactions on Embedded Computing Systems (TECS)},
  15(3):49, 2016.

\bibitem{farrington2000impact}
Rob arrington and John Rugh.
\newblock Impact of vehicle air-conditioning on fuel economy, tailpipe
  emissions, and electric vehicle range.
\newblock In {\em Earth technologies forum}, pages 1--6. NREL Washington, DC,
  2000.

\bibitem{bailis2011dimetrodon}
Peter Bailis, Vijay~Janapa Reddi, Sanjay Gandhi, David Brooks, and Margo
  Seltzer.
\newblock Dimetrodon: processor-level preventive thermal management via idle
  cycle injection.
\newblock In {\em 2011 48th ACM/EDAC/IEEE Design Automation Conference (DAC)},
  pages 89--94. IEEE, 2011.

\bibitem{bernat1999new}
Guillem Bernat and Alan Burns.
\newblock New results on fixed priority aperiodic servers.
\newblock In {\em IEEE Real-Time Systems Symposium (RTSS)}, 1999.

\bibitem{bevilacqua1999effect}
Oreste~M Bevilacqua.
\newblock {\em Effect of Air Conditioning on Regulated Emissions for In-use
  Vehicles: Phase I}.
\newblock Coordinating Research Council, Incorporated, 1999.

\bibitem{bletsas2018errata}
Konstantinos Bletsas, Neil Audsley, Wen-Hung Huang, Jian-Jia Chen, and Geoffrey
  Nelissen.
\newblock Errata for three papers (2004-05) on fixed-priority scheduling with
  self-suspensions.
\newblock {\em Leibniz Transactions on Embedded Systems}, 5(1):02--1, 2018.

\bibitem{chandarli2015response}
Youn{\`e}s Chandarli, Nathan Fisher, and Damien Masson.
\newblock Response time analysis for thermal-aware real-time systems under
  fixed-priority scheduling.
\newblock In {\em IEEE International Symposium on Real-Time Distributed
  Computing (ISORC)}, 2015.

\bibitem{4484694}
T.~Chantem, R.~P. Dick, and X.~S. Hu.
\newblock Temperature-aware scheduling and assignment for hard real-time
  applications on mpsocs.
\newblock In {\em 2008 Design, Automation and Test in Europe}, pages 288--293,
  March 2008.

\bibitem{chantem2009online}
Thidapat Chantem, X~Sharon Hu, and Robert~P Dick.
\newblock Online work maximization under a peak temperature constraint.
\newblock In {\em Proceedings of the 2009 ACM/IEEE international symposium on
  Low power electronics and design}, pages 105--110. ACM, 2009.

\bibitem{chantem2010temperature}
Thidapat Chantem, X~Sharon Hu, and Robert~P Dick.
\newblock Temperature-aware scheduling and assignment for hard real-time
  applications on mpsocs.
\newblock {\em IEEE Transactions on Very Large Scale Integration (VLSI)
  Systems}, 19(10):1884--1897, 2010.

\bibitem{chen2007minimization}
Jian-Jia Chen, Chia-Mei Hung, and Tei-Wei Kuo.
\newblock On the minimization for the instantaneous temperature for periodic
  real-time tasks.
\newblock In {\em 13th IEEE Real Time and Embedded Technology and Applications
  Symposium (RTAS)}, pages 236--248. IEEE, 2007.

\bibitem{chen2009proactive}
Jian-Jia Chen, Shengquan Wang, and Lothar Thiele.
\newblock Proactive speed scheduling for real-time tasks under thermal
  constraints.
\newblock In {\em 2009 15th IEEE Real-Time and Embedded Technology and
  Applications Symposium}, pages 141--150. IEEE, 2009.

\bibitem{choi2004fine}
Kihwan Choi, Ramakrishna Soma, and Massoud Pedram.
\newblock Fine-grained dynamic voltage and frequency scaling for precise energy
  and performance tradeoff based on the ratio of off-chip access to on-chip
  computation times.
\newblock {\em IEEE transactions on computer-aided design of integrated
  circuits and systems}, 24(1):18--28, 2004.

\bibitem{pascal}
{Pascal Tuning Guide.}
\newblock https://docs.nvidia.com/cuda/pascal-tuning-guide/index.html, 2018.

\bibitem{DSouza2017ThermalIO}
Sandeep~M D'souza and Ragunathan~Raj Rajkumar.
\newblock Thermal implications of energy-saving schedulers.
\newblock In {\em 29th Euromicro Conference on Real-Time Systems (ECRTS 2017)}.
  Schloss Dagstuhl-Leibniz-Zentrum fuer Informatik, 2017.

\bibitem{5456979}
T.~J.~A. {Eguia}, S.~X.~. {Tan}, R.~{Shen}, E.~H. {Pacheco}, and M.~{Tirumala}.
\newblock General behavioral thermal modeling and characterization for
  multi-core microprocessor design.
\newblock In {\em 2010 Design, Automation Test in Europe Conference Exhibition
  (DATE)}, pages 1136--1141, 2010.

\bibitem{elghool2017review}
Ali Elghool, Firdaus Basrawi, Thamir~Khalil Ibrahim, Khairul Habib, Hassan
  Ibrahim, and Daing Mohamad Nafiz~Daing Idris.
\newblock A review on heat sink for thermo-electric power generation:
  Classifications and parameters affecting performance.
\newblock {\em Energy conversion and management}, 134:260--277, 2017.

\bibitem{GPUSync}
G.~A. Elliott, B.~C. Ward, and J.~H. Anderson.
\newblock Gpusync: A framework for real-time gpu management.
\newblock In {\em IEEE Real-Time Systems Symposium(RTSS)}, 2014.

\bibitem{elliott2012globally}
Glenn~A Elliott and James~H Anderson.
\newblock Globally scheduled real-time multiprocessor systems with gpus.
\newblock {\em Real-Time Systems}, 48(1):34--74, 2012.

\bibitem{elliott2013optimal}
Glenn~A Elliott and James~H Anderson.
\newblock An optimal k-exclusion real-time locking protocol motivated by
  multi-gpu systems.
\newblock {\em Real-Time Systems}, 49(2):140--170, 2013.

\bibitem{exynos}
{Exynoss 5422.}
\newblock https://www.samsung.com/semiconductor/minisite/exynos
  /products/mobileprocessor/exynos-5-octa-5422, 2019.

\bibitem{6629322}
Ming Fan, Vivek Chaturvedi, Shi Sha, and Gang Quan.
\newblock An analytical solution for multi-core energy calculation with
  consideration of leakage and temperature dependency.
\newblock In {\em International Symposium on Low Power Electronics and Design
  (ISLPED)}, pages 353--358. IEEE, 2013.

\bibitem{fisher2009thermal}
Nathan Fisher, Jian-Jia Chen, Shengquan Wang, and Lothar Thiele.
\newblock Thermal-aware global real-time scheduling on multicore systems.
\newblock In {\em 2009 15th IEEE Real-Time and Embedded Technology and
  Applications Symposium}, pages 131--140. IEEE, 2009.

\bibitem{flir}
{FLIR A325sc.}
\newblock https://www.flir.com/products/a325sc, 2019.

\bibitem{fu2010feedback}
Yong Fu, Nicholas Kottenstette, Yingming Chen, Chenyang Lu, Xenofon~D
  Koutsoukos, and Hongan Wang.
\newblock Feedback thermal control for real-time systems.
\newblock In {\em 2010 16th IEEE Real-Time and Embedded Technology and
  Applications Symposium}, pages 111--120. IEEE, 2010.

\bibitem{fu2012feedback}
Yong Fu, Nicholas Kottenstette, Chenyang Lu, and Xenofon~D Koutsoukos.
\newblock Feedback thermal control of real-time systems on multicore
  processors.
\newblock In {\em Proceedings of the tenth ACM international conference on
  Embedded software}, pages 113--122. ACM, 2012.

\bibitem{gartner}
{Gartner.}
\newblock https://www.gartner.com/newsroom/id/3187134, 2016.

\bibitem{ghahremannezhad2018thermal}
Ali Ghahremannezhad and Kambiz Vafai.
\newblock Thermal and hydraulic performance enhancement of microchannel heat
  sinks utilizing porous substrates.
\newblock {\em International Journal of Heat and Mass Transfer},
  122:1313--1326, 2018.

\bibitem{ghahremannezhad2019effect}
Ali Ghahremannezhad, Huijin Xu, Mohammad~Alhuyi Nazari, Mohammad~Hossein
  Ahmadi, and Kambiz Vafai.
\newblock Effect of porous substrates on thermohydraulic performance
  enhancement of double layer microchannel heat sinks.
\newblock {\em International Journal of Heat and Mass Transfer}, 131:52--63,
  2019.

\bibitem{giannopoulou2013scheduling}
Georgia Giannopoulou, Nikolay Stoimenov, Pengcheng Huang, and Lothar Thiele.
\newblock Scheduling of mixed-criticality applications on resource-sharing
  multicore systems.
\newblock In {\em Proceedings of the Eleventh ACM International Conference on
  Embedded Software}, page~17. IEEE Press, 2013.

\bibitem{7904613}
Young-Ho Gong, Jae~Jeong Yoo, and Sung~Woo Chung.
\newblock Thermal modeling and validation of a real-world mobile ap.
\newblock {\em IEEE Design Test}, 35(1):55--62, Feb 2018.

\bibitem{herbert2007analysis}
Sebastian Herbert and Diana Marculescu.
\newblock Analysis of dynamic voltage/frequency scaling in
  chip-multiprocessors.
\newblock In {\em Proceedings of the 2007 international symposium on Low power
  electronics and design (ISLPED'07)}, pages 38--43. IEEE, 2007.

\bibitem{mehdi2020dynamic}
Seyedmehdi Hosseinimotlagh, Ali Ghahremannezhad, and Hyoseung Kim.
\newblock On dynamic thermal conditions in mixed-criticality systems.
\newblock In {\em 2020 IEEE Real-Time and Embedded Technology and Applications
  Symposium (RTAS)}, pages 336--349, 2020.

\bibitem{hosseinimotlagh2019thermal}
Seyedmehdi Hosseinimotlagh and Hyoseung Kim.
\newblock Thermal-aware servers for real-time tasks on multi-core
  gpu-integrated embedded systems.
\newblock In {\em 2019 IEEE Real-Time and Embedded Technology and Applications
  Symposium (RTAS)}, pages 254--266. IEEE, 2019.

\bibitem{hua1989generalized}
Yingbo Hua and Tapan~K Sarkar.
\newblock Generalized pencil-of-function method for extracting poles of an em
  system from its transient response.
\newblock {\em IEEE transactions on antennas and propagation}, 37(2):229--234,
  1989.

\bibitem{Huang2014}
Huang Huang, Vivek Chaturvedi, Gang Quan, Jeffrey Fan, and Meikang Qiu.
\newblock Throughput maximization for periodic real-time systems under the
  maximal temperature constraint.
\newblock {\em ACM Trans. Embed. Comput. Syst.}, 13(2s):70:1--70:22, January
  2014.

\bibitem{huang2006hotspot}
Wei Huang, Shougata Ghosh, Sivakumar Velusamy, Karthik Sankaranarayanan, Kevin
  Skadron, and Mircea~R Stan.
\newblock Hotspot: A compact thermal modeling methodology for early-stage vlsi
  design.
\newblock {\em IEEE Transactions on very large scale integration (VLSI)
  systems}, 14(5):501--513, 2006.

\bibitem{iranfar2017thespot}
Arman Iranfar, Mehdi Kamal, Ali Afzali-Kusha, Massoud Pedram, and David
  Atienza.
\newblock Thespot: Thermal stress-aware power and temperature management for
  multiprocessor systems-on-chip.
\newblock {\em IEEE Transactions on Computer-Aided Design of Integrated
  Circuits and Systems}, 2017.

\bibitem{jayaseelan2008temperature}
Ramkumar Jayaseelan and Tulika Mitra.
\newblock Temperature aware task sequencing and voltage scaling.
\newblock In {\em Proceedings of the 2008 IEEE/ACM International Conference on
  Computer-Aided Design}, pages 618--623. IEEE Press, 2008.

\bibitem{kato2011rgem}
Shinpei Kato, Karthik Lakshmanan, Aman Kumar, Mihir Kelkar, Yutaka Ishikawa,
  and Ragunathan Rajkumar.
\newblock Rgem: A responsive gpgpu execution model for runtime engines.
\newblock In {\em Real-Time Systems Symposium (RTSS), 2011 IEEE 32nd}, pages
  57--66. IEEE, 2011.

\bibitem{kato2011timegraph}
Shinpei Kato, Karthik Lakshmanan, Raj Rajkumar, and Yutaka Ishikawa.
\newblock Timegraph: Gpu scheduling for real-time multi-tasking environments.
\newblock In {\em Proc. USENIX ATC}, pages 17--30, 2011.

\bibitem{kato2012gdev}
Shinpei Kato, Michael McThrow, Carlos Maltzahn, and Scott~A Brandt.
\newblock Gdev: First-class gpu resource management in the operating system.
\newblock In {\em USENIX Annual Technical Conference}, pages 401--412. Boston,
  MA;, 2012.

\bibitem{8046309}
Hyoseung Kim, Pratyush Patel, Shige Wang, and Ragunathan~Raj Rajkumar.
\newblock A server-based approach for predictable gpu access control.
\newblock In {\em 2017 IEEE 23rd International Conference on Embedded and
  Real-Time Computing Systems and Applications (RTCSA)}, pages 1--10. IEEE,
  2017.

\bibitem{kim2018server}
Hyoseung Kim, Pratyush Patel, Shige Wang, and Ragunathan~Raj Rajkumar.
\newblock A server-based approach for predictable {GPU} access with improved
  analysis.
\newblock {\em Journal of Systems Architecture}, 88:97--109, 2018.

\bibitem{kim2017predictable}
Hyoseung Kim and Ragunathan Rajkumar.
\newblock Predictable shared cache management for multi-core real-time
  virtualization.
\newblock {\em ACM Transactions on Embedded Computing Systems (TECS)},
  17(1):1--27, 2017.

\bibitem{7010477}
Hyoseung Kim, Shige Wang, and Ragunathan Rajkumar.
\newblock {vMPCP}: A synchronization framework for multi-core virtual machines.
\newblock In {\em 2014 IEEE Real-Time Systems Symposium (RTSS)}, pages 86--95,
  Dec 2014.

\bibitem{kumar2011cool}
Pratyush Kumar and Lothar Thiele.
\newblock Cool shapers: shaping real-time tasks for improved thermal
  guarantees.
\newblock In {\em Proceedings of the 48th Design Automation Conference (DAC)},
  pages 468--473. ACM, 2011.

\bibitem{kumar2011system}
Pratyush Kumar and Lothar Thiele.
\newblock System-level power and timing variability characterization to compute
  thermal guarantees.
\newblock In {\em Proceedings of the seventh IEEE/ACM/IFIP international
  conference on Hardware/software codesign and system synthesis}, pages
  179--188. ACM, 2011.

\bibitem{kumar2011thermally}
Pratyush Kumar and Lothar Thiele.
\newblock Thermally optimal stop-go scheduling of task graphs with real-time
  constraints.
\newblock In {\em 16th Asia and South Pacific Design Automation Conference
  (ASP-DAC 2011)}, pages 123--128. IEEE, 2011.

\bibitem{kuroda2001cmos}
Tadahiro Kuroda.
\newblock Cmos design challenges to power wall.
\newblock In {\em Digest of Papers. Microprocesses and Nanotechnology 2001.
  2001 International Microprocesses and Nanotechnology Conference}, pages 6--7.
  IEEE, 2001.

\bibitem{8442110}
Ohchul Kwon, Wonjae Jang, Giyeon Kim, and Chang-Gun Lee.
\newblock Accurate thermal prediction for nans (n-app n-screen) services on a
  smart phone.
\newblock In {\em 2018 IEEE 13th International Symposium on Industrial Embedded
  Systems (SIES)}, pages 1--10. IEEE, 2018.

\bibitem{lasance2003thermally}
Clemens~JM Lasance.
\newblock Thermally driven reliability issues in microelectronic systems:
  status-quo and challenges.
\newblock {\em Microelectronics Reliability}, 43(12):1969--1974, 2003.

\bibitem{lee2013kernel}
JongHo Lee, Young-Woo Seo, Wende Zhang, and David Wettergreen.
\newblock Kernel-based traffic sign tracking to improve highway workzone
  recognition for reliable autonomous driving.
\newblock In {\em Intelligent Transportation Systems-(ITSC), 2013 16th
  International IEEE Conference on}, pages 1131--1136, 2013.

\bibitem{Youngmoon2018}
Youngmoon Lee, Hoonsung Chwa, Kang~G. Shin, and Shige Wang.
\newblock Thermal-aware resource management for embedded real-time systems.
\newblock In {\em Embedded Software (EMSOFT), 2018 International Conference
  on}. IEEE, 2018.

\bibitem{4212027}
Yongpan Liu, Robert~P Dick, Li~Shang, and Huazhong Yang.
\newblock Accurate temperature-dependent integrated circuit leakage power
  estimation is easy.
\newblock In {\em 2007 Design, Automation \& Test in Europe Conference \&
  Exhibition (DATE)}, pages 1--6. IEEE, 2007.

\bibitem{ma2015improving}
Yue Ma, Thidapat Chantem, X~Sharon Hu, and Robert~P Dick.
\newblock Improving lifetime of multicore soft real-time systems through global
  utilization control.
\newblock In {\em Proceedings of the 25th edition on Great Lakes Symposium on
  VLSI}, pages 79--82. ACM, 2015.

\bibitem{Mali}
{Mali OpenCl SDK.}
\newblock https://developer.arm.com/products/software/mali-sdks, 2016.

\bibitem{5753822}
S.~Murali, A.~Mutapcic, D.~Atienza, R.~Gupta, S.~Boyd, and G.~De Micheli.
\newblock Temperature-aware processor frequency assignment for mpsocs using
  convex optimization.
\newblock In {\em 2007 5th IEEE/ACM/IFIP International Conference on
  Hardware/Software Codesign and System Synthesis (CODES+ISSS)}, pages
  111--116, Sept 2007.

\bibitem{ODROIDXU4}
{ODROID-XU4.}
\newblock http://www.hardkernel.com/, 2016.

\bibitem{otterness2017evaluation}
Nathan Otterness, Ming Yang, Sarah Rust, Eunbyung Park, James~H Anderson,
  F~Donelson Smith, Alex Berg, and Shige Wang.
\newblock An evaluation of the nvidia tx1 for supporting real-time
  computer-vision workloads.
\newblock In {\em Real-Time and Embedded Technology and Applications Symposium
  (RTAS), 2017 IEEE}, pages 353--364. IEEE, 2017.

\bibitem{pagani2015matex}
Santiago Pagani, Jian-Jia Chen, Muhammad Shafique, and J{\"o}rg Henkel.
\newblock Matex: Efficient transient and peak temperature computation for
  compact thermal models.
\newblock In {\em 2015 Design, Automation \& Test in Europe Conference \&
  Exhibition (DATE)}, pages 1515--1520. IEEE, 2015.

\bibitem{pagani2015seboost}
Santiago Pagani, Muhammad Shafique, Heba Khdr, Jian-Jia Chen, and J{\"o}rg
  Henkel.
\newblock seboost: Selective boosting for heterogeneous manycores.
\newblock In {\em Proceedings of the 10th International Conference on
  Hardware/Software Codesign and System Synthesis}, pages 104--113. IEEE Press,
  2015.

\bibitem{park2010dynamic}
Sangyoung Park, Jian-Jia Chen, Donghwa Shin, Younghyun Kim, Chia-Lin Yang, and
  Naehyuck Chang.
\newblock Dynamic thermal management for networked embedded systems under harsh
  ambient temperature variation.
\newblock In {\em 2010 ACM/IEEE International Symposium on Low-Power
  Electronics and Design (ISLPED)}, pages 289--294. IEEE, 2010.

\bibitem{patel2017}
Pratyush Patel, Iljoo Baek, Hyoseung Kim, and Ragunathan Rajkumar.
\newblock Analytical enhancements and practical insights for mpcp with
  self-suspensions.
\newblock In {\em 2018 IEEE Real-Time and Embedded Technology and Applications
  Symposium (RTAS)}, pages 177--189. IEEE, 2018.

\bibitem{7372657}
Francesco Paterna and Tajana~{\v{S}}imunic Rosing.
\newblock Modeling and mitigation of extra-soc thermal coupling effects and
  heat transfer variations in mobile devices.
\newblock In {\em 2015 IEEE/ACM International Conference on Computer-Aided
  Design (ICCAD)}, pages 831--838. IEEE, 2015.

\bibitem{Prakash}
Alok Prakash, Hussam Amrouch, Muhammad Shafique, Tulika Mitra, and J\"{o}rg
  Henkel.
\newblock Improving mobile gaming performance through cooperative cpu-gpu
  thermal management.
\newblock In {\em Proceedings of the 53rd Annual Design Automation Conference},
  DAC '16, pages 47:1--47:6, New York, NY, USA, 2016. ACM.

\bibitem{rai2015calibration}
Devendra Rai and Lothar Thiele.
\newblock A calibration based thermal modeling technique for complex multicore
  systems.
\newblock In {\em 2015 Design, Automation \& Test in Europe Conference \&
  Exhibition (DATE)}, pages 1138--1143. IEEE, 2015.

\bibitem{rai2011worst}
Devendra Rai, Hoeseok Yang, Iuliana Bacivarov, Jian-Jia Chen, and Lothar
  Thiele.
\newblock Worst-case temperature analysis for real-time systems.
\newblock In {\em 2011 Design, Automation \& Test in Europe (DATE)}, pages
  1--6. IEEE, 2011.

\bibitem{rai2012power}
Devendra Rai, Hoeseok Yang, Iuliana Bacivarov, and Lothar Thiele.
\newblock Power agnostic technique for efficient temperature estimation of
  multicore embedded systems.
\newblock In {\em Proceedings of the 2012 international conference on
  Compilers, architectures and synthesis for embedded systems}, pages 61--70,
  2012.

\bibitem{rajkumar1990real}
Ragunathan Rajkumar.
\newblock Real-time synchronization protocols for shared memory
  multiprocessors.
\newblock In {\em Distributed Computing Systems, 1990. Proceedings., 10th
  International Conference on}, pages 116--123. IEEE, 1990.

\bibitem{rajkumar1988real}
Ragunathan Rajkumar, Lui Sha, and John~P Lehoczky.
\newblock Real-time synchronization protocols for multiprocessors.
\newblock In {\em Real-Time Systems Symposium (RTSS), 1988., Proceedings.},
  pages 259--269. IEEE, 1988.

\bibitem{saewong2002analysis}
Saowanee Saewong, Ragunathan~Raj Rajkumar, John~P Lehoczky, and Mark~H Klein.
\newblock Analysis of hierarchical fixed-priority scheduling.
\newblock In {\em Proceedings 14th Euromicro Conference on Real-Time Systems
  (ECRTS)}, 2002.

\bibitem{7746768}
Onur Sahin and Ayse~K Coskun.
\newblock Providing sustainable performance in thermally constrained mobile
  devices.
\newblock In {\em 2016 14th ACM/IEEE Symposium on Embedded Systems For
  Real-time Multimedia (ESTIMedia)}, pages 1--6, Oct 2016.

\bibitem{schor2012fast}
Lars Schor, Iuliana Bacivarov, Hoeseok Yang, and Lothar Thiele.
\newblock Fast worst-case peak temperature evaluation for real-time
  applications on multi-core systems.
\newblock In {\em 2012 13th Latin American Test Workshop (LATW)}, pages 1--6.
  IEEE, 2012.

\bibitem{schor2012worst}
Lars Schor, Iuliana Bacivarov, Hoeseok Yang, and Lothar Thiele.
\newblock Worst-case temperature guarantees for real-time applications on
  multi-core systems.
\newblock In {\em 2012 IEEE 18th Real Time and Embedded Technology and
  Applications Symposium}, pages 87--96. IEEE, 2012.

\bibitem{schor2011thermal}
Lars Schor, Hoeseok Yang, Iuliana Bacivarov, and Lothar Thiele.
\newblock Thermal-aware task assignment for real-time applications on
  multi-core systems.
\newblock In {\em International Symposium on Formal Methods for Components and
  Objects}, pages 294--313. Springer, 2011.

\bibitem{sha1986solutions}
Lui Sha, John Lehoczky, and Ragunathan Rajkumar.
\newblock Solutions for some practical problems in prioritized preemptive
  scheduling.
\newblock In {\em IEEE Real-Time Systems Symposium (RTSS)}. IEEE Computer
  Society Press, 1986.

\bibitem{shin2008compositional}
Insik Shin and Insup Lee.
\newblock Compositional real-time scheduling framework with periodic model.
\newblock {\em ACM Transactions on Embedded Computing Systems (TECS)}, 7(3):30,
  2008.

\bibitem{7092527}
Gaurav Singla, Gurinderjit Kaur, Ali~K Unver, and Umit~Y Ogras.
\newblock Predictive dynamic thermal and power management for heterogeneous
  mobile platforms.
\newblock In {\em 2015 Design, Automation \& Test in Europe Conference \&
  Exhibition (DATE)}, pages 960--965, March 2015.

\bibitem{Skadron}
Kevin Skadron, Mircea~R. Stan, Karthik Sankaranarayanan, Wei Huang, Sivakumar
  Velusamy, and David Tarjan.
\newblock Temperature-aware microarchitecture: Modeling and implementation.
\newblock {\em ACM Trans. Archit. Code Optim.}, 1(1):94--125, March 2004.

\bibitem{sprunt1989aperiodic}
Brinkley Sprunt, Lui Sha, and John Lehoczky.
\newblock Aperiodic task scheduling for hard-real-time systems.
\newblock {\em Real-Time Systems}, 1(1):27--60, 1989.

\bibitem{srinivasan2004impact}
Jayanth Srinivasan, Sarita~V Adve, Pradip Bose, and Jude~A Rivers.
\newblock The impact of technology scaling on lifetime reliability.
\newblock In {\em International Conference on Dependable Systems and Networks,
  2004}, pages 177--186. IEEE, 2004.

\bibitem{strosnider1995deferrable}
Jay~K. Strosnider, John~P. Lehoczky, and Lui Sha.
\newblock The deferrable server algorithm for enhanced aperiodic responsiveness
  in hard real-time environments.
\newblock {\em IEEE Transactions on Computers}, 44(1):73--91, 1995.

\bibitem{Thingy52}
{Nordic Semiconductor Thingy:52 IoT Sensor Development Kit.}
\newblock https://www.mouser.com/new/nordicsemiconductor/nordic-thingy-52/,
  2019.

\bibitem{tiwari1998reducing}
Vivek Tiwari, Deo Singh, Suresh Rajgopal, Gaurav Mehta, Rakesh Patel, and
  Franklin Baez.
\newblock Reducing power in high-performance microprocessors.
\newblock In {\em Proceedings of the 35th annual Design Automation conference},
  pages 732--737, 1998.

\bibitem{toribio2016fire}
Bryan~Anthony Toribio, Cameron~Duross Peterson, David~Parker Rubenstein, and
  Timothy~Philip Neilan.
\newblock Fire containment drone.
\newblock Technical report, Worcester Polytechnic Institute, 2016.

\bibitem{vestal2007preemptive}
Steve Vestal.
\newblock Preemptive scheduling of multi-criticality systems with varying
  degrees of execution time assurance.
\newblock In {\em 28th IEEE International Real-Time Systems Symposium (RTSS)},
  pages 239--243. IEEE, 2007.

\bibitem{viswanath2000thermal}
Ram Viswanath, Vijay Wakharkar, Abhay Watwe, Vassou Lebonheur, et~al.
\newblock Thermal performance challenges from silicon to systems.
\newblock {\em Intel Technology Journal}, Q3, 2000.

\bibitem{wang2010schedulability}
Shengquan Wang, Youngwoo Ahn, and Riccardo Bettati.
\newblock Schedulability analysis in hard real-time systems under thermal
  constraints.
\newblock {\em Real-Time Systems}, 46(2):160--188, 2010.

\bibitem{4032360}
Shengquan Wang and Riccardo Bettati.
\newblock Delay analysis in temperature-constrained hard real-time systems with
  general task arrivals.
\newblock In {\em 2006 27th IEEE International Real-Time Systems Symposium
  (RTSS)}, pages 323--334. IEEE, 2006.

\bibitem{wang2006delay}
Shengquan Wang and Riccardo Bettati.
\newblock Delay analysis in temperature-constrained hard real-time systems with
  general task arrivals.
\newblock In {\em 2006 27th IEEE International Real-Time Systems Symposium
  (RTSS)}, pages 323--334. IEEE, 2006.

\bibitem{wang2009temperature}
Yefu Wang, Kai Ma, and Xiaorui Wang.
\newblock Temperature-constrained power control for chip multiprocessors with
  online model estimation.
\newblock {\em ACM SIGARCH computer architecture news}, 37(3):314--324, 2009.

\bibitem{Xi_EMSOFT11}
Sisu Xi, Justin Wilson, Chenyang Lu, and Christopher Gill.
\newblock {RT-Xen}: towards real-time hypervisor scheduling in {Xen}.
\newblock In {\em International Conference on Embedded Software (EMSOFT)},
  2011.

\bibitem{xiang2010system}
Yun Xiang, Thidapat Chantem, Robert~P Dick, X~Sharon Hu, and Li~Shang.
\newblock System-level reliability modeling for mpsocs.
\newblock In {\em Proceedings of the eighth IEEE/ACM/IFIP international
  conference on Hardware/software codesign and system synthesis}, pages
  297--306. ACM, 2010.

\bibitem{yang2013real}
Hoeseok Yang, Iuliana Bacivarov, Devendra Rai, Jian-Jia Chen, and Lothar
  Thiele.
\newblock Real-time worst-case temperature analysis with temperature-dependent
  parameters.
\newblock {\em Real-Time Systems}, 49(6):730--762, 2013.

\bibitem{zhang2010thermal}
Sushu Zhang and Karam~S Chatha.
\newblock Thermal aware task sequencing on embedded processors.
\newblock In {\em Proceedings of the 47th Design Automation Conference}, pages
  585--590. ACM, 2010.

\bibitem{zhou2015gpes}
Husheng Zhou, Guangmo Tong, and Cong Liu.
\newblock Gpes: A preemptive execution system for gpgpu computing.
\newblock In {\em Real-Time and Embedded Technology and Applications Symposium
  (RTAS), 2015 IEEE}, pages 87--97. IEEE, 2015.

\bibitem{ziabari2014power}
Amirkoushyar Ziabari, Je-Hyoung Park, Ehsan~K Ardestani, Jose Renau, Sung-Mo
  Kang, and Ali Shakouri.
\newblock Power blurring: Fast static and transient thermal analysis method for
  packaged integrated circuits and power devices.
\newblock {\em IEEE Transactions on Very Large Scale Integration (VLSI)
  Systems}, 22(11):2366--2379, 2014.

\end{thebibliography}

\appendix

\end{document}